%% file: thesis.tex
\documentclass[twoside,11pt,oneandahalfspaced]{ut-thesis}
\usepackage[section]{mytheorem}
\usepackage{index,multicol,array}
\usepackage{amsmath,amssymb}

\degree{Doctor of Philosophy}
\department{Computer Science}
\gradyear{2000}
\author{Fran\c{c}ois Pitt}
\title{A Quantifier-Free String Theory for $\ALT$ Reasoning}

\let\seiresfb=\bfseries
\def\bfseries{\seiresfb\boldmath}
\makeatletter
\@addtoreset{equation}{section}
\@addtoreset{figure}{section}
\@addtoreset{table}{section}
\makeatother

\allowdisplaybreaks[2]
\input{macros}

\newtheorem{openproblem}{Open Problem}
\newtheorem{generalfact}{General Fact}
\newfact{derule}{Derived Rule}
\newcommand{\leftright}[2]{\text{(L)} \ #1 \qquad \text{(R)} \ #2}
\newcounter{count}
\let\scar=\bigstar
\newcommand{\tableline}[2]%
  {\( \text{\hphantom{$R_{i+1}$}\llap{$#1$}} = 
   \text{\rlap{$#2$}\hphantom{$R_i+2^{d-i-1}$}} \)}
\let\formaid=\widetilde
\makeatletter
\if@twoside
   \newcommand{\heading}[1]{\subsection*{#1}\markright{\textsc{#1}}}
\else
   \newcommand{\heading}[1]{\subsection*{#1}}
\fi
\makeatother
\makeatletter
\newenvironment{claim*}[1]%
  {\trivlist\@topsep\theorempreskipamount
   \@topsepadd\theorempostskipamount
   \item[\hskip\labelsep\normalfont\scshape
      Claim~#1\quad]\normalfont\slshape\ignorespaces}%
  {\endtrivlist}
\newenvironment{lemma*}[1]%
  {\trivlist\@topsep\theorempreskipamount
   \@topsepadd\theorempostskipamount
   \item[\hskip\labelsep\normalfont\scshape
      Lemma~#1\quad]\normalfont\slshape\ignorespaces}%
  {\endtrivlist}
\newenvironment{theorem*}[1]%
  {\trivlist\@topsep\theorempreskipamount
   \@topsepadd\theorempostskipamount
   \item[\hskip\labelsep\normalfont\scshape
      Theorem~#1\quad]\normalfont\slshape\ignorespaces}%
  {\endtrivlist}
\newenvironment{corollary*}[1]%
  {\trivlist\@topsep\theorempreskipamount
   \@topsepadd\theorempostskipamount
   \item[\hskip\labelsep\normalfont\scshape
      Corollary~#1\quad]\normalfont\slshape\ignorespaces}%
  {\endtrivlist}
\makeatother
\newcommand{\ndx}[5]%
  {\ifx\empty#5\empty
      \ifx\empty#1\empty
         \ifx\empty#3\empty
            \index[fn]{#2}
         \else
            \ifx\empty#4\empty
               \index[fn]{#2!#3}
            \else
               \index[fn]{#2!#3!#4}
            \fi
         \fi
      \else
         \ifx\empty#3\empty
            \index[fn]{#2*#1}
         \else
            \ifx\empty#4\empty
               \index[fn]{#2!#3*#1}
            \else
               \index[fn]{#2!#3!#4*#1}
            \fi
         \fi
      \fi
   \else
      \ifx\empty#1\empty
         \ifx\empty#3\empty
            \index[fn]{#5#2=#2}
         \else
            \ifx\empty#4\empty
               \index[fn]{#5#2=#2!#3}
            \else
               \index[fn]{#5#2=#2!#3!#4}
            \fi
         \fi
      \else
         \ifx\empty#3\empty
            \index[fn]{#5#2=#2*#1}
         \else
            \ifx\empty#4\empty
               \index[fn]{#5#2=#2!#3*#1}
            \else
               \index[fn]{#5#2=#2!#3!#4*#1}
            \fi
         \fi
      \fi
   \fi}
\newcommand{\ind}[4][]{\ndx{#1}{#2}{#3}{#4}{}}
\newcommand{\bitind}[4][]{\ndx{#1}{#2}{#3}{#4}{b}}

\newcommand{\numind}[4][]{\ndx{#1}{#2}{#3}{#4}{n}}

\newcommand{\fcnind}[4][]{\ndx{#1}{#2}{#3}{#4}{z}}
\newcommand{\bitfcnind}[4][]{\ndx{#1}{#2}{#3}{#4}{bz}}
\newcommand{\lenfcnind}[4][]{\ndx{#1}{#2}{#3}{#4}{lz}}
\newcommand{\numfcnind}[4][]{\ndx{#1}{#2}{#3}{#4}{nz}}

\newindex{fn}{idx}{ind}{Index of Function Symbols}

\begin{document}

\include{prelim}   

\clearemptydoublepage

\include{chapter1} 

\clearemptydoublepage

\include{chapter2} 

\clearemptydoublepage

\include{chapter3} 

\clearemptydoublepage

\include{chapter4} 

\clearemptydoublepage

\include{chapter5} 

\clearemptydoublepage

\include{chapter6} 

\clearemptydoublepage

\include{chapter7} 

\appendix 

\clearemptydoublepage

\include{appendix} 

\clearemptydoublepage

\include{biblio}   

\end{document}

%% file: macros.tex
\long\def\ignore#1{}   
\def\clap#1{\hbox to 0pt{\hss#1\hss}}
\renewcommand{\today}{\number\year\space\ifcase\month\or
   January\or February\or March\or April\or May\or June\or
   July\or August\or September\or October\or November\or December\fi
   \space\number\day}
\newcommand{\symrm}[1]{\textnormal{\rmfamily#1}}

\newcommand{\symbf}[1]{\textnormal{\bfseries#1}}

\newcommand{\symsl}[1]{\textnormal{\slshape#1}}
\newcommand{\symsc}[1]{\textnormal{\scshape#1}}
\newcommand{\dotminus}{\mathbin{\mathchoice%
{\mathop{\smash-}\limits^{\displaystyle\cdot}}%
{\mathop{\smash-}\limits^{\textstyle\cdot}}%
{\mathop{\smash-}\limits^{\scriptstyle\cdot}}%
{\mathop{\smash-}\limits^{\scriptscriptstyle\cdot}}%
}}
\newcommand{\bigand}{\mathop{\mathchoice%
{\rlap{$\displaystyle\bigwedge$}\bigwedge}%
{\rlap{$\textstyle\bigwedge$}\bigwedge}%
{\rlap{$\textstyle\wedge$}\,{\textstyle\wedge}}%
{\rlap{$\scriptstyle\wedge$}\,{\scriptstyle\wedge}}%
}}
\newcommand{\bigor}{\mathop{\mathchoice%
{\rlap{$\displaystyle\bigvee$}\bigvee}%
{\rlap{$\textstyle\bigvee$}\bigvee}%
{\rlap{$\textstyle\vee$}\,{\textstyle\vee}}%
{\rlap{$\scriptstyle\vee$}\,{\scriptstyle\vee}}%
}}
\newcommand{\bigOh}{\mathcal{O}}
\newcommand{\NN}{\mathbb{N}} 
\let\goesto=\rightarrow

\newcommand{\eg}{\emph{e.g.}}
\newcommand{\ie}{\emph{i.e.}}

\newcommand{\st}{\mathrel{:}} 
\newcommand{\restrict}{\mathord{\big|}}
\newcommand{\true}{\mathtt{true}}
\newcommand{\false}{\mathtt{false}}
\let\ltrue=\top
\let\lfalse=\bot

\newcommand{\limp}{\mathbin\rightarrow}
\newcommand{\liff}{\mathbin\leftrightarrow}
\let\lxor=\oplus
\newcommand{\lded}{\mathrel{\:\vdash\:}}

\newcommand{\phanbin}[1]{\mathbin{\phantom{#1}}}
\newcommand{\phanrel}[1]{\mathrel{\phantom{#1}}}
\newcommand{\pheq}{\phanrel{=}}

\newcommand{\NW}{\mathbb{W}} 
\newcommand{\mach}[1]{\mathsf{#1}}    
\newcommand{\func}[1]{\mathit{#1}}    
\newcommand{\syst}[1]{\mathit{#1}}    
\newcommand{\class}[1]{\mathit{#1}}   
\newcommand{\prob}[1]{\mathrm{#1}}    
\newcommand{\F}{\mathcal{F}}
\newcommand{\eF}{{\mathit{e}\mathcal{F}}}

\newcommand{\aL}{\mathcal{L}}

\newcommand{\B}{\syst{B}}
\newcommand{\No}{\syst{N}_0}
\newcommand{\PV}{\syst{PV}}

\newcommand{\ALV}{\syst{ALV}}
\newcommand{\TNC}{\syst{TNC}}

\newcommand{\PT}{\syst{PT}}
\newcommand{\QF}{\syst{QF}}
\newcommand{\AID}{\syst{AID}}
\newcommand{\ATM}{\mach{ATM}}

\newcommand{\co}{\class{co}}

\newcommand{\TC}{\class{TC}}
\newcommand{\ALT}{\class{ALOGTIME}}
\newcommand{\FALT}{\class{FALOGTIME}}
\newcommand{\NC}{\class{NC}}
\newcommand{\FNC}{\class{FNC}}

\newcommand{\DP}{\class{P}}
\newcommand{\FP}{\class{FP}}
\newcommand{\NP}{\class{NP}}

\newcommand{\TAUT}{\prob{TAUT}}
\newcommand{\BSVP}{\prob{BSVP}}

\newcommand{\PHP}{\prob{PHP}}
\newcommand{\BASE}{\text{BASE}}
\newcommand{\COMP}{\text{COMP}}

\newcommand{\CRN}{\text{CRN}}
\newcommand{\lCRN}{\text{$\ell$CRN}}
\newcommand{\rCRN}{\text{$r$CRN}}
\newcommand{\TRN}{\text{TRN}}
\newcommand{\STRN}{\text{STRN}}

\newcommand{\NIND}{\text{NIND}}
\newcommand{\TIND}{\text{TIND}}
\newcommand{\fhead}{\lambda}
\newcommand{\fdot}{\mathpunct{\mspace{1mu}\symrm{\Large.}\mspace{-1mu}}}
\newcommand{\LIND}[1][]{\text{\ifx\empty#1\empty\else$#1$-\fi LIND}}
\newcommand{\CA}[1][]{\text{\ifx\empty#1\empty\else$#1$-\fi CA}}
\newcommand{\E}{\varepsilon}
\newcommand{\cat}{\mathbin\cdot}
\newcommand{\I}{\mathcal{I}}
\newcommand{\C}{\mathbin?}
\newcommand{\lchop}{\mathbin\vartriangleright}
\newcommand{\rchop}{\mathbin\vartriangleleft}
\newcommand{\lhalf}{\mathord\blacktriangleleft}
\newcommand{\rhalf}{\mathord\blacktriangleright}
\newcommand{\ldel}{\mathord{\mathchoice%
{\raisebox{.2ex}{\protect\scriptsize$\gtrdot$}}%
{\raisebox{.2ex}{\protect\scriptsize$\gtrdot$}}%
{\raisebox{.1ex}{\protect\tiny$\gtrdot$}}%
{\raisebox{.1ex}{\protect\tiny$\gtrdot$}}%
}}
\newcommand{\rdel}{\mathord{\mathchoice%
{\raisebox{.2ex}{\protect\scriptsize$\lessdot$}}%
{\raisebox{.2ex}{\protect\scriptsize$\lessdot$}}%
{\raisebox{.1ex}{\protect\tiny$\lessdot$}}%
{\raisebox{.1ex}{\protect\tiny$\lessdot$}}%
}}
\newcommand{\lbit}{\mathord{{}^\backprime}}
\newcommand{\rbit}{\mathord{{}^\prime}}
\newcommand{\tinyR}{\symsl{\protect\tiny R}}
\newcommand{\tinyL}{\symsl{\protect\tiny L}}
\newcommand{\tinyB}{\symsl{\protect\tiny B}}
\newcommand{\tinyN}{\symsl{\protect\tiny N}}
\newcommand{\tinyS}{\symsl{\protect\tiny S}}
\newcommand{\tpl}[1]{\vec{#1}}
\newcommand{\cst}[1]{\widehat{#1}}
\newcommand{\fcn}[1]{\mathsf{#1}}
\newcommand{\len}[1]{{#1}^{\tinyL}}
\newcommand{\bit}[1]{{#1}^{\tinyB}}
\newcommand{\num}[1]{{#1}^{\tinyN}}
\newcommand{\str}[1]{{#1}^{\tinyS}}

\newcommand{\bitbin}[1]{\mathbin{\bit{#1}}}
\newcommand{\numbin}[1]{\mathbin{\num{#1}}}

\newcommand{\lenrel}[1]{\mathrel{\len{#1}}}
\newcommand{\bitrel}[1]{\mathrel{\bit{#1}}}
\newcommand{\numrel}[1]{\mathrel{\num{#1}}}
\newcommand{\strrel}[1]{\mathrel{\str{#1}}}
\newcommand{\lenfcn}[1]{\len{\fcn{#1}}}
\newcommand{\bitfcn}[1]{\bit{\fcn{#1}}}
\newcommand{\numfcn}[1]{\num{\fcn{#1}}}

\newcommand{\aux}{\fcn{aux\_}\!}
\newcommand{\bC}{\bitbin\C}
\newcommand{\zlC}{\mathbin{\C^{\symsl{\protect\tiny ZL}}}}
\newcommand{\elC}{\mathbin{\C^{\symsl{\protect\tiny EL}}}}
\newcommand{\lsg}{\mathord\approx}
\newcommand{\smsh}{\mathbin\#}
\newcommand{\tuple}[1]{\left\langle{#1}\right\rangle}
\newcommand{\proj}{\pi}
\newcommand{\pref}{\sqsubseteq}
\newcommand{\lpowCRN}{\text{$\ell$powCRN}}
\newcommand{\rpowCRN}{\text{$r$powCRN}}
\newcommand{\tlen}{\operatorname{len}} 
\newcommand{\latom}{\langle}
\newcommand{\ratom}{\rangle}
\newcommand{\atom}[1]{\latom#1\ratom}
\newcommand{\bigatom}[1]{\bigl\latom#1\bigr\ratom}
\newcommand{\lform}{\lbrack\!\lbrack}
\newcommand{\rform}{\rbrack\!\rbrack}
\newcommand{\form}[1]{\lform#1\rform}
\newcommand{\code}{\mathord\#}
\newcommand{\tlimp}{\mathbin{\rlap{$\rightarrow$}\rightarrow}}
\newcommand{\tlor}{\mathbin{\rlap{$\vee$}\,\vee}}
\newcommand{\tcomp}{\circ}
\let\term=\widehat
\newcommand{\anc}{\mathrel{\rhd}} 
\newcommand{\anceq}{\mathrel{\unrhd}} 
\newcommand{\lca}{\operatorname{lca}} 
\newcommand{\out}{\symsc{out}}
\newcommand{\undef}{\bot}
\newcommand{\mapl}[1]{\symsc{#1}}
\newcommand{\pair}[2]{\langle#1,#2\rangle}
\newcommand{\set}[1]{{[#1]}}
\newcommand{\una}[1]{\symsc{#1}}

%% file: prelim.tex
\begin{preliminary}

\maketitle

\clearemptydoublepage

\begin{abstract}
   The main contribution of this work is the definition of a
   quantifier-free string theory $T_1$ suitable for formalizing
   $\ALT$ reasoning.  After describing $L_1$---a new, simple,
   algebraic characterization of the complexity class $\ALT$ based on
   strings instead of numbers---the theory $T_1$ is defined (based on
   $L_1$), and a detailed formal development of $T_1$ is given.

   {} Then, theorems of $T_1$ are shown to translate into families of
   propositional tautologies that have uniform polysize Frege proofs,
   $T_1$ is shown to prove the soundness of a particular Frege system
   $\F$, and $\F$ is shown to provably $p$-simulate any proof system
   whose soundness can be proved in $T_1$.  Finally, $T_1$ is
   compared with other theories for $\ALT$ reasoning in the
   literature.

   {} To our knowledge, this is the first formal theory for $\ALT$
   reasoning whose basic objects are strings instead of numbers, and
   the first quantifier-free theory formalizing $\ALT$ reasoning in
   which a direct proof of the soundness of some Frege system has been
   given (in the case of first-order theories, such a proof was first
   given by Arai for his theory $\AID$).  Also, the polysize Frege
   proofs we give for the propositional translations of theorems of
   $T_1$ are considerably simpler than those for other theories, and
   so is our proof of the soundness of a particular $\F$-system in
   $T_1$.  Together with the simplicity of $T_1$'s recursion
   schemes, axioms, and rules these facts suggest that $T_1$ is one
   of the most natural theories available for $\ALT$ reasoning.
\end{abstract}

\clearemptydoublepage

\begin{acknowledgements}
   First and foremost, I would like to thank my advisor, Stephen Cook,
   without whose help, patience, and guidance this work would have
   been impossible; Alasdair Urquhart, Alan Borodin, and Charles
   Rackoff (the other members of my advisory committee) for all their
   suggestions; and Samuel Buss (my external appraiser) for taking the
   time to read the whole thesis and saying such nice things about it!

   {} I would also like to thank Stephen Bellantoni and Stephen Bloch,
   for some interesting and fruitful discussions on $\NC$ and $\NC^1$;
   Arnold Rosenbloom, for all the discussions; the staff of the
   Department of Computer Science, for all their help with
   administrative matters; and The Fields Institute, for their program
   on Computational Complexity.

   {} Last, but certainly not least, I would like to thank my parents
   and my wife, Marie-Jos\'ee, for their patience and encouragement
   over the last six years, as well as for everything else.
\end{acknowledgements}

\newpage

\clearemptydoublepage

\tableofcontents

\newpage

\clearemptydoublepage

\printindex[fn] 

\end{preliminary}

%% file: chapter1.tex
\chapter{Introduction}
\label{chap:intro}

 The starting point for this work is the following open problem in
 complexity theory, concerning propositional proof systems (for a good
 introduction to propositional proof systems, including the basic
 definitions, see Cook and Reckhow~\cite{CooRec79}).
\begin{openproblem}
\label{prob:FvseF}
   Are Frege (``$\F$'') and extended Frege (``$\eF$'') proof systems
   $p$-equivalent?
\end{openproblem}
 To provide some motivation for studying Open Problem~\ref{prob:FvseF}
 and to give an indication of its importance, note its connection to some
 major open questions in complexity theory through the following
 facts.
\begin{generalfact}
   If \(\NP\neq\co\NP\), then \(\DP\neq\NP\).
\end{generalfact}
\begin{generalfact}
   \(\NP=\co\NP\) if and only if \(\TAUT\in\NP\).
\end{generalfact}
\begin{generalfact}
   \(\TAUT\in\NP\) if and only if there exists a \emph{super} (\ie,
   polynomially-bounded) proof system for $\TAUT$.
\end{generalfact}
\begin{generalfact}
\label{fact:super}
   Given two proof systems $f_1$ and $f_2$, if $f_1$ is super and
   $f_2$ $p$-simulates $f_1$, then $f_2$ is also super.
\end{generalfact}
 From Cook and Reckhow's paper, we know that $p$-simulation imposes a
 partial order on proof systems.  Determining the relative position of
 particular proof systems in this order helps shed some light on their
 relative power and, because of General Fact~\ref{fact:super}, on such
 major open problems as \(\NP\stackrel{?}{=}\co\NP\) or
 \(\DP\stackrel{?}{=}\NP\).  From this point of view, determining the
 exact position of Frege systems relative to extended Frege systems in
 this order is one of the most important questions still open in this
 area.  For the rest of this chapter, I will give a short survey of
 the major results and issues connected with Open
 Problem~\ref{prob:FvseF}.

\section{$\eF$-systems and $\DP$}

 It is traditional in complexity theory to equate ``feasible'' with
 ``polynomial time''.  Moreover, there is a close association between
 polytime and $\eF$-systems since $\eF$-systems can be thought of as
 reasoning on uniform polysize circuits.  For the rest of this work, I
 will use the traditional notation ``$\DP$'' when referring to the
 class of polytime decidable languages, and ``$\FP$'' when referring
 to the class of polytime computable functions.  Over the years, many
 characterizations of the classes $\DP$ and $\FP$ have been given,
 most notably Cobham's ``$\aL$'' and Bellantoni and Cook's ``$B$''.
\begin{itemize}
\item
   Cobham's $\aL$~\cite{Cobh64} is the first machine-independent
   characterization of the class $\FP$ using a form of bounded
   recursion on notation.
\item
   Bellantoni and Cook's $\B$~\cite{BelCoo92} uses a tiered approach
   (\ie, it distinguishes between ``safe'' and ``normal'' parameters)
   in order to dispense with explicit bounds as in Cobham's scheme.
\end{itemize}
 Also, many logical theories have been proposed to capture polytime
 reasoning, most notably Cook's ``$\PV$'', Buss's ``$S^1_2$'',
 Leivant's ``$\PT(\NW)$'', and various second-order theories.
\begin{itemize}
\item
   Cook's $\PV$~\cite{Cook75,CooUrq93} is a free-variable equational
   theory based on Cobham's $\aL$.  Cook showed that every formula
   \(t=u\) provable in $\PV$ gives rise to a family of propositional
   tautologies which assert the equation and have uniform polysize
   $\eF$ proofs, that $\PV$ can define and prove the soundness of
   $\eF$, and that if the soundness of a propositional proof system
   $T$ is provable in $\PV$, then $\eF$ $p$-simulates $T$.
\item
   Buss's $S^1_2$~\cite{Buss86} is a system of Bounded Arithmetic that
   can define exactly the polytime functions.  Buss showed that
   $S^1_2$ is $\Sigma^b_1$-conservative over $\PV$ (when its language
   is suitably extended to include all the function symbols of $\PV$),
   which implies that $S^1_2$ proves the soundness of $\eF$ and that
   the $\Sigma^b_1$-theorems of $S^1_2$ can be translated into
   propositional tautologies that have uniform polysize $\eF$-proofs
   (by the corresponding results for $\PV$).
\item
   Leivant's $\PT(\NW)$~\cite{Leiv92b} has generative axioms for $\NW$
   (intuitively, binary strings) and instances of $\NW$-induction as
   its only axioms.  It proves the convergence of exactly the polytime
   functions over $\NW$ (when induction is restricted to positive
   existential formulas).  This formalization is conceptually and
   technically very simple because it does not rely on any particular
   initial functions, other than the algebra's constructors (in fact,
   the theory can talk about any computable function).
\item
   Buss's $V^1_1$ (studied by Razborov~\cite{Razb95}) and Leivant's
   $L_2(\QF^+)$~\cite{Leiv91} are two of the most notable examples of
   second-order theories for $\DP$.
\end{itemize}

\section{$\F$-systems and $\ALT$}

 The computational power of $\F$-systems seems to be captured by the
 uniform class $\NC^1$, since $\F$-systems can be thought of as
 reasoning on polysize formulas, which are the same as
 logarithmic-depth circuits.  Recall that $\NC^1$ is the class of
 languages decidable by families of logarithmic-depth circuits
 ($\FNC^1$ is the functional equivalent, using multi-output circuits),
 and by results of Ruzzo~\cite{Ruzz81},
 \(U_{E^*}\text{-uniform }\NC^1=\ALT\), where $\ALT$ is the class of
 languages decidable in logarithmic time by a random access
 alternating Turing machine.  The functional class $\FALT$ can be
 defined in two different ways: if functions are thought of as
 operating on integers in binary notation, we get a ``numerical''
 version of the class, whereas if functions are thought of as
 operating on strings of bits (which is closer to the circuit model),
 we get a ``string'' version of the class.  Fortunately, with a
 suitable interpretation of numbers as strings (or of strings as
 numbers), both versions are equivalent.

 {}Therefore, for the rest of this work, I will use $\ALT$ and $\NC^1$
 interchangeably, always referring to the uniform version of the class
 (unless otherwise specified).  Also, I will use ``$\FALT$'' (or
 ``$\FNC^1$'') to refer to the functional version of the class.
 Various characterizations of $\FALT$ have been given over the years,
 most notably Clote's ``$\No$'' and ``$\No'$'', and Bloch's string
 algebra.
\begin{itemize}
\item
   Clote's $\No$ and $\No'$~\cite{Clot89,Clot92a,Clot93} are
   ``numerical'' characterizations that use restricted forms of
   Cobham's recursion on notation.  Unfortunately, $\No$ includes a
   complete function for $\FALT$ as a base function, and $\No'$
   depends on Barrington's deep result about bounded-width branching
   programs~\cite{Barr89}, so neither algebra is as natural for
   $\FNC^1$ as Cobham's $\aL$ is for $\FP$.
\item
   Bloch's algebra~\cite{Bloc92,Bloc94} is a ``string''
   characterization that uses the ``safe'' versus ``normal'' parameter
   idea together with a form of recursion similar to Allen's ``divide
   and conquer recursion'' ({DCR})~\cite{Alle91}.  Bloch recognized
   that Allen's scheme of {DCR} (which Allen used to characterize
   uniform $\NC$) is particularly well-suited to characterizing
   uniform parallel complexity classes.  Combining this with the
   tiered approach allows him to dispense with explicit bounds on the
   rate of growth of functions and to give an elegant characterization
   that uses only simple base functions and one natural scheme of
   recursion.
\end{itemize}
 Based on Bloch's ideas but incorporating some of Clote's, I will
 introduce in Chapter~\ref{chap:L1} a new simple string algebra $L_1$
 that characterizes $\FALT$ using very few simple base functions and
 two simple schemes of recursion ($\CRN$ and $\TRN$, to be defined
 there).  It appears to us that $L_1$ is simpler than previous
 characterizations because it has fewer, simpler base functions, and
 no need for explicit bounds on the growth of functions or for
 different types of parameters.

 {}Based on the characterizations of $\ALT$ given above, a number of
 theories to capture $\ALT$ reasoning have been defined, most notably
 Clote's ``$\ALV$'' and ``$\ALV'$'', Takeuti and Clote's ``$\TNC^0$'',
 and Arai's ``$\AID$'' (all of which are based on ``numerical''
 characterizations of $\ALT$).
\begin{itemize}
\item
   Clote's $\ALV$ and $\ALV'$~\cite{Clot92a,Clot93} are free-variable
   equational theories based on his characterizations of $\ALT$
   mentioned above and on Cook's $\PV$.  Clote showed that theorems of
   $\ALV$ and $\ALV'$ give rise to families of tautologies which have
   polysize $\F$-proofs, but did not show that either of his theories
   can prove the soundness of $\F$-systems.  Also, the proof that the
   propositional translations of theorems of $\ALV$ or $\ALV'$ have
   polysize $\F$-proofs is fairly involved, and properties of even
   simple functions (such as ``parity'' or ``majority'') are difficult
   to prove.
\item
   Takeuti and Clote's $\TNC^0$~\cite{CloTak95} (first defined by
   Takeuti~\cite{Take94}) is a first-order theory similar to Buss's
   $S^1_2$ that was shown to be conservative over $\ALV'$ (when
   suitably extended to include every function symbol of $\ALV'$).
   Unfortunately, this theory needs to use a fairly complex form of
   inference called \emph{bounded successive nomination}, because of
   its implicit dependence on Barrington's result (through Clote's
   characterization of $\ALT$), which detracts greatly from its
   simplicity.
\item
   Arai's $\AID$~\cite{Arai98} is a system of bounded arithmetic
   inspired by Buss's consistency proof for
   $\F$-systems~\cite{Buss91}, which proves the soundness of $\F$ and
   whose $\Sigma_0^b$-theorems have polysize $\F$-proofs when suitably
   translated.  Moreover, Arai shows that $\AID$ is equivalent to a
   quantified version of Clote's $\ALV$, and hence that $\ALV$ can
   prove the soundness of $\F$.
\end{itemize}
 Unlike the situation for $\DP$, there is no quantifier-free theory
 for $\ALT$ which has the simplicity and naturalness of $\PV$.  I
 claim that $T_1$ fills that role, its axioms and induction schemes
 being based directly on $L_1$'s simple base functions and natural
 recursion operations.  Moreover, the proofs that propositional
 translations of the theorems of $T_1$ have uniform polysize
 $\F$-proofs and that $T_1$ can prove the soundness of $\F$-systems
 are much simpler than the corresponding proofs for other theories in
 the literature.

\section{Overview}

 Now that I have provided some context and motivation for studying
 Open Problem~\ref{prob:FvseF}, let me give a brief overview of the
 rest of the thesis.  In Chapter~2, I will introduce the string
 algebra $L_1$, followed in Chapter~3 by the quantifier-free theory
 $T_1$ (including a formal development of the theory, showing how to
 prove the pigeonhole principle in $T_1$).  In Chapter~4, I will
 define propositional translations for theorems of $T_1$ and show
 that they have polysize $\F$-proofs, while in Chapter~5, I will show
 that $T_1$ proves the soundness of $\F$, by formalizing an algorithm
 for the ``Boolean Sentence Value Problem'' ($\BSVP$) in $T_1$, and
 that $\F$ provably $p$-simulates any proof system whose soundness can
 be proved in $T_1$.  Finally, in Chapter~6, I will compare $T_1$
 with various other formalisms for $\ALT$ reasoning, most notably
 Arai's $\AID$.

%% file: chapter2.tex
\chapter{The String Algebra $L_1$}
\label{chap:L1}

 In this chapter, we define $L_1$ and show that it contains exactly
 the functions in $\FALT$.  We also give many examples of natural
 $L_1$ definitions for simple $\FALT$ functions.

\section{Basic definitions}

 The basic objects of the algebra are strings over the alphabet
 $\{0,1\}$.  The set of all such strings can be defined inductively:
 $\E$ (the empty string), $0$, $1$ are strings, and if $x$ and $y$ are
 strings, then so is $xy$.  Together with a wish for simplicity, this
 inductive definition motivates our choice of base functions.

 {}The reader should keep in mind that our definitions in this chapter
 are based on, and guided by, the idea of computation by uniform
 families of circuits.  In particular, all our functions will be
 \emph{length-determined}, \ie, the length of a function depends only
 on the lengths of the arguments, not their values.  Also, the
 starting point for our algebra $L_1$ is Bloch's paper~\cite{Bloc94},
 where he carries out a similar function-algebraic characterization of
 $\FALT$, so we will borrow many concepts and definitions from there.
 (We also borrow certain concepts and definitions from Clote's
 work~\cite{Clot89,Clot92a,Clot93}.)

 {}Now, we define the base functions and the basic operators that we
 will use to construct new functions.  We use $|x|$ to denote the
 length of $x$ (\ie, the number of symbols (bits) in the string $x$),
 $\tpl x_k$ to denote a $k$-tuple of variables, and $\tpl x$ to denote
 an arbitrary tuple of variables.

\begin{description}

\item[\BASE:]
   The set of \emph{base functions} consists of (in order of increasing
   arity):
   \begin{align*}
      \E\ind[ldef]{$\E$ (empty string)}{}{},0,1
      &= \text{empty string, 0-bit, and 1-bit (constants)}, \\
      \rhalf x\ind[ldef]{$\rhalf$ (right half)}{}{}
      &= \text{the \(\lceil|x|/2\rceil\) rightmost bits of $x$
         (``right half'')}, \\  
      x\rchop y\ind[ldef]{$\rchop$ (right chop)}{}{}
      &= \text{$x$ with \(|y|\) bits removed from the right
         (``right chop'')}, \\
      x\cat y\ind[ldef]{$\cat$ (concatenation)}{}{}
      &= \text{$x$ followed by $y$ (``concatenation'')}, \\
      x\C(y,z_0,z_1)\ind[ldef]{$\C$ (conditional)}{}{} &=
         \begin{cases}
            y   & \text{if \(x=\E\)}, \\
            z_0 & \text{if \(x=w\cat 0\) for some $w$}, \\
            z_1 & \text{if \(x=w\cat 1\) for some $w$}, 
         \end{cases} \quad\text{(``conditional'')} \\
      \I^n_k(x_1,\dots,x_n)\ind[ldef]{$\I^n_k$ (identity)}{}{}
      &= x_k \quad\text{for any } 1\leq k\leq n
         \quad\text{(``identity'' or ``projection'')}. 
   \end{align*}
   \begin{remark}
      In the definition of $x\C(y,z_0,z_1)$, it is assumed that
      \(|z_0|=|z_1|\).  If that is not the case, then the value
      returned will be padded on the left with as many $0$'s as are
      necessary to make $z_0$ and $z_1$ the same length (the length of
      $y$ does not change).
   \end{remark}

\item[\COMP:]
   $f$ is defined from $g$ and $h_1,\dots,h_k$ by \emph{composition}
   if
   \[ f(\tpl x) = g(h_1(\tpl x),\dots,h_k(\tpl x)). \]

\item[\CRN:]
   $f$ is defined from $h$ by \emph{concatenation recursion on
   notation} on $x$ if \(h(x,\tpl y)\in\{0,1\}\) for all $x,\tpl y$
   and
   \ind[ldef]{\CRN}{}{}
   \begin{align*}
      f(\E,\tpl y) &= \E, \\
      f(xi,\tpl y) &= f(x,\tpl y)\cat h(xi,\tpl y)
         \quad\text{for } i=0,1. 
   \end{align*}

\item[\TRN:]
   $f$ is defined from $g$, $h$, $h_\ell$, and $h_r$ by
   \emph{tree recursion on notation} on $x$ if
   \ind[ldef]{\TRN}{}{}
   \[
      f(x,z,\tpl y) =
         \begin{cases}
            g(x,z,\tpl y) & \text{if } x=\E,0,1, \\
            h\big(x,z,\tpl y,f(x\lhalf,h_\ell(z),\tpl y),
               f(\rhalf x,h_r(z),\tpl y)\big)
          & \text{otherwise}, 
         \end{cases} 
   \]
   where
   \( x\lhalf = x\rchop\rhalf x \)\ind[ldef]{$\lhalf$ (left half)}{}{}
   (the $\lfloor|x|/2\rfloor$ leftmost bits of $x$).  In what follows,
   we will omit the parameter $z$ when neither $g$ nor $h$ depend on
   it (in which case the functions $h_\ell$ and $h_r$ are irrelevant
   and will not be specified); we will refer to this form of $\TRN$ as
   \emph{simple $\TRN$}.

\end{description}

\begin{remark}
   Our ``right half'' function was called ``back half ($\func{Bh}$)''
   by Allen~\cite{Alle91} and Bloch~\cite{Bloc94}.  We introduce the
   new nomenclature because we feel that it is more representative of
   the action of the function, and the new notation to serve as a
   graphical reminder of that action (picture the black triangle
   cutting into the left part of $x$).  Similarly, our ``right chop''
   function was called ``chop'' by Cook~\cite{CooUrq93} and ``most
   significant part ($\func{Msp}$)'' by Allen and Bloch.  Our new
   notation should serve as a useful graphical mnemonic for the
   function's purpose and action (picture the bits of $y$ cutting into
   the bits of $x$ from the right---in the direction pointed to by the
   function symbol).  Our scheme of $\CRN$ is based on the operation
   of the same name in Clote's work~\cite{Clot89,Clot92a,Clot93},
   except that our version has been simplified by eliminating the
   function $g$ from the base case (without loss of generality since
   we can simply concatenate $g(\tpl y)$ to the left of our functions
   to get Clote's).  Our scheme of $\TRN$ is based on Bloch's ``very
   safe {DCR}'', which is itself based on Allen's ``{DCR}'' (for
   ``divide-and-conquer recursion''), except that our base case is
   simpler (defined for $x=\E,0,1$ instead of when \(|x|\leq|b|\) for
   some extra parameter $b$), and we have added the functions $h_\ell$
   and $h_r$ that allow parameter $z$ to vary during recursive calls
   (hence, $\TRN$ is technically a scheme of ``recursion with
   replacement'').
\end{remark}

\begin{definition}
\label{def:algebra}
   If we let $\TRN\restrict^L_{L'}$ represent the operation of $\TRN$
   restricted to functions \(g\in L\) and \(h,h_\ell,h_r\in L'\), for
   function classes $L$ and $L'$, then
   \begin{itemize}
   \item
      $L_0$ is the closure of $\BASE$ under $\COMP$ and $\CRN$;
   \item
      $L_1$ is the closure of $L_0$ under $\COMP$, $\CRN$, and
      $\TRN\restrict^{L_1}_{L_0}$, defined recursively.
   \end{itemize}
\end{definition}

 {}The next few sections contain mainly function definitions, where
 the following notational conventions will be used.

\begin{itemize}

\item
   For any constant string $c$, $\tpl c_k$ represents the tuple
   consisting of $k$ copies of $c$.

\item
   Unary functions have higher precedence than binary functions and
   binary functions have higher precedence than functions of higher
   arity (keep in mind that ``$\C$'' has arity $4$).  Concatenation
   has higher precedence than any other binary function when
   represented by juxtaposition; it has lower precedence than any
   other binary function when represented by ``$\cat$''.

\item
   $i$ and $j$ represent arbitrary fixed single bits, whereas $k$,
   $\ell$, $m$, and $n$ represent arbitrary fixed non-zero natural
   numbers.  When $2^k$ is used, $k$ ranges over all natural numbers
   (including zero), and similarly for $2^\ell$.

\item
   The notation $k\times x$ stands for $\overbrace{x\dotsm x}^k$ (\ie,
   $x$ concatenated with itself $k$ times).  We let \(0\times x=\E\)
   and use $\cst k$ as an abbreviation for $k\times 1$, \ie, the unary
   string representing $k$.
   \ind[ldef]{$\times$ (repeat)}{}{}
   \ind[ldef]{$\cst{\ }$ (unary constant)}{}{}

\end{itemize}

\section{Functions in $L_0$}

 In this section, we define many functions in $L_0$ and show that
 many useful generalizations of $\CRN$ can be simulated in $L_0$.  We
 are motivated by two goals: to define the machinery necessary to
 prove that $L_1$ contains all of $\FNC^1$, and to show that many
 useful functions have simple definitions in our algebra.

\subsection{Basic functions}

 First, we define a few simple variations on some of the $\BASE$
 functions.  The rightmost bit of $x$:
\( x\rbit = x\C(\E,0,1) \)\ind[ldef]{$\rbit$ (right bit)}{}{};
 $x$ with its rightmost bit removed:
\( x\rdel = x\rchop 1 \)\ind[ldef]{$\rdel$ (right delete)}{}{};
 the $\lfloor|x|/2\rfloor$ leftmost bits of $x$:
\( x\lhalf = x\rchop\rhalf x \).

 {}Next, a function that reverses the bits of $x$ can be defined by
 first using $\CRN$ to define a function $\fcn{reverse}(x,y)$, which
 returns the $|y|$ rightmost bits of $x$ reversed:
\begin{align*}
   \fcn{reverse}(x,\E) &= \E, \\*
   \fcn{reverse}(x,yi) &= \fcn{reverse}(x,y)\cat(x\rchop y)\rbit
      \quad\text{for } i=0,1. 
\end{align*}
 Then, \(\fcn{rev}(x)=\fcn{reverse}(x,x)\) returns the reverse of $x$.
 Using this function, we can now define symmetric counterparts to some
 of the earlier functions:
\[
   y\lchop x = \fcn{rev}(\fcn{rev}(x)\rchop\fcn{rev}(y)), \qquad
   \ind[ldef]{$\lchop$ (left chop)}{}{}
   \ldel x = \fcn{rev}(\fcn{rev}(x)\rdel), \qquad
   \ind[ldef]{$\ldel$ (left delete)}{}{}
   \lbit x = \fcn{rev}(x)\rbit. 
   \ind[ldef]{$\lbit$ (left bit)}{}{}
\]

 {}Now, let us introduce a generalization of $\CRN$: a function $f$ is
 defined from $h$ by \emph{left $\CRN$} (or \emph{reverse $\CRN$}) on
 $x$ if \(h(x,\tpl y)\in\{0,1\}\) for all $x,\tpl y$ and
\ind[ldef]{\CRN}{left \CRN}{}
\begin{align*}
   f(\E,\tpl y) &= \E, \\*
   f(ix,\tpl y) &= h(ix,\tpl y)\cat f(x,\tpl y)
      \quad\text{for } i=0,1. 
\end{align*}
 If $f$ is defined from $h$ by left $\CRN$ on $x$, then we can use
 $\CRN$ to define
\begin{align*}
   \aux f(\E,\tpl y) &= \E, \\*
   \aux f(xi,\tpl y) &= \aux f(x,\tpl y)\cat h(\fcn{rev}(xi),\tpl y)
      \quad\text{for } i=0,1, 
\end{align*}
 and \(f(x,\tpl y)=\fcn{rev}(\aux f(\fcn{rev}(x),\tpl y))\), using
 $\COMP$.  In what follows, we will use the notational conventions
 outlined before this section and we will no longer include the
 trivial base case \(f(\E,\tpl y)=\E\) or write ``for \(i=0,1\)'' when
 using $\CRN$ to define new functions.

\subsection{String manipulation functions}

 Now, we will define useful functions for manipulating strings.
 First, two simple functions that returns a string of the same length
 as its input, but consisting entirely of 0's or entirely of 1's:
\[
   {}_j(xi) = {}_jx\cat j \qquad \text{(by $\CRN$)}.
   \ind[ldef]{${}_0$ (all-zero)}{}{}
   \ind[ldef]{${}_1$ (all-one)}{}{}
\]
 Next, we can define a number of functions to compare the lengths of
 strings (these function symbols will be distinguished by putting a
 superscript ``${}^{\tinyL}$'' next to them).  First, it is useful to
 have a conditional that tests for the length of a string:
\( x\zlC(y,z) = x\C(y,z,z) \)
\ind[ldef]{$\C$ (conditional)}{$\zlC$ (zero-length conditional)}{}
 is equal to $y$ if $x$ is empty and equal to $z$ otherwise.  Because
 $\zlC$ distinguishes only between ``empty'' and ``non-empty'', we will
 define the ``length-relational'' functions below so that they return
 $\E$ when the relation holds and some fixed non-empty string (like
 ``$1$'') otherwise.  Accordingly, we define a simple signum function
 that returns $\E$ if its argument is empty and $1$ otherwise:
 \(\len\lsg x=x\zlC(\E,1)\) and a corresponding ``negation'':
 \(\len\lnot x=x\zlC(1,\E)\).  Now, we are ready to define the
 comparison functions.
\begin{gather*}
   \begin{alignedat}{3}
      x\lenrel\geq y &= \len\lsg(x\lchop y) &\qquad
         x\lenrel\leq y &= \len\lsg(x\rchop y) &\qquad
         x\lenrel=y &= \len\lsg((x\lchop y)\cat(x\rchop y)) \\*
      x\lenrel>y &= \len\lnot(x\lenrel\leq y) &\qquad
         x\lenrel<y &= \len\lnot(x\lenrel\geq y) &\qquad
         x\lenrel\neq y &= \len\lnot(x\lenrel=y) 
   \end{alignedat} \\
   \lenfcn{max}(x,y) = (x\lenrel\geq y)\zlC(x,y)
   \lenfcnind[ldef]{$\lenfcn{max}$}{}{}
      \qquad \lenfcn{max}_1(x) = x \\*
   \lenfcn{max}_{k+1}(x,\tpl x_k) =
      \lenfcn{max}\big(x,\lenfcn{max}_k(\tpl x_k)\big) 
   \lenfcnind[ldef]{$\lenfcn{max}$}{$\lenfcn{max}_k$}{}
\end{gather*}
 We can also define functions to manipulate the lengths of strings,
 namely $\lenfcn{div}_{2^k}(x)$ that returns a string of $1$'s whose
 length is $\lfloor|x|/2^k\rfloor$ and a corresponding
 $\lenfcn{mod}_{2^k}(x)$ function satisfying
\( {}_1x = 2^k\times\lenfcn{div}_{2^k}(x)\cat\lenfcn{mod}_{2^k}(x) \).
\begin{gather*}
   \lenfcn{div}_1(x) = {}_1x \qquad
      \lenfcn{div}_{2k}(x) = \lenfcn{div}_k(x\lhalf) \\*
   \lenfcn{mod}_{2^k}(x) =
      \big(2^k\times\lenfcn{div}_{2^k}(x)\big)\lchop{}_1x 
   \lenfcnind[ldef]{$\lenfcn{div}_{2^k},\lenfcn{mod}_{2^k}$}{}{}
\end{gather*}
 (Interestingly, there does not seem to be a way to define a
 $\lenfcn{div}_k$ function for arbitrary $k$ without using $\TRN$.)
 Following this, we define functions to perform simple bit
 manipulations on strings (extract single bits or substrings, pad to a
 certain length).
\begin{itemize}

\item
   ``Left bit'':
   \( \fcn{lb}(x,y) = y\zlC(\E,\lbit(\ldel y\lchop x)) \)
   \fcnind[ldef]{$\fcn{lb}$ (left bit)}{}{}
   returns bit number \(|y|\) of $x$ from the left; ``right bit'':
   \( \fcn{rb}(x,y) = y\zlC(\E,(x\rchop y\rdel)\rbit) \)
   \fcnind[ldef]{$\fcn{rb}$ (right bit)}{}{}
   returns bit
   number $|y|$ of $x$ from the right (both are equal to $\E$ if
   \(y=\E\) or \(|y|>|x|\)).  For convenience, we also define
   \( \bitfcn{lb}(x,y) = \fcn{lb}(x,y)\C(0,0,1) \)
   and
   \( \bitfcn{rb}(x,y) = \fcn{rb}(x,y)\C(0,0,1) \)
   which return $0$ or $1$ for all arguments.

\item
   ``Left cut'':
   \( \fcn{lc}(x,y) = x\rchop(y\lchop x) \)
   \fcnind[ldef]{$\fcn{lc}$ (left cut)}{}{}
   returns the $|y|$ leftmost bits of $x$; ``right cut'':
   \( \fcn{rc}(x,y) = (x\rchop y)\lchop x \)
   \fcnind[ldef]{$\fcn{rc}$ (right cut)}{}{}
   returns the $|y|$ rightmost bits of $x$ (both return $\E$ if
   \(y=\E\) and $x$ if \(|y|\geq|x|\)).

\item
   ``Left pad'':
   \( \fcn{lp}_j(x,y) = {}_j(y\rchop x)\cat x \)
   \fcnind[ldef]{$\fcn{lp}_0,\fcn{lp}_1$ (left pad)}{}{}
   returns $x$ padded on the left with $j$'s so that
   \(|\fcn{lp}_j(x,y)|\geq|y|\); ``right pad'':
   \( \fcn{rp}_j(x,y) = x\cat{}_j(x\lchop y) \)
   \fcnind[ldef]{$\fcn{rp}_0,\fcn{rp}_1$ (right pad)}{}{}
   returns $x$ padded on the right with $j$'s so that
   \(|\fcn{rp}_j(x,y)|\geq|y|\) (both return $x$ if \(|y|\leq|x|\)).

\item
   ``Left adjust'':
   \( \fcn{la}_j(x,y) = {}_j(y\rchop x)\cat((x\rchop y)\lchop x) \)
   \fcnind[ldef]{$\fcn{la}_0,\fcn{la}_1$ (left adjust)}{}{}
   returns $x$ either chopped or padded on the left so that
   \(|\fcn{la}_j(x,y)|=|y|\); ``right adjust'':
   \( \fcn{ra}_j(x,y) = (x\rchop(y\lchop x))\cat{}_j(x\lchop y) \)
   \fcnind[ldef]{$\fcn{ra}_0,\fcn{ra}_1$ (right adjust)}{}{}
   returns $x$ either chopped or padded on the right so that
   \(|\fcn{ra}_j(x,y)|=|y|\).

\end{itemize}
 Finally, we have all the functions we need to define a tuple function
 ($\tuple{\tpl x_k}_k$) and corresponding projection functions
 ($\proj^k_\ell(x)$).  To form tuples, we simply concatenate the
 arguments together after padding them on the left so that they all
 have the same length.  The projection functions are then defined
 easily using $\lhalf$ and $\rhalf$.  One small complication arises
 because we can only divide the length of a string by a power of 2, so
 we need to form tuples that always have a power of 2 elements even
 when there are fewer of them that are actually input values.  The
 definitions follow and are inspired by similar definitions in Bloch's
 paper~\cite{Bloc94}.  (The tuple function is defined in terms of an
 auxiliary function $\fcn{tuple}$ that has an extra parameter
 specifying the length to which each value should be padded.)
\begin{align*}
   \fcn{tuple}_1(x,z) &= \fcn{lp}_0(x,z) \\
   \fcn{tuple}_{2k}(\tpl x_k,\tpl y_{k},z) &=
      \fcn{tuple}_k(\tpl x_k,z)\cat\fcn{tuple}_k(\tpl y_k,z) \\
   \fcn{tuple}_{2k+1}(\tpl x_k,\tpl y_{k+1},z) &=
      \fcn{tuple}_{k+1}(\E,\tpl x_k,z)\cat
      \fcn{tuple}_{k+1}(\tpl y_{k+1},z) \\
   \tuple{\tpl x_k}_k &= \fcn{tuple}_k
      \big(\tpl x_k,\lenfcn{max}_k(\tpl x_k)\big)
   \ind[ldef]{$\tuple{\dotsc}_k$ (tuple)}{}{} \\
   \proj^1_1(y) &= y \\
   \proj^{2k}_\ell(y) &= \begin{cases}
         \proj^k_\ell(y\lhalf) & \text{if } \ell\leq k \\
         \proj^k_{\ell-k}(\rhalf y) & \text{if } \ell>k 
      \end{cases} \\
   \proj^{2k+1}_\ell(y) &= \begin{cases}
         \proj^{k+1}_{\ell+1}(y\lhalf) & \text{if } \ell\leq k \\
         \proj^{k+1}_{\ell-k}(\rhalf y) & \text{if } \ell>k 
      \end{cases} 
   \ind[ldef]{$\proj^k_\ell$ (projection)}{}{}
\end{align*}
 Note that these functions satisfy the following relations:
\begin{align*}
   \proj^{k}_\ell(\tuple{x_1,\dots,x_k}_k)
   &= \fcn{lp}_0(x_\ell,\lenfcn{max}_k(\tpl x_k)), \\
   \tuple{\proj^{2^k}_1(y),\dots,\proj^{2^k}_{2^k}(y)}_{2^k}
   &= y 
\end{align*}
 (unfortunately,
\( \tuple{\proj^k_1(y),\dots,\proj^k_k(y)}_k \neq y \)
 for arbitrary $k$ because of the way the tuple function is defined).

\subsection{Generalizations of $\CRN$}

 We now introduce a generalization of $\CRN$ where the recursion is
 defined on several variables at once.  (We assume that
 $x_1,\dots,x_m$ all have the same length, or are appropriately padded
 on the left with $0$'s to make them all the same length.)
\begin{definition}[$\CRN_m$]
   We say that $f$ is defined from $h$ by
   \emph{$\CRN_m$ on $x_1,\dots,x_m$} if
   \( h(\tpl x_m,\tpl y) \in \{0,1\} \) for all $\tpl x_m,\tpl y$ and
   \ind[ldef]{\CRN}{\CRN$_m$}{}
   \begin{align*}
      f(\tpl\E_m,\tpl y) &= \E, \\*
      f(x_1i_1,\dots,x_mi_m,\tpl y) &= f(x_1,\dots,x_m,\tpl y)\cat
         h(x_1i_1,\dots,x_mi_m,\tpl y) \quad
      \text{for } i_1=0,1;\dots;i_m=0,1. 
   \end{align*}
\end{definition}
 (We can also define \emph{left} $\CRN_m$ similarly to left $\CRN$.)
 If $f$ is defined from $h$ by $\CRN_m$ on $x_1,\dots,x_m$, then we
 can define $f$ using $\CRN$ as follows: We will define an auxiliary
 function $\aux f$ by $\CRN$ on a parameter $z$; this function will
 mimic the recursion on $x_1,\dots,x_m$ by using $\fcn{lc}$ to extract
 the correct substrings of $x_1,\dots,x_m$ based on the length of $z$.
 Then, $f$ is easily defined from $\aux f$ by $\COMP$.
\begin{align*}
   \aux f(zi,\tpl x_m,\tpl y) &= \aux f(z,\tpl x_m,\tpl y)\cat
      h\big(\fcn{lc}(x_1,zi),\dots,\fcn{lc}(x_m,zi),\tpl y\big) \\*
   f(\tpl x_m,\tpl y) &= \aux f\big(\lenfcn{max}_m(\tpl x_m),
      \fcn{lp}_0(x_1,\lenfcn{max}_m(\tpl x_m)),\dots,
      \fcn{lp}_0(x_m,\lenfcn{max}_m(\tpl x_m)),\tpl y\big) 
\end{align*}

 {}When a function $f(x,\tpl y)$ is defined by $\CRN$ on $x$ from $h$,
 every bit in the output corresponds to one bit from $x$.  Now, we
 will show how to define a function where every bit of $x$ corresponds
 to two bits in the output, and then generalize this to arbitrary
 values (where every group of $2^k$ bits in the input corresponds to a
 group of $2^n$ bits in the output, which we will call
 ``$2^k$-to-$2^n$-$\CRN$'', or ``$(2^k,2^n)$-$\CRN$'').

 {}Following the notation mentioned above, we say that a function $f$
 is defined from $h$ by \emph{$1$-to-$2$-$\CRN$ ($(1,2)$-$\CRN$) on
 $x$} if \(|h(x,\tpl y)|=2\) for all $x,\tpl y$ and
\begin{align*}
   f(\E,\tpl y) &= \E, \\*
   f(xi,\tpl y) &= f(x,\tpl y)\cat h(xi,\tpl y) \quad
   \text{for } i=0,1. 
\end{align*}
 (We can also define \emph{left} $(1,2)$-$\CRN$.)  If $f(x,\tpl y)$ is
 defined from $h$ by $(1,2)$-$\CRN$ on $x$, we can define $f$ using
 $\CRN$ as follows: We will first define an auxiliary function
 $\aux f(z,x,\tpl y)$ by $\CRN$ on $z$, to return the $|z|$ leftmost
 bits of $f(x,\tpl y)$ and then define $f$ from $\aux f$ by $\COMP$.
 Intuitively, $\aux f(z,x,\tpl y)$ uses $\lenfcn{div}_2(z)$ to
 determine which bits of $x$ to give as input to $h$ and
 $\lenfcn{mod}_2(z)$ to determine which bit of $h$ to output next.
\begin{align*}
   \aux f(zi,x,\tpl y) &= \aux f(z,x,\tpl y)\cat\bitfcn{lb}
      \big(h\big(\fcn{lc}(x,\lenfcn{div}_2(z)\cat i),\tpl y\big),
         \lenfcn{mod}_2(z)\cat i\big) \\*
   f(x,\tpl y) &= \aux f(x\cat x,x,\tpl y) 
\end{align*}

\noindent
 Now, we can introduce the generalization mentioned above.
\begin{definition}[$(2^k,2^n)\text{-}\CRN$]
   We say that $f$ is defined from $g$ and $h$ by
   \emph{$2^k$-to-$2^n$-$\CRN$ ($(2^k,2^n)$-$\CRN$) on $x$} if
   \(|h(x,\tpl y)|=2^n\) for all $x,\tpl y$ and
   \ind[ldef]{\CRN}{$(2^k,2^n)$-\CRN}{}
   \begin{alignat*}{2}
      f(x,\tpl y) &= g(x,\tpl y) &&\quad \text{if } |x|<2^k, \\*
      f(x\cat z,\tpl y) &= f(x,\tpl y)\cat h(x\cat z,\tpl y) &&\quad
         \text{for } z\in\{0,1\}^{2^k}. 
   \end{alignat*}
\end{definition}
 (As before, we can also define \emph{left} $(2^k,2^n)$-$\CRN$.)  If
 $f$ is defined from $g$ and $h$ by $(2^k,2^n)$-$\CRN$ on $x$, then we
 can define $f$ using $\CRN$ as follows.  (The intuition is similar to
 that for $(1,2)$-$\CRN$ given above.)

\pagebreak[0]
\vspace*{-2.25\baselineskip}
\begin{align*}
   \aux f(zi,x,\tpl y) &= \aux f(z,x,\tpl y)\cat\bitfcn{lb}\Bigl(
      h\bigl(\fcn{lc}\big(x,\lenfcn{mod}_{2^k}(x)\cat
         2^k\times(\lenfcn{div}_{2^n}(z)\cat i)\big),\tpl y\bigr),
      \lenfcn{mod}_{2^n}(z)\cat i\Bigr) \\
   f(x,\tpl y) &=
      g\big(\fcn{lc}(x,\lenfcn{mod}_{2^k}(x)),\tpl y\big)\cat
      \aux f\big(2^n\times\lenfcn{div}_{2^k}(x),x,\tpl y\big) 
\end{align*}

 {}By combining the two generalizations above, we can show that any
 function defined by ``$(2^k,2^n)$-$\CRN_m$'' can be defined using
 $\CRN$ and $\COMP$ alone, which gives us a relatively powerful way to
 define many more useful functions.

\subsection{Boolean functions}

 The next functions we will introduce are the Boolean operators, \ie,
 the standard connectives together with some useful functions for
 comparing bits (these function symbols will be distinguished by
 putting a superscript ``${}^{\tinyB}$'' next to them).  First, we
 will define a ``Boolean test'' function, which tests for the
 \emph{truth-value} of its argument (where a string's truth-value is
 determined by its rightmost bit by convention, with $1$ = $\true$ and
 $0$ = $\false$ --- $\E$ is treated the same way as $0$):
\( x\bC(y,z) = x\C(z,z,y) \)
\ind[ldef]{$\C$ (conditional)}{$\bC$ (boolean conditional)}{}
 is equal to $y$ if $x$ is ``true''; $z$ if $x$ is ``false''
 (according to the convention above).  Then, \(\bit\lsg x=x\bC(1,0)\)
 returns the truth-value of $x$ and we can define the boolean
 connectives in the usual way.
\begin{alignat*}{3}
   \bitind[(ldef]{$\bit\lnot,\bitbin\land,\bitbin\lor$, etc.}{}{}
   \bit\lnot x &= x\bC(0,1) &\qquad
      x\bitbin\land y &= x\bC(\bit\lsg y,0) &\qquad
      x\bitbin\lor y &= x\bC(1,\bit\lsg y) \\
   x\bitbin\limp y &= x\bC(\bit\lsg y,1) &\qquad
      x\bitbin\liff y &= x\bC(\bit\lsg y,\bit\lnot y) &\qquad
      x\bitbin\lxor y &= x\bC(\bit\lnot y,\bit\lsg y) \\
   x\bitrel\geq y &= y\bitbin\limp x &\qquad
      x\bitrel\leq y &= x\bitbin\limp y &\qquad
      x\bitrel=y &= x\bitbin\liff y \\
   x\bitrel<y &= \bit\lnot(x\bitrel\geq y) &\qquad
      x\bitrel>y &= \bit\lnot(x\bitrel\leq y) &\qquad
      x\bitrel\neq y &= \bit\lnot(x\bitrel=y) 
   \bitind[)]{$\bit\lnot,\bitbin\land,\bitbin\lor$, etc.}{}{}
\end{alignat*}
 Using $\CRN_m$, we can now easily define the following useful
 functions that perform bitwise operations on their arguments.
\begin{align*}
   \bitfcnind[(ldef]{$\bitfcn{not},\bitfcn{and},\bitfcn{or}$, etc.}{}{}
   \bitfcn{not}(xi) &= \bitfcn{not}(x)\cat\bit\lnot i \\
   \bitfcn{and}_k(x_1i_1,\dots,x_ki_k) &=
      \bitfcn{and}_k(x_1,\dots,x_k)\cat
         (i_1\bitbin\land\dotsb\bitbin\land i_k) \\
   \bitfcn{or}_k(x_1i_1,\dots,x_ki_k) &=
      \bitfcn{or}_k(x_1,\dots,x_k)\cat
         (i_1\bitbin\lor\dotsb\bitbin\lor i_k) \\
   \bitfcn{xor}_k(x_1i_1,\dots,x_ki_k) &=
      \bitfcn{xor}_k(x_1,\dots,x_k)\cat
         (i_1\bitbin\lxor\dotsb\bitbin\lxor i_k) \\
   \bitfcn{iff}_k(x_1i_1,\dots,x_ki_k) &=
      \bitfcn{iff}_k(x_1,\dots,x_k)\cat{}
      \big((i_1\bitbin\liff i_2)\bitbin\land\dotsb
         \bitbin\land(i_{k-1}\bitbin\liff i_k)\big) 
   \bitfcnind[)]{$\bitfcn{not},\bitfcn{and},\bitfcn{or}$, etc.}{}{}
\end{align*}

\noindent
 And following Buss~\cite{Buss87b}, we can define functions that
 implement \emph{carry-save addition}: $\fcn{CScar}$ to compute the
 carry bits and $\fcn{CSadd}$ to compute the addition bits.  Note that
 these functions are defined so that
\( \fcn{CScar}(x,y,z,w)+\fcn{CSadd}(x,y,z,w) = x+y+z+w \)
 (a fact that will be proved rigorously in Chapter~\ref{chap:T1}.)

\pagebreak[0]
\vspace*{-2.25\baselineskip}
\begin{align*}
   \fcn{CScar}_3(i_1x_1,i_2x_2,i_3x_3) &=
      \big((i_1\bitbin\land i_2)\bitbin\lor
         (i_2\bitbin\land i_3)\bitbin\lor
         (i_3\bitbin\land i_1)\big)\cat\fcn{CScar}_3(x_1,x_2,x_3) \\
   \fcn{CSadd}_3(x_1,x_2,x_3) &= \bitfcn{xor}_3(0x_1,0x_2,0x_3)
    = 0\cat\bitfcn{xor}_3(x_1,x_2,x_3) \\
   \fcn{CScar}(x_1,x_2,x_3,x_4) &= \fcn{CScar}_3\big(
      \fcn{CScar}_3(x_1,x_2,x_3)\cat 0,\fcn{CSadd}_3(x_1,x_2,x_3),
      0x_4\big)\cat 0 \\
   \fcn{CSadd}(x_1,x_2,x_3,x_4) &= \fcn{CSadd}_3\big(
      \fcn{CScar}_3(x_1,x_2,x_3)\cat 0,\fcn{CSadd}_3(x_1,x_2,x_3),
      0x_4\big) 
   \fcnind[ldef]{$\fcn{CScar},\fcn{CSadd}$}{}{}
\end{align*}

\section{Functions in $L_1$}

 In this section, we define many functions in $L_1$ and show that
 some useful generalizations of $\TRN$ can be simulated in $L_1$.
 Again, we are motivated by two goals: to define the machinery
 necessary to prove that $L_1$ contains all of $\FNC^1$, and to show
 that many useful functions have simple definitions in our algebra.

\subsection{Basic functions}

 Recall that the operation of $\TRN$ is restricted in $L_1$ so that
 we cannot define a function by $\TRN$ from functions that are
 themselves defined by $\TRN$.  Hence, it will be useful to be able to
 define more than one function simultaneously by $\TRN$.
\begin{definition}[$\TRN_k$]
   The functions $f_i$ (\(1\leq i\leq k\)) are defined from functions
   $g_i$, $h_i$, $h_\ell$, and $h_r$ by \emph{$\TRN_k$ on $x$} if
   \( |f_1(x,z,\tpl y)| = \dots = |f_k(x,z,\tpl y)| \)
   for all $x,z,\tpl y$, and for every \(1\leq i\leq k\),
   \ind[ldef]{\TRN}{\TRN$_k$}{}
   \[
      f_i(x,z,\tpl y) = \begin{cases}
         g_i(x,z,\tpl y) & \text{if } x=\E,0,1, \\
         h_i\big(x,z,f_1(x\lhalf,h_\ell(z),\tpl y),
            f_1(\rhalf x,h_r(z),\tpl y),\dots, \\
         \hphantom{h_i\big(x,z,\,}
            f_k(x\lhalf,h_\ell(z),\tpl y),
            f_k(\rhalf x,h_r(z),\tpl y),
            \tpl y\big) & \text{otherwise}. 
      \end{cases}
   \]
\end{definition}
 If $\tpl f_k$ are defined from $\tpl g_k$, $\tpl h_k$, $h_\ell$, and
 $h_r$ by $\TRN_k$, we can define the $k$-tuple
\( F(x,z,\tpl y) = \tuple{\smash[t]{\tpl f_k(x,z,\tpl y)}}_k \)
 by $\TRN$ as follows.
\[
   F(x,z,\tpl y) = \begin{cases}
      \tuple{\tpl g_k(x,z,\tpl y)}_k
       & \text{if } x=\E,0,1, \\
      \Big\langle\tpl h_k\big(x,z,
         \proj^k_1(F(x\lhalf,h_\ell(z),\tpl y)),
         \proj^k_1(F(\rhalf x,h_r(z),\tpl y)),\dots, \\
      \hphantom{\big\langle\tpl h_k\big(x,z,\,}
         \proj^k_k(F(x\lhalf,h_\ell(z),\tpl y)),
         \proj^k_k(F(\rhalf x,h_r(z),\tpl y)),
         \tpl y\big)\Big\rangle_k & \text{otherwise}. 
   \end{cases}
\]
 Then, a simple composition gives
\( f_i(x,z,\tpl y) = \proj^k_i(F(x,z,\tpl y)) \)
 for \(1\leq i\leq k\).

 {}Now, we can define some functions by $\TRN$ (actually, by simple
 $\TRN$).  The first two perform Boolean operations on all the bits of
 their input; $|x|$ returns the length of $x$, expressed as a binary
 number; $x\smsh y$ returns $|x|$ copies of $y$ concatenated together
 (so that \(|x\smsh y|=|x|\times|y|\)); $\lenfcn{div}_k(x)$ returns a
 string whose length is equal to $\lfloor|x|/k\rfloor$ (where we use
 the notation ``$\lenfcn{mod}_k(x)$'' as a shorthand for the term
 $(k\times\lenfcn{div}_k(x))\lchop{}_1x$); and the last two functions
 are defined by $\TRN_2$ and will be used to count the number of $1$'s
 in the string $x$, using the carry-save technique of
 Buss~\cite{Buss87b} (this will be done below).
\begin{gather*}
\begin{aligned}
   \fcn{AND}(x) &= \begin{cases}
      x & \text{if } x=\E,0,1 \\
      \fcn{AND}(x\lhalf)\bitbin\land\fcn{AND}(\rhalf x)
       & \text{otherwise} 
   \end{cases}
   \fcnind[ldef]{$\fcn{AND}$}{}{} \\
   \fcn{OR}(x) &= \begin{cases}
      x & \text{if } x=\E,0,1 \\
      \fcn{OR}(x\lhalf)\bitbin\lor\fcn{OR}(\rhalf x)
       & \text{otherwise} 
   \end{cases}
   \fcnind[ldef]{$\fcn{OR}$}{}{} \\
   |x| &= \begin{cases}
      {}_1x & \text{if } x=\E,0,1 \\
      x\lhalf\lchop\rhalf x\zlC
         \big(|x\lhalf|\cat 0,|x\lhalf|\cat 1\big)
       & \text{otherwise} 
   \end{cases}
   \ind[ldef]{$|\mathord\cdot|$ (length)}{}{} \\
   x\smsh y &= \begin{cases}
      x\zlC(\E,y) & \text{if } x=\E,0,1 \\
      (x\lhalf\smsh y)\cat(\rhalf x\smsh y) & \text{otherwise} 
   \end{cases}
   \ind[ldef]{$\smsh$ (smash)}{}{}
\end{aligned} \\
\begin{aligned}
   \lenfcn{div}_k(x) &= \begin{cases}
       x\lchop\cst k\zlC({}_1x,\E) & \text{if } x=\E,0,1 \\
       \lenfcn{div}_k(x\lhalf)\cat\lenfcn{div}_k(\rhalf x)\cat
       \big((\lenfcn{mod}_k(x\lhalf)\cat\lenfcn{mod}_k(\rhalf x))
          \lchop\cst k\zlC(1,\E)\big) & \text{otherwise}
   \end{cases}
   \lenfcnind[ldef]{$\lenfcn{div}_k,\lenfcn{mod}_k$}{}{} \\
   \fcn{CAR}(x) &= \begin{cases}
      {}_0x & \text{if } x=\E,0,1 \\
      \fcn{CScar}(\fcn{CAR}(x\lhalf),\fcn{CAR}(\rhalf x),
                  \fcn{ADD}(x\lhalf),\fcn{ADD}(\rhalf x))
       & \text{otherwise} 
   \end{cases}
   \fcnind[ldef]{$\fcn{CAR},\fcn{ADD}$}{}{} \\
   \fcn{ADD}(x) &= \begin{cases}
      x & \text{if } x=\E,0,1 \\
      \fcn{CSadd}(\fcn{CAR}(x\lhalf),\fcn{CAR}(\rhalf x),
                  \fcn{ADD}(x\lhalf),\fcn{ADD}(\rhalf x))
       & \text{otherwise} 
   \end{cases} 
\end{aligned}
\end{gather*}
 Note that \(|\fcn{CAR}(x)|=|\fcn{ADD}(x)|\) (easy to show
 inductively) so the definition by $\TRN_2$ is correct.  Also note
 that with $\lenfcn{div}_k$ and $\lenfcn{mod}_k$, we can now define
 ``$(k,\ell)$-$\CRN$'' similarly to $(2^k,2^\ell)$-$\CRN$, but for
 blocks of bits of arbitrary fixed lengths.  Interestingly, it does
 not seem possible to define a more general $\lenfcn{div}$ function
 that would take two parameters, and thus to define a general form of
 $\CRN$ where the lengths of the input and output blocks of bits are
 specified by extra parameters (we discuss this issue further in
 Chapter~\ref{chap:T1}).

\subsection{Numerical functions}

 Unfortunately, the fact that functions in $L_1$ are
 length-determined makes it harder to define ``numerical'' functions,
 \ie, functions that treat their inputs as binary notation for numbers
 (ignoring leading 0's).  For example, the definition of $|x|$ given
 above is quite simple whereas the definition of $|x\num|$ given below
 relies on some more complex functions.

 {}Now, we will define a number of ``numerical'' functions
 (distinguished by putting a superscript ``${}^{\tinyN}$'' next to
 them).  We start with an equality operator for numbers, and also one
 for strings.
\[
   x\numrel=y = \fcn{AND}(1\cat\bitfcn{iff}_2(x,y))
   \numind[ldef]{$\numrel"=$}{}{}
   \qquad x\strrel=y = \bit\lnot(x\lenrel=y)\bitbin\land(x\numrel=y)
\]
 Note that the extra ``$1$'' is necessary in the definition of
 $\numrel=$ for $\E\numrel=\E$ to be true, and the value of
 $x\numrel=y$ is independent of the lengths of $x$ and $y$, \ie, the
 function really does behave as though its string inputs were binary
 representations of numbers.  Also note that ``$\bit\lnot$'' is
 necessary in the definition of $\strrel=$ because of our convention
 that $\lenrel=$ returns $\E$ for ``true'' and $1$ for ``false''.

 {}Next, we define a successor and a predecessor function, both by
 left $\CRN$.  The successor function is defined in terms of an
 auxiliary function that simply replaces each bit by its negation
 until it encounters a $0$, which it replaces by $1$, and then outputs
 each bit unchanged (\eg, $11010$ becomes $11011$, $1011$ becomes
 $1100$, $111$ becomes $000$).  The successor function first adds a
 $0$ to the front (left) of its argument before calling the auxiliary
 function, in case the string consists of all $1$'s (\eg, $111$
 correctly becomes $1000$ and not just $000$).  The predecessor
 function performs a similar computation, except replacing bits by
 their negation until it encounters a $1$, and first checking that the
 string does not consist of all $0$'s before calling the auxiliary
 function.
\begin{gather*}
   \numfcn{aux\_succ}(ix) = \big(\fcn{AND}(1x)\bC(\bit\lnot i,i)\big)
      \cat\numfcn{aux\_succ}(x) \qquad
   \numfcn{succ}(x) = \numfcn{aux\_succ}(0x)
   \numfcnind[ldef]{$\numfcn{succ}$}{}{} \\
   \numfcn{aux\_pred}(ix) = \big(\fcn{OR}(x)\bC(i,\bit\lnot i)\big)
      \cat\numfcn{aux\_pred}(x) \qquad
   \numfcn{pred}(x) = \fcn{OR}(x)\bC(\numfcn{aux\_pred}(x),x) 
   \numfcnind[ldef]{$\numfcn{pred}$}{}{}
\end{gather*}
 Note that one unfortunate side-effect of the fact that the functions
 are length-determined is that the successor function always appends a
 bit to the left of its argument.  So starting from $\E$ and applying
 $\numfcn{succ}$ repeatedly, we get a series of strings that represent
 $0,1,2,\dotsc$ in binary, but whose lengths are also $0,1,2,\dotsc$
 Next, we define the numerical predicate $\numrel<$, which together
 with $\numrel=$ allows us to define all other relational operators on
 numbers using the Boolean connectives.  Note that to define
 $\numfcn{less}$, we use Clote's ``programming trick''~\cite{Clot89}
 of making a sweep through the bits of the strings $x$ and $y$,
 appending a 1 when some condition is met so that the final
 composition with $\fcn{OR}$ yields 1 iff the condition was met at
 some position.  Using $\fcn{AND}$, we could similarly define
 functions that test for some condition on every bit of their inputs.
 Also recall that functions defined by $\CRN_2$ (such as
 $\numfcn{less}$ below) first pad their arguments on the left with
 $0$'s so they have the same length.
\begin{gather*}
   \numfcn{less}(xi,yj) = \numfcn{less}(x,y)\cat
      \big((i\bitrel<j)\bitbin\land(x\numrel=y)\big) \qquad
   x\numrel<y = \fcn{OR}(\numfcn{less}(x,y)) 
   \numind[ldef]{$\numrel<$}{}{}
\end{gather*}

 {}The next function we want to define is $\numfcn{bit}(x,z)$, which
 returns bit number $z$ of $x$, starting at $1$ and counting from the
 right, where $z$ is interpreted as a binary number.  The easiest way
 to do this is by defining a function $\numfcn{pow}(z,x)$ that returns
 a string of length $|x|$ consisting entirely of 0's except at bit
 position $z$ (from the right), if \(1\numrel\leq z\numrel\leq|x|\).
 Then, we define a function $\numfcn{maskbit}(x,y)$ that treats $y$ as
 a mask to determine which bit of $x$ to return.
\begin{gather*}
   \numfcn{pow}(z,ix) =
      \big(|ix|\numrel=z\big)\cat\numfcn{pow}(z,x) \\
   \numfcn{maskbit}(x,y) = \fcn{OR}(\bitfcn{and}_2(x,y)) \\
   \numfcn{bit}(x,z) = \numfcn{maskbit}(x,\numfcn{pow}(z,x)) 
   \numfcnind[ldef]{$\numfcn{bit}$}{}{}
\end{gather*}
 Next, we want to define addition.  This will require only a few more
 definitions.  First, in order to simulate a function that strips
 leading ones (or zeros) from $x$, we can define functions
 $\fcn{first}_j(x)$ that return a mask which is 1 on the leftmost bit
 of $x$ equal to $j$ and 0 elsewhere.  We can also define a function
 that returns a mask which is 1 on every significant bit of $x$ (\ie,
 every bit to the right of the first ``1'' in $x$) and 0 elsewhere.
\begin{gather*}
   \fcn{first}_1(xi) = \fcn{first}_1(x)\cat
      \big(\fcn{OR}(x)\bC(0,i)\big) \qquad
   \fcn{first}_0(xi) = \fcn{first}_0(x)\cat
      \big(\fcn{AND}(1x)\bC(\bit\lnot i,0)\big)
   \fcnind[ldef]{$\fcn{first}_0,\fcn{first}_1$}{}{} \\
   \numfcn{mask}(xi) = \numfcn{mask}(x)\cat
      \big(\fcn{OR}(x)\bC(1,i)\big) 
   \numfcnind[ldef]{$\numfcn{mask}$}{}{}
\end{gather*}
 Then, we can define a function which computes the carry bits and an
 addition function, as follows.
\begin{align*}
   \numfcn{carry}(ix,jy) &= \numfcn{maskbit}\big(
      \bitfcn{and}_2(ix,jy),\fcn{first}_0(\bitfcn{xor}_2(ix,jy))
      \big)\cat\numfcn{carry}(x,y)
   \numfcnind[ldef]{$\numfcn{carry}$}{}{} \\
   x\numbin+y &= \bitfcn{xor}_3(\numfcn{carry}(x,y)\cat 0,x,y) 
   \numind[ldef]{$\numbin+$}{}{}
\end{align*}
 Finally, using the addition function, we can define a function that
 counts the number of ones in a string:
\( \fcn{sum}(x) = \fcn{CAR}(x)\numbin+\fcn{ADD}(x) \)
\fcnind[ldef]{$\fcn{sum}$}{}{}
 and using this function, define the ``numerical'' length of $x$:
\( |x\num| = \fcn{sum}(\numfcn{mask}(x)) \).
\ind[ldef]{$|\mathord\cdot|$ (length)}{$|\mathord\cdot\num|$}{}
 (Note that we could also have defined \(|x|=\fcn{sum}({}_1x)\)
 instead of directly using $\TRN$.)

\section{$L_1$ and $\FALT$}

 In this section, we prove the following claim.
\begin{claim}
\label{claim:L1}
   \( L_1 = \FALT \) (= uniform $\FNC^1$).
\end{claim}
 To be precise, we say that a $k$-ary function $f$ belongs to $\FALT$
 if there exists an integer polynomial $p_f$ such that
\( |f(\tpl x_k)| \leq p_f(|x_1|,\dots,|x_k|) \)
 for every $\tpl x_k$, and if the language
\( \{ \langle\tpl x_k,i,b\rangle \st
      \text{the $i$-th bit of $f(\tpl x_k)$ is equal to $b$} \} \)
 is recognizable by an $\ATM$ running in time
 $\bigOh\big(\max\{\log|x_1|,\dots,\log|x_k|\}\big)$.

\subsection{$\FALT$ is contained in $L_1$}

 To prove that \( \FALT \subseteq L_1 \), we show how to simulate the
 computation of an $\ATM$ using functions in $L_1$.  Then, if
 \(f\in\FALT\), there exists a term \(t_f\in L_1\) such that
 \(|t_f(\tpl x_k)|=p_f(|\tpl x_k|)\), so we can use $\CRN$ on
 $t_f(\tpl x_k)$ to compute each bit of $f$ by simulating the $\ATM$
 on the appropriate input.  (Technically speaking, this works only if
 \(|f(\tpl x_k)|=|t_f(\tpl x_k)|\) for all $\tpl x_k$, but if that is
 not the case, we can simply use Bloch's idea of ``length masks'' to
 compute a mask $b_f(\tpl x_k)$ of length $t_f(\tpl x_k)$ that has a
 $1$ in every bit position where $f(\tpl x_k)$ is defined and a $0$
 elsewhere.)

 {}Now, without loss of generality, let the $\ATM$ have the following
 properties.
\begin{enumerate}

\item
   There is a function \(t\in L_1\) such that the $\ATM$ runs for no
   more than \(||t(\tpl x)||=\bigOh(\log|t(\tpl x)|)\) steps on inputs
   $\tpl x$ (always possible when the $\ATM$ runs in logarithmic time
   since $t$ can use $\smsh$ and $\cat$ to output a string whose
   length is an arbitrary polynomial in the lengths of the inputs).
   Also, universal states of the $\ATM$ are given \emph{even} numbers
   and existential states, \emph{odd} numbers.  Moreover, we assume
   that the function $t$ is defined so that $|t(\tpl x)|$ is always a
   power of $2$ and $2^{2|t(\tpl x)|}$ is greater than the number of
   states of the $\ATM$ for any inputs $\tpl x$ (in other words, a
   string of length $2|t(\tpl x)|$ is long enough to encode the state
   of the $\ATM$).

\item
   The $\ATM$ has $n$ read-only input tapes represented by strings
   $x_1,\dots,x_n$ and $k$ worktapes represented by pairs of strings
   $y^\ell_1,y^r_1,\dots,y^\ell_k,y^r_k$, each of length exactly
   $2|t(\tpl x)|$, where $y^\ell_i$ represents the content of tape
   number $i$ to the left of the tape head and $y^r_i$ represents the
   content to the right, with the head scanning the rightmost symbol
   of $y^\ell_i$.  Each of the three possible worktape symbols (1, 0,
   or blank) is encoded using two bits (11 for 1, 10 for 0, and 00 for
   blank).  Initially, the worktapes are blank.

\item
   The computation tree of the $\ATM$ is a complete binary tree (each
   non-leaf node has exactly two successor configurations, a left
   successor and a right successor, and every leaf occurs at the same
   level).

\item
   Access to the input occurs only at the leaves of the computation
   tree and is of the form ``accept iff symbol number $y^\ell_i$
   (interpreted as a binary number) on input tape number $j$ is equal
   to $b$'', where $i$, $j$, and $b$ are encoded in the current state
   of the $\ATM$.

\end{enumerate}
 Then, if we let
\( \symsc{con} = \tuple{s,y^\ell_1,y^r_1,\dots,y^\ell_k,y^r_k}_{2k+1} \)
 represent a configuration of the $\ATM$ when in state $s$, we can
 define $l\symsc{con}$ and $r\symsc{con}$, the left and right
 successor configurations of $\symsc{con}$, as follows:
\begin{align*}
   \ell\symsc{con} &= \tuple{\fcn{lstate}(\symsc{con}),
      \fcn{ltape}^\ell_1(\symsc{con}),\fcn{ltape}^r_1(\symsc{con}),\dots,
      \fcn{ltape}^\ell_k(\symsc{con}),\fcn{ltape}^r_k(\symsc{con})}_{2k+1}, \\
   r\symsc{con} &= \tuple{\fcn{rstate}(\symsc{con}),
      \fcn{rtape}^\ell_1(\symsc{con}),\fcn{rtape}^r_1(\symsc{con}),\dots,
      \fcn{rtape}^\ell_k(\symsc{con}),\fcn{rtape}^r_k(\symsc{con})}_{2k+1}, 
\end{align*}
 where each of \( \fcn{lstate},\fcn{ltape}^\ell_i,\fcn{ltape}^r_i,
    \fcn{rstate},\fcn{rtape}^\ell_i,\fcn{rtape}^r_i \) is easily seen
 to be in $L_0$, involving only simple string manipulations and
 finite table lookup on the state $s$ (for example,
\( \fcn{ltape}^\ell_i(\symsc{con}) = d_i\bC\big(
   00\lchop((y^\ell_i\rchop 00\cat b_i)\cat\fcn{lc}(y^r_i,00)),
   (00\cat y^\ell_i)\rchop 00\big) \)
 computes the contents of tape $i$ to the left of the head in the left
 successor of $\symsc{con}$, where $d_i$ (the direction of movement
 for head $i$) and $b_i$ (the tape symbol to write on tape~$i$) are
 obtained from the state and tape contents of $\symsc{con}$ using the
 conditional function).  Moreover, if we let $\fcn{select}$ be defined
 by $(2,1)$-$\CRN$ to output every second bit of its input string, the
 function
\( \fcn{input}(\symsc{con},\tpl x) =
   \numfcn{bit}(x_j,\fcn{select}(y^\ell_i))\bitbin\liff b \)
 (where $i$, $j$, and $b$ are extracted from the state $s$) is equal
 to the accept state of the given input configuration and is in
 $L_1$.  Finally, we let
\( \symsc{con}_0 = \tuple{s_0,t(\tpl x)\smsh 00,\dots,
      t(\tpl x)\smsh 00}_{2k+1} \)
 denote the initial configuration, where $s_0$ is the initial state of
 the $\ATM$.

 {}Now, we can easily use $\TRN$ to define a function $\fcn{eval}$
 that evaluates the computation tree of the $\ATM$, so that the result
 of the entire computation is given by
 $\fcn{eval}(t(\tpl x),\symsc{con}_0,\tpl x)$:
\[
   \fcn{eval}(z,\symsc{con},\tpl x) =
   \begin{cases}
      \fcn{input}(\symsc{con},\tpl x) & \text{if } z=\E,0,1, \\
      \proj^{2k+1}_1(\symsc{con})'\bC\big(
         \fcn{eval}(z\lhalf,\ell\symsc{con},\tpl x)\bitbin\lor
         \fcn{eval}(\rhalf z,r\symsc{con},\tpl x), \\
      \hphantom{\proj^{2k+1}_1(\symsc{con})'\bC\big( }
         \fcn{eval}(z\lhalf,\ell\symsc{con},\tpl x)\bitbin\land
         \fcn{eval}(\rhalf z,r\symsc{con},\tpl x)\big)
       & \text{otherwise}. 
   \end{cases}
\]
 Note that in the recursive call, $\proj^{2k+1}_1(\symsc{con})$ simply
 extracts the current state from the given configuration, and the
 rightmost bit of the state number is used to determine whether the
 configuration is universal or existential.  Also note that this half
 of the proof is considerably simpler than the corresponding proofs in
 Bloch~\cite{Bloc94} and Clote~\cite{Clot92a,Clot93}.  This seems to
 be because our scheme of $\TRN$ encapsulates the sort of computation
 carried out by $\ATM$'s more directly than the schemes considered by
 Bloch and Clote, especially by its use of the parameter replacement
 functions $h_\ell$ and $h_r$.

\subsection{$L_1$ is contained in $\FALT$}

 To prove that \( L_1 \subseteq \FALT \), we argue that every
 function in $L_0$ can be computed by a family of circuits in uniform
 $\FNC^0$, and that every function in $L_1$ can be computed by a
 family of circuits in uniform $\FNC^1$, where we use Bloch's notion
 of \emph{mapping-uniformity}, defined in~\cite{Bloc94}, which
 generalizes $U_{E^*}$-uniformity to make sense for circuits of
 constant depth.  As will be seen, the facts that functions in $L_0$
 have constant depth circuits and that functions in $L_1$ have
 logarithmic depth circuits are quite simple to prove; the technical
 difficulties arise mainly from uniformity considerations.

 {}First, we give bounds on the rate of growth of functions in $L_0$
 and $L_1$.

\begin{lemma}
\label{lem:L0}
   For every $n$-ary function \(f\in L_0\), there exist constants
   \(a^f_0,a^f_1,\dots,a^f_n\in\NN\) such that
   \( |f(x_1,\dots,x_n)| \leq a^f_0+a^f_1|x_1|+\dots+a^f_n|x_n| \)
   for all strings $x_1,\dots,x_n$.
\end{lemma}
\begin{proof}
   The result is proved by induction on the definition of $f$.
   \begin{itemize}
   \item
      If \(f=\E\), then \(a^f_0=0\) since \(|\E|=0\).
   \item
      If \(f=0\) or \(f=1\), then \(a^f_0=1\) since \(|0|=|1|=1\).
   \item
      If \(f(x)=\rhalf x\), then \(a^f_0=0,a^f_1=1\) since
      \(|\rhalf x|=\lceil|x|/2\rceil\leq|x|\).
   \item
      If \(f(x,y)=x\rchop y\), then \(a^f_0=0,a^f_1=1,a^f_2=0\) since
      \(|x\rchop y|=|x|\dotminus|y|\leq|x|\).
   \item
      If \(f(x,y)=x\cat y\), then \(a^f_0=0,a^f_1=1,a^f_2=1\) since
      \(|x\cat y|=|x|+|y|\).
   \item
      If \(f(x,y,z,w)=x\C(y,z,w)\), then
      \(a^f_0=0,a^f_1=0,a^f_2=1,a^f_3=1,a^f_4=1\) since \\
      \(|x\C(y,z,w)|\leq\max\{|y|,|z|,|w|\}\leq|y|+|z|+|w|\).
   \item
      If \(f(x_1,\dots,x_n)=\I^n_k(x_1,\dots,x_n)\), then
      \(a^f_0=0,\dots,a^f_{k-1}=0,a^f_k=1,a^f_{k+1}=0,\dots,a^f_n=0\)
      since \(|\I^n_k(x_1,\dots,x_n)|=|x_k|\).
   \item
      If $f$ is defined by $\CRN$ from $h$, then
      \( |f(x,\tpl y)| = |x| \) so
      \(a^f_0=0,a^f_1=1,a^f_2=0,\dots,a^f_{n+1}=0\).
   \item
      If $f$ is defined by $\COMP$ from $g$ and $h_1,\dots,h_k$, then
      \begin{align*}
         |f(\tpl x)|
         &= |g(h_1(\tpl x),\dots,h_k(\tpl x))| \\
         &\leq a^g_0+a^g_1|h_1(\tpl x)|+\dots+a^g_k|h_k(\tpl x)| \\
         &\leq a^g_0+\sum_{1\leq i\leq k}
               a^g_i\Big(a^{h_i}_0+
                  \sum_{1\leq j\leq n}a^{h_i}_j|x_j|\Big) \\
         &\leq \Big(a^g_0+\sum_{1\leq i\leq k}a^g_ia^{h_i}_0\Big)+
               \sum_{1\leq j\leq n}
                  \Big(\sum_{1\leq i\leq k}a^g_ia^{h_i}_j\Big)|x_j| 
      \end{align*}
      so
      \( a^f_0 = a^g_0+a^g_1a^{h_1}_0+\dots+a^g_ka^{h_k}_0 \),
      and
      \( a^f_i = a^g_1a^{h_1}_i+\dots+a^g_ka^{h_k}_i \)
      for \(1\leq i\leq n\).
      \quad\QED
   \end{itemize}
\end{proof}

\begin{lemma}
\label{lem:L1}
   For every $n$-ary function \(f\in L_1\), there exists a polynomial
   \(p_f\in\NN[x_1,\dots,x_n]\) such that
   \( |f(x_1,\dots,x_n)| \leq p_f(|x_1|,\dots,|x_n|) \)
   for all strings $x_1,\dots,x_n$.
\end{lemma}
\begin{proof}
   The result is proved by induction on the definition of $f$, where
   we use the notation $|\tpl x_n|$ to stand for the list
   $|x_1|,\dots,|x_n|$.
   \begin{itemize}
   \item
      If \(f\in L_0\), then by the preceding lemma,
      \( p_f(|\tpl x_n|) = a^f_0+a^f_1|x_1|+\dots+a^f_n|x_n| \).
   \item
      If $f$ is defined by $\CRN$ from $h$, then as in the preceding
      lemma, \( p_f(|x|,|\tpl y|) = |x| \).
   \item
      If $f$ is defined by $\COMP$ from $g$ and $h_1,\dots,h_k$, then
      \begin{align*}
         |f(\tpl x_n)|
         &= |g(h_1(\tpl x_n),\dots,h_k(\tpl x_n))| \\
         &\leq p_g\big(|h_1(\tpl x_n)|,\dots,|h_k(\tpl x_n)|\big) \\
         &\leq p_g\big(p_{h_1}(\tpl x_n),\dots,p_{h_k}(\tpl x_n)\big) 
      \end{align*}
      (since polynomials in $\NN[\tpl x]$ are non-decreasing), so
      \( p_f(|\tpl x_n|) =
         p_g\big(p_{h_1}(|\tpl x_n|),\dots,p_{h_k}(|\tpl x_n|)\big) \).
   \item
      If $f$ is defined by $\TRN$ form $g,h,h_\ell,h_r$, where we
      assume without loss of generality that
      \begin{itemize}
      \item
         \( |g(x,z,\tpl y)| \leq p_g(|x|,|z|,|\tpl y|) \),
      \item
         \( |h_\ell(z)| \leq c|z| \) \quad and \quad
         \( |h_r(z)| \leq c|z| \),
      \item
         \( |h(x,z,\tpl y,v_\ell,v_r)| \leq a_0+a_1|x|+a_2|z|+
               \sum(b_i|y_i|)+a(|v_\ell|+|v_r|) \),
      \end{itemize}
      then intuitively, at each level of the recursion, the length of
      the second argument is multiplied by $c$ so that at the bottom
      level (after $\lg|x|$ steps), the second argument has length
      $\bigOh(c^{\lg|x|}|z|)$.  At the same time, the lengths of each
      recursive call to $f$ are multiplied by $a$, which means that
      the total length of $f$ (bounded by the length of $g$ in the
      base case) is multiplied by $a^{\lg|x|}$.  More precisely, we
      show that
      \begin{align*}
         p_f(|x|,|z|,|\tpl y|)
         &= \left(a_0+\sum(b_i|y_i|)\right)\cdot
               \sum_{j=0}^{\lceil\lg|x|\rceil-1}(2a)^j +
            a_1|x|\cdot\sum_{j=0}^{\lceil\lg|x|\rceil-1}a^j \\
         &\pheq + a_2|z|\cdot\sum_{j=0}^{\lceil\lg|x|\rceil-1}(2ac)^j
            +(2a)^{\lceil\lg|x|\rceil}\cdot
               p_g\big(1,c^{\lceil\lg|x|\rceil}|z|,|\tpl y|\big) \\
         &\leq 2|x|^{\lceil\lg a\rceil+1}\left[a_0+\sum(b_i|y_i|)+
               a_1|x|+a_2|z||x|^{\lceil\lg c\rceil} \right.\\
         &\phanrel{\leq} \hphantom{2|x|^{\lceil\lg a\rceil+1}\Big[}
            \left. +a\cdot p_g\big(1,c|z||x|^{\lceil\lg c\rceil},
                  |\tpl y|\big)\right]. 
      \end{align*}
      Technically speaking, we need to use $\max\{1,|z|\}$ everywhere
      that $|z|$ appears in this expression, but this does not change
      the proof substantially besides making it longer to write down.
      Also, we need to deal separately with special cases such as when
      \(a=0\) or when the lengths of $h_\ell$ and $h_r$ are constants
      independent of $|z|$, but all of these cases simplify the proof
      so we present the general case only.

      Now, if \(x=\E,0,1\), then
      \( |f(x,z,\tpl y)|
          = |g(x,z,\tpl y)|
         \leq p_g(|x|,|z|,|\tpl y|)
         \leq p_g(1,|z|,|\tpl y|)
         \leq a_0+a_1|x|+a_2|z|+\sum(b_i|y_i|)+p_g(1,|z|,|\tpl y|)
          = p_f(1,|z|,|\tpl y|) \).
      If \(|x|>1\), then we consider two subcases.  If $|x|$ is even,
      then \( \lceil\lg(|x|/2)\rceil = \lceil\lg|x|\rceil-1 \), so
      \begin{align*}
         |f(x,z,\tpl y)|
         &= \big|h\big(x,z,\tpl y,f(x\lhalf,h_\ell(z),\tpl y),
            f(\rhalf x,h_r(z),\tpl y)\big)\big| \\
         &\leq a_0+a_1|x|+a_2|z|+\sum(b_i|y_i|) \\*
         &\pheq +a\big(|f(x\lhalf,h_\ell(z),\tpl y)|
            +|f(\rhalf x,h_r(z),\tpl y)|\big) \\
         &\leq a_0+a_1|x|+a_2|z|+\sum(b_i|y_i|) \\*
         &\pheq +a\big(p_f(|x\lhalf|,|h_\ell(z)|,|\tpl y|)
            +p_f(|\rhalf x|,|h_r(z)|,|\tpl y|)\big) \\
         &\leq a_0+a_1|x|+a_2|z|+\sum(b_i|y_i|)
            +2a\,p_f(|x|/2,c|z|,|\tpl y|) \\
         &\leq a_0+a_1|x|+a_2|z|+\sum(b_i|y_i|) \\*
         &\pheq +2a\left[\left(a_0+\sum(b_i|y_i|)\right)\cdot
               \sum_{j=0}^{\lceil\lg|x|\rceil-2}(2a)^j
            +a_1\frac{|x|}{2}\cdot\sum_{j=0}^{\lceil\lg|x|\rceil-2}a^j
            \right.\\*
         &\pheq \hphantom{+2a\left[\right. } \left.
            +a_2c|z|\cdot\sum_{j=0}^{\lceil\lg|x|\rceil-2}(2ac)^j
            +(2a)^{\lceil\lg|x|\rceil-1}\cdot
               p_g\big(1,c^{\lceil\lg|x|\rceil-1}c|z|,|\tpl y|\big)
            \right] \\
         &\leq p_f(|x|,|z|,|\tpl y|). 
      \end{align*}
      If $|x|$ is odd, then
      \( \lceil\lg((|x|-1)/2)\rceil \leq \lceil\lg|x|\rceil-1 \) and
      \( \lceil\lg((|x|+1)/2)\rceil = \lceil\lg|x|\rceil-1 \), so
      \begin{align*}
         |f(x,z,\tpl y)|
         &= \big|h\big(x,z,\tpl y,f(x\lhalf,h_\ell(z),\tpl y),
            f(\rhalf x,h_r(z),\tpl y)\big)\big| \\
         &\leq a_0+a_1|x|+a_2|z|+\sum(b_i|y_i|) \\*
         &\pheq +a\big(|f(x\lhalf,h_\ell(z),\tpl y)|
            +|f(\rhalf x,h_r(z),\tpl y)|\big) \\
         &\leq a_0+a_1|x|+a_2|z|+\sum(b_i|y_i|) \\*
         &\pheq +a\big(p_f(|x\lhalf|,|h_\ell(z)|,|\tpl y|)
            +p_f(|\rhalf x|,|h_r(z)|,|\tpl y|)\big) \\
         &\leq a_0+a_1|x|+a_2|z|+\sum(b_i|y_i|) \\*
         &\pheq +a\big(p_f((|x|-1)/2,c|z|,|\tpl y|)
            +p_f((|x|+1)/2,c|z|,|\tpl y|)\big) \\
         &\leq a_0+a_1|x|+a_2|z|+\sum(b_i|y_i|) \\
         &\pheq +a\left[\left(a_0+\sum(b_i|y_i|)\right)\cdot
               \sum_{j=0}^{\lceil\lg|x|\rceil-2}(2a)^j
            +a_1\frac{|x|-1}{2}\cdot
               \sum_{j=0}^{\lceil\lg|x|\rceil-2}a^j \right.\\*
         &\pheq \hphantom{+a\left[\right. } \left.
            +a_2c|z|\cdot\sum_{j=0}^{\lceil\lg|x|\rceil-2}(2ac)^j
            +(2a)^{\lceil\lg|x|\rceil-1}\cdot
               p_g\big(1,c^{\lceil\lg|x|\rceil-1}c|z|,|\tpl y|\big)
            \right] \\
         &\pheq +a\left[\left(a_0+\sum(b_i|y_i|)\right)\cdot
               \sum_{j=0}^{\lceil\lg|x|\rceil-2}(2a)^j
            +a_1\frac{|x|+1}{2}\cdot
               \sum_{j=0}^{\lceil\lg|x|\rceil-2}a^j \right.\\*
         &\pheq \hphantom{+a\left[\right. } \left.
            +a_2c|z|\cdot\sum_{j=0}^{\lceil\lg|x|\rceil-2}(2ac)^j
            +(2a)^{\lceil\lg|x|\rceil-1}\cdot
               p_g\big(1,c^{\lceil\lg|x|\rceil-1}c|z|,|\tpl y|\big)
            \right] \\
         &\leq a_0+a_1|x|+a_2|z|+\sum(b_i|y_i|) \\*
         &\pheq +\left(a_0+\sum(b_i|y_i|)\right)\cdot
               \sum_{j=1}^{\lceil\lg|x|\rceil-1}(2a)^j
            +a_1|x|\cdot\sum_{j=1}^{\lceil\lg|x|\rceil-1}a^j \\*
         &\pheq +a_2|z|\cdot\sum_{j=1}^{\lceil\lg|x|\rceil-1}(2ac)^j
            +(2a)^{\lceil\lg|x|\rceil}\cdot
               p_g\big(1,c^{\lceil\lg|x|\rceil}|z|,|\tpl y|\big) \\*
         &\pheq +\frac{-a_1}{2}\sum_{j=1}^{\lceil\lg|x|\rceil-1}a^j
            +\frac{a_1}{2}\sum_{j=1}^{\lceil\lg|x|\rceil-1}a^j 
          \leq p_f(|x|,|z|,|\tpl y|). 
         \quad\QED
      \end{align*}
   \end{itemize}
\end{proof}

 {}Now, we are ready to discuss circuits.  For the sake of
 completeness, we summarize here Bloch's definitions and results,
 suitably modified to apply to our setting.  We will be working with
 circuit families that compute \emph{functions} instead of relations,
 \ie, circuits will generally have multiple output gates.  In this
 setting, we define $\FNC^0$ to be the class of functions computed by
 constant depth circuit families (because the circuits have multiple
 output gates, this class contains interesting functions, unlike the
 relational counterpart $\NC^0$), and $\FNC^1$ to be the class of
 functions computed by logarithmic depth circuit families.

 {}We assume that the gate set for our circuits consists of constants
 $0$ and $1$, unary identity ($\lsg$) and negation ($\lnot$), and
 binary conjunction ($\land$), disjunction ($\lor$), left projection
 ($\pi_{\tinyL}$), and right projection ($\pi_{\tinyR}$) (this could be
 reduced at the cost of longer proofs).  Given a circuit family
 composed of such gates, we identify gates in the circuits by pairs
 $\pair{\out}{\sigma}$, where $\out$ is the number of an output gate
 of the circuit (in binary) and \(\sigma\in\{L,R,S\}^*\) represents a
 path in the circuit from the given output gate, where $L$ and $R$
 indicate the left and right inputs of a binary gate, respectively,
 and $S$ indicates the only input of a unary gate.  (Note that we
 number the output gates from right to left, starting with $1$, and
 similarly for the input gates of each input.)

 {}Now, we want to work with \emph{uniform} families of circuits.
 Unfortunately, the standard notion of $U_{E^*}$-uniformity defined by
 Ruzzo~\cite{Ruzz81} does not make sense for constant depth circuits
 (it would require the extended connection language for the circuits
 to be recognizable by an $\ATM$ in constant time, which is not even
 enough time for the $\ATM$ to examine a gate number or an input
 length).  To remedy this, Bloch defines a notion of
 \emph{mapping-uniform} circuits, where the uniformity computation is
 divided in two phases.  The main purpose of the uniformity
 computation is to be able to recognize connections in the circuit
 (\ie, given a gate and a path, what gate is at the end of the path?)
 and gate information (\ie, given a gate number, what type is that
 gate?).  Because of our numbering scheme for gates, determining the
 descendant of a gate $\pair{\out}{\sigma}$ along a path $\sigma'$ is
 easy: the answer is simply $\pair{\out}{\sigma\sigma'}$.  Determining
 the type of an internal gate is also not difficult, as we will see.
 The hard part, requiring the ``two-phase'' approach, is to determine
 which input bit is tied to an input gate (which is necessary to fully
 specify the ``type'' of that input gate).

 {}Here is some intuition behind the two phases of the uniformity
 computation.  Essentially, given an input gate $\pair{\out}{\sigma}$
 in a circuit, the first phase must use $\sigma$ to identify
 \emph{how} to compute the number of the input bit as a function of
 the output gate number, but without carrying out that computation.
 This will be done by a deterministic Turing machine running in time
 proportional to $|\sigma|$, which is bounded by the depth of the
 circuit, and it is this phase of the computation that will be
 composed or iterated for functions defined by composition or
 recursion. The second phase of the computation will only be carried
 out once, by an $\ATM$ that combines the information from the first
 phase together with the rest of the information about the circuit.

 {}More precisely, the first phase will output a term $t_\sigma$ that
 may be thought of as mapping output bit positions to input bit
 positions that ``affect'' that output bit, \ie, given an input gate
 $\pair{\out}{\sigma}$ in the circuit tied to bit number $r$ of some
 input parameter, we want \(t_\sigma(\out)=r\).  For this purpose, we
 introduce the \emph{mapping language} of a family of circuits, which
 consists of a set of functions that encapsulate all the ``primitive''
 dependencies that may exist between output and input bit numbers.
 Besides the natural numbers used to represent bit positions, we also
 use the symbol ``$\undef$'' to indicate that a given output bit does
 not depend on any input bit.  For a circuit with $k$ input parameters
 $y_1,\dots,y_k$, the mapping language contains the following function
 symbols (each function is implicitly defined to be equal to $\undef$
 when its argument is $\undef$).
\begin{multicols}{3}
\begin{itemize}
\item
   \( \mapl{undef}(x) = \undef \)
\item
   \( \mapl{one}(x) = 1 \)
\item
   \( \mapl{add}_j(x) = x+|y_j| \)
\item
   \( \mapl{sub}_j(x) = x\dotminus|y_j| \)
\item
   \( \mapl{min}_j(x) = \min\{x,|y_j|\} \)
\end{itemize}
\end{multicols}

 {}Before we move back to circuits, we argue that terms $t_\sigma$ in
 the mapping language can be computed in alternating logarithmic time
 (as a function of \(m=\max\{|t_\sigma|,\out,|y_1|,\dots,|y_k|\}\)).
 Given $t_\sigma,\out,|y_1|,\dots,|y_k|$, an $\ATM$ can check in
 parallel if $t_\sigma$ contains $\mapl{undef}$ and output $\undef$
 immediately if this is the case; otherwise, the $\ATM$ guesses the
 position in $t_\sigma$ where the last ``$\mapl{one}$'' appears, and
 replaces the subsequent part of the term with $1$.  Once these simple
 checks are done, the $\ATM$ can construct lists of numbers from
 $|y_1|,\dots,|y_k|$ and subterms of $t_\sigma$ for each block of
 functions of the form
 $\mapl{min}_{j_1}(\dots\mapl{min}_{j_\ell}(t')\dots)$ in $t_\sigma$,
 and evaluate each of these blocks in parallel.  The rest of the
 subterms will contain only $\mapl{add}_j$ and $\mapl{sub}_j$
 functions, and these lists of terms and numbers from
 $|y_1|,\dots,|y_k|$ can be added and subtracted (using two's
 complement) with standard carry-save techniques.

 {}Now, following Bloch, we say that a circuit family is
 \emph{mapping-uniform} if there exist a deterministic multi-tape
 Turing machine $P$ and an $\ATM$ $Q$ such that for every gate
 $\pair{\out}{\sigma}$ in a circuit of the family, the following two
 conditions hold.
\begin{enumerate}
\item
   If $\pair{\out}{\sigma}$ is an input gate tied to bit number $r$ of
   some input parameter, then $P$ on input $\sigma$ runs in time
   $\bigOh(|\sigma|)$ and outputs a term $t_\sigma$ in the mapping
   language such that \(t_\sigma(\out)=r\).
\item
   Machine $Q$ on input $\out,\sigma,|y_1|,\dots,|y_k|,\tau$ runs in
   logarithmic time, \ie, in alternating time
   $\bigOh\big(\log(\max\{\out,|\sigma|,|y_1|,\dots,|y_k|\})\big)$
   and accepts iff $\pair{\out}{\sigma}$ is an internal gate of type
   $\tau$ or $\pair{\out}{\sigma}$ is an input gate tied to bit number
   $r$ of $y_j$ and \(\tau=\pair{j}{r}\) (in some standard encoding).
\end{enumerate}
 A direct argument shows that any mapping-uniform family of circuits
 of at least logarithmic depth is also $U_{E^*}$-uniform.  Now, we are
 ready to show that \( L_1 \subseteq \FNC^1 \).

\begin{lemma}
\label{lem:base}
   For every \(f\in\BASE\), $f$ can be computed by a uniform circuit
   family in $\FNC^0$.
\end{lemma}
\begin{proof}\mbox{}
\begin{description}
\item[{$\E$}:]
   The empty circuit is the only one computing this function.  Hence,
   on input \(\sigma=\E\), $P$ outputs $\mapl{undef}(x)$; on input
   $\out,\sigma,\tau$, $Q$ accepts iff \(\out=0\), \(\sigma=\E\), and
   \(\tau=\E\).
\item[{$0,1$}:]
   Circuits for these functions consist of a single output gate, one
   of the ``constant'' gates $0$ or $1$, appropriately.  Hence, on
   input \(\sigma=\E\), $P$ outputs $\mapl{undef}(x)$; on input
   $\out,\sigma,\tau$, $Q$ accepts iff \(\out=1\), \(\sigma=\E\), and
   \(\tau=0\) or \(\tau=1\), respectively.
\item[{$\rhalf y_1$}:]
   Circuits for this function connect output gates to input gates
   directly, using unary identity gates $\lsg$.  Hence, on input
   \(\sigma=S\), $P$ outputs simply $x$; on input
   $\out,\sigma,|y_1|,\tau$, $Q$ accepts iff
   \(1\leq\out\leq\lceil|y_1|/2\rceil\), \(\sigma=\E\) and
   \(\tau=\lsg\), or \(\sigma=S\) and \(\tau=\pair{1}{\out}\).
\item[{$y_1\rchop y_2$}:]
   Again, circuits for this function connect output gates to input
   gates directly, using unary identity gates $\lsg$.  Hence, on input
   \(\sigma=S\), $P$ outputs $\mapl{add}_2(x)$; on input
   $\out,\sigma,|y_1|,|y_2|,\tau$, $Q$ accepts iff
   \(1\leq\out\leq|y_1|\dotminus|y_2|\), \(\sigma=\E\) and
   \(\tau=\lsg\), or \(\sigma=S\) and \(\tau=\pair{1}{r}\) for
   \(r=\mapl{add}_2(\out)\).
\item[{$y_1\cat y_2$}:]
   The simplest circuits to compute this function would use unary
   identity gates connected directly to the input bits, as in the last
   two cases.  Unfortunately, this would not allow $P$ to know from
   $\sigma$ alone which term to output.  Therefore, we do something
   slightly different, as depicted in Figure~\ref{fig:cat}.

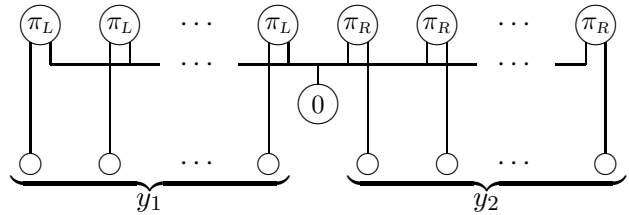
\begin{figure}[!ht]
\centering\setlength{\unitlength}{.75pt}
   \begin{picture}(310,105)(-15,-95)
      \put(0,0){\makebox(0,0){\small$\pi_{\tinyL}$}}
      \put(0,0){\circle{20}}
      \put(-5,-8.66){\line(0,-1){56.34}}
      \put(-5,-70){\circle{10}}
      \put(5,-8.66){\line(0,-1){11.34}}
      \put(40,0){\makebox(0,0){\small$\pi_{\tinyL}$}}
      \put(40,0){\circle{20}}
      \put(35,-8.66){\line(0,-1){56.34}}
      \put(35,-70){\circle{10}}
      \put(45,-8.66){\line(0,-1){11.34}}
      % ...
      \put(80,0){\makebox(0,0){$\dotsm$}}
      \put(80,-70){\makebox(0,0){$\dotsm$}}
      \put(120,0){\makebox(0,0){\small$\pi_{\tinyL}$}}
      \put(120,0){\circle{20}}
      \put(115,-8.66){\line(0,-1){56.34}}
      \put(115,-70){\circle{10}}
      \put(125,-8.66){\line(0,-1){11.34}}
      \put(140,-40){\makebox(0,0){\small$0$}}
      \put(140,-40){\circle{20}}
      \put(140,-30){\line(0,1){10}}
      \put(140,-20){\line(-1,0){40}}
      \put(80,-20){\makebox(0,0){$\dotsm$}}
      \put(60,-20){\line(-1,0){55}}
      \put(140,-20){\line(1,0){80}}
      \put(240,-20){\makebox(0,0){$\dotsm$}}
      \put(260,-20){\line(1,0){15}}
      \put(160,0){\makebox(0,0){\small$\pi_{\tinyR}$}}
      \put(160,0){\circle{20}}
      \put(155,-8.66){\line(0,-1){11.34}}
      \put(165,-8.66){\line(0,-1){56.34}}
      \put(165,-70){\circle{10}}
      \put(200,0){\makebox(0,0){\small$\pi_{\tinyR}$}}
      \put(200,0){\circle{20}}
      \put(195,-8.66){\line(0,-1){11.34}}
      \put(205,-8.66){\line(0,-1){56.34}}
      \put(205,-70){\circle{10}}
      % ...
      \put(240,0){\makebox(0,0){$\dotsm$}}
      \put(240,-70){\makebox(0,0){$\dotsm$}}
      \put(280,0){\makebox(0,0){\small$\pi_{\tinyR}$}}
      \put(280,0){\circle{20}}
      \put(275,-8.66){\line(0,-1){11.34}}
      \put(285,-8.66){\line(0,-1){56.34}}
      \put(285,-70){\circle{10}}
      \put(55,-75){\makebox(0,0)[t]
        {$\underbrace{\begin{picture}(140,0)\end{picture}}$}}
      \put(55,-85){\makebox(0,0)[t]{$y_1$}}
      \put(225,-75){\makebox(0,0)[t]
        {$\underbrace{\begin{picture}(140,0)\end{picture}}$}}
      \put(225,-85){\makebox(0,0)[t]{$y_2$}}
   \end{picture}
\caption{Uniform circuits for the concatenation function ``$\cat$''.}
\label{fig:cat}
\end{figure}

   Now, on input \(\sigma=L\), $P$ outputs $\mapl{sub}_2(x)$, while on
   input \(\sigma=R\), $P$ outputs simply $x$; on input
   $\out,\sigma,|y_1|,|y_2|,\tau$, $Q$ accepts iff
   \begin{multicols}{2}
   \begin{itemize}
   \item
      \(1\leq\out\leq|y_2|\) and
      \begin{itemize}
      \item
         \(\sigma=\E\), \(\tau=\pi_{\tinyR}\), or
      \item
         \(\sigma=L\), \(\tau=0\), or
      \item
         \(\sigma=R\), \(\tau=\pair{2}{\out}\); or
      \item[]
      \end{itemize}
   \item
      \(|y_2|+1\leq\out\leq|y_2|+|y_1|\) and
      \begin{itemize}
      \item
         \(\sigma=\E\), \(\tau=\pi_{\tinyL}\), or
      \item
         \(\sigma=R\), \(\tau=0\), or
      \item
         \(\sigma=L\), \(\tau=\pair{1}{r}\) \\
         for \(r=\mapl{sub}_2(\out)\).
      \end{itemize}
   \end{itemize}
   \end{multicols}
\item[{$y_1\C(y_2,y_3,y_4)$}:]
   This is the only base function requiring circuits of depth greater
   than one.  There is one consideration making the circuits slightly
   more complicated than it would seem necessary at first: the shorter
   of the last two input parameters must be ``padded'' to the same
   length as the longer, requiring some extra gates.  So, the circuits
   for $\C$ are of two different kinds: when \(|y_1|=0\), the circuits
   simply use unary identity gates $\lsg$ for output, connected
   directly to the input gates of $y_2$, while if \(|y_1|>0\), the
   circuits are depicted in Figure~\ref{fig:C} (we illustrate the case
   when \(0<|y_3|<|y_4|\); the other cases are identical except for
   the obvious modifications to the types of the projection gates).

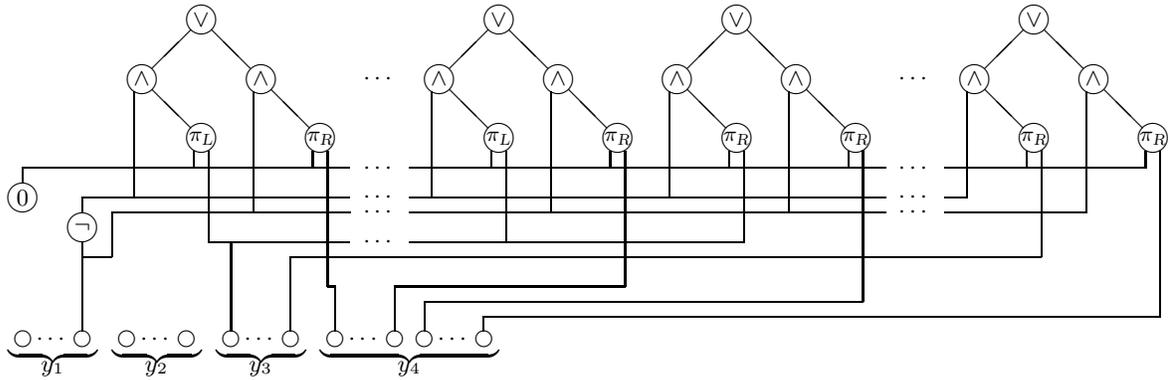
\begin{figure}[!ht]
\centering\footnotesize\setlength{\unitlength}{.5625pt}
   \begin{picture}(780,250)(-690,-240)
      \put(0,0){\makebox(0,0){$\lor$}}
      \put(0,0){\circle{20}}
      \put(-7.07,-7.07){\line(-1,-1){25.86}}
      \put(7.07,-7.07){\line(1,-1){25.86}}
      \put(-40,-40){\makebox(0,0){$\land$}}
      \put(-40,-40){\circle{20}}
      \put(-45,-48.66){\line(0,-1){71.34}}
      \put(-32.93,-47.07){\line(1,-1){25.86}}
      \put(40,-40){\makebox(0,0){$\land$}}
      \put(40,-40){\circle{20}}
      \put(35,-48.66){\line(0,-1){81.34}}
      \put(47.07,-47.07){\line(1,-1){25.86}}
      \put(0,-80){\makebox(0,0){\scriptsize$\pi_{\tinyR}$}}
      \put(0,-80){\circle{20}}
      \put(-5,-88.66){\line(0,-1){11.34}}
      \put(5,-88.66){\line(0,-1){71.34}}
      \put(80,-80){\makebox(0,0){\scriptsize$\pi_{\tinyR}$}}
      \put(80,-80){\circle{20}}
      \put(75,-88.66){\line(0,-1){11.34}}
      \put(85,-88.66){\line(0,-1){111.34}}
      % ...
      \put(-80,-40){\makebox(0,0){$\dotsm$}}
      \put(-200,0){\makebox(0,0){$\lor$}}
      \put(-200,0){\circle{20}}
      \put(-207.07,-7.07){\line(-1,-1){25.86}}
      \put(-192.93,-7.07){\line(1,-1){25.86}}
      \put(-240,-40){\makebox(0,0){$\land$}}
      \put(-240,-40){\circle{20}}
      \put(-245,-48.66){\line(0,-1){71.34}}
      \put(-232.93,-47.07){\line(1,-1){25.86}}
      \put(-160,-40){\makebox(0,0){$\land$}}
      \put(-160,-40){\circle{20}}
      \put(-165,-48.66){\line(0,-1){81.34}}
      \put(-152.93,-47.07){\line(1,-1){25.86}}
      \put(-200,-80){\makebox(0,0){\scriptsize$\pi_{\tinyR}$}}
      \put(-200,-80){\circle{20}}
      \put(-205,-88.66){\line(0,-1){11.34}}
      \put(-195,-88.66){\line(0,-1){61.34}}
      \put(-120,-80){\makebox(0,0){\scriptsize$\pi_{\tinyR}$}}
      \put(-120,-80){\circle{20}}
      \put(-125,-88.66){\line(0,-1){11.34}}
      \put(-115,-88.66){\line(0,-1){101.34}}
      \put(-360,0){\makebox(0,0){$\lor$}}
      \put(-360,0){\circle{20}}
      \put(-367.07,-7.07){\line(-1,-1){25.86}}
      \put(-352.93,-7.07){\line(1,-1){25.86}}
      \put(-400,-40){\makebox(0,0){$\land$}}
      \put(-400,-40){\circle{20}}
      \put(-405,-48.66){\line(0,-1){71.34}}
      \put(-392.93,-47.07){\line(1,-1){25.86}}
      \put(-320,-40){\makebox(0,0){$\land$}}
      \put(-320,-40){\circle{20}}
      \put(-325,-48.66){\line(0,-1){81.34}}
      \put(-312.93,-47.07){\line(1,-1){25.86}}
      \put(-360,-80){\makebox(0,0){\scriptsize$\pi_{\tinyL}$}}
      \put(-360,-80){\circle{20}}
      \put(-365,-88.66){\line(0,-1){11.34}}
      \put(-355,-88.66){\line(0,-1){61.34}}
      \put(-280,-80){\makebox(0,0){\scriptsize$\pi_{\tinyR}$}}
      \put(-280,-80){\circle{20}}
      \put(-285,-88.66){\line(0,-1){11.34}}
      \put(-275,-88.66){\line(0,-1){91.34}}
      % ...
      \put(-440,-40){\makebox(0,0){$\dotsm$}}
      \put(-560,0){\makebox(0,0){$\lor$}}
      \put(-560,0){\circle{20}}
      \put(-567.07,-7.07){\line(-1,-1){25.86}}
      \put(-552.93,-7.07){\line(1,-1){25.86}}
      \put(-600,-40){\makebox(0,0){$\land$}}
      \put(-600,-40){\circle{20}}
      \put(-605,-48.66){\line(0,-1){71.34}}
      \put(-592.93,-47.07){\line(1,-1){25.86}}
      \put(-520,-40){\makebox(0,0){$\land$}}
      \put(-520,-40){\circle{20}}
      \put(-525,-48.66){\line(0,-1){81.34}}
      \put(-512.93,-47.07){\line(1,-1){25.86}}
      \put(-560,-80){\makebox(0,0){\scriptsize$\pi_{\tinyL}$}}
      \put(-560,-80){\circle{20}}
      \put(-565,-88.66){\line(0,-1){11.34}}
      \put(-555,-88.66){\line(0,-1){61.34}}
      \put(-480,-80){\makebox(0,0){\scriptsize$\pi_{\tinyR}$}}
      \put(-480,-80){\circle{20}}
      \put(-485,-88.66){\line(0,-1){11.34}}
      \put(-475,-88.66){\line(0,-1){91.34}}
      \put(-680,-120){\makebox(0,0){$0$}}
      \put(-680,-120){\circle{20}}
      \put(-680,-110){\line(0,1){10}}
      \put(-680,-100){\line(1,0){220}}
      \put(-440,-100){\makebox(0,0){$\dotsm$}}
      \put(-420,-100){\line(1,0){320}}
      \put(-80,-100){\makebox(0,0){$\dotsm$}}
      \put(-60,-100){\line(1,0){135}}
      \put(-680,-215){\circle{10}}
      \put(-660,-215){\makebox(0,0){$\dotsm$}}
      \put(-640,-215){\circle{10}}
      \put(-660,-220){\makebox(0,0)[t]
        {$\underbrace{\begin{picture}(60,0)\end{picture}}$}}
      \put(-660,-230){\makebox(0,0)[t]{$y_1$}}
      \put(-640,-210){\line(0,1){60}}
      \put(-640,-160){\line(1,0){20}}
      \put(-620,-160){\line(0,1){30}}
      \put(-620,-130){\line(1,0){160}}
      \put(-440,-130){\makebox(0,0){$\dotsm$}}
      \put(-420,-130){\line(1,0){320}}
      \put(-80,-130){\makebox(0,0){$\dotsm$}}
      \put(-60,-130){\line(1,0){95}}
      \put(-640,-140){\makebox(0,0){$\lnot$}}
      \put(-640,-140){\circle{20}}
      \put(-640,-130){\line(0,1){10}}
      \put(-640,-120){\line(1,0){180}}
      \put(-440,-120){\makebox(0,0){$\dotsm$}}
      \put(-420,-120){\line(1,0){320}}
      \put(-80,-120){\makebox(0,0){$\dotsm$}}
      \put(-60,-120){\line(1,0){15}}
      \put(-610,-215){\circle{10}}
      \put(-590,-215){\makebox(0,0){$\dotsm$}}
      \put(-570,-215){\circle{10}}
      \put(-590,-220){\makebox(0,0)[t]
        {$\underbrace{\begin{picture}(60,0)\end{picture}}$}}
      \put(-590,-230){\makebox(0,0)[t]{$y_2$}}
      \put(-540,-215){\circle{10}}
      \put(-520,-215){\makebox(0,0){$\dotsm$}}
      \put(-500,-215){\circle{10}}
      \put(-520,-220){\makebox(0,0)[t]
        {$\underbrace{\begin{picture}(60,0)\end{picture}}$}}
      \put(-520,-230){\makebox(0,0)[t]{$y_3$}}
      \put(-540,-210){\line(0,1){60}}
      \put(-540,-150){\line(-1,0){15}}
      \put(-540,-150){\line(1,0){80}}
      \put(-440,-150){\makebox(0,0){$\dotsm$}}
      \put(-420,-150){\line(1,0){225}}
      \put(-500,-210){\line(0,1){50}}
      \put(-500,-160){\line(1,0){505}}
      \put(-470,-215){\circle{10}}
      \put(-450,-215){\makebox(0,0){$\dotsm$}}
      \put(-430,-215){\circle{10}}
      \put(-410,-215){\circle{10}}
      \put(-390,-215){\makebox(0,0){$\dotsm$}}
      \put(-370,-215){\circle{10}}
      \put(-420,-220){\makebox(0,0)[t]
        {$\underbrace{\begin{picture}(120,0)\end{picture}}$}}
      \put(-420,-230){\makebox(0,0)[t]{$y_4$}}
      \put(-470,-210){\line(0,1){30}}
      \put(-470,-180){\line(-1,0){5}}
      \put(-430,-210){\line(0,1){30}}
      \put(-430,-180){\line(1,0){155}}
      \put(-410,-210){\line(0,1){20}}
      \put(-410,-190){\line(1,0){295}}
      \put(-370,-210){\line(0,1){10}}
      \put(-370,-200){\line(1,0){455}}
   \end{picture}
\caption{Uniform circuits for the conditional function ``$\C$''.}
\label{fig:C}
\end{figure}

   Now, there are only a constant number of possibilities for $\sigma$
   that machine $P$ needs to check.  We list each one and the
   corresponding output for $P$, as well as a brief explanation
   indicating which bit of which input parameter is designated by the
   given path $\sigma$, in Table~\ref{table:C}.

\begin{table}[!ht]
\centering
   \begin{tabular}{|l|l|l|}
   \hline
      $\sigma$ & output            & explanation \\
   \hline
      $S$      & $x$               & same bit of $y_2$ \\
      $RL$     & $\mapl{one}(x)$   & first bit of $y_1$ \\
      $LLS$    & $\mapl{one}(x)$   & first bit of $y_1$ (negated) \\
      $LRR$    & $\mapl{min}_3(x)$ & bit of $y_3$ (padded) \\
      $RRR$    & $\mapl{min}_4(x)$ & bit of $y_4$ (padded) \\
   \hline
   \end{tabular}
\caption{Behaviour of machine $P$ for the conditional function.}
\label{table:C}
\end{table}

   Next, on input $\out,\sigma,|y_1|,|y_2|,|y_3|,|y_4|,\tau$, $Q$
   accepts iff
   \begin{itemize}
   \item
      \(|y_1|=0\), \(1\leq\out\leq|y_2|\), and
      \begin{itemize}
      \item
         \(\sigma=\E\), \(\tau=\lsg\), or
      \item
         \(\sigma=S\), \(\tau=\pair{2}{\out}\); or
      \end{itemize}
   \item
      \(|y_1|>0\), \(1\leq\out\leq\max\{|y_3|,|y_4|\}\), and
      \begin{itemize}
      \item
         \(\sigma=\E\), \(\tau=\lor\), or
      \item
         \(\sigma=L\), \(\tau=\land\), or
      \item
         \(\sigma=R\), \(\tau=\land\), or
      \item
         \(\sigma=LL\), \(\tau=\lnot\), or
      \item
         \(\sigma=LR\), \(\tau=\pi_{\tinyR}\) if \(\out\leq|y_3|\),
         \(\tau=\pi_{\tinyL}\) if \(|y_3|<\out\), or
      \item
         \(\sigma=RL\), \(\tau=\pair{1}{1}\), or
      \item
         \(\sigma=RR\), \(\tau=\pi_{\tinyR}\) if \(\out\leq|y_4|\),
         \(\tau=\pi_{\tinyL}\) if \(|y_4|<\out\), or
      \item
         \(\sigma=LLS\), \(\tau=\pair{1}{1}\), or
      \item
         \(\sigma=LRL\), \(\tau=0\), or
      \item
         \(\sigma=LRR\), \(\tau=\pair{3}{r}\) for
         \(r=\mapl{min}_3(\out)\), or
      \item
         \(\sigma=RRL\), \(\tau=0\), or
      \item
         \(\sigma=RRR\), \(\tau=\pair{4}{r}\) for
         \(r=\mapl{min}_4(\out)\).
      \end{itemize}
   \end{itemize}
\item[{$\I^n_k(y_1,\dots,y_n)$}:]
   Circuits will use unary identity gates $\lsg$ directly connected to
   the proper input bits.  Hence, on input \(\sigma=S\), $P$ outputs
   simply $x$; on input $\out,\sigma,|y_1|,\dots,|y_n|,\tau$, $Q$
   accepts iff \(1\leq\out\leq|y_k|\), \(\sigma=\E\) and
   \(\tau=\lsg\), or \(\sigma=S\) and \(\tau=\pair{k}{\out}\).
   \quad\QED
\end{description}
\end{proof}

 {}Next, we want to show that functions defined by $\CRN$ also have
 uniform circuit families.  For technical reasons (\ie, to simplify
 the proof), we actually show the result for \emph{left} $\CRN$.
 (Since $L_0$ and $L_1$ remain the same whether $\CRN$ or left
 $\CRN$ is used to define them, this is sufficient.)
\begin{lemma}
\label{lem:crn}
   If $f(y_1,y_2,\dots,y_n)$ is defined from $h$ by left $\CRN$ on
   $y_1$, where $h$ has uniform circuits of depth
   $d_h(|y_1|,|y_2|,\dots,|y_n|)$, then $f$ has uniform circuits of
   depth
   \[ \max\big\{d_h(1,|y_2|,\dots,|y_n|),d_h(2,|y_2|,\dots,|y_n|),
      \dots,d_h(|y_1|,|y_2|,\dots,|y_n|)\big\}. \]
\end{lemma}
\begin{proof}
   A natural circuit for $f$ consists of a series of $h$-circuits in
   parallel, one for each output bit of $f$, where the $i$-th
   $h$-circuit is connected to the first $i$ bits of $y_1$ (as well as
   to every other input parameter).  Clearly, the depth of this
   circuit is as stated above.  Moreover, given a path $\sigma$,
   machine $P$ simply simulates $P_h$ to get a term $t^h_\sigma$ and
   outputs \(t_\sigma(x)=t^h_\sigma(\mapl{one}(x))\), while machine
   $Q$ accepts $\out,\sigma,|y_1|,|y_2|,\dots,|y_n|,\tau$ iff
   \(|y_1|=0\), \(\out=0\), \(\sigma=\E\), \(\tau=\E\), or $Q_h$
   accepts $\out'=1,\sigma,|y_1'|=\out,|y_2|,\dots,|y_n|,\tau$.
\end{proof}

 {}Following this, we need to show that the composition of functions
 computed by uniform families of circuits is also computable by
 uniform families of circuits, of the right depth.

\begin{lemma}
\label{lem:comp}
   If $f(y_1,\dots,y_n)$ is defined from $g$ and $h_1,\dots,h_k$ by
   $\COMP$, where $g$ has uniform circuits of depth
   $d_g(|z_1|,\dots,|z_k|)$ and $h_i$ has uniform circuits of depth
   $d_{h_i}(|y_1|,\dots,|y_n|)$ (for \(1\leq i\leq k\)), then $f$ has
   uniform circuits of depth
   \[
      d_g(|h_1(\tpl y_n)|,\dots,|h_k(\tpl y_n)|)
       + \max\{d_{h_i}(|\tpl y_n|)\} + \lceil\lg(k)\rceil.
   \]
\end{lemma}
\begin{proof}
   A natural circuit for $f$ consists of a circuit for $g$ whose input
   gates are connected to the output gates of the corresponding
   circuits for $h_i$ directly.  Unfortunately, machine $P$ cannot
   tell what term to output just from a path in such a circuit,
   because $P$ does not have enough time to determine which $h_i$ the
   path leads into (this would require determining the number of the
   input gate of $g$ through which the path passes).  Therefore, we
   construct a slightly more complicated circuit by adding a layer of
   selection gates of depth $\lceil\lg(k)\rceil$ between the circuit
   for $g$ and the $h_i$ circuits (somewhat like what was done for the
   concatenation function), in such a way that machine $P$ can easily
   determine from a path $\sigma$ which $h_i$ the path goes through.
   Clearly, the depth of such a circuit is as stated above.  Moreover,
   given a path $\sigma$, machine $P$ can break it up into $\sigma_g$
   through $g$ (getting a term $t_{\sigma_g}$), followed by $\sigma'$
   through the selection subcircuit (which gives the index $i$ of the
   function $h_i$ feeding into the $g$ circuit), followed by a final
   part $\sigma_i$ through $h_i$ (getting a term $t_{\sigma_i}$).  $P$
   then outputs $t_{\sigma_i}(t_{\sigma_g}(x))$, in time linear in
   $|\sigma|$.  Also, on input $\out,\sigma,|y_1|,\dots,|y_n|,\tau$,
   machine $Q$ accepts iff $\pair{\out}{\sigma}$ is a gate of type
   $\tau$ in $g$ (by simulating $Q_g$), or $\pair{\out}{\sigma}$ is a
   gate within $\lceil\lg(k)\rceil$ steps of an input of $g$ and
   $\tau$ is the correct type of projection gate (computing
   $t_{\sigma_g}(\out)$ to figure out which input bit of $g$ the path
   $\sigma$ goes through, and then checking the constantly many
   possibilities for the part of $\sigma$ through the selection
   subcircuit), or $\pair{\out}{\sigma}$ is a gate of type $\tau$ or
   an input gate \(\tau=\pair{j}{r}\) for
   \(r=t_{\sigma_i}(t_{\sigma_g}(\out))\) within an $h_i$ circuit
   (computing $t_{\sigma_g}(\out)$ and tracking the path through the
   selection subcircuit to figure out the index $i$, and then
   simulating $Q_{h_i}$ to verify $\tau$).  All this can be done in
   logarithmic time.
\end{proof}

 {}Finally, we show that functions defined by $\TRN$ can be computed
 by uniform families of circuits.

\begin{lemma}
\label{lem:trn}
   If $f(y_1,y_2,y_3,\dots,y_n)$ is defined from $g$, $h$, $h_\ell$,
   and $h_r$ by $\TRN$, where $g$ has uniform circuits of depth
   $d_g(|y_1|,|y_2|,|y_3|,\dots,|y_n|)$, $h$ has uniform circuits of
   depth $d_h$ (a constant), and $h_\ell$ and $h_r$ have uniform
   circuits of depth $d_\ell$ and $d_r$ (constants), then $f$ has
   uniform circuits of depth
   \[
      \bigOh\big(d_g(1,|y_2||y_1|^c,|y_3|,\dots,|y_n|)+
         \log|y_1|\cdot(d_h+d_\ell+d_r)\big)
   \]
   for some constant \(c\in\NN\).
\end{lemma}
\begin{proof}
   A natural circuit for $f$ consists of a binary tree of
   $h$-subcircuits connected to the appropriate bits of the first
   input and to each other, with a layer of $g$ circuits at the
   bottom, where the second input of the $h$ and $g$ circuits is
   connected to subtrees of $h_\ell$ and $h_r$ circuits.  As in the
   proof for composition, we use layers of selection gates to ``glue''
   together the different subcircuits (between successive $h$
   circuits, between $h$ and $g$ circuits, between $h$ or $g$ circuits
   and the circuits to compute left and right halves of the first
   input or left and right functions of the second input, as well as
   between successive ``half'' functions for the first and second
   inputs), so that $P$ can tell from the path $\sigma$ alone which
   subcircuit a path leads to.  The total depth of such a circuit is
   obviously as stated.  Moreover, given a path $\sigma$, $P$ can
   divide the path into portions through $h$ circuits, accumulating
   the terms for each one, and the final portion of the path through
   an $h$, $g$, $h_\ell$, $h_r$, or ``half'' circuit, outputting the
   composition of each term.  This can obviously be done in linear
   time in the length of $\sigma$.  On input
   $\out,\sigma,|y_1|,|y_2|,|y_3|,\dots,|y_n|,\tau$, machine $Q$ can
   break up the path $\sigma$ into a first part through some number of
   $h$ circuits and a final part $\sigma'$ entirely contained inside
   some $h$, $g$, $h_\ell$, $h_r$, or ``half'' subcircuit.  $Q$ can
   compute the term corresponding to the first part of the path to
   figure out which output bit of the subcircuit $\sigma$ passes
   through, and then simulate the $\ATM$ for that subcircuit to verify
   $\tau$, all in logarithmic time.
\end{proof}

 {}Now, we can put all of these results together.
\begin{theorem}
   For all \(f\in L_0\), $f$ can be computed by a uniform family of
   circuits in $\FNC^0$ (\ie, of constant depth).
\end{theorem}
\begin{proof}
   By induction on the definition of $f$: Lemma~\ref{lem:base} shows
   that functions in $\BASE$ have uniform constant depth circuits, and
   Lemmas~\ref{lem:crn} and~\ref{lem:comp} show that $\CRN$ and
   $\COMP$ preserve uniform constant depth.
\end{proof}

\begin{theorem}
   For all \(f\in L_1\), $f$ can be computed by a uniform family of
   circuits in $\FNC^1$ (\ie, of logarithmic depth).
\end{theorem}
\begin{proof}
   By induction on the definition of $f$.
   \begin{itemize}
   \item
      If \(f\in L_0\), then the preceding theorem shows the result.
   \item
      If $f$ is defined by $\CRN$ from \(h\in L_1\), then
      Lemma~\ref{lem:crn} shows that $f$ can be computed by uniform
      circuits of the same depth as that of the circuits for $h$,
      which shows the result.
   \item
      If $f$ is defined by $\COMP$ from \(g,h_1,\dots,h_k\in L_1\),
      then Lemma~\ref{lem:comp} and Lemma~\ref{lem:L1} (on the length
      of functions in $L_1$) show that $f$ can be computed by uniform
      circuits of depth
      \begin{multline*}
         d_g(|h_1(\tpl y)|,\dots,|h_k(\tpl y)|)+
         \max\{d_{h_i}(|\tpl y|)\}+\lceil\lg(k)\rceil \\*
          = \bigOh\Big(\log\big(\max\{p_{h_1}(|\tpl y|),\dots,
            p_{h_k}(|\tpl y|)\}\big)+\log(\max\{|\tpl y|\})\Big) \\*
          = \bigOh\big(\log(\max\{|\tpl y|\})\big). 
      \end{multline*}
   \item
      If $f$ is defined by $\TRN$ from \(g\in L_1\) and
      \(h,h_\ell,h_r\in L_0\), then Lemma~\ref{lem:trn}, together with
      Lemmas~\ref{lem:L0} and~\ref{lem:L1} on the length of functions
      in $L_0$ and $L_1$, shows that $f$ can be computed by uniform
      circuits of depth
      \begin{multline*}
         \bigOh\big(d_g(1,|y_2||y_1|^c,|y_3|,\dots,|y_n|)+
            \log|y_1|\cdot(d_h+d_\ell+d_r)\big) \\*
          = \bigOh\big(\log(\max\{|y_2|,\dots,|y_n|\})+\log|y_1|\big) \\*
          = \bigOh\big(\log(\max\{|y_1|,|y_2|,\dots,|y_n|\})\big). 
         \quad\QED
      \end{multline*}
   \end{itemize}
\end{proof}

%% file: chapter3.tex
\chapter{The Quantifier-Free Theory $T_1$}
\label{chap:T1}

 In this chapter, we will define the theory $T_1$ and give its formal
 development, including many proofs of simple properties of functions
 of $T_1$, as well as many derived rules, and concluding with an
 illustrative example by proving the pigeonhole principle.

\section{Definitions}

 The theory $T_1$ that we now describe is a \emph{quantifier-free}
 system, \ie, a free-variable theory with propositional connectives,
 modeled after Cook's $\PV$~\cite{Cook75} but based on the algebra
 $L_1$.  The language of $T_1$ consists of the function symbols
\[
   \{\E,0,1,{}_0,{}_1,\lhalf,\rhalf,\cat,\lchop,\rchop,\C\},
\]
 the function constructors $\{\fhead,\lCRN,\rCRN,\TRN\}$, the
 predicate symbol $\{=\}$, and the usual propositional connectives
 $\{\lnot,\land,\lor,\limp,\liff\}$.  More precisely, we have the
 following definitions (where we use the informal notation $x0$ for
 $(x\cat 0)$ and $0x$ for $(0\cat x)$---similarly for $x1$ and $1x$).

\begin{definition}
   The \emph{function symbols} and \emph{terms} of $T_1$ are defined
   as follows.  (The intended interpretation of each function symbol
   is as given in Chapter~\ref{chap:L1}, where $\lCRN$ represents
   ``left'' (or ``reverse'') $\CRN$ and $\rCRN$ represents ``right''
   (or ``plain'') $\CRN$, and we use the notation introduced there
   instead of the more formal prefix notation.  Also, each function
   symbol and each term has a \emph{rank} of either $0$ or $1$---that
   intuitively indicates which one of $L_0$ or $L_1$ the function
   symbol or term belongs to.)
   \begin{enumerate}

   \item
      Each variable $x_0,x_1,x_2,\dotsc$ is a term of rank $0$.

   \item
      If $f$ is an $n$-place function symbol and $t_1,\dots,t_n$ are
      terms, then $f(t_1,\dots,t_n)$ is a term whose rank is the
      maximum of the ranks of $f,t_1,\dots,t_n$ (\ie, the rank of
      $f(t_1,\dots,t_n)$ is $0$ iff the rank of each one of
      $f,t_1,\dots,t_n$ is $0$).

   \item
      \ind[tdef]{$\E$ (empty string)}{}{}
      $\E,0,1$ are $0$-place function symbols (constants) of rank $0$.

   \item
      \ind[tdef]{${}_0$ (all-zero)}{}{}
      \ind[tdef]{${}_1$ (all-one)}{}{}
      \ind[tdef]{$\lhalf$ (left half)}{}{}
      \ind[tdef]{$\rhalf$ (right half)}{}{}
      ${}_0,{}_1,\lhalf,\rhalf$ are $1$-place function symbols of rank
      $0$.

   \item
      \ind[tdef]{$\cat$ (concatenation)}{}{}
      \ind[tdef]{$\lchop$ (left chop)}{}{}
      \ind[tdef]{$\rchop$ (right chop)}{}{}
      $\cat,\lchop,\rchop$ are $2$-place function symbols of rank $0$.

   \item
      \ind[tdef]{$\C$ (conditional)}{}{}
      $\C$ is a $3$-place function symbol of rank $0$.

   \item
      If $t$ is a term and $x_1,\dots,x_n$ is a list of variables
      including all the variables in $t$, then
      $[\fhead x_1\dots x_n\fdot t]$ is an $n$-place function symbol
      of the same rank as that of $t$.

   \item
      \ind[tdef]{\CRN}{\lCRN,\rCRN}{}
      If $h$ is an $(n+1)$-place function symbol, then $\lCRN[h]$ and
      $\rCRN[h]$ are $(n+1)$-place function symbols whose rank is that
      of $h$.

   \item
      \ind[tdef]{\TRN}{}{}
      If $g$ is an $(n+2)$-place function symbol, $h$ is an
      $(n+4)$-place function symbol of rank $0$, and $h_\ell$ and
      $h_r$ are $1$-place function symbols of rank $0$, then
      $\TRN[g,h,h_\ell,h_r]$ is an $(n+2)$-place function symbol of
      rank $1$.

   \end{enumerate}
\end{definition}

\begin{definition}
   The \emph{axioms} of $T_1$ are as follows (except for the
   propositional and equality axioms, they simply define the function
   symbols).
   \begin{enumerate}\setcounter{enumi}{-1}

   \item
      Any standard, complete set of axioms for the propositional
      calculus (with equations of the form \(x=y\) in place of
      propositional atoms, for arbitrary variables $x$ and $y$).

   \item
      \begin{enumerate}

      \item
         \( x=x \)

      \item
         \( x=y \ \limp \ y=x \)

      \item
         \( (x=y\land y=z) \ \limp \ x=z \)

      \item
      \label{axiom:fcneq}
         \( (x_1=y_1\land\dots\land x_k=y_k) \ \limp
             \ f(x_1,\dots,x_k)=f(y_1,\dots,y_k) \) \\
         (for all $k$-ary function symbols $f$, for all $k\geq 1$)

      \end{enumerate}

   \item 
   \label{axiom:E01}
      \ind{$\E$ (empty string)}{}{}
      \( \E \neq 0 \ \land \ 0 \neq 1 \ \land \ 1 \neq \E \)

   \item 
      \begin{enumerate}
      \ind[(]{$\cat$ (concatenation)}{}{}

      \item
      \label{axiom:cat}
         \( x\cat\E = x \ \land
          \ x\cat y0 = (x\cat y)\cat 0 \ \land
          \ x\cat y1 = (x\cat y)\cat 1 \)

      \item
      \label{axiom:cat=E}
         \( x\cat y=\E \ \limp \ (x=\E\land y=\E) \)

      \item
         \( x\cat y=0 \ \limp \ (x=\E\land y=0)\lor(x=0\land y=\E) \) \\
         \( x\cat y=1 \ \limp \ (x=\E\land y=1)\lor(x=1\land y=\E) \)

      \ind[)]{$\cat$ (concatenation)}{}{}
      \end{enumerate}

   \item 
      \begin{enumerate}
      \ind[(]{$\lchop$ (left chop)}{}{}

      \item
      \label{axiom:lchop}
         \( \E\lchop x = x \ \land
          \ 0y\lchop x = 0\lchop(y\lchop x) \ \land
          \ 1y\lchop x = 1\lchop(y\lchop x) \)

      \item
      \label{axiom:ilchop}
         \( 0\lchop\E = \E \ \land
          \ 0\lchop 0x = x \ \land
          \ 0\lchop 1x = x \) \\
         \( 1\lchop\E = \E \ \land
          \ 1\lchop 0x = x \ \land
          \ 1\lchop 1x = x \)

      \item
      \label{axiom:lchopE}
         \( y\lchop x=\E \ \liff
          \ x\lchop y0\neq\E \ \liff
          \ x\lchop y1\neq\E \)

      \ind[)]{$\lchop$ (left chop)}{}{}
      \end{enumerate}

   \item 
      \begin{enumerate}
      \ind[(]{$\rchop$ (right chop)}{}{}

      \item
      \label{axiom:rchop}
         \( x = x\rchop\E \ \land
          \ (x\rchop y)\rchop 0 = x\rchop y0 \ \land
          \ (x\rchop y)\rchop 1 = x\lchop y1 \)

      \item
      \label{axiom:rchopi}
         \( \E = \E\rchop 0 \ \land
          \ x = x0\rchop 0 \ \land
          \ x = x1\rchop 0 \) \\
         \( \E = \E\rchop 1 \ \land
          \ x = x0\rchop 1 \ \land
          \ x = x1\rchop 1 \)

      \item
      \label{axiom:rchopE}
         \( \E\neq 1y\rchop x \ \liff
          \ \E\neq 0y\rchop x \ \liff
          \ \E=x\rchop y \)

      \ind[)]{$\rchop$ (right chop)}{}{}
      \end{enumerate}

   \item 
      \begin{enumerate}

      \item
      \label{axiom:_0}
         \ind{${}_0$ (all-zero)}{}{}
         \( {}_0\E = \E \ \land
          \ {}_0(x0) = {}_0x\cat 0 \ \land
          \ {}_0(x1) = {}_0x\cat 0 \)

      \item
      \label{axiom:_1}
         \ind{${}_1$ (all-one)}{}{}
         \( {}_1\E = \E \ \land
          \ {}_1(x0) = {}_1x\cat 1 \ \land
          \ {}_1(x1) = {}_1x\cat 1 \)

      \end{enumerate}

   \item 
   \label{axiom:C}
      \ind{$\C$ (conditional)}{}{}
      \( \E\C(x,y,z) = x \ \land
       \ w0\C(x,y,z) = {}_0(z\rchop y)\cat y \ \land
       \ w1\C(x,y,z) = {}_0(y\rchop z)\cat z \)

   \item 
      \begin{enumerate}
      \ind[(]{$\lhalf$ (left half)}{}{}
      \ind[(]{$\rhalf$ (right half)}{}{}

      \item
         \( (x\lhalf)\cat(\rhalf x) = x \)

      \item
      \label{axiom:halfl}
         \( (x\lhalf\rchop\rhalf x) = \E \ \land
          \ 1\lchop(x\lhalf\lchop\rhalf x) = \E \)

      \ind[)]{$\rhalf$ (right half)}{}{}
      \ind[)]{$\lhalf$ (left half)}{}{}
      \end{enumerate}

   \item 
   \label{axiom:halves}
      We use ``$x\elC(y,z)$'' as a shorthand notation for
      $(x\lhalf\lchop\rhalf x)\C(y,z,z)$ (which equals $y$ if the
      length of $x$ is even, $z$ if the length of $x$ is odd).
      \begin{enumerate}

      \item
         \( (x0)\lhalf = x\elC\big(x\lhalf,x\lhalf\cat
               ((\rhalf x\cat 0)\rchop\rhalf x)\big) \ \land
          \ (x1)\lhalf = x\elC\big(x\lhalf,x\lhalf\cat
               ((\rhalf x\cat 1)\rchop\rhalf x)\big) \)

      \item
         \( \rhalf(x0) = x\elC\big(\rhalf x\cat 0,
            1\lchop(\rhalf x\cat 0)\big) \ \land
          \ \rhalf(x1) = x\elC\big(\rhalf x\cat 1,
            1\lchop(\rhalf x\cat 1)\big) \)

      \item
         \( (0x)\lhalf = x\elC\big((0\cat x\lhalf)\rchop 1,
            0\cat x\lhalf\big) \ \land
          \ (1x)\lhalf = x\elC\big((1\cat x\lhalf)\rchop 1,
            1\cat x\lhalf\big) \)

      \item
         \( \rhalf(0x) = x\elC\big((x\lhalf\lchop(0\cat x\lhalf))
               \cat\rhalf x,\rhalf x\big) \ \land
          \ \rhalf(1x) = x\elC\big((x\lhalf\lchop(1\cat x\lhalf))
               \cat\rhalf x,\rhalf x\big) \)

      \end{enumerate}

   \item 
      \( [\fhead x_1\dots x_n\fdot t](x_1,\dots,x_n) = t \)

   \item 
      \begin{enumerate}
      \ind[(]{\CRN}{\lCRN,\rCRN}{}

      \item
         \( \lCRN[h](\E,\tpl y) = \E \) \\
         \( \land \quad \lCRN[h](0x,\tpl y) =
            \big((h(0x,\tpl y)\cat 0)\rchop h(0x,\tpl y)\big)
            \cat\lCRN[h](x,\tpl y) \) \\
         \( \land \quad \lCRN[h](1x,\tpl y) =
            \big((h(1x,\tpl y)\cat 0)\rchop h(1x,\tpl y)\big)
            \cat\lCRN[h](x,\tpl y) \)

      \item
         \( \rCRN[h](\E,\tpl y) = \E \) \\
         \( \land \quad \rCRN[h](x0,\tpl y) = \rCRN[h](x,\tpl y)\cat
            \big(h(x0,\tpl y)\lchop(0\cat h(x0,\tpl y))\big) \) \\
         \( \land \quad \rCRN[h](x1,\tpl y) = \rCRN[h](x,\tpl y)\cat
            \big(h(x1,\tpl y)\lchop(0\cat h(x1,\tpl y))\big) \)

      \ind[)]{\CRN}{\lCRN,\rCRN}{}
      \end{enumerate}

   \item 
      \ind{\TRN}{}{}
      \( \TRN[g,h,h_\ell,h_r](x,z,\tpl y) =
         x\rchop 1\C(g(x,z,\tpl y),t,t) \) \\
      where \( t = h\bigl(x,z,\tpl y,
         \TRN[g,h,h_\ell,h_r](x\lhalf,h_\ell(z),\tpl y),
         \TRN[g,h,h_\ell,h_r](\rhalf x,h_r(z),\tpl y)\bigr) \)

   \end{enumerate}
\end{definition}

\begin{remark}
   By Claim~\ref{claim:L1}, every function in $\FALT$ is represented
   by some function symbol in $T_1$, and every function symbol in
   $T_1$ represents a function in $\FALT$.
\end{remark}

\begin{definition}
\label{def:T1rules}
   The \emph{rules of inference} of $T_1$ are as follows.
   \begin{enumerate}\setcounter{enumi}{-1}

   \item
      Any standard, complete set of rules for the propositional
      calculus.

   \item
   \label{rule:sub}
      \emph{Substitution} for an arbitrary formula $A$, variable $x$,
      and term $t$:
      \[
         A \lded A[t/x]
      \]

   \item
      \emph{Induction on Notation} ($\NIND$) for an arbitrary formula
      $A$ and variable $x$:
      \begin{enumerate}

      \item   (``left'' version) \ 
         \( A[\E],A[x]\limp A[0x],A[x]\limp A[1x] \lded A \)

      \item   (``right'' version) \ 
         \( A[\E],A[x]\limp A[x0],A[x]\limp A[x1] \lded A \)

      \end{enumerate}

   \item
      \emph{Tree Induction} ($\TIND$) for an arbitrary formula $A$,
      variables $x,z$, and unary function symbols $h_\ell,h_r$ of rank
      $0$:
      \[
         A[\E,z],A[0,z],A[1,z],\big(A[x\lhalf,h_\ell(z)]\land
         A[\rhalf x,h_r(z)]\big)\limp A[x,z] \lded A
      \]

   \end{enumerate}
\end{definition}

\section{Developing the theory}

 In this section, we give a formal development of $T_1$, starting
 with a few simple theorems and working our way towards multi-variable
 versions of $\CRN$ and $\NIND$.  These will be used to define binary
 addition and ``counting'' functions, and to prove their properties.

\subsection{Basic definitions and theorems}

\begin{claim}
   \( \E\cat x = x \)
\end{claim}
\begin{proof}
   By $\NIND$ on $x$:
   \( \E\cat\E = \E \) (by Axiom~\ref{axiom:cat} and
   Rule~\ref{rule:sub}),
   \begin{alignat*}{2}
      \E\cat x0
      &= (\E\cat x)\cat 0 &\qquad
         &\ \text{(Axiom~\ref{axiom:cat}, Rule~\ref{rule:sub})} \\
      &= x\cat 0 &\qquad
         &\ \text{(Induction Hypothesis, Axiom~\ref{axiom:fcneq},
            Rule~\ref{rule:sub})}, \\
      \E\cat x1
      &= (\E\cat x)\cat 1 &\qquad
         &\ \text{(Axiom~\ref{axiom:cat}, Rule~\ref{rule:sub})} \\
      &= x\cat 1 &\qquad
         &\ \text{(Induction Hypothesis, Axiom~\ref{axiom:fcneq},
            Rule~\ref{rule:sub})}. 
      \quad\QED
   \end{alignat*}

\end{proof}

\begin{claim}
\label{claim:catassoc}
   \( x\cat(y\cat z) = (x\cat y)\cat z \)
\end{claim}
\begin{proof}
   By $\NIND$ on $z$:
   \( x\cat(y\cat\E) = x\cat y = (x\cat y)\cat\E \)
   (by Axiom~\ref{axiom:cat} and Rule~\ref{rule:sub}),
   \begin{alignat*}{2}
      x\cat(y\cat z0)
      &= x\cat((y\cat z)\cat 0) &\qquad
         &\ \text{(Axioms~\ref{axiom:cat} and~\ref{axiom:fcneq},
            Rule~\ref{rule:sub})} \\
      &= (x\cat(y\cat z))\cat 0 &\qquad
         &\ \text{(Axiom~\ref{axiom:cat} and Rule~\ref{rule:sub})} \\
      &= ((x\cat y)\cat z)\cat 0 &\qquad
         &\ \text{(Induction Hypothesis)} \\
      &= (x\cat y)\cat z0 &\qquad
         &\ \text{(Axiom~\ref{axiom:cat} and Rule~\ref{rule:sub})}, 
   \end{alignat*}
   and a similar proof shows \( x\cat(y\cat z1) = (x\cat y)\cat z1 \).
\end{proof}

 {}Note that in the proofs that follow, we will not mention explicitly
 the application of particular axioms, of the induction hypothesis, or
 of the substitution rule when they are self-evident.  Also, when
 proving a statement by $\NIND$, the cases for $x0$ and $x1$ will
 often be almost identical (as above) so we will prove both cases at
 once using ``$i$'' to stand for either $0$ or $1$.

\begin{remark}
   Be advised that the rest of this section contains a large number of
   technical claims, together with their proofs, which are included
   here for the sake of completeness.  Most of these claims are of
   limited interest in themselves, apart from illustrating the style
   of proofs in $T_1$ and giving basic properties of functions which
   will be used in later proofs.  For this reason, we recommend that
   on a first reading, the reader focus mainly on the
   \textsc{Definition}s, \textsc{Theorem}s, and \textsc{Derived
   Rule}s, which contain the essential results.
\end{remark}

\subsubsection{On ``$\lchop$'' and ``$\rchop$''}

 We start by defining two functions that will serve as a convenient
 shorthand notation throughout the rest of this chapter, and prove
 basic properties of these functions.

\begin{definition}
   \ind[tdef]{$\ldel$ (left delete)}{}{}
   \ind[tdef]{$\rdel$ (right delete)}{}{}
   \leftright
  {\( \ldel = [\fhead x\fdot 1\lchop x] \)}
  {\( \rdel = [\fhead x\fdot x\rchop 1] \)}
\end{definition}
 (To make the notation more consistent with previous usage, we will
 write ``$\ldel x$'' and ``$x\rdel$'' instead of the more formal
 ``$\ldel(x)$'' and ``$\rdel(x)$'', respectively.)

\begin{claim}
   \leftright
  {\( 0\lchop x = 1\lchop x \)}
  {\( x\rchop 1 = x\rchop 0 \)}
\end{claim}
\begin{proof}
   (L): Immediate from Axiom~\ref{axiom:ilchop}.
   (R): Immediate from Axiom~\ref{axiom:rchopi}.
\end{proof}

\begin{claim}
\label{claim:del}
   \ind{$\ldel$ (left delete)}{}{}
   \ind{$\rdel$ (right delete)}{}{}
   \leftright
  {\( \ldel\E=\E \land \ldel(0x)=x \land \ldel(1x)=x \)}
  {\( \E\rdel=\E \land (x0)\rdel=x \land (x1)\rdel=x \)}
\end{claim}
\begin{proof}
   (L): This is just a restatement of Axiom~\ref{axiom:ilchop}.
   (R): This is just a restatement of Axiom~\ref{axiom:rchopi}.
\end{proof}

\begin{claim}
   \leftright{\( y\lchop\E = \E \)}{\( \E = \E\rchop y \)}
\end{claim}
\begin{proof}
   (L) By $\NIND$ on $y$:
   \( \E\lchop\E = \E \),
   \( iy\lchop\E = i\lchop(y\lchop\E) = i\lchop\E = \E \).
   (R) By $\NIND$ on $y$:
   \( \E\lchop\E = \E \),
   \( \E\rchop yi = (\E\rchop y)\rchop i = \E\rchop i = \E \).
\end{proof}
 Many of the theorems that follow will be similar to the ones above in
 having a ``left'' and ``right'' version, both of which can be proved
 in the same way (by using the appropriate version of $\NIND$ when
 necessary).  Hence, to avoid unnecessary repetition, we will only
 give the proof of one version from now on.

\begin{claim}
   \leftright
  {\( \ldel(y\lchop x) = y\lchop\ldel x \)}
  {\( (x\rchop y)\rdel = x\rdel\rchop y \)}
\end{claim}
\begin{proof}
   (L) By $\NIND$ on $y$:
   \( \ldel(\E\lchop x) = \ldel x = \E\lchop\ldel x \),
   \( \ldel(iy\lchop x) = \ldel\ldel(y\lchop x)
       = \ldel(y\lchop\ldel x) = iy\lchop\ldel x \).
\end{proof}

\begin{claim}
   \leftright
  {\( zy\lchop zx = y\lchop x \)}
  {\( xz\rchop yz = x\rchop y \)}
\end{claim}
\begin{proof}
   (L) By $\NIND$ on $z$:
   \( \E y\lchop\E x = y\lchop x \),
   \( (iz)y\lchop(iz)x = \ldel(zy\lchop(iz)x)
       = zy\lchop\ldel(iz)x = zy\lchop zx = y\lchop x \).
\end{proof}

\begin{corollary}
\label{cor:chopnil}
   \leftright{\( x\lchop xy = y \)}{\( y = yx\rchop x \)}
\end{corollary}

\begin{corollary}
   \leftright{\( x\lchop x = \E \)}{\( \E = x\rchop x \)}
\end{corollary}

 {}Now, we are ready to define two more functions that will also be
 used as a convenient shorthand notation for the rest of the chapter.

\begin{definition}
   \ind[tdef]{$\lbit$ (left bit)}{}{}
   \ind[tdef]{$\rbit$ (right bit)}{}{}
   \leftright
  {\( \lbit = [\fhead x\fdot x\rchop\ldel x] \)}
  {\( \rbit = [\fhead x\fdot x\rdel\lchop x] \)}
\end{definition}
 (To make the notation more consistent with previous usage, we will
 write ``$\lbit x$'' and ``$x\rbit$'' instead of the more formal
 ``$\lbit(x)$'' and ``$\rbit(x)$'', respectively.)

\begin{claim}
\label{claim:bit}
   \ind{$\lbit$ (left bit)}{}{}
   \ind{$\rbit$ (right bit)}{}{}
   \leftright
  {\( \lbit\E=\E \land \lbit(0x)=0 \land \lbit(1x)=1 \)}
  {\( \E\rbit=\E \land (x0)\rbit=0 \land (x1)\rbit=1 \)}
\end{claim}
\begin{proof}
   (L) From the definition, by Claim~\ref{claim:del} and by
   Corollary~\ref{cor:chopnil}:
   \( \lbit\E = \E\rchop\ldel\E = \E\rchop\E = \E \),
   \( \lbit(ix) = ix\rchop\ldel(ix) = ix\rchop x = i \).
\end{proof}

\begin{claim}
   \leftright
  {\( \lbit x\cat\ldel x = x \)}
  {\( x = x\rdel\cat x\rbit \)}
\end{claim}
\begin{proof}
   (L) By $\NIND$ on $x$:
   \( \lbit\E\cat\ldel\E = \E\cat\E = \E \),
   \( \lbit(ix)\cat\ldel(ix) = i\cat x = ix \).
\end{proof}

\subsubsection{On ``$\cat$''}

\begin{claim}
\label{claim:Eneqcati}
   \leftright{\( \E \neq ix \)}{\( xi \neq \E \)}
\end{claim}
\begin{proof}
   (L) By Axioms~\ref{axiom:E01} and~\ref{axiom:cat=E}, and by taking
   the contrapositive:
   \( xi=\E \limp x=\E\land i=\E \limp i=\E \).
\end{proof}

\begin{claim}
\label{claim:cat0neqcat1}
   \leftright{\( 0x \neq 1y \)}{\( x0 \neq y1 \)}
\end{claim}
\begin{proof}
   (L) By Axiom~\ref{axiom:E01}, Claim~\ref{claim:bit},
   Axiom~\ref{axiom:fcneq}, and by taking the contrapositive:
   \( 0x = 1y \limp \lbit(0x) = \lbit(1y) \limp 0 = 1 \).
\end{proof}

\begin{claim}
   \leftright
  {\( \ldel x=\E \liff x=\E\lor x=0\lor x=1 \)}
  {\( x=\E\lor x=0\lor x=1 \liff \E=x\rdel \)}
\end{claim}
\begin{proof}
   (L) By $\NIND$ on $x$:
   \( \ldel\E=\E \liff \E=\E \),
   \( \ldel(ix)=\E \liff x=\E \liff ix=i\E \liff ix=i \).
\end{proof}

\begin{theorem}
\label{thm:cases}
   \leftright
  {\( x=\E \lor x=0\cat\ldel x \lor x=1\cat\ldel x \)}
  {\( x=\E \lor x=x\rdel\cat 0 \lor x=x\rdel\cat 1 \)}
\end{theorem}
\begin{proof}
   (L) By $\NIND$ on $x$: \( \E = \E \), \( ix = i\cat\ldel(ix) \).
\end{proof}
 Note that by Claims~\ref{claim:Eneqcati} and~\ref{claim:cat0neqcat1},
 we can show that exactly one of the disjuncts holds (\ie, that
 \( x=\E \limp x\neq 0\cat\ldel x \land x\neq 1\cat\ldel x \) and
 \( x=0\cat\ldel x \limp x\neq\E \land x\neq 1\cat\ldel x \) and
 \( x=1\cat\ldel x \limp x\neq 0\cat\ldel x \land x\neq\E \)).

\begin{corollary}
   \leftright
  {\( x\neq\E \liff x=0\cat\ldel x\lor x=1\cat\ldel x \)}
  {\( x\neq\E \liff x=x\rdel\cat 0\lor x=x\rdel\cat 1 \)}
\end{corollary}

 {}Note that Theorem~\ref{thm:cases} can easily be generalized to
 show, for example,
\( x=\E \lor x=0 \lor x=1 \lor x=00\cat\ldel\ldel x \lor
   x=01\cat\ldel\ldel x \lor x=10\cat\ldel\ldel x \lor
   x=11\cat\ldel\ldel x \),
 or, by substituting various terms for $x$,
\( xi=i \lor xi=0\cat(\ldel x)i \lor xi=1\cat(\ldel x)i \),
 etc.  Because 
\( (A\lor B)\land(A\limp C)\land(B\limp C) \limp C \)
 is a theorem of $T_1$, we can use Theorem~\ref{thm:cases} together
 with substitution to define an entire family of ``derived rules'' in
 $T_1$, like the following.

\begin{derule}
\label{derule:cases}\mbox{}\nopagebreak
   \begin{enumerate}
   \item
      \leftright
     {\( A[\E],A[0x],A[1x] \lded A \)}
     {\( A[\E],A[x0],A[x1] \lded A \)}
   \item
      \( A[\E],A[0],A[1],A[0x0],A[0x1],A[1x0],A[1x1] \lded A \)
   \end{enumerate}
\end{derule}

\noindent
 As an example of application of Derived Rule~\ref{derule:cases}, we
 prove the following simple claim.

\begin{claim}
   \( (\ldel x)\rdel = \ldel(x\rdel) \)
\end{claim}
\begin{proof}
   By Derived Rule~\ref{derule:cases}:
   \( (\ldel\E)\rdel = \E = \ldel(\E\rdel) \),
   \( (\ldel i)\rdel = \E\rdel = \ldel\E = \ldel(i\rdel) \),
   \( (\ldel(ixj))\rdel = (xj)\rdel = x = \ldel(ix)
       = \ldel((ixj)\rdel) \).
\end{proof}

\subsubsection{On ``${}_0$'' and ``${}_1$''}

\begin{claim}
\label{claim:_jcatdist}
   \( {}_0(x\cat y) = {}_0x\cat{}_0y \)
\end{claim}
\begin{proof}
   By $\NIND$ on $y$:
   \( {}_0(x\cat\E) = {}_0x = {}_0x\cat\E = {}_0x\cat{}_0\E \),
   \( {}_0(x\cat yi) = {}_0(x\cat y)\cat 0 = {}_0x\cat{}_0y\cat 0
       = {}_0x\cat{}_0(yi) \).
\end{proof}
 Note that an identical theorem can be proved with ${}_1x$ in place of
 ${}_0x$.  In what follows, we will often need to prove theorems in
 which ``${}_0$'' or ``${}_1$'' appear, where the particular function
 used does not matter.  We will indicate this by using ``${}_j$'' to
 stand for either of the above functions.

\begin{claim}
\label{claim:_jcomm}
   \( {}_jx\cat j = j\cat{}_jx \)
\end{claim}
\begin{proof}
   By $\NIND$ on $x$:
   \( {}_j\E\cat j = \E\cat j = j = j\cat\E = j\cat{}_j\E \),
   \( {}_j(xi)\cat j = ({}_jx\cat j)\cat j = (j\cat{}_jx)\cat j
       = j\cat({}_jx\cat j) = j\cat{}_j(xi) \).
\end{proof}

\begin{claim}
\label{claim:_jcatcomm}
   \( {}_jx\cat{}_jy = {}_jy\cat{}_jx \)
\end{claim}
\begin{proof}
   By $\NIND$ on $y$, and by Claims~\ref{claim:_jcatdist}
   and~\ref{claim:_jcomm}:
   \( {}_jx\cat{}_j\E = {}_jx\cat\E = {}_jx = \E\cat{}_jx
       = {}_j\E\cat{}_jx \),
   \( {}_jx\cat{}_j(yi) = {}_jx\cat{}_jy\cat j = {}_jy\cat{}_jx\cat j
       = {}_j(yx)\cat j = j\cat{}_j(yx) = j\cat{}_jy\cat{}_jx
       = {}_jy\cat j\cat{}_jx = {}_j(yi)\cat{}_jx \).
\end{proof}

\begin{corollary}
   \( {}_j(xy) = {}_j(yx) \)
\end{corollary}

\begin{claim}
   \( {}_jx = {}_j({}_0x) \) \qquad and \qquad
   \( {}_jx = {}_j({}_1x) \)
\end{claim}
\begin{proof}
   (We will prove only the first property, the second one being almost
   identical.)  By $\NIND$ on $x$:
   \( {}_j\E = {}_j({}_0\E) \),
   \( {}_j(xi) = {}_jx\cat j = {}_j({}_0x)\cat j = {}_j({}_0x\cat 0)
       = {}_j({}_0(xi)) \).
\end{proof}

\begin{corollary}
   \( {}_0x={}_0y \liff {}_1x={}_1y \)
\end{corollary}

\begin{claim}
\label{claim:_jdelcomm}
   \( \ldel{}_jx = {}_jx\rdel \)
\end{claim}
\begin{proof}
   By $\NIND$ on $x$, and by Claim~\ref{claim:_jcomm}:
   \( \ldel{}_j\E = \ldel\E = \E = \E\rdel = {}_j\E\rdel \),
   \( \ldel{}_j(xi) = \ldel({}_jx\cat j) = \ldel(j\cat{}_jx)
       = {}_jx = ({}_jx\cat j)\rdel = {}_j(xi)\rdel \).
\end{proof}

\begin{claim}
\label{claim:_jdeldist}
   \leftright
  {\( \ldel({}_jx) = {}_j(\ldel x) \)}
  {\( {}_j(x\rdel) = ({}_jx)\rdel \)}
\end{claim}
\begin{proof}
   (L) By $\NIND$ on $x$, and by Claim~\ref{claim:_jcatdist}:
   \( \ldel({}_j\E) = \ldel\E = \E = {}_j\E = {}_j(\ldel\E) \),
   \( \ldel({}_j(ix)) = \ldel(j\cat{}_jx) = {}_jx
       = {}_j(\ldel(ix)) \).
\end{proof}

\begin{claim}
   \( x=\E \liff {}_jx=\E \)
\end{claim}
\begin{proof}
   By $\NIND$ on $x$:
   \( \E=\E \liff {}_j\E=\E \),
   \( xi=\E \liff {}_jx\cat j=\E \).
\end{proof}

\begin{claim}
\label{claim:_jchop}
   \leftright
  {\( x\lchop y = {}_jx\lchop y \)}
  {\( y\rchop x = y\rchop{}_jx \)}
\end{claim}
\begin{proof}
   (L) By $\NIND$ on $x$:
   \( \E\lchop y = y = {}_j\E\lchop y \),
   \( ix\lchop y = \ldel(x\lchop y) = \ldel({}_jx\lchop y)
       = j\cat{}_jx\lchop y = {}_j(ix)\lchop y \).
\end{proof}

\begin{claim}
   \leftright
  {\( x\lchop{}_jy = {}_j(x\lchop y) \)}
  {\( {}_jy\rchop x = {}_j(y\rchop x) \)}
\end{claim}
\begin{proof}
   (L) By $\NIND$ on $x$, and by Claim~\ref{claim:_jdeldist}:
   \( \E\lchop{}_jy = {}_jy = {}_j(\E\lchop y) \),
   \( ix\lchop{}_jy = \ldel(x\lchop{}_jy) = \ldel{}_j(x\lchop y)
       = {}_j(\ldel(x\lchop y)) = {}_j(ix\lchop y) \).
\end{proof}

\begin{claim}
   \( {}_j(x\lchop y) = {}_j(y\rchop x) \)
\end{claim}
\begin{proof}
   By $\NIND$ on $x$, and by Claims~\ref{claim:_jdeldist}
   and~\ref{claim:_jdelcomm}:
   \( {}_j(\E\lchop y) = {}_jy = {}_j(y\rchop\E) \),
   \( {}_j(ix\lchop y) = {}_j(\ldel(x\lchop y))
       = \ldel({}_j(x\lchop y)) = \ldel({}_j(y\rchop x))
       = ({}_j(y\rchop x))\rdel = {}_j((y\rchop x)\rdel)
       = {}_j(y\rchop ix) \).
\end{proof}

\begin{corollary}
   \( x\lchop{}_jy = {}_jy\rchop x \)
\end{corollary}

\begin{claim}
   \leftright
  {\( \lbit x = x\rchop x\rdel \)}
  {\( \ldel x\lchop x = x\rbit \)}
\end{claim}
\begin{proof}
   (L) By $\NIND$ on $x$, and by Corollary~\ref{cor:chopnil} and
   Claims~\ref{claim:_jchop} and~\ref{claim:_jdelcomm}:
   \( \lbit\E = \E = \E\rchop\E\rdel \),
   \( \lbit(ix) = i = ix\rchop x = ix\rchop{}_0x
       = ix\rchop\ldel(0{}_0x) = ix\rchop\ldel{}_0(ix)
       = ix\rchop{}_0(ix)\rdel = ix\rchop(ix)\rdel \).
\end{proof}

\begin{corollary}
   \leftright
  {\( \lbit(xi) = xi\rchop x \)}
  {\( x\lchop ix = (ix)\rbit \)}
\end{corollary}

\begin{claim}
\label{claim:chops}\mbox{}\nopagebreak
   \begin{enumerate}
   \item
      \leftright
     {\( y\lchop xi\neq\E \liff y\lchop xi=y\lchop x\cat i \)}
     {\( \E\neq ix\rchop y \liff i\cat x\rchop y=ix\rchop y \)}
   \item
      \leftright
     {\( y\lchop x\neq\E \limp y\lchop xi\neq\E \)}
     {\( \E\neq x\rchop y \limp \E\neq ix\rchop y \)}
   \item
      \leftright
     {\( {}_0(y\cat(y\lchop x)) = {}_0(x\cat(x\lchop y)) \)}
     {\( {}_0((x\rchop y)\cat y) = {}_0((y\rchop x)\cat x) \)}
   \end{enumerate}
\end{claim}
\begin{proof}
\mbox{}\nopagebreak
   \begin{enumerate}
   \item
      (L) The first direction is proved by Claim~\ref{claim:Eneqcati}:
      \( y\lchop xi = y\lchop x\cat i \limp y\lchop xi \neq\E \).
      The other direction is proved by $\NIND$ on $y$:
      \( \E\lchop xi\neq\E \limp \E\lchop xi = \E\lchop x\cat i \),
      \( jy\lchop xi\neq\E \limp \ldel(y\lchop xi)\neq\E
         \limp y\lchop xi\neq\E
         \limp y\lchop xi = y\lchop x\cat i
         \limp \ldel(y\lchop xi) = \ldel(y\lchop x\cat i)
                = y\lchop x\zlC(\E,\ldel(y\lchop x)\cat i)
         \limp jy\lchop xi = jy\lchop xi\zlC(\E,jy\lchop x\cat i)
                = jy\lchop x\cat i \).
   \item
      (L) By $\NIND$ on $y$ and the preceding claim:
      \( \E\lchop x = x\neq\E \limp \E\lchop xi = xi\neq\E \),
      \( jy\lchop x = \ldel(y\lchop x) \neq \E
         \limp y\lchop x \neq \E
         \limp y\lchop xi \neq \E
         \limp y\lchop xi = y\lchop x\cat i
         \limp \ldel(y\lchop xi) = \ldel(y\lchop x)\cat i
                = jy\lchop x\cat i \neq \E \).
   \item
      (L) The claim is proved first under the assumption that
      \(x\lchop y=\E\) (which implies by Axioms~\ref{axiom:lchopE}
      and~\ref{axiom:rchopE}, and by the preceding claims, that
      \(y\lchop xi=y\lchop x\cat i\), and also implies that
      \(xi\lchop y=\ldel(x\lchop y)=\E\)), and then under the
      assumption that \(x\lchop y\neq\E\) (which implies by
      Axioms~\ref{axiom:lchopE} and~\ref{axiom:rchopE}, and by the
      preceding claims, that \(y\lchop xi=y\lchop x=\E\)).  Then, a
      simple application of modus ponens with the tautology
      \(x\lchop y=\E \lor x\lchop y\neq\E\) yields the claim.

      By $\NIND$ on $x$, and under the assumption that
      \(x\lchop y=\E\):
      \( {}_0(y\cat(y\lchop\E)) = {}_0y = {}_0(\E\cat(\E\lchop y)) \),
      \( {}_0(y\cat(y\lchop xi))
          = {}_0(y\cat(y\lchop x)\cat i)
          = {}_0(y\cat(y\lchop x))\cat 0
          = {}_0(x\cat(x\lchop y))\cat 0
          = {}_0x\cat 0
          = {}_0(xi\cat\E)
          = {}_0(xi\cat(xi\lchop y)) \).

      By $\NIND$ on $x$, and under the assumption that
      \(x\lchop y\neq\E\):
      \( {}_0(y\cat(y\lchop\E)) = {}_0y = {}_0(\E\cat(\E\lchop y)) \),
      \( {}_0(y\cat(y\lchop xi))
          = {}_0(y\cat\E)
          = {}_0(y\cat(y\lchop x))
          = {}_0(x\cat(x\lchop y))
          = {}_0x\cat{}_0(x\lchop y)
          = {}_0x\cat 0\cat\ldel{}_0(x\lchop y)
          = {}_0(xi)\cat{}_0(xi\lchop y)
          = {}_0(xi\cat(xi\lchop y)) \).
      \quad\QED
   \end{enumerate}
\end{proof}

\subsubsection{On ``$\C$'' and related functions}

 Now, we will prove a group of theorems about the conditional function
 ``$\C$''.  Note that in the statement of some of the theorems below,
 we will need to express the fact that terms $t$ and $u$ have the same
 length, something which can be done by the equation \({}_jt={}_ju\).

 {}First, we introduce two new functions defined in terms of ``$\C$''
 that will be used throughout the rest of this chapter for notational
 convenience.  Whereas the conditional function ``$\C$'' performs a
 three-way test on its first argument, the ``zero-length conditional''
 function ``$\zlC$'' tests whether the length of its first argument is
 zero or not, and the ``even-length conditional'' function ``$\elC$''
 tests whether the length of its first argument is even or odd ($\elC$
 has already been introduced informally in Axiom~\ref{axiom:halves}).
\begin{definition}
   \ind[tdef]{$\C$ (conditional)}{$\zlC$ (zero-length conditional)}{}
   \( \zlC = [\fhead xyz\fdot x\C(y,z,z)] \)
\end{definition}
\begin{definition}
   \ind[tdef]{$\C$ (conditional)}{$\elC$ (even-length conditional)}{}
   \( \elC = [\fhead xyz\fdot (x\lhalf\lchop\rhalf x)\zlC(y,z)] \)
\end{definition}
 (To make the notation more consistent with previous usage, we will
 write ``$x\zlC(y,z)$'' and ``$x\elC(y,z)$'' instead of the more formal
 ``$\zlC(x,y,z)$'' and ``$\elC(x,y,z)$'', respectively.)

\begin{claim}
   \( w\C(x,y,z) =
      w\C\big(x,{}_0(z\rchop y)\cat y,{}_0(y\rchop z)\cat z\big) \)
\end{claim}
\begin{proof}
   Immediate from Axiom~\ref{axiom:C}.
\end{proof}

\begin{claim}
   \( w\C(x,y,z) = w\rbit\C(x,y,z) \)
\end{claim}
\begin{proof}
   By $\NIND$ on $w$:
   \( \E\C(x,y,z) = x = \E\rbit\C(x,y,z) \),
   \( w0\C(x,y,z) = {}_0(z\rchop y)\cat y = 0\C(x,y,z)
       = (w0)\rbit\C(x,y,z) \),
   \( w1\C(x,y,z) = {}_0(y\rchop z)\cat z = 1\C(x,y,z)
       = (w1)\rbit\C(x,y,z) \).
\end{proof}

\begin{claim}
   \( wi\zlC(x,y) = y \)
\end{claim}
\begin{proof}
   By Corollary~\ref{cor:chopnil}:
   \( wi\zlC(x,y) = wi\C(x,y,y) = {}_0(y\rchop y)\cat y = {}_0\E\cat y
       = y \).
\end{proof}

\begin{corollary}
\label{cor:lC}
   \( iw\zlC(x,y) = y \)
\end{corollary}

\begin{corollary}
   \( w\zlC(x,y) = {}_jw\zlC(x,y) \)
\end{corollary}

\begin{corollary}
\label{cor:lCid}
   \( w\zlC(y,y) = w\C(y,y,y) = y \)
\end{corollary}

\begin{theorem}
\label{thm:Cdist}
   For any $k$-ary function symbol $f$,
   \[
      \big({}_0y_1={}_0z_1\land\dots\land{}_0y_k={}_0z_k\big)
      \limp w\C\big(f(\tpl x_k),f(\tpl y_k),f(\tpl z_k)\big)
             = f\big(w\C(x_1,y_1,z_1),\dots,w\C(x_n,y_n,z_n)\big).
   \]
\end{theorem}
\begin{proof}
   By $\NIND$ on $w$, and under the assumption that
   \( {}_0y_1={}_0z_1 \land \dots \land {}_0y_k={}_0z_k \):
   \( \E\C\big(f(\tpl x_k),f(\tpl y_k),f(\tpl z_k)\big) = f(\tpl x_k)
       = f\big(\E\C(x_1,y_1,z_1),\dots,\E\C(x_k,y_k,z_k)\big) \),
   \( w0\C\big(f(\tpl x_k),f(\tpl y_k),f(\tpl z_k)\big) = f(\tpl y_k)
       = f\big(w0\C(x_1,y_1,z_1),\dots,w0\C(x_k,y_k,z_k)\big) \),
   \( w1\C\big(f(\tpl x_k),f(\tpl y_k),f(\tpl z_k)\big) = f(\tpl z_k)
       = f\big(w1\C(x_1,y_1,z_1),\dots,w1\C(x_k,y_k,z_k)\big) \).
\end{proof}

\begin{claim}
\label{claim:Ccomp}
   \[
      w\C\big(w\C(x_1,y_1,z_1),w\C(x_2,y_2,z_2),w\C(x_3,y_3,z_3)\big)
       = w\C\big(x_1,{}_0(z_2\rchop y_2)\cat y_2,
                     {}_0(y_3\rchop z_3)\cat z_3\big)
   \]
\end{claim}
\begin{proof}
   By $\NIND$ on $w$ and by Claim~\ref{claim:chops}:
   \( \E\C\big(\E\C(x_1,y_1,z_1),\E\C(x_2,y_2,z_2),
         \E\C(x_3,y_3,z_3)\big) = \E\C(x_1,x_2,x_3) = x_1
       = \E\C\big(x_1,{}_0(z_2\rchop y_2)\cat y_2,
         {}_0(y_3\rchop z_3)\cat z_3\big) \),
   \( w0\C\big(w0\C(x_1,y_1,z_1),w0\C(x_2,y_2,z_2),
         w0\C(x_3,y_3,z_3)\big)
       = w0\C({}_0(z_1\rchop y_1)\cat y_1,{}_0(z_2\rchop y_2)\cat y_2,
         {}_0(z_3\rchop y_3)\cat y_3)
       = {}_0\big(({}_0(z_3\rchop y_3)\cat y_3)\rchop
            ({}_0(z_2\rchop y_2)\cat y_2)\big)\cat
         {}_0(z_2\rchop y_2)\cat y_2
       = {}_0\big(({}_0(y_3\rchop z_3)\cat z_3)\rchop
            ({}_0(z_2\rchop y_2)\cat y_2)\big)\cat
         {}_0(z_2\rchop y_2)\cat y_2
       = w0\C\big(x_1,{}_0(z_2\rchop y_2)\cat y_2,
         {}_0(y_3\rchop z_3)\cat z_3\big) \),
   and similarly for $w1$.
\end{proof}

\begin{claim}
   \( w\C(x_0,y_0,z_0)\C(x_1,y_1,z_1) = \)
   \[
      w\C\big(x_0\C(x_1,y_1,z_1),
         ({}_0(z_0\rchop y_0)\cat y_0)\C(x_1,y_1,z_1),
         ({}_0(y_0\rchop z_0)\cat z_0)\C(x_1,y_1,z_1)\big)
   \]
\end{claim}
\begin{proof}
   A straightforward $\NIND$ on $w$, very similar to the proof of
   Claim~\ref{claim:Ccomp}.
\end{proof}

\begin{corollary}
   \( w\zlC(x_0,y_0)\zlC(x_1,y_1)
       = w\zlC\big(x_0\zlC(x_1,y_1),y_0\zlC(x_1,y_1)\big) \)
\end{corollary}

\begin{claim}
   \( x\zlC\big(y\zlC(z_0,w_0),y\zlC(z_1,w_1)\big)
       = y\zlC\big(x\zlC(z_0,z_1),x\zlC(w_0,w_1)\big) \)
\end{claim}
\begin{proof}
   By $\NIND$ on $x$:
   \( \E\zlC\big(y\zlC(z_0,w_0),y\zlC(z_1,w_1)\big) = y\zlC(z_0,w_0)
       = y\zlC\big(\E\zlC(z_0,z_1),\E\zlC(w_0,w_1)\big) \),
   \( xi\zlC\big(y\zlC(z_0,w_0),y\zlC(z_1,w_1)\big) = y\zlC(z_1,w_1)
       = y\zlC\big(xi\zlC(z_0,z_1),xi\zlC(w_0,w_1)\big) \).
\end{proof}

\begin{claim}
   \( w = w\C(\E,w\rdel\cat 0,w\rdel\cat 1) \)
\end{claim}
\begin{proof}
   By $\NIND$ on $w$:
   \( \E = \E\C(\E,\E\rdel\cat 0,\E\rdel\cat 1) \),
   \( w0 = w0\C(\E,w0,w1)
         = w0\C(\E,(w0)\rdel\cat 0,(w0)\rdel\cat 1) \),
   \( w1 = w1\C(\E,w0,w1)
         = w1\C(\E,(w1)\rdel\cat 0,(w1)\rdel\cat 1) \).
\end{proof}

\begin{corollary}
\label{cor:lCcases}
   \( w = w\zlC(\E,w) \)
\end{corollary}

\begin{theorem}
   \( w\C(x,y_0,y_1)=z \liff (w=\E\land x=z)\lor
      (w=w\rdel\cat 0\land {}_0(y_1\rchop y_0)\cat y_0=z)\lor
      (w=w\rdel\cat 1\land {}_0(y_0\rchop y_1)\cat y_1=z) \)
\end{theorem}
\begin{proof}
   By $\NIND$ on $w$:
   \( \E\C(x,y_0,y_1)=z \liff x=z \liff \E=\E\land x=z \),
   \( w0\C(x,y_0,y_1)=z \liff {}_0(y_1\rchop y_0)\cat y_0=z
      \liff w0=(w0)\rdel\cat 0\land{}_0(y_1\rchop y_0)\cat y_0=z \),
   \( w1\C(x,y_0,y_1)=z \liff {}_0(y_0\rchop y_1)\cat y_1=z
      \liff w1=(w1)\rdel\cat 1\land{}_0(y_0\rchop y_1)\cat y_1=z \).
\end{proof}

\begin{corollary}
   \( w\zlC(x,y)=z \liff (w=\E\land x=z)\lor(w\neq\E\land y=x) \)
\end{corollary}

\begin{theorem}
   For any term $u$, \( w\zlC(u[\E/w],u) = u \).
\end{theorem}
\begin{proof}
   By induction on the structure of $u$: if \(u=w\), then
   \( w\zlC(u[\E/w],u) = w\zlC(\E,w) = w \)
   by Corollary~\ref{cor:lCcases}; if \(u=x\neq w\), then
   \( w\zlC(u[\E/w],u) = w\zlC(x,x) = x \)
   by Corollary~\ref{cor:lCid}; if \(u=f(t_1,\dots,t_n)\), then
   \( w\zlC(u[\E/w],u) = w\zlC\big(f(t_1[\E/w],\dots,t_n[\E/w]),
         f(t_1,\dots,t_n)\big) = f\big(w\zlC(t_1[\E/w],t_1),\dots,
         w\zlC(t_n[\E/w],t_n)\big) = f(t_1,\dots,t_n) = u \)
   by Theorem~\ref{thm:Cdist} and the induction hypothesis.
\end{proof}

\begin{claim}
   \leftright
  {\( \lbit(zx) = z\zlC(\lbit x,\lbit z) \)}
  {\( (xz)\rbit = z\zlC(x\rbit,z\rbit) \)}
\end{claim}
\begin{proof}
   (L) By $\NIND$ on $z$, and by Corollary~\ref{cor:lC}:
   \( \lbit(\E x) = \lbit x = \E\zlC(\lbit x,\lbit\E) \),
   \( \lbit((iz)x) = i = iz\zlC(\lbit x,\lbit(iz)) \).
\end{proof}

\begin{claim}
   \leftright
  {\( \ldel(zx) = z\zlC(\ldel x,\ldel z\cat x) \)}
  {\( (xz)\rdel = z\zlC(x\rdel,x\cat z\rdel) \)}
\end{claim}
\begin{proof}
   (L) By $\NIND$ on $z$, and by Corollary~\ref{cor:lC}:
   \( \ldel(\E x) = \ldel x = \E\zlC(\ldel x,\ldel\E\cat x) \),
   \( \ldel((iz)x) = zx = iz\zlC(\ldel x,\ldel(iz)\cat x) \).
\end{proof}

\begin{claim}
   \( v\zlC(u,t)=u \liff (v\neq\E\limp t=u) \)
\end{claim}
\begin{proof}
   By $\NIND$ on $v$:
   \( \E\zlC(u,t)=u \liff u=u \liff (\E\neq\E\limp u=u) \),
   \( vi\zlC(u,t)=u \liff t=u \liff (vi\neq\E\limp t=u) \).
\end{proof}

\subsubsection{On ``$\lhalf$'' and ``$\rhalf$''}

\begin{claim}
   \leftright
  {\( {}_j(x\lhalf) = ({}_jx)\lhalf \)}
  {\( {}_j(\rhalf x) = \rhalf({}_jx) \)}
\end{claim}
\begin{proof}
   (L) By $\NIND$ on $x$, Axioms~\ref{axiom:halves}, and various
   theorems proved above:
   \( {}_j(\E\lhalf) = \E = ({}_j\E)\lhalf \),
   \begin{align*}
      {}_j((xi)\lhalf)
      &= {}_j\big((x\lhalf\lchop\rhalf x)\zlC(x\lhalf,
            x\lhalf\cat\lbit(\rhalf x\cat i))\big) \\
      &= {}_j(x\lhalf\lchop\rhalf x)\zlC\big({}_j(x\lhalf),
            {}_j(x\lhalf\cat\lbit(\rhalf x\cat i))\big) \\
      &= ({}_j(x\lhalf)\lchop{}_j(\rhalf x))\zlC\big(({}_jx)\lhalf,
            {}_j(x\lhalf)\cat\lbit{}_j(\rhalf x\cat i)\big) \\
      &= (({}_jx)\lhalf\lchop\rhalf({}_jx))\zlC\big(({}_jx)\lhalf,
            ({}_jx)\lhalf\cat\lbit(\rhalf({}_jx)\cat j)\big) \\
      &= ({}_jx\cat j)\lhalf 
       = ({}_j(xi))\lhalf 
      \quad\QED
   \end{align*}

\end{proof}

 {}Basic properties of ``$\elC$'' can easily be obtained from the
 basic properties of ``$\zlC$'', on which it is based.  In order to
 prove properties particular to ``$\elC$'', we will need the following
 lemmas.

 {}But first, a few reminders.
\begin{itemize}
\item
   \( {}_j(\ldel(\rhalf x\cat i)) = \ldel(\rhalf{}_jx\cat j)
       = \ldel(j\cat\rhalf{}_jx) = {}_j(\rhalf x) \)
\item
   \( (z\cat i)\lchop y = {}_j(z\cat i)\lchop y
       = (j\cat{}_jz)\lchop y = \ldel({}_jz\lchop y)
       = \ldel(z\lchop y) \)
\end{itemize}

\begin{lemma}
\label{lem:elCparity1}
   \( {}_0x\lhalf\lchop(\rhalf{}_0x\cat 0)
       = ({}_0x\lhalf\lchop\rhalf{}_0x)\cat 0 \)
\end{lemma}
\begin{proof}
   By Axioms~\ref{axiom:lchopE}, \ref{axiom:rchopE},
   and~\ref{axiom:halfl}, and by Claim~\ref{claim:chops}:
   \( {}_0x\lhalf\rchop\rhalf{}_0x=\E \limp
      {}_0x\lhalf\lchop(\rhalf{}_0x\cat 0)\neq\E \limp
      {}_0x\lhalf\lchop(\rhalf{}_0x\cat 0)
       = ({}_0x\lhalf\lchop\rhalf{}_0x)\cat 0 \).
\end{proof}

\begin{lemma}
\label{lem:elCparity2}
   \( {}_0(xi)\lhalf\lchop\rhalf{}_0(xi)
       = ({}_0x\lhalf\lchop\rhalf{}_0x)\zlC(0,\E) \)
\end{lemma}
\begin{proof}
   By Lemma~\ref{lem:elCparity1}:
   \begin{align*}
      {}_0(xi)\lhalf\lchop\rhalf{}_0(xi)
      &= x\elC\big({}_0x\lhalf\lchop(\rhalf{}_0x\cat 0),
            ({}_0x\lhalf\cat\lbit(\rhalf{}_0x\cat 0))\lchop
               \ldel(\rhalf{}_0x\cat 0)\big) \\
      &= x\elC\big(({}_0x\lhalf\lchop\rhalf{}_0x)\cat 0,
            ({}_0x\lhalf\cat 0)\lchop\rhalf{}_0x\big) \\
      &= (x\lhalf\lchop\rhalf x)\zlC\big(
            {}_0(x\lhalf\lchop\rhalf x)\cat 0,
            \ldel{}_0(x\lhalf\lchop\rhalf x)\big) \\
      &= ({}_0x\lhalf\lchop\rhalf{}_0x)\zlC\big(0,\E) 
      \quad\QED
   \end{align*}

\end{proof}

\begin{theorem}
\label{thm:elCparity}
   \( xi\elC(y,z) = x\elC(z,y) \)
\end{theorem}
\begin{proof}
   By Lemmas~\ref{lem:elCparity1} and~\ref{lem:elCparity2}:
   \begin{align*}
      xi\elC(y,z)
      &= {}_0(xi)\elC(y,z) \\
      &= ({}_0(xi)\lhalf\lchop\rhalf{}_0(xi))\zlC(y,z) \\
      &= ({}_0x\lhalf\lchop\rhalf{}_0x)\zlC(0,\E)\zlC(y,z) \\
      &= ({}_0x\lhalf\lchop\rhalf{}_0x)\zlC
         \big(0\zlC(y,z),\E\zlC(y,z)\big) \\
      &= ({}_0x\lhalf\lchop\rhalf{}_0x)\zlC(z,y) \\
      &= {}_0x\elC(z,y) 
       = x\elC(z,y) 
      \quad\QED
   \end{align*}

\end{proof}

\begin{claim}
\label{claim:cathalfcat}
   \leftright
  {\( (ixj)\lhalf = i\cat x\lhalf \)}
  {\( \rhalf(ixj) = \rhalf x\cat j \)}
\end{claim}
\begin{proof}
   (L) By Axioms~\ref{axiom:halves} and Theorem~\ref{thm:elCparity}:
   \begin{align*}
      (ixj)\lhalf
      &= xj\elC\big((i(xj)\lhalf)\rdel,i(xj)\lhalf\big) \\
      &= x\elC\big(i(xj)\lhalf,(i(xj)\lhalf)\rdel\big) \\
      &= x\elC\big(i\cat x\lhalf,
            (i\cat x\lhalf\cat\lbit\rhalf x)\rdel\big) \\
      &= x\elC\big(i\cat x\lhalf,i\cat x\lhalf\big) \\
      &= i\cat x\lhalf 
      \quad\QED
   \end{align*}

\end{proof}

\begin{claim}
   \leftright
  {\( (\ldel x\rdel)\lhalf = \ldel(x\lhalf) \)}
  {\( \rhalf(\ldel x\rdel) = (\rhalf x)\rdel \)}
\end{claim}
\begin{proof}
   (L) By Derived Rule~\ref{derule:cases} and
   Claim~\ref{claim:cathalfcat}:
   \( (\ldel\E\rdel)\lhalf = \E = \ldel(\E\lhalf) \),
   \( (\ldel i\rdel)\lhalf = \E = \ldel(i\lhalf) \),
   \( (\ldel(ixj)\rdel)\lhalf = x\lhalf = \ldel(i\cat x\lhalf)
       = \ldel((ixj)\lhalf) \).
\end{proof}

\subsection{Further definitions and theorems}

 In this section, we define many functions in $T_1$ and prove their
 basic properties.  We also give (and prove) a number of useful
 derived rules for $T_1$.

 {}From now on, we will not give proofs that consist only in a
 straightforward application of $\NIND$.  Proof sketches will be given
 for more complex theorems, and complete proofs are provided in
 Appendix~\ref{app:proofs} for most of the theorems below.

\subsubsection{On generalizations of $\NIND$}

 First, we define some generalizations of $\NIND$ based on Derived
 Rule~\ref{derule:cases}.

\begin{derule}
\label{derule:NIND2}\mbox{}\\*\nopagebreak
   \leftright
  {\( A[\E],A[0],A[1],A[x]\limp
      A[00x]\land A[01x]\land A[10x]\land A[11x] \lded A \)\\}
  {\( A[\E],A[0],A[1],A[x]\limp
      A[x00]\land A[x01]\land A[x10]\land A[x11] \lded A \)}
\end{derule}
\begin{proof}
   (We will prove only (R), the case for (L) being almost identical.)
   Let us define a formula \(EL[x]:{}_0(x\lhalf\lchop\rhalf x)=\E\).
   By Lemma~\ref{lem:elCparity2}, we immediately get that
   \( EL[xi] \liff \lnot EL[x] \liff EL[ix] \).
   To prove that $A$ is true under the given hypotheses, we will show
   that the hypotheses imply the following two statements:
   \begin{gather}
      (EL[x]\limp A[x]) \land (\lnot EL[x]\limp A[x\rdel]),
   \label{eq:EL1} \\
      (\lnot EL[x]\limp A[x]) \land (EL[x]\limp A[x\rdel]).
   \label{eq:EL2} 
   \end{gather}
   Together with the fact that \(EL[x]\lor\lnot EL[x]\), this will
   imply that \(A[x]\land A[x\rdel]\), \ie, \(A[x]\) is true.

   We can prove statement~\ref{eq:EL1} by $\NIND$ on $x$:
   \( (EL[\E]\limp A[\E]) \land (\lnot EL[\E]\limp A[\E\rdel]) \)
   is trivially true since \(A[\E]\) is true by assumption, while
   \[
      (EL[xi]\limp A[xi])\land(\lnot EL[xi]\limp A[(xi)\rdel])
      \liff (\lnot EL[x]\limp A[xi])\land(EL[x]\limp A[x])
   \]
   is true since the second conjunct is true by the induction
   hypothesis, and so is \((\lnot EL[x]\limp A[x\rdel])\), which,
   together with the assumption that \(A[x]\limp A[xji]\), implies
   that \(A[x\rdel]\limp A[x\rdel\cat ji] \limp A[xi]\).

   The same reasoning applies to statement~\ref{eq:EL2}, which
   concludes the proof.
\end{proof}
 Note that this proof can easily be modified to get a similar derived
 rule for \(A[x]\limp A[ixj]\), and it can easily be extended to cover
 other variations of Theorem~\ref{thm:cases}.

 {}Next, we want to define simultaneous $\NIND$ on two variables.
 Before we can do this, we need to define a few functions and prove
 their basic properties.

\begin{definition}
   \fcnind[tdef]{$\fcn{lb}$ (left bit)}{}{}
   \fcnind[tdef]{$\fcn{rb}$ (right bit)}{}{}
   \leftright
  {\( \fcn{lb} =
      \big[\fhead xy\fdot y\zlC(\E,\lbit(\ldel y\lchop x))\big] \)}
  {\( \fcn{rb} =
      \big[\fhead xy\fdot y\zlC(\E,(x\rchop y\rdel)\rbit)\big] \)}
\end{definition}
\begin{definition}
   \fcnind[tdef]{$\fcn{lc}$ (left cut)}{}{}
   \fcnind[tdef]{$\fcn{rc}$ (right cut)}{}{}
   \leftright
  {\( \fcn{lc} = \big[\fhead xy\fdot x\rchop(y\lchop x)\big] \)}
  {\( \fcn{rc} = \big[\fhead xy\fdot (x\rchop y)\lchop x\big] \)}
\end{definition}
\begin{definition}
   \lenfcnind[tdef]{$\lenfcn{min}$}{}{}
   \lenfcnind[tdef]{$\lenfcn{max}$}{}{}
   \( \lenfcn{min} = \big[\fhead xy\fdot x\rchop y\zlC(x,y)\big] \)
   \qquad
   \( \lenfcn{max} = \big[\fhead xy\fdot x\lchop y\zlC(x,y)\big] \)
\end{definition}

\begin{claim}
\label{claim:cuts}\mbox{}\nopagebreak
   \begin{enumerate}
   \item
      \leftright
     {\( z\lchop yx = z\lchop y\cat(z\rchop y)\lchop x \)}
     {\( xy\rchop z = x\rchop(y\lchop z)\cat y\rchop z \)}
   \item
      \leftright
     {\( y\rdel\lchop x = \fcn{lb}(x,y)\cat y\lchop x \)}
     {\( x\rchop\ldel y = x\rchop y\cat\fcn{rb}(x,y) \)}
   \item
      \leftright
     {\( y\lchop((x\rchop y)\lchop x) = \E \)}
     {\( (x\rchop(y\lchop x))\rchop y = \E \)}
   \item
      \leftright
     {\( \fcn{lc}(\fcn{rc}(x,y),y) = \fcn{rc}(x,y) \)}
     {\( \fcn{rc}(\fcn{lc}(x,y),y) = \fcn{lc}(x,y) \)}
   \item
      \leftright
     {\( \lbit(y\lchop x) =
         y\lchop x\zlC\big(\E,(x\rchop\ldel(y\lchop x))\rbit\big) \)}
     {\( (x\rchop y)\rbit =
         x\rchop y\zlC\big(\E,\lbit((x\rchop y)\rdel\lchop x)\big) \)}
   \item
      \leftright
     {\( \fcn{lc}(x,yi) = \fcn{lc}(x,y)\cat\fcn{lb}(x,yi) \)}
     {\( \fcn{rc}(x,iy) = \fcn{rb}(x,iy)\cat\fcn{rc}(x,y) \)}
   \item
      \leftright
     {\( \fcn{lc}(x,y)\cat y\lchop x = x \)}
     {\( x = x\rchop y\cat\fcn{rc}(x,y) \)}
   \end{enumerate}
\end{claim}

\noindent
 Now, we can state and prove a derived rule for simultaneous $\NIND$
 on two variables.

\begin{derule}
\label{derule:2NIND}
\begin{align*}
   (LL)\ &\ A[\E,y],A[x,\E],A[x,y]\limp
      A[0x,0y]\land A[0x,1y]\land A[1x,0y]\land A[1x,1y] \lded A \\
   (LR)\ &\ A[\E,y],A[x,\E],A[x,y]\limp
      A[0x,y0]\land A[0x,y1]\land A[1x,y0]\land A[1x,y1] \lded A \\
   (RL)\ &\ A[\E,y],A[x,\E],A[x,y]\limp
      A[x0,0y]\land A[x0,1y]\land A[x1,0y]\land A[x1,1y] \lded A \\
   (RR)\ &\ A[\E,y],A[x,\E],A[x,y]\limp
      A[x0,y0]\land A[x0,y1]\land A[x1,y0]\land A[x1,y1] \lded A 
\end{align*}
\end{derule}
\begin{proof}
   (We will prove only (RR), the other cases being almost identical.)
   Under the given assumptions, we will prove
   \( A[x_L\cat\fcn{lc}(x_R,z),y_L\cat\fcn{lc}(y_R,z)] \)
   by $\NIND$ on $z$, where
   \begin{alignat*}{2}
      x_L &= x\rchop\lenfcn{min}(x,y) &\qquad
      x_R &= \fcn{rc}(x,\lenfcn{min}(x,y)) \\
      y_L &= y\rchop\lenfcn{min}(x,y) &\qquad
      y_R &= \fcn{rc}(y,\lenfcn{min}(x,y)) 
   \end{alignat*}

   Base case:
   \( A[x_L\cat\fcn{lc}(x_R,\E),y_L\cat\fcn{lc}(y_R,\E)]
       = A[x_L,y_L] \).
   By the definition of $\lenfcn{min}$, we know that
   \( \lenfcn{min}(x,y)=x \lor \lenfcn{min}(x,y)=y \),
   which means that
   \( x_L=\E \lor y_L=\E \),
   which implies
   \( (A[x_L,y_L]\liff A[x_L,\E]) \lor (A[x_L,y_L]\liff A[\E,y_L]) \), 
   so we know that \(A[x_L,y_L]\) is true by the assumptions.

   Induction Step: we have that
   \[
      A[x_L\cat\fcn{lc}(x_R,zi),y_L\cat\fcn{lc}(y_R,zi)]
       = A[x_L\cat\fcn{lc}(x_R,z)\cat\fcn{lb}(x_R,zi),
         y_L\cat\fcn{lc}(y_R,z)\cat\fcn{lb}(y_R,zi)],
   \]
   which follows directly from the induction hypothesis by the
   assumptions.

   Finally, we know that
   \( A[x_L\cat\fcn{lc}(x_R,\lenfcn{min}(x,y)),
      y_L\cat\fcn{lc}(y_R,\lenfcn{min}(x,y))]
       = A[x_L\cat x_R,y_L\cat y_R] = A[x,y] \),
   which completes the proof.
\end{proof}
 Note that this rule, and its proof, can easily be extended to more
 than two variables, giving us a very useful form of $\NIND$ on many
 variables.

\subsubsection{On propositional reasoning}

 {}Now, we will show how to formalize propositional connectives in
 $T_1$.  (The definitions are identical to those for $L_1$, and we
 use ``$x\bC(y,x)$'' instead of the more formal ``$\bC(x,y,z)$''.)
\begin{definition}
\mbox{}\nopagebreak
   \begin{enumerate}
   \bitind[(tdef]{$\bit\lnot,\bitbin\land,\bitbin\lor$, etc.}{}{}
   \item
      \ind[tdef]{$\C$ (conditional)}{$\bC$ (boolean conditional)}{}
      \( \bC = [\fhead xyz\fdot x\C(z,z,y)] \)
   \item
      \( \bit\lsg = [\fhead x\fdot x\bC(1,0)] \)
   \item
      \( \bit\lnot = [\fhead x\fdot x\bC(0,1)] \)
   \item
      \( \bitbin\land = [\fhead xy\fdot x\bC(\bit\lsg y,0)] \)
   \item
      \( \bitbin\lor = [\fhead xy\fdot x\bC(1,\bit\lsg y)] \)
   \item
      \( \bitbin\limp = [\fhead xy\fdot x\bC(\bit\lsg y,1)] \)
   \item
      \( \bitbin\liff = [\fhead xy\fdot x\bC(\bit\lsg y,\bit\lnot y)] \)
   \item
      \( \bitbin\lxor = [\fhead xy\fdot x\bC(\bit\lnot y,\bit\lsg y)] \)
   \bitind[)]{$\bit\lnot,\bitbin\land,\bitbin\lor$, etc.}{}{}
   \end{enumerate}
\end{definition}
 The properties of ``$\C$'' already proven immediately extend to $\bC$
 in the obvious way, and the following theorem follows directly from
 these properties.
\begin{theorem}
\label{thm:prop}\mbox{}\nopagebreak
   \begin{enumerate}
   \bitind[(]{$\bit\lnot,\bitbin\land,\bitbin\lor$, etc.}{}{}
   \item
      \( \bit\lsg x=1 \lor \bit\lsg x=0 \)
   \item
      \( \bit\lsg\bit\lsg x = \bit\lsg x \)
   \item
      \( \bit\lnot x=1 \liff \lnot(\bit\lsg x=1) \)
   \item
      \( x\bitbin\land y=1 \liff (\bit\lsg x=1\land\bit\lsg y=1) \)
   \item
      \( x\bitbin\lor y=1 \liff (\bit\lsg x=1\lor\bit\lsg y=1) \)
   \item
      \( x\bitbin\limp y=1 \liff (\bit\lsg x=1\limp\bit\lsg y=1) \)
   \item
      \( x\bitbin\liff y=1 \liff (\bit\lsg x=1\liff\bit\lsg y=1) \)
   \item
      \( x\bitbin\lxor y=1 \liff (\bit\lsg x=1\lxor\bit\lsg y=1) \)
   \bitind[)]{$\bit\lnot,\bitbin\land,\bitbin\lor$, etc.}{}{}
   \end{enumerate}
\end{theorem}
 This theorem gives us direct proofs of the usual properties of the
 defined connectives, from the corresponding properties of the
 connectives in $T_1$, and it allows us to introduce the following
 notation: we will write ``$t$'' instead of ``$t=1$'' for $T_1$-terms
 $t$.  (For example, we could state that
 ``$\bit\lsg x\bitbin\lxor\bit\lnot x$'' is a theorem.)

\subsubsection{On variations of $\TRN$}

 To define functions by ``simple'' $\TRN$, we will use $\STRN[g,h]$ as
 shorthand for
\[
   \big[\fhead x\tpl y\fdot
   \TRN[\fhead xz\tpl y\fdot g(x,\tpl y),
        \fhead xz\tpl yv_\ell v_r\fdot h(x,\tpl y,v_\ell,v_r),
        \fhead z\fdot z,\fhead z\fdot z](x,\E,\tpl y)\big].
\]
 The following property is then a direct consequence of the axiom for
 $\TRN$.
\begin{claim}
\mbox{}\nopagebreak
\[
   \STRN[g,h](x,\tpl y) = x\rdel\zlC\Bigl(g(x,\tpl y),
      h\bigl(x,\tpl y,\STRN[g,h](x\lhalf,\tpl y),
         \STRN[g,h](\rhalf x,\tpl y)\bigr)\Bigr)
\]
\end{claim}

\subsubsection{On ``$\fcn{AND}$'' and ``$\fcn{OR}$''}

\begin{definition}
   \fcnind[tdef]{$\fcn{AND}$}{}{}
   \( \fcn{AND} = \STRN[\fhead x\fdot x,
         \fhead xv_\ell v_r\fdot v_\ell\bitbin\land v_r] \)
\end{definition}
\begin{definition}
   \fcnind[tdef]{$\fcn{OR}$}{}{}
   \( \fcn{OR} = \STRN[\fhead x\fdot x,
         \fhead xv_\ell v_r\fdot v_\ell\bitbin\lor v_r] \)
\end{definition}
 We can use $\TIND$ to prove the following simple theorem (we give the
 proof here to illustrate the use of $\TIND$).
\begin{theorem}
   \( \fcn{AND}({}_1x) = x\zlC(\E,1) \)
\end{theorem}
\begin{proof}
   By $\TIND$ on $x$:
   \( \fcn{AND}({}_1\E) = \fcn{AND}(\E) = \E = \E\zlC(\E,1) \),
   \( \fcn{AND}({}_1i) = \fcn{AND}(1) = 1 = i\zlC(\E,1) \),
   \( \fcn{AND}({}_1x)
       = \fcn{AND}(({}_1x)\lhalf)\bitbin\land\fcn{AND}(\rhalf({}_1x))
       = \fcn{AND}({}_1(x\lhalf))\bitbin\land\fcn{AND}({}_1(\rhalf x))
       = x\lhalf\zlC(\E,1)\bitbin\land\rhalf x\zlC(\E,1)
       = 1\bitbin\land 1 = 1 \).
\end{proof}
 Similarly, we can prove that \( \fcn{OR}({}_0x) = x\zlC(\E,0) \).

 {}Now, we want to prove some more basic properties of $\fcn{AND}$ and
 $\fcn{OR}$, the most important of which being
\( \fcn{AND}(xy) = \fcn{AND}(x)\bitbin\land\fcn{AND}(y) \)
 (and similarly for $\fcn{OR}$).  This property would naturally be
 proved using $\NIND$ since it involves the concatenation of two
 variables, but $\fcn{AND}$ is defined by $\TRN$ which makes it more
 natural to use $\TIND$.  In fact, we will use $\TIND$ to prove the
 property but because of the messy interaction between concatenation
 recursion and tree recursion, the proof will unfortunately not be as
 simple as one might expect.

 {}Before we get started, note that it is a simple matter to extend
 Derived Rule~\ref{derule:NIND2} to give us rules similar to the
 following ones.
\begin{derule}
\label{derule:NINDE}
   \( A[0],A[1],y\neq\E\land A[y]\limp A[y0]\land A[y1] \lded
      x\neq\E\limp A[x] \)
\end{derule}
\begin{derule}
\label{derule:TINDE}\mbox{}\nopagebreak
   \begin{align*}
      &\phanrel{\lded} A[00],A[01],A[10],A[11],A[000],\dots,A[111],
      y\lhalf\rdel\neq\E\land A[y\lhalf]\land A[\rhalf y]\limp A[y] \\
      &\lded x\neq\E\land x\neq 0\land x\neq 1 \limp A[x] 
   \end{align*}
\end{derule}

\noindent
 These rules can then be used to prove the following claim and
 theorem.

\begin{lemma}
\label{lem:AND}\mbox{}\nopagebreak
   \begin{enumerate}
   \item
      \leftright
     {\( x\neq\E\land x\neq 0\land x\neq 1 \limp
         (\ldel x\cat j)\lhalf = \ldel(x\lhalf)\cat\lbit\rhalf x \)\\}
     {\( x\neq\E\land x\neq 0\land x\neq 1 \limp
         \rhalf(\ldel x\cat j) = \ldel\rhalf x\cat j \)}
   \item
      \leftright
     {\( \lbit x\bitbin\land\fcn{AND}(\ldel x\cat j)
          = \fcn{AND}(x)\bitbin\land j \)}
     {\( j\bitbin\land\fcn{AND}(x)
          = \fcn{AND}(j\cat x\rdel)\bitbin\land x\rbit \)}
   \item
      \( \lbit x\bitbin\land\fcn{AND}(\ldel x) = \fcn{AND}(x)
          = \fcn{AND}(x\rdel)\bitbin\land x\rbit \) \quad
      for \(x\neq\E,0,1\)
   \end{enumerate}
\end{lemma}

\begin{theorem}
   \( \fcn{AND}(xy) = \fcn{AND}(x)\bitbin\land\fcn{AND}(y) \) \quad
   for \(x,y\neq\E\)
\end{theorem}
\begin{proof}
   By Lemma~\ref{lem:AND} and by Derived Rule~\ref{derule:NINDE} on
   $y$:
   \( \fcn{AND}(xi) = \fcn{AND}(x)\bitbin\land i
       = \fcn{AND}(x)\bitbin\land\fcn{AND}(i) \),
   \( \fcn{AND}(x(yi)) = \fcn{AND}(xy)\bitbin\land i
       = \fcn{AND}(x)\bitbin\land\fcn{AND}(y)\bitbin\land i
       = \fcn{AND}(x)\bitbin\land\fcn{AND}(yi) \).
\end{proof}
 Note that a similar lemma and theorem can be used to show
\( \fcn{OR}(xy) = \fcn{OR}(x)\bitbin\lor\fcn{OR}(y) \)
 for \(x,y\neq\E\).

\subsubsection{On generalizations of $\CRN$---part I}

 Now, we will define simultaneous concatenation recursion on notation
 for many variables, prove its basic properties, and define a few more
 useful functions based on this generalized $\CRN$.  But first, we
 must prove a number of technical lemmas.

\begin{theorem}
\label{thm:_jchop}
   \( {}_jx={}_jy \liff x\lchop y=\E=x\rchop y \)
\end{theorem}

\begin{claim}
\label{claim:_jCRN}
   For \(f=\rCRN[h]\),
   \begin{enumerate}
   \item
      \( {}_j(f(x,\tpl y)) = {}_jx \)
   \item
      \( f(x,\tpl y)\rchop z = f(x\rchop z,\tpl y) \)
   \item
      \( \fcn{lb}(f(x,\tpl y),z) = x\lchop z\zlC
         \big((0\cat h(\fcn{lc}(x,z),\tpl y))\rbit,\E\big) \)
      \quad for \(x,z\neq\E\)
   \item
      \( \fcn{lc}(f(x,\tpl y),z) = f(\fcn{lc}(x,z),\tpl y) \)
   \end{enumerate}
\end{claim}
 (A similar claim can be proved about $\lCRN$.)

\begin{definition}
   \fcnind[tdef]{$\fcn{lp}_0,\fcn{lp}_1$ (left pad)}{}{}
   \fcnind[tdef]{$\fcn{rp}_0,\fcn{rp}_1$ (right pad)}{}{}
   \leftright
  {\( \fcn{lp}_j = \big[\fhead xy\fdot{}_j(y\rchop x)\cat x\big] \)}
  {\( \fcn{rp}_j = \big[\fhead xy\fdot x\cat{}_j(x\lchop y)\big] \)}
\end{definition}

\begin{definition}
   \begin{align*}
      \lenfcnind[(tdef]{$\lenfcn{max}$}{$\lenfcn{max}_k$}{}
      \lenfcn{max}_1 &= [\fhead x\fdot x] \\
      \lenfcn{max}_{k+1} &= \big[\fhead x\tpl x_k\fdot
         \lenfcn{max}\big(x,\lenfcn{max}_k(\tpl x_k)\big)\big] 
      \lenfcnind[)]{$\lenfcn{max}$}{$\lenfcn{max}_k$}{}
   \end{align*}
\end{definition}

\begin{lemma}
\label{lem:chopchop}\mbox{}\nopagebreak
   \begin{enumerate}
   \item
      \leftright
     {\( y\lchop x = x\lchop y\zlC(y\lchop x,\E) \)}
     {\( x\rchop y = x\lchop y\zlC(x\rchop y,\E) \)}
   \item
      \leftright
     {\( (y\rchop(x\lchop y))\lchop x = y\lchop x \)}
     {\( x\rchop((y\rchop x)\lchop y) = x\rchop y \)}
   \end{enumerate}
\end{lemma}

\begin{claim}
\label{claim:lpmaxl}\mbox{}\nopagebreak
   \begin{enumerate}
   \item
      \leftright
     {\( \fcn{lp}_j(x,\lenfcn{max}(x,y)) = \fcn{lp}_j(x,y) \)}
     {\( \fcn{rp}_j(x,\lenfcn{max}(x,y)) = \fcn{rp}_j(x,y) \)}
   \item
      \leftright
     {\( \fcn{lb}\big(\fcn{lp}_0(xi,\lenfcn{max}(xi,yj)),
      \lenfcn{max}(xi,yj)\big) = i \)\\}
     {\( \fcn{rb}\big(\fcn{rp}_0(ix,\lenfcn{max}(ix,jy)),
      \lenfcn{max}(ix,jy)\big) = i \)}
   \end{enumerate}
\end{claim}
 (Note that this can easily be generalized to more than two
 variables.)

\noindent
 Now, we are ready to define $\lCRN_m$ and $\rCRN_m$.
\begin{definition}
   \begin{align*}
      \ind[(tdef]{\CRN}{\lCRN$_m$,\rCRN$_m$}{}
      \lCRN_m[h] &= \Big[\fhead \tpl x_m\tpl y\fdot
         \lCRN\big[\fhead z\tpl x_m\tpl y\fdot
            h(\fcn{rc}(x_1,z),\dots,\fcn{rc}(x_m,z),\tpl y) \big] \\*
      &\pheq \hphantom{\Big[\fhead \tpl x_m\tpl y\fdot }
         \big(\lenfcn{max}_m(\tpl x_m),
         \fcn{lp}_0(x_1,\lenfcn{max}_m(\tpl x_m)),\dots,
         \fcn{lp}_0(x_m,\lenfcn{max}_m(\tpl x_m)),\tpl y \big) \Big] \\
      \rCRN_m[h] &= \Big[\fhead \tpl x_m\tpl y\fdot
         \rCRN\big[\fhead z\tpl x_m\tpl y\fdot
            h(\fcn{lc}(x_1,z),\dots,\fcn{lc}(x_m,z),\tpl y) \big] \\*
      &\pheq \hphantom{\Big[\fhead \tpl x_m\tpl y\fdot }
         \big(\lenfcn{max}_m(\tpl x_m),
         \fcn{rp}_0(x_1,\lenfcn{max}_m(\tpl x_m)),\dots,
         \fcn{rp}_0(x_m,\lenfcn{max}_m(\tpl x_m)),\tpl y \big) \Big] 
      \ind[)]{\CRN}{\lCRN$_m$,\rCRN$_m$}{}
   \end{align*}
\end{definition}
 A simple application of Derived Rule~\ref{derule:2NIND} together with
 Claim~\ref{claim:lpmaxl}, both generalized to the $m$ variables
 $x_1,\dots,x_m$, suffices to show the following basic theorem about
 $\lCRN_m$ and $\rCRN_m$.

\begin{theorem}
\label{thm:CRNm}\mbox{}\nopagebreak
   \begin{align*}
      \ind[(]{\CRN}{\lCRN$_m$,\rCRN$_m$}{}
      &\!\lCRN_m[h](x_1,\dots,x_m,\tpl y) \\*
      &= x_1\cat\dotsb\cat x_m\zlC\Big(\E,\lbit
         \big(h\big(\fcn{lp}_0(x_1,\lenfcn{max}_m(\tpl x_m)),\dots,
         \fcn{lp}_0(x_m,\lenfcn{max}_m(\tpl x_m)),\tpl y\big)\cat
         0\big) \\*
      &\pheq \hphantom{x_1\cat\dotsb\cat x_m\zlC\Big(\E, } \cat
         \lCRN_m[h]\big(\ldel\fcn{lp}_0(x_1,\lenfcn{max}_m(\tpl x_m)),
         \dots,\ldel\fcn{lp}_0(x_m,\lenfcn{max}_m(\tpl x_m)),
         \tpl y\big)\Big) \\
      &\!\rCRN_m[h](x_1,\dots,x_m,\tpl y) \\*
      &= x_1\cat\dotsb\cat x_m\zlC\Big(\E,\rCRN_m[h]
         \big(\fcn{rp}_0(x_1,\lenfcn{max}_m(\tpl x_m))\rdel,\dots,
         \fcn{rp}_0(x_m,\lenfcn{max}_m(\tpl x_m))\rdel,\tpl y\big) \\*
      &\pheq \hphantom{x_1\cat\dotsb\cat x_m\zlC\Big(\E, } \cat
         \big(0\cat h\big(\fcn{rp}_0(x_1,\lenfcn{max}_m(\tpl x_m)),
         \dots,\fcn{rp}_0(x_m,\lenfcn{max}_m(\tpl x_m)),
         \tpl y\big)\big)\rbit\Big) 
      \ind[)]{\CRN}{\lCRN$_m$,\rCRN$_m$}{}
   \end{align*}
\end{theorem}

\noindent
 Using $\lCRN_m$, we can now define some useful functions and prove
 their basic properties.

\begin{definition}
\mbox{}\nopagebreak
   \begin{enumerate}
   \bitfcnind[(tdef]{$\bitfcn{not},\bitfcn{and},\bitfcn{or}$, etc.}{}{}
   \item
      \( \bitfcn{not} = \lCRN[\fhead x\fdot\bit\lnot(\lbit x)] \)
   \item
      \( \bitfcn{and}_m = \lCRN_m\big[\fhead\tpl x_m\fdot
         (\lbit x_1)\bitbin\land\dotsb\bitbin\land(\lbit x_m)\big] \)
   \item
      \( \bitfcn{or}_m = \lCRN_m\big[\fhead\tpl x_m\fdot
         (\lbit x_1)\bitbin\lor\dotsb\bitbin\lor(\lbit x_m)\big] \)
   \item
      \( \bitfcn{xor}_m = \lCRN_m\big[\fhead\tpl x_m\fdot
         (\lbit x_1)\bitbin\lxor\dotsb\bitbin\lxor(\lbit x_m)\big] \)
   \item
      \( \bitfcn{iff}_m = \lCRN_m\big[\fhead\tpl x_m\fdot
         ((\lbit x_1)\bitbin\liff(\lbit x_2))\bitbin\land\dotsb
         \bitbin\land((\lbit x_{m-1})\bitbin\liff(\lbit x_m))\big] \)
   \bitfcnind[)]{$\bitfcn{not},\bitfcn{and},\bitfcn{or}$, etc.}{}{}
   \end{enumerate}
\end{definition}

\begin{claim}
\label{claim:bitand}\mbox{}\nopagebreak
   \begin{enumerate}
   \item
      \( \bitfcn{and}_m(x_1i_1,\dots,x_mi_m) =
         \bitfcn{and}_m(\tpl x_m)\cat\bitfcn{and}_m(\tpl i_m) \)
   \item
      \( {}_jx_1=\dots={}_jx_m\land{}_jy_1=\dots={}_jy_m \limp
         \bitfcn{and}_m(x_1y_1,\dots,x_my_m)=
         \bitfcn{and}_m(\tpl x_m)\cat\bitfcn{and}_m(\tpl y_m) \).
   \item
      \( \bitfcn{and}_m(\tpl x_m)\rdel =
         \bitfcn{and}_m(x_1\rdel,\dots,x_m\rdel) \)
   \item
      \( \bitfcn{and}_m(\tpl x_m)\rchop y =
         \bitfcn{and}_m(x_1\rchop y,\dots,x_m\rchop y) \)
   \end{enumerate}
\end{claim}
 And similarly for $\bitfcn{or}_m$, $\bitfcn{iff}_m$, and
 $\bitfcn{not}$.  We can now prove a theorem relating the functions
 $\fcn{AND}$ and $\fcn{OR}$ to each other.

\begin{theorem}
\label{thm:ANDnotOR}
   \( \bit\lnot\fcn{AND}(x) = \fcn{OR}(\bitfcn{not}(x)) \)
   \quad and \quad
   \( \bit\lnot\fcn{OR}(x) = \fcn{AND}(\bitfcn{not}(x)) \)
   \quad for \(x\neq\E\)
\end{theorem}

\subsubsection{On generalizations of $\CRN$---part II}

 Finally, we are ready to generalize $\CRN$ to operate on blocks of
 bits instead of single bits.  For technical reasons to be discussed
 below, this will be done only for variable-length blocks of bits
 whose lengths are powers of two.

 {}In order to get such a version of $\CRN$, we first need to define a
 length-division function.  Ideally, we would like to define a
 function $\lenfcn{div}(x,y)$ whose length would be equal to
 $\lfloor|x|/|y|\rfloor$, but this seems to be impossible in $T_1$.
 Instead, we can define a function $\lenfcn{powdiv}(x,y)$ whose length
 is equal to $\lfloor|x|/2^{\lceil\lg|y|\rceil}\rfloor$, \ie, the
 function divides the length of $x$ by the smallest power of $2$
 larger than or equal to the length of $y$.  Luckily, this will be
 sufficient for our purposes, as will be seen in
 Chapter~\ref{chap:soundness}.

 {}The first functions we define are a function that returns a string
 whose length is the smallest power of two larger than or equal to the
 length of its input, and a function that tests whether or not the
 length of its input is a power of two.
\begin{definition}
   \lenfcnind[tdef]{$\lenfcn{pow}$}{}{}
   \( \lenfcn{pow} = \STRN[{}_1,\fhead yv_\ell v_r\fdot v_r\cat v_r] \)
\end{definition}
\begin{definition}
   \( \lenfcn{ispow} = \STRN[\fhead y\fdot\E,
         \fhead yv_\ell v_r\fdot y\elC(v_r,1)] \)
\end{definition}
 (Note that $\lenfcn{ispow}$ returns $\E$ if the length of its input
 is a power of two and $1$ otherwise.)  The basic properties of
 $\lenfcn{pow}$ and $\lenfcn{ispow}$ are now easy to prove by $\TIND$.
\begin{claim}
\label{claim:pow}\mbox{}\nopagebreak
   \begin{enumerate}
   \item
      \( \rhalf\lenfcn{pow}(y) = \lenfcn{pow}(\rhalf y)
          = \lenfcn{pow}(y)\lhalf \)
      \quad (for \(y\neq\E,0,1\))
   \item
      \( {}_1\lenfcn{pow}(y) = \lenfcn{pow}({}_1y)
          = \lenfcn{pow}(y) \)
   \item
      \( \lenfcn{pow}(\lenfcn{pow}(y)) = \lenfcn{pow}(y) \)
   \item
      \( \lenfcn{pow}(y)\lchop y = \E \)
   \item
      \( \lenfcn{ispow}(\lenfcn{pow}(y)) = \E \)
   \end{enumerate}
\end{claim}

\noindent
 Before we can define the length-division function, we need to define
 a ``length multiplication'' function (this is just the ``smash''
 function).
\begin{definition}
   \ind[tdef]{$\smsh$ (smash)}{}{}
   \( \smsh = \STRN[\fhead xy\fdot x\zlC(\E,y),
         \fhead xyv_\ell v_r\fdot v_\ell\cat v_r] \)
\end{definition}
 These properties of $\smsh$ can then be proven with simple
 applications of $\TIND$ and $\NIND$.
\begin{claim}
\label{claim:smsh}\mbox{}\nopagebreak
   \begin{enumerate}
   \item
      \( \E\smsh y = \E = x\smsh\E \)
   \item
      \leftright
     {\( {}_1(xi\smsh y) = {}_1(x\smsh y)\cat{}_1y \)}
     {\( {}_1(x\smsh yi) = {}_1(x\smsh y)\cat{}_1x \)}
   \item
      \leftright
     {\( {}_1((x\cat y)\smsh z) = {}_1(x\smsh z)\cat{}_1(y\smsh z) \)}
     {\( {}_1(x\smsh(y\cat z)) = {}_1(x\smsh y)\cat{}_1(x\smsh z) \)}
   \item
      \( {}_1(x\smsh y) = {}_1(y\smsh x) \)
   \end{enumerate}
\end{claim}

\noindent
 Now, we can define the $\lenfcn{powdiv}$ function, and a
 corresponding $\lenfcn{powmod}$ function.
\begin{definition}
   \begin{align*}
      \lenfcnind[(tdef]{$\lenfcn{powdiv},\lenfcn{powmod}$}{}{}
      \lenfcn{powdiv} &= \TRN\big[\fhead yx\fdot y\zlC(\E,{}_1x),
         \fhead yxv_\ell v_r\fdot v_r,\lhalf,\lhalf\big] \\
      \lenfcn{powmod} &= \big[\fhead xy\fdot
         (\lenfcn{pow}(y)\smsh\lenfcn{powdiv}(x,y))\lchop{}_1x\big] 
      \lenfcnind[)]{$\lenfcn{powdiv},\lenfcn{powmod}$}{}{}
   \end{align*}
\end{definition}
 Straightforward applications of $\TIND$ then prove these basic
 properties.
\begin{claim}
\label{claim:powdiv}\mbox{}\nopagebreak
   \begin{enumerate}
   \item
      \( \lenfcn{powdiv}(x,y) = \lenfcn{powdiv}({}_1x,{}_1y) \)
   \item
      \( \lenfcn{powdiv}(x,y) = \lenfcn{powdiv}(x,\lenfcn{pow}(y)) \)
   \item
      \( x\lchop\lenfcn{pow}(y)\neq\E \limp \lenfcn{powdiv}(x,y)=\E \)
   \item
      \( \lenfcn{powdiv}(\lenfcn{pow}(y)\smsh z,y) = y\zlC(\E,{}_1z) \)
   \end{enumerate}
\end{claim}
 And properties of $\lenfcn{powmod}$ follow directly from the
 properties of $\lenfcn{powdiv}$.
\begin{corollary}
\label{cor:powmod}\mbox{}\nopagebreak
   \begin{enumerate}
   \item
      \( \lenfcn{powmod}(x,y) = \lenfcn{powmod}({}_1x,{}_1y) \)
   \item
      \( \lenfcn{powmod}(x,y) = \lenfcn{powmod}(x,\lenfcn{pow}(y)) \)
   \item
      \( x\lchop\lenfcn{pow}(y)\neq\E \limp
         \lenfcn{powmod}(x,y)={}_1x \)
   \item
      \( \lenfcn{powmod}(\lenfcn{pow}(y)\smsh z,y) = \E \)
   \end{enumerate}
\end{corollary}

 {}We need just a few more technical lemmas about $\lenfcn{powdiv}$
 and $\lenfcn{powmod}$ before we can define generalized $\CRN$ and
 prove its properties.
\begin{claim}
\label{claim:powdivmod}\mbox{}\nopagebreak
   \begin{enumerate}
   \item
      \( x\lchop(\lenfcn{pow}(y)\smsh\lenfcn{powdiv}(x,y)) = \E \)
   \item
      \( y\neq\E \limp
         \lenfcn{powmod}(x,y)\lchop\lenfcn{pow}(y)\neq\E \)
   \item
      \( y\neq\E \limp \lenfcn{powdiv}(x1,y)=\lenfcn{powdiv}(x,y)\cat
         \big((\lenfcn{powmod}(x,y)\cat 1)\lchop\lenfcn{pow}(y)
            \zlC(1,\E)\big) \) \\
      \( y\neq\E \limp \lenfcn{powmod}(x1,y)=
         (\lenfcn{powmod}(x,y)\cat 1)\lchop\lenfcn{pow}(y)\zlC
         \big(\E,\lenfcn{powmod}(x,y)\cat 1\big) \)
   \item
      \( \lenfcn{powdiv}((\lenfcn{pow}(y)\smsh z)\cat x,y) =
         y\zlC(\E,{}_1z)\cat\lenfcn{powdiv}(x,y) \land{} \) \\
      \( \lenfcn{powmod}((\lenfcn{pow}(y)\smsh z)\cat x,y) =
         \lenfcn{powmod}(x,y) \)
   \item
      \( y\neq\E \land x\lchop\lenfcn{pow}(y)=\E \limp
         \lenfcn{powdiv}(x,y)=
            \lenfcn{powdiv}(x\rchop\lenfcn{pow}(y),y)\cat 1 \) \\
      \( y\neq\E \land x\lchop\lenfcn{pow}(y)=\E \limp
         \lenfcn{powmod}(x,y)=
            \lenfcn{powmod}(x\rchop\lenfcn{pow}(y),y) \)
   \end{enumerate}
\end{claim}

 {}At last, we are ready to define the generalized versions of $\lCRN$
 and $\rCRN$ which we will name ``$\lpowCRN$'' and ``$\rpowCRN$''.
 Functions defined by $\lpowCRN$ or $\rpowCRN$ take two extra
 parameters $u,v$ as input, and essentially perform $\CRN$ on their
 first input by replacing blocks of $|\lenfcn{pow}(u)|$ bits with
 blocks of $|\lenfcn{pow}(v)|$ bits.  Just like $(2^k,2^\ell)$-$\CRN$
 in Chapter~\ref{chap:L1}, we will simulate these generalized forms of
 $\CRN$ by using $\lenfcn{powdiv}$ and $\lenfcn{powmod}$ to extract
 the correct substring of the first input to pass to $h$ and to output
 the correct bits of $h$ in sequence.  Intuitively, $\lpowCRN[g,h]$
 and $\rpowCRN[g,h]$ will behave as follows (for all strings $z$ such
 that \(|z|=|\lenfcn{pow}(u)|\)).
\begin{align*}
   \lpowCRN[g,h](x,u,v,\tpl y) &= g(x,u,v,\tpl y)
      \qquad \text{(if } |x|<|\lenfcn{pow}(u)| \text{)} \\*
   \lpowCRN[g,h](z\cat x,u,v,\tpl y) &= \fcn{la}_0
      \big(h(z\cat x,u,v,\tpl y),\lenfcn{pow}(v)\big)
         \cat\lpowCRN[g,h](x,u,v,\tpl y) \\
   \rpowCRN[g,h](x,u,v,\tpl y) &= g(x,u,v,\tpl y)
      \qquad \text{(if } |x|<|\lenfcn{pow}(u)| \text{)} \\*
   \rpowCRN[g,h](x\cat z,u,v,\tpl y) &= \rpowCRN[g,h](x,u,v,\tpl y)
      \cat\fcn{ra}_0\big(h(x\cat z,u,v,\tpl y),\lenfcn{pow}(v)\big) 
\end{align*}
 Where we have used the functions $\fcn{la}_j$ and $\fcn{ra}_j$, whose
 definitions and basic properties (easily proved by Derived
 Rule~\ref{derule:2NIND}) appear below.
\begin{definition}
\mbox{}\\*\nopagebreak
   \leftright
  {\( \fcn{la}_j = \big[\fhead xy\fdot
         {}_j(y\rchop x)\cat((x\rchop y)\lchop x)\big] \)}
  {\( \fcn{ra}_j = \big[\fhead xy\fdot
         (x\rchop(y\lchop x))\cat{}_j(x\lchop y)\big] \)}
\end{definition}
\begin{claim}
\mbox{}\nopagebreak
   \begin{enumerate}
   \item
      \leftright
     {\( {}_1(\fcn{la}_j(x,y)) = {}_1y \)}
     {\( {}_1y = {}_1(\fcn{ra}_j(x,y)) \)}
   \item
      \leftright
     {\( \fcn{lc}(\fcn{la}_j(x,y),y) = \fcn{la}_j(x,y) \)}
     {\( \fcn{ra}_j(x,y) = \fcn{rc}(\fcn{ra}_j(x,y),y) \)}
   \end{enumerate}
\end{claim}

\begin{definition}
\mbox{}\nopagebreak
   \begin{align*}
      \ind[(tdef]{\CRN}{\lpowCRN,\rpowCRN}{}
      \lpowCRN[g,h] &= \Big[\fhead xuv\tpl y\fdot
         \lCRN\Big[\fhead zxuv\tpl y\fdot \\*
      &\pheq \hphantom{\Big[ }
         \fcn{rb}\Big(\fcn{la}_0\Big(h\big(\fcn{rc}\big(x,
            (\lenfcn{pow}(u)\smsh(1\cat\lenfcn{powdiv}(\ldel z,v)))
            \cat\lenfcn{powmod}(x,u)\big),u,v,\tpl y\big), \\*
      &\pheq \hphantom{\Big[\fcn{rb}\Big(\fcn{la}_0\Big( }
         \lenfcn{pow}(v)\Big),1\cat\lenfcn{powmod}(\ldel z,v)\Big)
         \Big](\lenfcn{pow}(v)\smsh\lenfcn{powdiv}(x,u),x,u,v,\tpl y) \\*
      &\pheq \hphantom{\Big[ } \rule{200pt}{0pt} \cat
         g\big(\fcn{rc}(x,\lenfcn{powmod}(x,u)),u,v,\tpl y\big)\Big] \\
      \rpowCRN[g,h] &= \Big[\fhead xuv\tpl y\fdot
         g\big(\fcn{lc}(x,\lenfcn{powmod}(x,u)),u,v,\tpl y\big)\cat
         \rCRN\Big[\fhead zxuv\tpl y\fdot \\*
      &\pheq \hphantom{\Big[ } \fcn{lb}\Big(\fcn{ra}_0\Big(
         h\big(\fcn{lc}\big(x,\lenfcn{powmod}(x,u)\cat
         (\lenfcn{pow}(u)\smsh(\lenfcn{powdiv}(z\rdel,v)\cat 1))\big),
            u,v,\tpl y\big), \\*
      &\pheq \hphantom{\Big[\fcn{rb}\Big(\fcn{la}_0\Big( }
         \lenfcn{pow}(v)\Big),\lenfcn{powmod}(z\rdel,v)\cat 1\Big)
         \Big](\lenfcn{pow}(v)\smsh\lenfcn{powdiv}(x,u),x,u,v,\tpl y)
         \Big] 
      \ind[)]{\CRN}{\lpowCRN,\rpowCRN}{}
   \end{align*}
\end{definition}

\begin{theorem}
   For \(u\neq\E\) and \(v\neq\E\),
   \begin{align*}
      \ind[(]{\CRN}{\lpowCRN,\rpowCRN}{}
      &\lpowCRN[g,h](x,u,v,\tpl y) = \\*
      &\ x\lchop\lenfcn{pow}(u)\zlC\Big(
            \fcn{la}_0(h(x,u,v,\tpl y),\lenfcn{pow}(v))\cat
            \lpowCRN[g,h](\lenfcn{pow}(u)\lchop x,u,v,\tpl y),
            g(x,u,v,\tpl y)\Big), \\
      &\rpowCRN[g,h](x,u,v,\tpl y) = \\*
      &\ x\lchop\lenfcn{pow}(u)\zlC\Big(
            \rpowCRN[g,h](x\rchop\lenfcn{pow}(u),u,v,\tpl y)
            \cat\fcn{ra}_0(h(x,u,v,\tpl y),\lenfcn{pow}(v)),
            g(x,u,v,\tpl y)\Big). 
      \ind[)]{\CRN}{\lpowCRN,\rpowCRN}{}
   \end{align*}
\end{theorem}
\begin{proof}
   We prove the theorem for $\rpowCRN$ only, the case for $\lpowCRN$
   being almost identical.  To start with, if
   \(x\lchop\lenfcn{pow}(u)\neq\E\), then Claim~\ref{claim:powdiv} and
   Corollary~\ref{cor:powmod} give us the result immediately since
   \(\lenfcn{powdiv}(x,u)=\E\) and \(\lenfcn{powmod}(x,u)={}_1x\).

   Next, suppose that \(x\lchop\lenfcn{pow}(u)=\E\).  Then,
   Claim~\ref{claim:powdivmod} implies that
   \(\lenfcn{powdiv}(x,u)\neq\E\), which shows that
   \( \lenfcn{pow}(v)\smsh\lenfcn{powdiv}(x,u) =
      (\lenfcn{pow}(v)\smsh\lenfcn{powdiv}(x,u)\rdel)\cat
         \lenfcn{pow}(v) =
      (\lenfcn{pow}(v)\smsh\lenfcn{powdiv}(x\rchop\lenfcn{pow}(u),u))
         \cat\lenfcn{pow}(v) \).
   The following facts are then direct consequences of preceding
   claims, and hold for all strings $z$ such that \( z\neq\E \land
      z\rchop\lenfcn{pow}(v)=\E \):
   \begin{itemize}
   \item
      \( \fcn{lc}(x,y\cat z) =
         \fcn{lc}(x,y)\cat\fcn{lc}(y\lchop x,z) \)
      (easy to prove by $\NIND$ on $z$),
   \item
      \( \lenfcn{powmod}\big((\lenfcn{pow}(v)\smsh
            \lenfcn{powdiv}(x,u)\rdel)\cat{}_1z\rdel,
         \lenfcn{pow}(v)\big)\cat 1 =
         \lenfcn{powmod}({}_1z\rdel,\lenfcn{pow}(v))\cat 1 =
         {}_1z\rdel\cat 1 = {}_1z \),
   \item
      \( \fcn{lc}\Big(x,\lenfcn{powmod}(x,u)\cat
            \Big(\lenfcn{pow}(u)\smsh\big(\lenfcn{powdiv}\big(
               (\lenfcn{pow}(v)\smsh\lenfcn{powdiv}(x,u)\rdel)\cat
               z\rdel,v\big)\cat 1\big)\Big)\Big)
          = \fcn{lc}\big(x,\lenfcn{powmod}(x,u)\cat(\lenfcn{pow}(u)
               \smsh(\lenfcn{powdiv}(x,u)\rdel\cat 1))\big)
          = \fcn{lc}(x,{}_1x) = x \).
   \end{itemize}
   These facts can be used to prove by $\NIND$ on $z$ that
   \begin{multline*}
      x\lchop\lenfcn{pow}(u)=\E \land z\rchop\lenfcn{pow}(v)=\E
      \limp f\big((\lenfcn{pow}(v)\smsh\lenfcn{powdiv}(x,u)\rdel)
         \cat{}_1z,x,u,v,\tpl y\big) = \\*
      f\big(\lenfcn{pow}(v)\smsh\lenfcn{powdiv}
            (x\rchop\lenfcn{pow}(u),u),x,u,v,\tpl y\big)\cat
         \fcn{rc}\big(\fcn{ra}_0(h(x,u,v,\tpl y),
            \lenfcn{pow}(v)),z\big) 
   \end{multline*}
   (where we use ``$f$'' to denote the function defined by $\rCRN$ in
   the definition of $\rpowCRN$), and putting \(z=\lenfcn{pow}(v)\) in
   this last fact gives us the theorem.
\end{proof}

\subsection{Numerical definitions and theorems}

 In this section, we will give definitions for numerical predicates
 and functions (\ie, ones that treat their string arguments as
 encoding binary numbers) and prove their properties.

\subsubsection{On ``$\numrel=$'' and ``$\numrel<$''}

 The definitions of ``$\numrel=$'' and ``$\numrel<$'' inside $T_1$
 are the same as in $L_1$.  Intuitively, $x\numrel=y$ if the two
 strings are equal when padded on the left with 0's to the same
 length.
\begin{definition}
   \numind[tdef]{$\numrel"=$}{}{}
   \( \numrel= =
      \big[\fhead xy\fdot\fcn{AND}(1\cat\bitfcn{iff}_2(x,y))\big] \)
\end{definition}
 In a similar way, $x\numrel<y$ if there is a bit position where $x$
 has a $0$, $y$ has a $1$, and the portions of $x$ and $y$ to the left
 of that position are numerically equal.
\begin{definition}
\mbox{}\nopagebreak
   \numind[tdef]{$\numrel<$}{}{}
   \[
      \numrel< = \big[\fhead xy\fdot\fcn{OR}\big(\rCRN_2
         [\fhead xy\fdot(x\rbit\bitrel<y\rbit)\bitbin\land
         (x\rdel\numrel=y\rdel)]
         (\fcn{lp}_0(x,y),\fcn{lp}_0(y,x))\big)\big]
   \]
\end{definition}
 (Where we used ``$i\bitrel<j$'' as shorthand for
 ``$\bit\lnot i\bitbin\land j$''.)

 {}Now, we prove basic properties of the two predicates just defined.
 A simple $\NIND$ suffices to show the following theorem.
\begin{theorem}
\label{thm:=n}\mbox{}\nopagebreak
   \begin{enumerate}
   \item
      \( x\numrel=\E \liff x={}_0x \)
   \item
      \( x\numrel=y \liff \fcn{lp}_0(x,y)=\fcn{lp}_0(y,x) \)
   \end{enumerate}
\end{theorem}
 This immediately implies that ``$\numrel=$'' is an equivalence
 relation.  If we define
\( \lenrel= = [\fhead xy\fdot (x\lchop y)\cat(x\rchop y)\zlC(\E,1)] \)
 and
\( \strrel= = [\fhead xy\fdot
      \bit\lnot(x\lenrel=y)\bitbin\land x\numrel=y] \),
 then the theorem we just proved immediately implies that
\( x\strrel=y \liff x=y \).  Together with the facts about
 propositional connectives proved in Theorem~\ref{thm:prop}, this
 means that for any formula $A$ of $T_1$, there exists a term
 $\term{A}$ of $T_1$ such that $T_1$ can prove \(A\liff\term{A}\)
 (with our usual convention whereby ``$\term{A}$'' stands for the
 formula $\term{A}=1$).

 {}Now, we prove more properties of $\numrel=$ and $\numrel<$.  By
 Theorem~\ref{thm:CRNm}, we have the following two theorems.
\begin{claim}
\label{claim:=ncat}\mbox{}\nopagebreak
   \begin{enumerate}
   \item
      \( x0\numrel=y0 \liff x\numrel=y \liff x1\numrel=y1 \)
   \item
      \( \bit\lnot(x0\numrel=y1) \)
   \item
      \( \bit\lnot(x1\numrel=y0) \)
   \end{enumerate}
\end{claim}

\pagebreak[0]

\begin{claim}
\label{claim:<ncat}\mbox{}\nopagebreak
   \begin{enumerate}
   \item
      \( x0\numrel<y0
          = x\numrel<y\bitbin\lor(x\numrel=y\bitbin\land 0\bitrel<0)
          = x\numrel<y \)
   \item
      \( x0\numrel<y1
          = x\numrel<y\bitbin\lor(x\numrel=y\bitbin\land 0\bitrel<1)
          = x\numrel\leq y \)
   \item
      \( x1\numrel<y0
          = x\numrel<y\bitbin\lor(x\numrel=y\bitbin\land 1\bitrel<0)
          = x\numrel<y \)
   \item
      \( x1\numrel<y1
          = x\numrel<y\bitbin\lor(x\numrel=y\bitbin\land 1\bitrel<1)
          = x\numrel<y \)
   \end{enumerate}
\end{claim}
  Simple proofs by $\NIND$ now suffice to show the following lemma.
\begin{claim}
\label{claim:<n}\mbox{}\nopagebreak
   \begin{enumerate}
   \item
      \( \bit\lnot(x\numrel<\E) \)
   \item
      \( \bit\lnot(x\numrel<x) \)
   \item
      \( \bit\lnot(\E\numrel<{}_0x) \)
   \end{enumerate}
\end{claim}
 Using the notation ``$x\numrel>y$'' for $y\numrel<x$,
 ``$x\numrel\leq y$'' for $x\numrel<y\bitbin\lor x\numrel=y$, and
 ``$\numrel\geq$'' similarly defined, we have the following theorem.
 (We give its proof here because it is representative of the kind of
 proof that will be used for most theorems concerning numerical
 functions.)
\begin{theorem}
   \begin{align*}
      &\phanbin{\bitbin\lor} (x\numrel<y \bitbin\land
         \bit\lnot(x\numrel=y) \bitbin\land \bit\lnot(x\numrel>y)) \\
      &\bitbin\lor (\bit\lnot(x\numrel<y) \bitbin\land
         x\numrel=y \bitbin\land \bit\lnot(x\numrel>y)) \\
      &\bitbin\lor (\bit\lnot(x\numrel<y) \bitbin\land
         \bit\lnot(x\numrel=y) \bitbin\land x\numrel>y) 
   \end{align*}
\end{theorem}
\begin{proof}
   By Derived Rule~\ref{derule:2NIND}, and the lemma above:
   When $y=\E$, the statement of the theorem reduces to
   \( (x\numrel=\E \bitbin\land \bit\lnot(x\numrel>\E)) \bitbin\lor
      (\bit\lnot(x\numrel=\E) \bitbin\land x\numrel>\E) \),
   which can be proved by regular $\NIND$ on $x$:
   \( \E\numrel=\E \bitbin\land \bit\lnot(\E\numrel>\E) \),
   \( (x0\numrel=\E \bitbin\land \bit\lnot(x0\numrel>\E)) \bitbin\lor
      (\bit\lnot(x0\numrel=\E) \bitbin\land x0\numrel>\E)
       = (x\numrel=\E \bitbin\land \bit\lnot(x\numrel>\E)) \bitbin\lor
         (\bit\lnot(x\numrel=\E)\bitbin\land x\numrel>\E) \),
   \( (x1\numrel=\E \bitbin\land \bit\lnot(x1\numrel>\E)) \bitbin\lor
      (\bit\lnot(x1\numrel=\E) \bitbin\land x1\numrel>\E)
       = \E\numrel<x \bitbin\lor \E\numrel=x = \E\numrel\leq x \).
   We can show that the statement holds when $x=\E$ in the same way.
   Next, we have four cases to consider:
   \begin{align*}
      \phanbin{\bitbin\lor} (x0\numrel<y0 \bitbin\land
         \bit\lnot(x0\numrel=y0) \bitbin\land \bit\lnot(x0\numrel>y0)) \\*
      \bitbin\lor (\bit\lnot(x0\numrel<y0) \bitbin\land
         x0\numrel=y0 \bitbin\land \bit\lnot(x0\numrel>y0)) \\*
      \bitbin\lor (\bit\lnot(x0\numrel<y0) \bitbin\land
         \bit\lnot(x0\numrel=y0) \bitbin\land x0\numrel>y0)
      &= \phanbin{\bitbin\lor} (x\numrel<y \bitbin\land
         \bit\lnot(x\numrel=y) \bitbin\land \bit\lnot(x\numrel>y)) \\*
      &\pheq \bitbin\lor (\bit\lnot(x\numrel<y)
         \bitbin\land x\numrel=y \bitbin\land \bit\lnot(x\numrel>y)) \\*
      &\pheq \bitbin\lor (\bit\lnot(x\numrel<y)
         \bitbin\land \bit\lnot(x\numrel=y) \bitbin\land x\numrel>y) 
   \end{align*}
   and similarly for $x1,y1$,
   \begin{align*}
      \phanbin{\bitbin\lor} (x0\numrel<y1 \bitbin\land
         \bit\lnot(x0\numrel=y1) \bitbin\land \bit\lnot(x0\numrel>y1)) \\*
      \bitbin\lor (\bit\lnot(x0\numrel<y1) \bitbin\land
         x0\numrel=y1 \bitbin\land \bit\lnot(x0\numrel>y1)) \\*
      \bitbin\lor (\bit\lnot(x0\numrel<y1) \bitbin\land
         \bit\lnot(x0\numrel=y1) \bitbin\land x0\numrel>y1)
      &= \phanbin{\bitbin\lor} (x\numrel\leq y \bitbin\land 1
         \bitbin\land \bit\lnot(x\numrel>y)) \\*
      &\pheq \bitbin\lor (\bit\lnot(x\numrel\leq y)
         \bitbin\land 0 \bitbin\land \bit\lnot(x\numrel>y)) \\*
      &\pheq \bitbin\lor (\bit\lnot(x\numrel\leq y)
         \bitbin\land 1 \bitbin\land x\numrel>y) \\
      &= \phanbin{\bitbin\lor} (x\numrel<y \bitbin\land
         \bit\lnot(x\numrel>y)) \\*
      &\pheq \bitbin\lor (x\numrel=y \bitbin\land
         \bit\lnot(x\numrel>y)) \\*
      &\pheq \bitbin\lor (\bit\lnot(x\numrel<y)
         \bitbin\land \bit\lnot(x\numrel=y) \bitbin\land x\numrel>y) 
   \end{align*}
   and similarly for $x1,y0$.
\end{proof}

\begin{corollary}
   \( \E \numrel\leq x \)
\end{corollary}
\begin{corollary}
   \( x\numrel=y = x\numrel\leq y\bitbin\land x\numrel\geq y \)
\end{corollary}

 {}Next, from the fact that
\( \fcn{lp}_0(0x,y) = \fcn{lp}_0(x,y) \ \lor
 \ \fcn{lp}_0(0x,y) = 0\cat\fcn{lp}_0(x,y) \)
 (which can easily be proved by cases depending on the length of
 $x\lchop y$), simple proofs by Derived Rule~\ref{derule:2NIND} show
 the following lemma.
\begin{lemma}
\label{lem:cat=n}\mbox{}\nopagebreak
   \begin{enumerate}
   \item
      \( x\numrel=y = 0x\numrel=y = x\numrel=0y = 0x\numrel=0y \)
   \item
      \( x\numrel<y = 0x\numrel<0y \)
   \item
      \( x\numrel<y = 0x\numrel<y = x\numrel<0y \)
   \end{enumerate}
\end{lemma}
 This lemma can be used, with a generalization of Derived
 Rule~\ref{derule:2NIND} to three variables, to show the following
 theorems and their corollaries (from Theorem~\ref{thm:=n}).
\begin{theorem}
\label{thm:=n<ntrans}
   \( x\numrel=y \land y\numrel<z \limp x\numrel<z \)
   \quad and \quad
   \( x\numrel=y \land y\numrel>z \limp x\numrel>z \)
\end{theorem}
\begin{corollary}
\label{cor:=nleqntrans}
   \( x\numrel=y \land y\numrel\leq z \limp x\numrel\leq z \)
   \quad and \quad
   \( x\numrel=y \land y\numrel\geq z \limp x\numrel\geq z \)
\end{corollary}
\begin{theorem}
\label{thm:<ntrans}
   \( x\numrel<y \land y\numrel<z \limp x\numrel<z \)
   \quad and \quad
   \( x\numrel>y \land y\numrel>z \limp x\numrel>z \)
\end{theorem}
\begin{corollary}
\label{cor:leqn<ntrans}
   \( x\numrel\leq y \land y\numrel<z \limp x\numrel<z \)
   \quad and \quad
   \( x\numrel\geq y \land y\numrel>z \limp x\numrel>z \)
\end{corollary}
\begin{corollary}
\label{cor:leqntrans}
   \( x\numrel\leq y \land y\numrel\leq z \limp x\numrel\leq z \)
   \quad and \quad
   \( x\numrel\geq y \land y\numrel\geq z \limp x\numrel\geq z \)
\end{corollary}

\subsubsection{On ``$|\mathord\cdot|$'' and ``$\numfcn{succ}$''}

 Now, we define the binary length function ``$|\mathord\cdot|$'' and
 the numerical successor function ``$\numfcn{succ}$'' as in $L_1$,
 and prove some of their basic properties.
\begin{definition}
   \begin{align*}
      \numfcnind[(tdef]{$\numfcn{succ}$}{}{}
      \numfcn{cuss} &= \lCRN\big[\fhead x\fdot
         \fcn{AND}(1\ldel x)\bC(\bit\lnot\lbit x,\lbit x)\big] \\
      \numfcn{succ} &= [\fhead x\fdot\numfcn{cuss}(0x)] 
      \numfcnind[)]{$\numfcn{succ}$}{}{}
   \end{align*}
\end{definition}
 Simple proofs by $\NIND$ show the following theorem (proving the
 relevant properties first for the auxiliary function $\numfcn{cuss}$,
 and then for $\numfcn{succ}$).
\begin{claim}
\label{claim:succcat}\mbox{}\nopagebreak
   \begin{align*}
      \numfcn{succ}(\E) &= 1 \\
      \numfcn{succ}(x0) &= 0x1 \\
      \numfcn{succ}(x1) &= \numfcn{succ}(x)\cat 0 
   \end{align*}
\end{claim}
 Using this theorem, a simple $\NIND$ will now prove the following
 properties.
\begin{claim}
\label{claim:catsucc}\mbox{}\nopagebreak
   \begin{align*}
      \numfcn{succ}(0x) &= 0\cat\numfcn{succ}(x) \\
      \numfcn{succ}(1x) &= \lbit\numfcn{succ}(x)\cat
         \bit\lnot\lbit\numfcn{succ}(x)\cat\ldel\numfcn{succ}(x) \\
      \lbit\numfcn{succ}(x) &= \fcn{AND}(1x) \\
      \numfcn{succ}(x) &= \fcn{AND}(1x)\bC
         \big(1\cat{}_0x,0\cat\ldel\numfcn{succ}(x)\big) 
   \end{align*}
\end{claim}
 Now, we can prove a few theorems involving $\numfcn{succ}$ together
 with some of the other numerical functions already defined.
\begin{theorem}
\label{thm:<nsucc>n}\mbox{}\nopagebreak
   \begin{enumerate}
   \item
      \( x \numrel< \numfcn{succ}(x) \)
   \item
      \( x\numrel>y = x\numrel\geq\numfcn{succ}(y) \)
   \end{enumerate}
\end{theorem}

\noindent
 The binary length function is defined in the same way as in $L_1$,
 as follows.
\begin{definition}
   \ind[tdef]{$|\mathord\cdot|$ (length)}{}{}
   \( |\mathord\cdot| = \STRN[{}_1,\fhead xv_\ell v_r\fdot
         x\elC(v_\ell\cat 0,v_\ell\cat 1)] \)
\end{definition}
 (To be consistent with previous notation, we will write ``$|x|$''
 instead of the more formal ``$||(x)$''.)
\begin{claim}
\label{claim:|_j|}
   \( |x| = |{}_jx| \)
\end{claim}

\begin{theorem}
   \ind{$|\mathord\cdot|$ (length)}{}{}
   \( {}_jx={}_jy \liff |x|=|y| \)
\end{theorem}
\begin{proof}
   One direction (\({}_jx={}_jy\limp|x|=|y|\)) is immediate from the
   preceding claim.  The other is proved by $\TIND$ on $y$ (with
   $h_\ell=\lhalf$ and $h_r=\rhalf$):
   \( |x|=|\E| \limp |x|=\E \limp x=\E \limp {}_jx={}_j\E \),
   \( |x|=|i| \limp |x|=1 \limp x=i' \limp {}_jx={}_ji \),
   and assuming that
   \( |x\lhalf|=|y\lhalf| \limp {}_jx\lhalf={}_jy\lhalf \) and
   \( |\rhalf x|=|\rhalf y| \limp {}_j\rhalf x={}_j\rhalf y \),
   we have that
   \begin{align*}
      |x|=|y|
      &\limp x\elC\big(|x\lhalf|\cat 0,|x\lhalf|\cat 1\big)
             = y\elC\big(|y\lhalf|\cat 0,|y\lhalf|\cat 1\big) \\
      &\limp \big( x\lhalf\lchop\rhalf x=\E \land
            y\lhalf\lchop\rhalf y=\E \land
            |x\lhalf|\cat 0=|y\lhalf|\cat 0 \big) \lor \\*
      &\phanbin{\limp} \big( x\lhalf\lchop\rhalf x\neq\E \land
            y\lhalf\lchop\rhalf y\neq\E \land
            |x\lhalf|\cat 1=|y\lhalf|\cat 1 \big) \\
      &\limp \big( x\lhalf\lchop\rhalf x=\E \land
            y\lhalf\lchop\rhalf y=\E \land
            {}_jx\lhalf={}_jy\lhalf \big) \lor \\*
      &\phanbin{\limp} \big(x\lhalf\lchop\rhalf x=i \land
            y\lhalf\lchop\rhalf y=i' \land
            {}_jx\lhalf={}_jy\lhalf \big) \\
      &\limp {}_jx={}_jy 
   \end{align*}
   (where the two cases for \(|x\lhalf|\cat 0=|y\lhalf|\cat 1\) and
   \(|x\lhalf|\cat 1=|y\lhalf|\cat 0\) were not included in the
   disjunction on the second and third lines since they are known to
   be false).
\end{proof}

 The following theorem can be proved with an easy $\TIND$ and its
 corollaries are immediate from previously proved theorems.
\begin{theorem}
\label{thm:succ||}
   \( |xi| \numrel= \numfcn{succ}(|x|) \)
\end{theorem}
\begin{corollary}
   \( |x| \numrel< |xi| \)
\end{corollary}
\begin{corollary}
   \( x\numrel>|y\rdel| \limp x\numrel\geq|y| \)
\end{corollary}

\subsubsection{On ``masking'' functions}

 In order to define binary addition, and to prove its properties, we
 will need ``masking'' functions like the ones that were defined in
 $L_1$.  We give their definition and basic properties here.

\begin{definition}
\mbox{}\nopagebreak
   \begin{align*}
      \fcnind[(tdef]{$\fcn{first}_0,\fcn{first}_1$}{}{}
      \fcn{first}_0
      &= \rCRN[\fhead x\fdot\fcn{AND}(1x\rdel)\bC(\bit\lnot x\rbit,0)] \\
      \fcn{first}_1
      &= \rCRN[\fhead x\fdot\fcn{OR}(x\rdel)\bC(0,x\rbit)] 
      \fcnind[)]{$\fcn{first}_0,\fcn{first}_1$}{}{}
   \end{align*}
\end{definition}
\begin{definition}
   \( \fcn{maskbit}
       = \big[\fhead xy\fdot\fcn{OR}(\bitfcn{and}_2(x,y))\big] \)
\end{definition}
\begin{definition}
   \( \fcn{delfirst}_1 = \big[\fhead x\fdot
      \bitfcn{and}_2\big(x,\bitfcn{not}(\fcn{first}_1(x))\big)\big] \)
\end{definition}
 The basic theorem below, as well as its corollary, can both be proved
 with a simple $\NIND$.
\begin{theorem}
\label{thm:firstj}\mbox{}\nopagebreak
   \begin{alignat*}{2}
      \fcnind[(]{$\fcn{first}_0,\fcn{first}_1$}{}{}
      \fcn{first}_0(0x) &= 1\cat{}_0x &\qquad
      \fcn{first}_0(1x) &= 0\cat\fcn{first}_0(x) \\
      \fcn{first}_1(0x) &= 0\cat\fcn{first}_1(x) &\qquad
      \fcn{first}_1(1x) &= 1\cat{}_0x 
      \fcnind[)]{$\fcn{first}_0,\fcn{first}_1$}{}{}
   \end{alignat*}
\end{theorem}
\begin{corollary}
\label{cor:first01}
   \( \fcn{first}_0(x) = \fcn{first}_1(\bitfcn{not}(x)) \)
\end{corollary}

\subsubsection{On binary addition}

 Before we define binary addition and prove its properties, let us
 make a remark about ``numerical'' functions.  If a formula $A$
 contains only terms made up of functions $f$ with the property that
\( f(x_1,\dots,x_m) = f\big(\fcn{lp}_0(x_1,\lenfcn{max}_m(\tpl x_m)),
   \dots,\fcn{lp}_0(x_m,\lenfcn{max}_m(\tpl x_m))\big) \)
 (which happens to be the case for the numerical functions), then
\( A[x_1,\dots,x_m] \liff
   \big({}_jx_1=\dots={}_jx_m\limp A[x_1,\dots,x_m]\big) \).
 Thus, we can use the following special form of Derived
 Rule~\ref{derule:2NIND} to prove any such formula $A$ (the rule is
 stated only for two variables but can easily be extended to more).

\begin{derule}
\label{derule:len2NIND}
   \( A[\E,\E], {}_jx={}_jy\land A[x,y]\limp
         A[0x,0y]\land A[0x,1y]\land A[1x,0y]\land A[1x,1y] \) \\
   \( \lded {}_jx={}_jy\limp A[x,y] \)
\end{derule}
 (The conclusion of the rule can easily be proved from the antecedent
 by a simple application of Derived Rule~\ref{derule:2NIND}.)

 {}Now, binary addition is defined just as in $L_1$, as follows.
\begin{definition}
   \numfcnind[tdef]{$\numfcn{carry}$}{}{}
   \( \numfcn{carry} = \lCRN_2\big[\fhead xy\fdot
      \fcn{maskbit}\big(\bitfcn{and}_2(x,y),
         \fcn{first}_0(\bitfcn{xor}_2(x,y))\big)\big] \)
\end{definition}
\begin{definition}
   \numind[tdef]{$\numbin+$}{}{}
   \( \numbin+ = \big[\fhead xy\fdot
      \bitfcn{xor}_3(\numfcn{carry}(x,y)\cat 0,x,y)\big] \)
\end{definition}
 (To make the notation consistent with previous usage, we will write
 ``$x\numbin+y$'' instead of the more formal ``$\numbin+(x,y)$''.)  The
 commutativity of ``$\numbin+$'' is a direct result of the commutativity
 of each function involved in its definition.
\begin{theorem}
\label{thm:+comm}
   \( x\numbin+y = y\numbin+x \)
\end{theorem}
 Proving the associativity of ``$\numbin+$'' will be slightly more
 complicated.  First, we relate the functions $\numbin+$ and
 $\numfcn{succ}$ through the following lemma and theorem.
\begin{lemma}
\label{lem:carry1}\mbox{}\nopagebreak
   \begin{align*}
      \numfcn{carry}(x0,1)
      &= \numfcn{carry}(x,\E)\cat 0 = {}_0x0 \\
      \numfcn{carry}(x1,1)
      &= x\zlC\big(1,\numfcn{carry}(x,1)\cat 1\big) 
   \end{align*}
\end{lemma}
\begin{theorem}
\label{thm:succ=+1}
   \( x\numbin+1 =
      x\zlC\big(0\cat\numfcn{succ}(x),\numfcn{succ}(x)\big)
      \numrel= \numfcn{succ}(x) \)
\end{theorem}
 Next, we can state certain facts about the carry function.
\begin{claim}
\label{claim:catcarry}
   For ${}_jx={}_jy$,
   \begin{align*}
      \numfcn{carry}(x,\E) &= {}_0x \\
      \numfcn{carry}(x,x) &= x \\
      \numfcn{carry}(0x,0y) &= 0\cat\numfcn{carry}(x,y) \\
      \numfcn{carry}(1x,0y)
      &= \lbit\numfcn{carry}(x,y)\cat\numfcn{carry}(x,y) \\
      \numfcn{carry}(1x,1y) &= 1\cat\numfcn{carry}(x,y) 
   \end{align*}
\end{claim}
 Note that we omitted the property
\( \numfcn{carry}(0x,1y) =
   \lbit\numfcn{carry}(x,y)\cat\numfcn{carry}(x,y) \)
 from this theorem since it follows directly by the commutativity of
 $\numfcn{carry}$.  This will be the case for many of the theorems and
 proofs about $\numbin+$ that we will now present: for the sake of
 brevity, we will omit statements and proofs that follow directly from
 previous ones by commutativity.  The following claim follows directly
 from the corresponding properties for $\numfcn{carry}$.
\begin{claim}
\label{claim:cat+}
   For ${}_jx={}_jy$,
   \begin{align*}
      x\numbin+\E &= 0x \\
      x\numbin+x &= x0 \\
      0x\numbin+0y &= 0\cat(x\numbin+y) \\
      1x\numbin+0y &= \lbit(x\numbin+y)\cat\bit\lnot\lbit(x\numbin+y)\cat
         \ldel(x\numbin+y) \\
      1x\numbin+1y &= 1\cat(x\numbin+y) 
   \end{align*}
\end{claim}
 Now, although we can use Claim~\ref{claim:cat+} to prove theorems
 about $\numbin+$ by Derived Rule~\ref{derule:len2NIND}, we will also
 have need of the following theorem further on.
\begin{claim}
\label{claim:+cat}\mbox{}\nopagebreak
   \begin{align*}
      x0\numbin+y0 &= (x\numbin+y)\cat 0 \\
      x1\numbin+y0 &= (x\numbin+y)\cat 1 \\
      x1\numbin+y1 &= \ldel\numfcn{succ}(x\numbin+y)\cat 0 
   \end{align*}
\end{claim}

 {}With the help of this theorem, we can now prove the following
 important properties of $\numbin+$ with a version of Derived
 Rule~\ref{derule:len2NIND} that concatenates bits to the right
 instead of to the left.

\pagebreak[2]

\begin{theorem}
\label{thm:+}\mbox{}\nopagebreak
   \begin{enumerate}
   \item
      \( x\numbin+\numfcn{succ}(y) \numrel= \numfcn{succ}(x\numbin+y) \)
   \item
      \( x\numbin+(y\numbin+z) \numrel= (x\numbin+y)\numbin+z \)
   \item
      \( y\numrel<z \liff x\numbin+y\numrel<x\numbin+z \)
   \item
      \( x\numrel=y \land z\numrel=w \limp x\numbin+z\numrel=y\numbin+w \)
   \item
      \( x\numrel=y \land z\numrel<w \limp x\numbin+z\numrel<y\numbin+w \)
   \item
      \( x\numrel<y \land z\numrel<w \limp x\numbin+z\numrel<y\numbin+w \)
   \end{enumerate}
\end{theorem}

\subsubsection{On iterated sums}

 The last functions we need to define are iterated sums, defined as in
 $L_1$ using Buss's ``carry-save'' technique.

\begin{definition}
\mbox{}\nopagebreak
   \begin{align*}
      \fcn{CScar}_3 &= \lCRN_3\big[\fhead x_1x_2x_3\fdot
         ((\lbit x_1\bitbin\land\lbit x_2)\bitbin\lor
          (\lbit x_2\bitbin\land\lbit x_3)\bitbin\lor
          (\lbit x_3\bitbin\land\lbit x_1))\big] \\
      \fcn{CSadd}_3 &= \big[\fhead x_1x_2x_3\fdot
         \bitfcn{xor}_3(0x_1,0x_2,0x_3)\big] \\
      \fcnind[(tdef]{$\fcn{CScar},\fcn{CSadd}$}{}{}
      \fcn{CScar} &= \big[\fhead x_1x_2x_3x_4\fdot
         \fcn{CScar}_3\big(\fcn{CScar}_3(x_1,x_2,x_3)\cat 0,
            \fcn{CSadd}_3(x_1,x_2,x_3),0x_4\big)\cat 0\big] \\
      \fcn{CSadd} &= \big[\fhead x_1x_2x_3x_4\fdot
         \fcn{CSadd}_3\big(\fcn{CScar}_3(x_1,x_2,x_3)\cat 0,
            \fcn{CSadd}_3(x_1,x_2,x_3),0x_4\big)\big] 
      \fcnind[)]{$\fcn{CScar},\fcn{CSadd}$}{}{}
   \end{align*}
\end{definition}
 The following properties are a direct consequence of these
 definitions.
\begin{claim}
\label{claim:CS}\mbox{}\nopagebreak
   \begin{gather*}
      \fcn{CScar}_3(x0,y0,z0) = \fcn{CScar}_3(x,y,z)\cat 0 \\
      \fcn{CScar}_3(x1,y0,z0) = \fcn{CScar}_3(x0,y1,z0)
       = \fcn{CScar}_3(x1,y0,z1) = \fcn{CScar}_3(x,y,z)\cat 0 \\
      \fcn{CScar}_3(x0,y1,z1) = \fcn{CScar}_3(x1,y0,z1)
       = \fcn{CScar}_3(x1,y1,z0) = \fcn{Cscar}_3(x,y,z)\cat 1 \\
      \fcn{CScar}_3(x1,y1,z1) = \fcn{CScar}_3(x,y,z)\cat 1 \\
      {}_0\fcn{CScar}_3(x,y,z)\cat 0 = {}_0\fcn{CSadd}_3(x,y,z)
       = 0\cat{}_0\lenfcn{max}_3(x,y,z)
       = 0\cat\lenfcn{max}_3({}_0x,{}_0y,{}_0z) \\
      0\cat\lenfcn{max}_3({}_0x,{}_0y,{}_0z)
       = \fcn{CScar}_3({}_0x,{}_0y,{}_0z)\cat 0
       = \fcn{CSadd}_3({}_0x,{}_0y,{}_0z) \\
      {}_0\fcn{CScar}(x,y,z,w) = {}_0\fcn{CSadd}(x,y,z,w)
       = 00\cat{}_0\lenfcn{max}_4(x,y,z,w) 
       = 00\cat\lenfcn{max}_4({}_0x,{}_0y,{}_0z,{}_0w) \\
      00\cat\lenfcn{max}_4({}_0x,{}_0y,{}_0z,{}_0w)
       = \fcn{CScar}({}_0x,{}_0y,{}_0z,{}_0w)
       = \fcn{CSadd}({}_0x,{}_0y,{}_0z,{}_0w) 
   \end{gather*}
\end{claim}

 {}We can now prove one main lemma and one main theorem about the
 ``carry-save'' addition functions.

\begin{lemma}
\label{lem:CSsucc}\mbox{}\nopagebreak
   \begin{multline*}
      (\fcn{CScar}_3(\numfcn{succ}(x),y,z)\cat 0)\numbin+
         \fcn{CSadd}_3(\numfcn{succ}(x),y,z) \\*
      \numrel= \numfcn{succ}\big((\fcn{CScar}_3(x,y,z)\cat 0)\numbin+
         \fcn{CSadd}_3(x,y,z)\big) 
   \end{multline*}
\end{lemma}
\begin{theorem}
\label{thm:CS+}
   \( \fcn{CScar}(x,y,z,w)\numbin+\fcn{CSadd}(x,y,z,w)
      \numrel= x\numbin+y\numbin+z\numbin+w \)
\end{theorem}

\noindent
 Finally, we can define the function ``$\fcn{sum}$'', that adds all
 the bits of its argument, as in $L_1$.
\begin{definition}
\mbox{}\nopagebreak
   \begin{align*}
      \fcnind[(tdef]{$\fcn{CAR},\fcn{ADD}$}{}{}
      \fcn{CARADD} &= \STRN\big[\fhead x\fdot{}_0x\cat x,
         \fhead xv_\ell v_r\fdot
         \fcn{CScar}(v_\ell\lhalf,\rhalf v_\ell,v_r\lhalf,\rhalf v_r)
         \cat
         \fcn{CSadd}(v_\ell\lhalf,\rhalf v_\ell,v_r\lhalf,\rhalf v_r)
         \big] \\
      \fcn{CAR} &= \big[\fhead x\fdot\fcn{CARADD}(x)\lhalf\big] \qquad
      \fcn{ADD} = \big[\fhead x\fdot\rhalf\fcn{CARADD}(x)\big] \\
      \fcnind[)]{$\fcn{CAR},\fcn{ADD}$}{}{}
      \fcnind[tdef]{$\fcn{sum}$}{}{}
      \fcn{sum} &= \big[\fhead x\fdot
         \fcn{CAR}(x)\numbin+\fcn{ADD}(x)\big] 
   \end{align*}
\end{definition}
 The following basic properties of $\fcn{CARADD}$ will be used to
 prove results about $\fcn{sum}$ and can be proved easily from
 previous theorems.
\begin{claim}
\label{claim:CARADD_0}\mbox{}\nopagebreak
   \begin{enumerate}
   \item
     \( \fcn{CARADD}({}_0x) \numrel= 0 \)
   \item
     \( \fcn{sum}({}_0x) = \fcn{CAR}({}_0x)\numbin+\fcn{ADD}({}_0x)
        \numrel= 0\numbin+0 \numrel= 0 \)
   \end{enumerate}
\end{claim}

\begin{theorem}
\label{thm:sumhalf}
   \( \fcn{sum}(x) \numrel=
      \fcn{sum}(x\lhalf)\numbin+\fcn{sum}(\rhalf x) \)
\end{theorem}
 From this theorem, it is possible to prove that
\( \fcn{sum}(xy) \numrel= \fcn{sum}(x)\numbin+\fcn{sum}(y) \)
 with a sequence of lemmas and theorems similar to the ones used to
 show that
\( \fcn{AND}(xy) = \fcn{AND}(x)\bitbin\land\fcn{AND}(y) \).
 In particular, we have that
\( \fcn{sum}(x0) \numrel= \fcn{sum}(x)\numbin+0
   \numrel= \fcn{sum}(x) \)
 and
\( \fcn{sum}(x1) \numrel= \fcn{sum}(x)\numbin+1
   \numrel= \numfcn{succ}(\fcn{sum}(x)) \).

 {}Simple proofs by $\NIND$ now show the following theorem.
\begin{theorem}
\label{thm:sum_1}
   \( \fcn{sum}(x) \numrel\leq \fcn{sum}({}_1x) \numrel= |x| \)
\end{theorem}

\section{Proving the pigeonhole principle in $T_1$}

 In this section, we will be working with the following form of the
 pigeonhole principle, denoted $\PHP_n(f)$ (or simply $\PHP$ when $n$
 and $f$ are clear from the context): ``no map
 \(f:\set{n+1}\goesto\set{n}\) is injective'', or equivalently ``if
 $f$ is a map from $\set{n+1}$ to $\set{n}$, then there exist
 \(i\neq j\in\set{n+1}\) such that \(f(i)=f(j)\)''.  (Note that we are
 using the common notation ``$\set{n}$'' to represent the set
 $\{1,2,\dots,n\}$, for any positive integer $n$, and in what follows,
 we will use the term \emph{map} to mean a (possibly) multi-valued
 function.)

 {}Informally, the proof of $\PHP$ goes as follows: Assume for a
 contradiction that $f$ is a map from $\set{n+1}$ to $\set{n}$ and
 that $f$ is injective (\ie, for every \(i\neq j\in \set{n+1}\),
 \(f(i)\neq f(j)\)).  Define
\[
   \symrm{count}(k,\ell) = \big|\{x\in\set{\ell}\st f(y)=x
      \text{ for some }y\in\set{k}\}\big|,
\]
 \ie, $\symrm{count}(k,\ell)$ is the number of elements in $\set{\ell}$
 mapped onto from elements in $\set{k}$ by $f$.  Then, the following
 facts are easy to prove.
\begin{enumerate}
\item\label{fact01}
   \(\symrm{count}(n+1,\ell)\leq\ell\) for any \(1\leq\ell\leq n\)
   (since there are $\ell$ elements in $\set{\ell}$).
\item\label{fact02}
   \(\symrm{count}(1,n)\geq 1\) (since \(f(1)\in\set{n}\)).
\item\label{fact03}
   \(\symrm{count}(k+1,n)>\symrm{count}(k,n)\)
   for \(1\leq k\leq n\) (since $f(k+1)$ must be different from
   $f(1),\dots,f(k)$ by the assumption that $f$ is injective).
\end{enumerate}
 Combining facts~\ref{fact02} and~\ref{fact03}, we get that
 \(\symrm{count}(k,n)\geq k\) for all \(1\leq k\leq n+1\).  But then,
 \(n+1\leq\symrm{count}(n+1,n)\leq n\), \ie, \(n+1\leq n\), which is a
 contradiction.  Hence, $\PHP$ is true.

 {}In the rest of this section, we will show how the informal proof
 given above can be formalized in $T_1$, in a top-down manner.  Also,
 we adopt the following notational convention: when a function is
 defined through an auxiliary function that is of no interest in
 itself, the name of the auxiliary function will consist of the
 function's name spelled backwards (\eg, we will define below a
 function ``$\fcn{map}$'' in terms of an auxiliary function
 ``$\fcn{pam}$'').

\subsection{Representation of $\PHP$ in $T_1$}

 The first step in the formalization will be to use the machinery
 given above to write down $L_1$-functions that define $\PHP$.
 Formally, \(\PHP=\PHP_n(f)\) depends on two parameters: $n$ and $f$;
 moreover, given $n$, $f$ can be described by an $[n\times(n+1)]$
 binary array whose $(i,j)$-th entry is equal to $1$ if \(i=f(j)\) and
 $0$ otherwise, \ie, each row corresponds to one hole and each column
 to one pigeon.  This is essentially the representation we will use to
 encode the problem.

 {}More precisely, given $n$, every bit string $a$ can be seen as
 encoding an $[n\times(n+1)]$ binary array (and therefore a partial
 map \(f_a^n:\set{n+1}\goesto\set{n}\)) by either padding $a$ on the
 right with $0$'s or chopping off enough bits from the right of $a$ so
 that its length is $n\cdot(n+1)$, and then reading the array in
 row-major order, so that the first $n+1$ bits of $a$ (from the left)
 represent the first row of the array, and so on.  For example, the
 string $1000110010$ represents at the same time
\[
   \text{the } [2\times 3] \text{ binary array }
      \begin{bmatrix}
         1 & 0 & 0 \\
         0 & 1 & 1 
      \end{bmatrix}, \qquad
   \text{the } [3\times 4] \text{ binary array }
      \begin{bmatrix}
         1 & 0 & 0 & 0 \\
         1 & 1 & 0 & 0 \\
         1 & 0 & 0 & 0 
      \end{bmatrix}, \qquad
   \text{etc.}
\]
 Then, every such $[n\times(n+1)]$ binary array $a$ represents a
 partial map \(f_a^n:\set{n+1} \goesto\set{n}\).

 {}Now, given a bit string $\una{n}$ such that \(n=|\una{n}|\), we can
 define a function $\fcn{adj}(a,\una{n})$ that ``adjusts'' the length
 of the bit string $a$ so that
 \(|\fcn{adj}(a,\una{n})|=|\una{n}|\cdot(|\una{n}|+1)\), \ie,
 $\fcn{adj}(a,\una{n})$ is exactly the $[n\times(n+1)]$ binary matrix
 encoded by $a$, written out in row-major order:
\[
   \fcn{adj}(a,\una{n}) = a\rchop((\una{n}1\smsh\una{n})\lchop a)
      \cat{}_0(a\lchop(\una{n}1\smsh\una{n})).
\]

 {}Next, given a column number $\una{k}$ and a row number $\una{l}$ in
 unary, we can easily define a function $\fcn{entry}$ that extracts a
 single entry (bit) of the matrix:
\[
   \fcn{entry}(a,\una{n},\una{k},\una{l}) = \big(
      ((\una{n}1\smsh\una{l}\rdel)\cat\una{k}\rdel)\lchop
      \fcn{adj}(a,\una{n})\big)\rbit.
\]
 Note that the value returned by this function is meaningless unless
 \(1\leq|\una{k}|=k\leq n+1\) and \(1\leq|\una{l}|=\ell\leq n\).  In
 order to simplify the presentation, we will implicitly assume that
 $k$ and $\ell$ fall within this range for the rest of the section,
 where \(k=|\una{k}|\) and \(\ell=|\una{l}|\) by convention (\ie,
 functions are implicitly defined by cases to be equal to $\E$ for
 values outside the meaningful range).

 {}Once we have the function $\fcn{entry}$, it is easy to define
 functions $\fcn{col}$ and $\fcn{row}$ that extract columns or rows of
 the matrix, by $\CRN$:
\begin{align*}
   \fcn{loc}(a,\una{n},\una{k},\una{l}) &=
      \fcn{loc}(a,\una{n},\una{k},\una{l}\rdel)\cat
      \fcn{entry}(a,\una{n},\una{k},\una{l})
      \quad \text{(for $\una{l}\neq\E$)} \\*
   \fcn{col}(a,\una{n},\una{k}) &=
      \fcn{loc}(a,\una{n},\una{k},\una{n}); \\
   \fcn{wor}(a,\una{n},\una{k},\una{l}) &=
      \fcn{wor}(a,\una{n},\una{k}\rdel,\una{l})\cat
      \fcn{entry}(a,\una{n},\una{k},\una{l})
      \quad \text{(for $\una{k}\neq\E$)} \\*
   \fcn{row}(a,\una{n},\una{l}) &=
      \fcn{wor}(a,\una{n},\una{n}1,\una{l}). 
\end{align*}
 So that $\fcn{col}(a,\una{n},\una{k})$ is the $k$-th column of $a$
 and $\fcn{row}(a,\una{n},\una{l})$ is the $\ell$-th row of $a$.
 Moreover, it is easy to prove in $T_1$ that
\begin{align*}
   \fcn{lc}(\fcn{col}(a,\una{n},\una{k}),\una{l}) &=
      \fcn{loc}(a,\una{n},\una{k},\una{l}), \\
   \fcn{lc}(\fcn{row}(a,\una{n},\una{l}),\una{k}) &=
      \fcn{wor}(a,\una{n},\una{k},\una{l}), 
\end{align*}
 directly from the properties of $\CRN$.

\noindent
 Using these functions, we can now define two functions needed to
 represent $\PHP_n(f)$:
\begin{align*}
   \fcn{map}(a,\una{n}) &=
      \begin{cases}
         1 & \text{if $f_a^n$ is a map, \ie, every column of $a$
                contains at least one $1$}, \\
         0 & \text{otherwise}; 
      \end{cases} \\
   \fcn{inj}(a,\una{n}) &=
      \begin{cases}
         1 & \text{if $f_a^n$ is injective, \ie, every row of $a$
                contains at most one $1$}, \\
         0 & \text{otherwise}. 
      \end{cases} 
\end{align*}
 To compute $\fcn{map}$ (resp.\ $\fcn{inj}$), we will first define a
 function $\fcn{pam}$ (resp.\ $\fcn{jni}$) that returns a bit string
 with one bit for each column (resp.\ row) of $a$ indicating whether
 the constraint is satisfied or not for that column (resp.\ row);
 then, we simply take the conjunction of all the bits to get the
 answer:
\begin{align*}
   \fcn{pam}(a,\una{n},x) &=
      \fcn{pam}(a,\una{n},x\rdel)\cdot
      \fcn{OR}\big(\fcn{col}(a,\una{n},x)\big)
      \quad \text{(for $x\neq\E$)} \\*
   \fcn{map}(a,\una{n}) &=
      \fcn{AND}\big(\fcn{pam}(a,\una{n},\una{n}1)\big), \\
   \fcn{jni}(a,\una{n},x) &=
      \fcn{jni}(a,\una{n},x\rdel)\cdot\bit\lnot\fcn{OR}
      \big(\fcn{delfirst}_1(\fcn{row}(a,\una{n},x))\big)
      \quad \text{(for $x\neq\E$)} \\*
   \fcn{inj}(a,\una{n}) &=
      \fcn{AND}\big(\fcn{jni}(a,\una{n},\una{n})\big). 
\end{align*}
 Finally, we can easily define a function that represents $\PHP$:
\begin{equation*}
   \fcn{php}(a,\una{n}) = \fcn{map}(a,\una{n})\bitbin\limp
      \bit\lnot\fcn{inj}(a,\una{n}).
\end{equation*}

\subsection{The $T_1$-proof of $\PHP$}

 First, let us define the function
 $\fcn{count}(a,\una{n},\una{k},\una{l})$, which returns the number of
 elements from $\set{\ell}$ that are mapped onto by elements from
 $\set{k}$ according to $f_a^n$.  We do this by first defining
 $\fcn{tnuoc}(a,\una{n},\una{k},\una{l})$, which returns a string of
 $\ell$ bits, one for each of the first $\ell$ rows of $a$, where bit
 $j$ is set to $1$ iff row $j$ contains at least one $1$ in the first
 $k$ columns:
\begin{equation*}
   \fcn{tnuoc}(a,\una{n},\una{k},\una{l}) =
      \fcn{tnuoc}(a,\una{n},\una{k},\una{l}\rdel)\cat
      \fcn{OR}\big(\fcn{wor}(a,\una{n},\una{k},\una{l})\big)
      \quad \text{(for $\una{l}\neq\E$)} 
\end{equation*}
 Then, \(\fcn{count}(a,\una{n},\una{k},\una{l})=
   \fcn{sum}\big(\fcn{tnuoc}(a,\una{n},\una{k},\una{l})\big)\).
 We can depict the situation as follows, where we have represented the
 submatrix of $a$ consisting of the first $k$ columns for each of the
 first $\ell$ rows, and where the value of
 $\fcn{tnuoc}(a,\una{n},\una{k},\una{l})$ can be read bit by bit, one
 for each row:
\[
   \begin{array}{r@{}ccl}
       & \overbrace{\hphantom{\begin{array}{ccccc}
            1&1&1&\dotsm&1\end{array}}}^{\textstyle k}
       & \quad
       & \fcn{tnuoc}(a,\una{n},\una{k},\una{l}): \\
      \ell\left\{\vphantom{\begin{array}{c}1\\1\\\vdots\\1\end{array}}
         \right.
       & \begin{array}{ccccc}
            1 & 0 & 0 & \dotsm & 1 \\
            0 & 0 & 0 & \dotsm & 0 \\
            \vdots & \vdots & \vdots & \vdots & \vdots \\
            0 & 0 & 0 & \dotsm & 1 
         \end{array}
       & \begin{array}{c}
            \longrightarrow \\
            \longrightarrow \\
            \vdots \\
            \longrightarrow 
         \end{array}
       & \begin{array}{ccl}
            1 & = & \fcn{OR}(\fcn{wor}(a,\una{n},\una{k},1)) \\
            0 & = & \fcn{OR}(\fcn{wor}(a,\una{n},\una{k},11)) \\
            \vdots & \vdots & \quad\vdots \\
            1 & = & \fcn{OR}(\fcn{wor}(a,\una{n},\una{k},\una{l})) 
         \end{array} 
   \end{array}
\]

 {}Now, we give a formalization in $T_1$ of the high-level proof of
 $\PHP$ outlined at the beginning of this section.  To make the
 notation easier to read, all theorems are conditional to the fact
 that $k$ and $\ell$ are within meaningful range.

 {}Recall the general outline of the proof: under the assumption that
 \(\bit\lnot\fcn{php}(a,\una{n})\), \ie, that \(\fcn{map}(a,\una{n})\)
 and \(\fcn{inj}(a,\una{n})\), it is possible to prove the following
 two facts.
\begin{fact}
\label{fact1}
   \( \fcn{count}(a,\una{n},\una{n}1,\una{l}) \numrel\leq \ell \)
   \quad (where \(\ell=|\una{l}|\), by convention)
\end{fact}
\begin{fact}
\label{fact2}
   \( k \numrel\leq \fcn{count}(a,\una{n},\una{k},\una{n}) \)
   \quad (where \(k=|\una{k}|\), by convention)
\end{fact}
 Then, we get that
 \(|\una{n}1|\numrel\leq\fcn{count}(a,\una{n},\una{n}1,\una{n})
   \numrel\leq|\una{n}|\),
 so that \(|\una{n}1|\numrel\leq|\una{n}|\) (by transitivity of
 $\numrel\leq$).  But since we know that
 \(\bit\lnot\big(|x1|\numrel\leq|x|\big)\),
 we get \(\fcn{php}(a,\una{n})\) by contradiction.

 {}Now, we can prove fact~\ref{fact1}:
\[ \fcn{count}(a,\una{n},\una{n}1,\una{l})
    = \fcn{sum}\big(\fcn{tnuoc}(a,\una{n},\una{n}1,\una{l})\big)
   \numrel\leq \big|\fcn{tnuoc}(a,\una{n},\una{n}1,\una{l})\big|
   \numrel\leq |\una{l}| = \ell
\]
 since \(\fcn{sum}(x)\numrel\leq|x|\) for any string $x$ and
 \(\big|\rCRN[h](x,\tpl y)\big|=|x|\) for any $L_1$-function $h$ and
 any strings $x$, $\tpl y$.

 {}To prove fact~\ref{fact2}, we will first show that it is possible
 to prove the following two facts (corresponding to facts~\ref{fact02}
 and~\ref{fact03} in the informal proof).
\begin{fact}
\label{fact2.1}
   \( \fcn{count}(a,\una{n},\E,\una{n}) \numrel= 0 \)
\end{fact}
\begin{fact}
\label{fact2.2}
   \( \fcn{count}(a,\una{n},\una{k},\una{n}) \numrel>
      \fcn{count}(a,\una{n},\una{k}\rdel,\una{n}) \)
\end{fact}
 Then, we can use $\NIND$ to show that
 \(\fcn{count}(a,\una{n},\una{k},\una{n})\numrel\geq|\una{k}|\):
 \(\fcn{count}(a,\una{n},\E,\una{n})\numrel\geq|\E|\)
 by fact~\ref{fact2.1} and
 \(\fcn{count}(a,\una{n},\una{k},\una{n})\numrel>
   \fcn{count}(a,\una{n},\una{k}\rdel,\una{n})\numrel\geq
   |\una{k}\rdel|\)
 by fact~\ref{fact2.2} and the induction hypothesis, so that
 \(\fcn{count}(a,\una{n},\una{k},\una{n})\numrel\geq|\una{k}|\).

 {}Next, to prove fact~\ref{fact2.1}, we use $\NIND$ together with the
 fact that \(\fcn{wor}(a,\una{n},\E,\una{l})=\E\) (by definition) to
 conclude that
 \(\fcn{tnuoc}(a,\una{n},\E,\una{l})={}_0\una{l}\):
 \(\fcn{tnuoc}(a,\una{n},\E,\E)=\E={}_0\E\),
 and assuming that
 \(\fcn{tnuoc}(a,\una{n},\E,\una{l}\rdel)={}_0(\una{l}\rdel)\),
 then
 \(\fcn{tnuoc}(a,\una{n},\E,\una{l})=
   \fcn{tnuoc}(a,\una{n},\E,\una{l}\rdel)\cat
      \fcn{OR}\big(\fcn{wor}(a,\una{n},\E,\una{l})\big)=
   {}_0(\una{l}\rdel)\cat\fcn{OR}(\E)=
   {}_0(\una{l}\rdel)\cat 0={}_0(\una{l})\).
 Finally, we use the fact that \(\fcn{sum}({}_0x)\numrel=0\) for any
 string $x$ to conclude that fact~\ref{fact2.1} holds.

 {}To prove fact~\ref{fact2.2}, we have to show that
 \(\fcn{sum}\big(\fcn{tnuoc}(a,\una{n},\una{k},\una{n})\big)
   \numrel>\fcn{sum}\big(\fcn{tnuoc}(a,\una{n},\una{k}\rdel,\una{n})
   \big)\).
 Intuitively, this will be true iff
 $\fcn{tnuoc}(a,\una{n},\una{k},\una{n})$ contains more $1$'s than
 $\fcn{tnuoc}(a,\una{n},\una{k}\rdel,\una{n})$.  Formally, for any two
 strings $x$ and $y$, the term
 \(\fcn{AND}\big(\bitfcn{or}_2(\bitfcn{not}(x),y)\big)\)
 expresses the fact that $y$ has a $1$ in every position where $x$ has
 a $1$, and the term
 \(\fcn{OR}\big(\bitfcn{and}_2(\bitfcn{not}(x),y)\big)\)
 expresses the fact that there is a position where $x$ has a $0$ but
 $y$ has a $1$.  Now, since we can prove that
 \(\fcn{AND}\big(\bitfcn{or}_2(\bitfcn{not}(x),y)\big)\bitbin\land
   \fcn{OR}\big(\bitfcn{and}_2(\bitfcn{not}(x),y)\big)\bitbin\limp
   \fcn{sum}(x)\numrel<\fcn{sum}(y)\)
 (the proof is given in Appendix~\ref{app:proofs}), we only need to
 show the following facts to complete the proof of fact~\ref{fact2.2}.
\begin{fact}
\label{fact2.2.1}
   \( \fcn{AND}\big(\bitfcn{or}_2\big(
      \bitfcn{not}(\fcn{tnuoc}(a,\una{n},\una{k}\rdel,\una{n})),
      \fcn{tnuoc}(a,\una{n},\una{k},\una{n})\big)\big) \)
\end{fact}
\begin{fact}
\label{fact2.2.2}
   \( \fcn{OR}\big(\bitfcn{and}_2\big(
      \bitfcn{not}(\fcn{tnuoc}(a,\una{n},\una{k}\rdel,\una{n})),
      \fcn{tnuoc}(a,\una{n},\una{k},\una{n})\big)\big) \)
\end{fact}

 {}It is relatively easy to prove fact~\ref{fact2.2.1} by $\NIND$ on
 the last argument:
\( \fcn{tnuoc}(a,\una{n},\una{k}\rdel,1) =
   \fcn{OR}(\fcn{wor}(a,\una{n},\una{k}\rdel,1)) \) and
\( \fcn{tnuoc}(a,\una{n},\una{k},1) =
   \fcn{OR}(\fcn{wor}(a,\una{n},\una{k},1)) \) so
\begin{align*}
   & \fcn{AND}\big(\bitfcn{or}_2\big(
      \bitfcn{not}(\fcn{tnuoc}(a,\una{n},\una{k}\rdel,\E)),
      \fcn{tnuoc}(a,\una{n},\una{k},\E)\big)\big) \\
   &\ = \bit\lnot\fcn{OR}(\fcn{wor}(a,\una{n},\una{k}\rdel,1))
        \bitbin\lor\fcn{OR}(\fcn{wor}(a,\una{n},\una{k},1)) \\
   &\ = \bit\lnot\fcn{OR}(\fcn{wor}(a,\una{n},\una{k}\rdel,\una{l}))
         \bitbin\lor\fcn{OR}\big(\fcn{wor}(a,\una{n},\una{k}\rdel,\una{l})
         \cat\fcn{entry}(a,\una{n},\una{k},\una{l})\big) \\
   &\ = \bit\lnot\fcn{OR}(\fcn{wor}(a,\una{n},\una{k}\rdel,\una{l}))
         \bitbin\lor\fcn{OR}(\fcn{wor}(a,\una{n},\una{k}\rdel,\una{l}))
         \bitbin\lor\fcn{entry}(a,\una{n},\una{k},\una{l}) \\
   &\ = 1\bitbin\lor\fcn{entry}(a,\una{n},\una{k},\una{l}) = 1. 
\end{align*}
 Also, by the definition of $\fcn{tnuoc}$, we get that
\begin{align*}
   & \fcn{AND}\big(\bitfcn{or}_2\big(
      \bitfcn{not}(\fcn{tnuoc}(a,\una{n},\una{k}\rdel,\una{l})),
      \fcn{tnuoc}(a,\una{n},\una{k},\una{l})\big)\big) \\
   &\ = \fcn{AND}\Big(\bitfcn{or}_2\Big(\bitfcn{not}\big(
      \fcn{tnuoc}(a,\una{n},\una{k}\rdel,\una{l}\rdel)\cat
         \fcn{OR}(\fcn{wor}(a,\una{n},\una{k}\rdel,\una{l}))\big),
      \Big.\Big.\\*
   &\ \hphantom{= \fcn{AND}\Big(\bitfcn{or}_2\Big(\Big.\Big.}
      \Big.\Big.\fcn{tnuoc}(a,\una{n},\una{k},\una{l}\rdel)\cat
         \fcn{OR}(\fcn{wor}(a,\una{n},\una{k},\una{l}))\Big)\Big) \\
   &\ = \fcn{AND}\big(\bitfcn{or}_2\big(\bitfcn{not}(
            \fcn{tnuoc}(a,\una{n},\una{k}\rdel,\una{l}\rdel)),
         \fcn{tnuoc}(a,\una{n},\una{k},\una{l}\rdel)\big)\big) \\*
   &\ \hphantom{ = }\ \bitbin\land\big(\bit\lnot
         \fcn{OR}(\fcn{wor}(a,\una{n},\una{k}\rdel,\una{l}))\bitbin\lor
         \fcn{OR}(\fcn{wor}(a,\una{n},\una{k},\una{l}))\big) \\
   &\ = 1\bitbin\land\big(\bit\lnot
         \fcn{OR}(\fcn{wor}(a,\una{n},\una{k}\rdel,\una{l}))\bitbin\lor
         \fcn{OR}(\fcn{wor}(a,\una{n},\una{k},\una{l}))\big) = 1 
\end{align*}
 (where the third equality holds by the induction hypothesis).

 {}The proof of fact~\ref{fact2.2.2} is the most involved so far.
 First, by the definition of $\fcn{tnuoc}$ and properties of $\CRN$,
 we know that
\[
   \fcn{tnuoc}(a,\una{n},\una{k},\una{n}) =
      \fcn{tnuoc}(a,\una{n},\una{k},\una{l}\rdel)\cat
      \fcn{OR}\big(\fcn{wor}(a,\una{n},\una{k},\una{l})\big)\cat
      \big(\una{l}\lchop\fcn{tnuoc}(a,\una{n},\una{k},\una{n})\big).
\]
 Because
 \(\fcn{OR}(xy)=\fcn{OR}(x)\bitbin\lor\fcn{OR}(y)\)
 and
 \(\bitfcn{and}_2(xy,wz)=
   \bitfcn{and}_2(x,w)\cat\bitfcn{and}_2(y,z)\)
 when \(|x|=|w|\) and \(|y|=|z|\), we get easy proofs in $T_1$ that
\begin{gather*}
   \bit\lnot\fcn{OR}\big(\fcn{wor}(a,\una{n},\una{k}\rdel,\una{l})\big)
   \bitbin\land\fcn{OR}\big(\fcn{wor}(a,\una{n},\una{k},\una{l})\big)
   \bitbin\limp \\
   \fcn{OR}\big(\bitfcn{and}_2\big(\bitfcn{not}
      (\fcn{tnuoc}(a,\una{n},\una{k}\rdel,\una{n})),
      \fcn{tnuoc}(a,\una{n},\una{k},\una{n})\big)\big).
\end{gather*}
 Hence, we can prove fact~\ref{fact2.2.2} by showing that there must
 exist some value $\una{l}$ for which
 \(\bit\lnot\fcn{OR}\big(\fcn{wor}(a,\una{n},\una{k}\rdel,\una{l})\big)
   \bitbin\land\fcn{OR}\big(\fcn{wor}(a,\una{n},\una{k},\una{l})
   \big)\).
 Now, we can prove that
 \(\bit\lnot\fcn{OR}(\fcn{delfirst}_1(x))\bitbin\land\fcn{lb}(x,y)
   \bitbin\limp\bit\lnot\fcn{OR}(\fcn{lc}(x,y\rdel))\)
 (see Appendix~\ref{app:proofs}), and since
 \(\fcn{wor}(a,\una{n},\una{k}\rdel,\una{l})=
   \fcn{lc}(\fcn{row}(a,\una{n},\una{l}),\una{k}\rdel)\)
 and
 \(\bit\lnot\fcn{OR}\big(\fcn{delfirst}_1
   (\fcn{row}(a,\una{n},\una{l}))\big)\)
 by the assumption that $\fcn{inj}(a,\una{n})$, we only need to show
 that there is some $\una{l}$ for which
 \(\fcn{lb}\big(\fcn{row}(a,\una{n},\una{l}),\una{k}\big)\),
 which is equivalent to showing that there is a value of $\una{l}$ for
 which \(\fcn{entry}(a,\una{n},\una{k},\una{l})\).

 {}Unfortunately, we do not have quantifiers to reason with so to show
 the existence of $\una{l}$, we have to construct it explicitly, \ie,
 to define a function $\fcn{pos}(a,\una{n},\una{k})$ that gives the
 value of $\una{l}$.  Because all functions definable in $T_1$ are
 length-determined, $\fcn{pos}$ will have to return a \emph{bitmask}
 to the position of $\una{l}$, and this bitmask cannot be used
 directly with the current definition of $\fcn{row}$ to prove what we
 need.  So, we will define an alternate function $\fcn{mrow}$ whose
 last argument is a bitmask instead of a unary string and for which we
 can show
 \(\fcn{row}(a,\una{n},\una{l})=\fcn{mrow}\big(a,\una{n},
      (1\cat{}_0(\una{l}\lchop\una{n}1))\rdel\big)\):
\begin{align*}
   \fcn{mwor}(a,\una{n},\una{k},m) &=
      \fcn{mwor}(a,\una{n},\una{k}\rdel,m)\cat
      \numfcn{maskbit}(\fcn{col}(a,\una{n},\una{k}),m)
      \quad \text{(for $\una{k}\neq\E$)} \\*
   \fcn{mrow}(a,\una{n},m) &= \fcn{mwor}(a,\una{n},\una{n}1,m) 
\end{align*}
 With this definition, $\fcn{pos}$ can easily be defined as
 \(\fcn{pos}(a,\una{n},\una{k})=
   \fcn{first}_1(\fcn{col}(a,\una{n},\una{k}))\).
 By the assumption that $\fcn{map}(a,\una{n})$, we know that
 \(\fcn{OR}(\fcn{col}(a,\una{n},\una{k}))\)
 and since we can prove that
 \(\fcn{OR}(x)\bitbin\liff\numfcn{maskbit}(x,\fcn{first}_1(x))\)
 in $T_1$ (see Appendix~\ref{app:proofs}), we get that
 \(\numfcn{maskbit}\big(\fcn{col}(a,\una{n},\una{k}),
      \fcn{pos}(a,\una{n},\una{k})\big)\),
 which implies immediately that
 \(\fcn{lb}\big(\fcn{mrow}(a,\una{n},
      \fcn{pos}(a,\una{n},\una{k})),\una{k}\big)\).

 {}Now, because $\fcn{row}$ and $\fcn{mrow}$ are both defined by
 $\CRN$ on the same parameter $\una{k}$, it is sufficient to show that
 \(\fcn{entry}(a,\una{n},\una{k},\una{l})=
   \numfcn{maskbit}\big(\fcn{col}(a,\una{n},\una{k}),
      (1\cat{}_0(\una{l}\lchop\una{n}1))\rdel\big)\)
 in order to prove that
 \(\fcn{row}(a,\una{n},\una{l})=\fcn{mrow}\big(a,\una{n},
      (1\cat{}_0(\una{l}\lchop\una{n}1))\rdel\big)\).
 And because we can prove in $T_1$ that
 \(\bitfcn{lb}(x,y)=
   \numfcn{maskbit}\big(x,(1\cat{}_0(y\lchop x1))\rdel\big)\)
 (see Appendix~\ref{app:proofs}), this is equivalent to showing that
 \(\fcn{entry}(a,\una{n},\una{k},\una{l})=
   \fcn{lb}(\fcn{col}(a,\una{n},\una{k}),\una{l})=
   \fcn{loc}(a,\una{n},\una{k},\una{l})\rbit\),
 a fact which is immediate by the definition of $\fcn{loc}$.

\noindent
 Finally, we can redefine $\fcn{tnuoc}$ using $\fcn{mwor}$ instead
 of $\fcn{wor}$, as follows:
\[
   \fcn{mtnuoc}(a,\una{n},\una{k},\una{l}) =
      \fcn{mtnuoc}(a,\una{n},\una{k},\una{l}\rdel)\cat
      \fcn{OR}\big(\fcn{mwor}\big(a,\una{n},\una{k},
         (1\cat{}_0(\una{l}\lchop\una{n}1))\rdel\big)\big)
      \quad \text{(for $\una{l}\neq\E$)} 
\]
 and using the reasoning given above, it is possible to prove that
\[
   \fcn{OR}\big(\bitfcn{and}_2\big(
   \bitfcn{not}(\fcn{mtnuoc}(a,\una{n},\una{k}\rdel,\una{n})),
   \fcn{mtnuoc}(a,\una{n},\una{k},\una{n})\big)\big).
\]
 Moreover, because of the equivalence between $\fcn{wor}$ and
 $\fcn{mwor}$ given above, this implies the corresponding result for
 $\fcn{tnuoc}$, which completes the proof.

\begin{remark}
   Based on the $T_1$-proof of the pigeonhole principle just given,
   it should be possible to prove other, similar combinatorial
   statements in $T_1$.  One example is Tutte's theorem, which states
   that a graph has no perfect matching iff it satisfies a certain
   simple form of decomposition.  This would give an alternative proof
   that ``perfect matching'' tautologies have short $\F$-proofs, and
   maybe provide a more precise estimate of the size of these proofs
   by the results of Chapter~\ref{chap:polyproofs}.  (The ``perfect
   matching'' tautologies were first discussed in a paper by
   Impagliazzo, Pitassi, and Urquhart \cite{ImPiUr94}, where it was
   shown that they have polysize $\F$-proofs---note that those proofs
   were \emph{non}-uniform, unlike the proofs we would obtain through
   $T_1$.)
\end{remark}

%% file: chapter4.tex
\chapter{Theorems of $T_1$ Have Polysize $\F$-proofs}
\label{chap:polyproofs}

 For every term $t$ of $T_1$ with free variables $x_1,\dots,x_k$, we
 define a \emph{length function} $\tlen_t(m_1,\dots,m_k)$ that gives
 the exact length of $t$ as a function of the lengths of
 $x_1,\dots,x_k$ (this function is well-defined because functions in
 $L_1$ are length-determined).  Then, we define a family of
 propositional \emph{term formulas}
 $\atom{t}^{\tpl m}_1,\dots,\atom{t}^{\tpl m}_{\tlen_t(\tpl m)}$
 that describe the bits of $t$ in terms of the bits of $x_1,\dots,x_k$
 (where \(\atom{t}^{\tpl m}_1\) describes the leftmost bit of $t$),
 \ie, given any truth-value assignment to the atoms representing the
 bits of $x_1,\dots,x_n$, the truth value of \(\atom{t}^{\tpl m}_i\)
 represents the correct value for bit number $i$ of term $t$.
 Finally, for any formula $A$ of $T_1$, we define a family of
 \emph{propositional translations} \(\form{A}^{\tpl m}\), where
 $\tpl m$ lists the lengths of all free variables in $A$, and show
 that there are short $\F$-proofs of \(\form{A}^{\tpl m}\) whenever
 $A$ is a theorem of $T_1$.

\section{Length functions}

 The length functions are defined inductively as follows (where
 ``$\operatorname{sg}$'' is the signum function).
\addtocounter{footnote}{1}
\footnotetext
  {Where $\tpl m_i$ represents the lengths of the variables that occur
   in $t_i$.}
\addtocounter{footnote}{-1}
\begin{align*}
   \tlen_{x}(m) = &\ m \\
   \tlen_{f(t_1,\dots,t_k)}(\tpl m) = &\ 
      \tlen_{f(x_1,\dots,x_k)}
      \bigl(\tlen_{t_1}(\tpl m_1),\dots,\tlen_{t_k}(\tpl m_k)\bigr)
      \footnotemark \\*
   &\ \text{( where $x_1,\dots,x_k$ occur in none of
      $t_1,\dots,t_k$ )} \\
   \tlen_\E = &\ \rlap{$0$}\hspace{1.25in}\clap{\(\tlen_0 = 1\)}
      \hspace{1.25in}\llap{$\tlen_1$} = 1 \\
   \tlen_{{}_0x}(m) = &\ \rlap{$m$}\hspace{2.5in}
      \llap{$\tlen_{{}_1x}(m)$} = m \\
   \tlen_{x\lhalf}(m) = &\ \rlap{$\lfloor m/2\rfloor$}\hspace{2.5in}
      \llap{$\tlen_{\rhalf x}(m)$} = \lceil m/2\rceil \\
   \tlen_{x\rchop y}(m,n) = &\ \rlap{$m\dotminus n$}\hspace{2.5in}
      \llap{$\tlen_{y\lchop x}(n,m)$} = m\dotminus n \\
   \tlen_{x\cat y}(m,n) = &\ m+n \\
   \tlen_{x\C(y,z_0,z_1)}(m,n,p_0,p_1) = &\ \symbf{if } m=0
      \symbf{ then } n \symbf{ else } \max\{p_0,p_1\} \\
   \tlen_{[\fhead\tpl x\fdot t](\tpl y)}(\tpl n) = &\ 
      \tlen_{t[\tpl y/\tpl x]}(\tpl n')\footnotemark \\ 
   \tlen_{\lCRN[h](x,\tpl y)}(m,\tpl n) = &\ \rlap{$m$}\hspace{2.5in}
      \llap{$\tlen_{\rCRN[h](x,\tpl y)}(m,\tpl n)$} = m \\
   \tlen_{\TRN[g,h,h_\ell,h_r](x,z,\tpl y)}(m,p,\tpl n) = &\ 
      \symbf{if } m\leq 1 \symbf{ then }
      \tlen_{g(x,z,\tpl y)}(m,p,\tpl n) \symbf{ else} \\*
   &\ \tlen_{h(x,z,\tpl y,v_0,v_1)}\Big(m,p,\tpl n,
      \tlen_{\TRN[g,h,h_\ell,h_r](x\lhalf,h_\ell(z),\tpl y)}
         (m,p,\tpl n), \\*
   &\ \hphantom{\tlen_{h(x,z,\tpl y,v_0,v_1)}\Big(m,p,\tpl n, }\:
      \tlen_{\TRN[g,h,h_\ell,h_r](\rhalf x,h_r(z),\tpl y)}
         (m,p,\tpl n)\Big) 
\end{align*}
\footnotetext
  {Where $\tpl n'$ represents the lengths of the variables from
   $\tpl y$ that actually occur in $t[\tpl y/\tpl x]$.}
\vspace*{-\baselineskip}

\section{Term formulas}

 To every variable $x$ of $T_1$ are associated propositional atoms
 \(\atom{x}^m_1,\dots,\atom{x}^m_m\).  For other terms of $T_1$, the
 term formulas are defined inductively as follows.  (When subscript
 $i$ is used without specifying its range in the definition of
 \(\atom{t}^{\tpl m}_i\), it is implicitly assumed that
 \(1\leq i\leq\tlen_t(\tpl m)\).)
\begin{align*}
   \atom{f(t_1,\dots,t_k)}^{\tpl m}_i = &\ 
      \atom{f(x_1,\dots,x_k)}
      ^{\tlen_{t_1}(\tpl m_1),\dots,\tlen_{t_k}(\tpl m_k)}_i
      \Bigl[\atom{t_j}^{\tpl m_j}_{i_j} /
      \atom{x_j}^{\tlen_{t_j}(\tpl m_j)}_{i_j}\Bigr]_{\substack
        {1\leq j\leq k \\ 1\leq i_j\leq \tlen_{t_j}(\tpl m_j)}} \\[-.25ex]
   &\ \text
      {( where $x_1,\dots,x_k$ occur in none of $t_1,\dots,t_k$ )} 
\end{align*}
\begin{align*}
   \atom{0}_1 = &\ \rlap{\(\lfalse\)}\hspace{2.5in}
      \llap{\(\atom{1}_1\)} = \ltrue \\
   \atom{{}_0x}^m_i = &\ \rlap{\(\lfalse\)}\hspace{2.5in}
      \llap{\(\atom{{}_1x}^m_i\)} = \ltrue \\
   \atom{x\lhalf}^m_i = &\ \rlap{\(\atom{x}^m_i\)}\hspace{2.5in}
      \llap{\(\atom{\rhalf x}^m_i\)} =
         \atom{x}^m_{i+\lfloor m/2\rfloor} \\
   \atom{x\rchop y}^{m,n}_i = &\ \rlap{\(\atom{x}^m_i\)}\hspace{2.5in}
      \llap{\(\atom{y\lchop x}^{m,n}_i\)} = \atom{x}^m_{i+n} \\
   \atom{x\cat y}^{m,n}_i = &\ \begin{cases}
         \atom{x}^m_i & \text{if } i\leq m \\
         \atom{y}^n_{i-m} & \text{if } m<i 
      \end{cases} \\
   \atom{x\C(y,z_0,z_1)}^{m,n,p_0,p_1}_i = &\ 
      \begin{cases}
         \atom{y}^n_i & \text{if } m=0 \\
         \bigl(\lnot\atom{x}^m_m\land
            \atom{{}_0(z_1\rchop z_0)\cat z_0}^{p_0,p_1}_i\bigr) \\
          \lor \bigl(\atom{x}^m_m\land
            \atom{{}_0(z_0\rchop z_1)\cat z_1}^{p_0,p_1}_i\bigr)
          & \text{if } m>0 
      \end{cases} \\
   \atom{[\fhead\tpl x\fdot t](\tpl y)}^{\tpl n}_i = &\ 
      \atom{t[\tpl y/\tpl x]}^{\tpl n'}_i \\
   \atom{\lCRN[h](x,\tpl y)}^{m,\tpl n}_i = &\ 
      \atom{h(z,\tpl y)\cat 0}^{m-i+1,\tpl n}_1
      \big[\atom{x}^m_{j+i-1}/\atom{z}^{m-i+1}_j\big]
         _{1\leq j\leq m-i+1} \\*
      &\ \text{( where $z$ does not occur in $h(x,\tpl y)$ )}
         \quad \text{for $m>0$} \\
   \atom{\rCRN[h](x,\tpl y)}^{m,\tpl n}_i = &\ 
      \atom{0\cat h(z,\tpl y)}^{i,\tpl n}
         _{\tlen_{h(z,\tpl y)}(i,\tpl n)+1}
      \big[\atom{x}^m_j/\atom{z}^i_j\big]_{1\leq j\leq i} \\*
      &\ \text{( where $z$ does not occur in $h(x,\tpl y)$ )}
         \quad \text{for $m>0$} \\
   \bigatom{\TRN[g,h,h_\ell,h_r](x,z,\tpl y)}^{m,p,\tpl n}_i = &\ 
      \begin{cases}
         \atom{g(x,z,\tpl y)}^{m,p,\tpl n}_i & \text{if } m\leq 1 \\
         \Big\latom h\big(x,z,\tpl y,
            \TRN[g,h,h_\ell,h_r](x\lhalf,h_\ell(z),\tpl y), \\
            \hphantom{\Big\latom h\big(x,z,\tpl y, }
            \TRN[g,h,h_\ell,h_r](\rhalf x,h_r(z),\tpl y)
            \big)\Big\ratom^{m,p,\tpl n}_i & \text{if } 1<m 
      \end{cases} 
\end{align*}

\begin{remark}
   Note that ``$\C$'' is the only primitive function symbol that has
   non-trivial term formulas (because it is the only function that
   depends directly on the values of its arguments), so that any
   non-trivial term formula must depend on $\C$ in some way.
\end{remark}

\section{Propositional translations}

 The propositional translations of formulas of $T_1$ are defined
 inductively as follows (where $\odot$ stands for any one of the
 binary propositional connectives, and $\tpl m_0$ and $\tpl m_1$
 represent the lengths of the variables that occur in $A$ and $B$ (or
 $t$ and $u$), respectively.)
\begin{align*}
   \form{t=u}^{\tpl m} &= \begin{cases}
         \bigand\limits_{1\leq i\leq \tlen_t(\tpl m_0)}
            \atom{t}^{\tpl m_0}_i\liff\atom{u}^{\tpl m_1}_i
          & \text{if } \tlen_t(\tpl m_0)=\tlen_u(\tpl m_1), \\
         \lfalse & \text{otherwise}. 
      \end{cases} \\
   \form{\lnot A}^{\tpl m} &= \lnot\form{A}^{\tpl m} \\
   \form{A\odot B}^{\tpl m} &=
      \form{A}^{\tpl m_0}\odot\form{B}^{\tpl m_1} 
\end{align*}

\section{The simulation result}

 Now, we can prove the following theorem.

\begin{theorem}
\label{thm:simulation}
   If $A$ is provable in $T_1$, then for any $\tpl m$,
   \(\form{A}^{\tpl m}\) has uniform polysize $\F$-proofs.
\end{theorem}

\begin{proof}
   The proof is by induction on the number of inferences in the proof
   of $A$.  If $A$ is an axiom, then Subsection~\ref{sec:sim_axioms}
   below shows that \(\form{A}^{\tpl m}\) has linear-size $\F$-proofs.
   If $A$ is obtained by a derivation, then by the induction
   hypothesis, the propositional translations of the premises of the
   last inference all have short $\F$-proofs.
   Subsection~\ref{sec:sim_rules} below shows that in this case also,
   \(\form{A}^{\tpl m}\) has short $\F$-proofs.  Moreover, all these
   $\F$-proofs are uniform, in the sense that there exists a specific
   function that takes a theorem of $T_1$ and the lengths of its
   variables into a $\F$-proof of the translation of the theorem.
   This function is described implicitly in the sections that follow,
   but it can be formalized in $T_1$ itself, using techniques similar
   to those developed in Chapter~\ref{chap:soundness} (we do not
   expect any technical difficulties in doing this but time
   constraints prevent us from working out the details).
\end{proof}

\subsection{Axioms}
\label{sec:sim_axioms}

 For most axioms of the form \(t=u\), just writing down the
 definitions of $\atom{t}_i$ and $\atom{u}_i$ is enough to see that
 the axiom is a theorem with short proofs since
 \(\atom{t}_i=\atom{u}_i\).  We give a more detailed argument only for
 a few axioms.
\begin{enumerate}\setcounter{enumi}{-1}

\item
   The axioms for the propositional calculus can obviously be
   simulated by any $\F$-system.

\item
   \begin{enumerate}

   \item
      By reflexivity, there are linear-size $\F$-proofs of
      \(\bigand\atom{x}^m_i\liff\atom{x}^m_i\)
      for any variable $x$.

   \item
      If \(m\neq n\), then the antecedent of the axiom translates to
      $\lfalse$ so the translation of the axiom is a trivial theorem.
      If \(m=n\), then the commutativity of $\liff$ gives linear-size
      $\F$-proofs of
      \( \big(\bigand\atom{x}^m_i\liff\atom{y}^n_i\big) \limp
         \big(\bigand\atom{y}^n_i\liff\atom{x}^m_i\big) \).

   \item
      If \(m\neq n\) or \(n\neq p\), then one of the antecedents of
      the axiom translates to $\lfalse$ so the translation of the
      axiom is a trivial theorem.  If \(m=n=p\), then the transitivity
      of $\liff$ gives linear-size $\F$-proofs of
      \( \big(\big(\bigand\atom{x}^m_i\liff\atom{y}^n_i\big)\land
         \big(\bigand\atom{y}^n_i\liff\atom{z}^p_i\big)\big) \limp
         \big(\bigand\atom{x}^m_i\liff\atom{z}^p_i\big) \).

   \item
      An easy induction on the structure of the function symbol $f$,
      together with properties of $\liff$, is sufficient to show that
      there are short $\F$-proofs of
      \( \big(\bigand\atom{x_1}^{m_1}_i\liff\atom{y_1}^{n_1}_i\land
         \dots\land\bigand\atom{x_k}^{m_k}_i\liff\atom{y_k}^{n_k}_i\big)
         \limp \bigand\atom{f(x_1,\dots,x_k)}^{m_1,\dots,m_k}_i\liff
            \atom{f(y_1,\dots,y_k)}^{n_1,\dots,n_k}_i \)
      when \(m_1=n_1,\dots,m_k=n_k\) (the axiom's translation becoming
      a trivial theorem otherwise as one of the antecedents translates
      to $\lfalse$).  The size of these proofs is linear if $f$ does
      not contain any function defined by $\TRN$; it is polynomial
      otherwise (since the size of the term formulas can be polynomial
      in the lengths of the variables).

   \end{enumerate}

\item 
   \( \atom{0}_1 = \lfalse \) and \( \atom{1}_1 = \ltrue \).

\item 
   \begin{enumerate}

   \item
      For all \(1\leq i\leq m+0\),
      \( \atom{x\cat\E}^m_i = \atom{x}^m_i \),
      and for all \(1\leq i\leq m+n+1\),
      \begin{align*}
         \atom{x\cat y0}^{m,n}_i &=
            \begin{cases}
               \atom{x}^m_i & \text{if } i\leq m \\
               \atom{y\cat 0}^{n}_{i-m} = \begin{cases}
                     \atom{y}^n_{i-m} & \text{if } i-m\leq n \\
                     \atom{0}_{i-m-n} & \text{if } n<i-m 
                  \end{cases} & \text{if } m<i 
            \end{cases} \\
         \atom{(x\cat y)\cat 0}^{m,n}_i &=
            \begin{cases}
               \atom{x\cat y}^{m,n}_i = \begin{cases}
                     \atom{x}^m_i & \text{if } i\leq m \\
                     \atom{y}^n_{i-m} & \text{if } m<i 
                  \end{cases} & \text{if } i\leq m+n \\
               \atom{0}_{i-(m+n)} & \text{if } m+n<i 
            \end{cases} 
      \end{align*}
      (similarly for \( x\cat y1 = (x\cat y)\cat 1 \)).

   \item
      \(\form{x\cat y=\E}^{m,n}\) holds iff \(m+n=0\), and
      \(\form{x=\E\land y=\E}^{m,n}\) holds iff \(m=0\) and \(n=0\).

   \item
      \(\form{x\cat y=0}^{m,n}\) holds iff \(m+n=1\) and
      \(\atom{x}^1_1=\lfalse\) or \(\atom{y}^1_1=\lfalse\).
      Similarly for \(x\cat y=1\).

   \end{enumerate}

\item 
   \begin{enumerate}

   \item
   \label{Fproof:lchop}
      For \(1\leq i\leq m\),
      \( \atom{\E\lchop x}^m_i = \atom{x}^m_{i+0} \)
      by definition; also, for \(1\leq i\leq m-(n+1)\),
      \(
         \atom{(0y)\lchop x}^{m,n}_i
          = \atom{x}^m_{i+n+1}
          = \atom{y\lchop x}^{m,n}_{i+1}
          = \atom{0\lchop(y\lchop x)}^{m,n}_i,
      \)
      and the same reasoning applies to
      \(\atom{(1y)\lchop x}^{m,n}_i\).

   \item
   \label{Fproof:ilchop}
      \( \tlen_{0\lchop\E} = 0\dotminus 1 = 0 = \tlen_\E \),
      and for \(1\leq i\leq m+1\dotminus 1\),
      \( \atom{0\lchop(0x)}^m_i = \atom{0x}^m_{i+1}
          = \atom{x}^m_{(i+1)-1} = \atom{x}^m_i \)
      (similarly for \(\atom{0\lchop(1x)}^m_i\)).  The same reasoning
      applies with ``$1\lchop\dots$'' in place of ``$0\lchop\dots$''.

   \item
   \label{Fproof:lchopE}
      \( \tlen_{y\lchop x}(m,n) = m\dotminus n \)
      and
      \( \tlen_{x\lchop y0}(m,n) = \tlen_{x\lchop y1}(m,n)
          = (n+1)\dotminus m \),
      and
      \( m\dotminus n = 0 \liff m-n\leq 0 \liff m\leq n \liff
         m<n+1 \liff 0<(n+1)-m \liff (n+1)\dotminus m \neq 0 \).

   \end{enumerate}

\item 
   \begin{enumerate}

   \item
      Similarly to~\ref{Fproof:lchop}.

   \item
      Similarly to~\ref{Fproof:ilchop}.

   \item
      Similarly to~\ref{Fproof:lchopE}.

   \end{enumerate}

\item 
   \begin{enumerate}

   \item
   \label{Fproof:_0}
      \( \form{{}_0\E=\E} = \ltrue \) since
      \( \tlen_{{}_0\E} = \tlen_\E = 0 \), and for \(1\leq i\leq m+1\),
      \( \atom{{}_0(x0)}^m_i = \atom{{}_0(x1)}^m_i = \lfalse \) and
      \(
         \atom{{}_0x\cat 0}^m_i = \begin{cases}
            \atom{{}_0x}^m_i & \text{if } i\leq m, \\
            \atom{0}_{i-m} & \text{if } m<i. 
         \end{cases}
      \)

   \item
      Similarly to~\ref{Fproof:_0}.

   \end{enumerate}

\item 
   \( \atom{\E\C(x,y,z)}^{m,n,p}_i = \atom{x}_i \)
   for \(1\leq i\leq m\), by definition; also, for
   \(1\leq i\leq\max\{n,p\}\),
   \begin{equation*}
   \begin{split}
      \atom{(w0)\C(x,y,z)}^{k,m,n,p}_i = &\ 
         \bigl(\lnot\atom{w0}^k_{k+1}\land
            \atom{{}_0(z\rchop y)\cat y}^{n,p}_i\bigr) \\
      &\ \lor \bigl(\atom{w0}^k_{k+1}\land
            \atom{{}_0(y\rchop z)\cat z}^{n,p}_i\bigr) \\
       = &\ \bigl(\lnot\lfalse\land
            \atom{{}_0(z\rchop y)\cat y}^{n,p}_i\bigr)
         \lor \bigl(\lfalse\land
            \atom{{}_0(y\rchop z)\cat z}^{n,p}_i\bigr) \\
      \liff &\ \atom{{}_0(z\rchop y)\cat y}^{n,p}_i, \\
      \atom{(w1)\C(x,y,z)}^{k,m,n,p}_i = &\ 
         \bigl(\lnot\atom{w1}^k_{k+1}\land
            \atom{{}_0(z\rchop y)\cat y}^{n,p}_i\bigr) \\
      &\ \lor \bigl(\atom{w1}^k_{k+1}\land
            \atom{{}_0(y\rchop z)\cat z}^{n,p}_i\bigr) \\
       = &\ \bigl(\lnot\ltrue\land
            \atom{{}_0(z\rchop y)\cat y}^{n,p}_i\bigr)
         \lor \bigl(\ltrue\land
            \atom{{}_0(y\rchop z)\cat z}^{n,p}_i\bigr) \\
      \liff &\ \atom{{}_0(y\rchop z)\cat z}^{n,p}_i. 
   \end{split}
   \end{equation*}

\item 
   \begin{enumerate}

   \item
      For all \(1\leq i\leq m\), \\
      \(
         \atom{(x\lhalf)\cat(\rhalf x)}^m_i = \begin{cases}
            \atom{x\lhalf}^m_i = \atom{x}^m_i
          & \text{if } i\leq\lfloor m/2\rfloor \\
            \atom{\rhalf x}^m_{i-\lfloor m/2\rfloor} =
               \atom{x}^m_{i-\lfloor m/2\rfloor+\lfloor m/2\rfloor}
          & \text{if } \lfloor m/2\rfloor<i
         \end{cases}
      \)

   \item
      \( x\lhalf\rchop\rhalf x = \E \) since
      \( \tlen_{x\lhalf}(m)\dotminus\tlen_{\rhalf x}(m) =
         \lfloor m/2\rfloor\dotminus\lceil m/2\rceil = 0 \); \\
      \( 1\lchop(x\lhalf\lchop\rhalf x) = \E \) since
      \( (\tlen_{\rhalf x}(m)\dotminus\tlen_{x\lhalf}(m))\dotminus 1
          = \lceil m/2\rceil\dotminus\lfloor m/2\rfloor\dotminus 1
          = 0 \).

   \end{enumerate}

\setcounter{count}{\value{enumi}}
\end{enumerate}
\begin{slshape}
 The last four axioms are easy to prove if we note the following two
 facts.
\begin{itemize}
\item[\textup{(a)}]
   For any term $t$, \(t\rchop t = \E\) and therefore
   \( \atom{{}_0(t\rchop t)\cat t}^{\tpl m}_i = \atom{t}^{\tpl m}_i \)
   for \(1\leq i\leq \tlen_t(\tpl m)\).
\item[\textup{(b)}]
   Within any $\F$-system, the identity
   \((\lnot p\land q)\lor(p\land q) \liff q\) has linear-size proofs.
\end{itemize}
 Together, these facts show that
\[
   \atom{x\C(y,z,z)}^{m,n,p}_i \ \liff
    \ \begin{cases}
         \atom{y}^n_i & \text{if } m=0, \\
         \atom{z}^p_i & \text{if } m>0. 
      \end{cases}
\]
\end{slshape}
\begin{enumerate}
\setcounter{enumi}{\value{count}}

\item 
   \begin{enumerate}

   \item
   \label{Fproof:halves1}
      Note that
      \( \tlen_{(x0)\lhalf}(m)
          = \lfloor(m+1)/2\rfloor = \lceil m/2\rceil \)
      and
      \begin{align*}
         &\!\tlen_{x\lhalf\lchop\rhalf x\C(x\lhalf,
            x\lhalf\cat((\rhalf x\cat 0)\rchop\rhalf x),
            x\lhalf\cat((\rhalf x\cat 0)\rchop\rhalf x))}(m) \\
         &= \symbf{if } \lceil m/2\rceil\dotminus\lfloor m/2\rfloor=0
            \symbf{ then } \lfloor m/2\rfloor
            \symbf{ else } \lfloor m/2\rfloor+
               (\lceil m/2\rceil+1-\lceil m/2\rceil) \\
         &= \lceil m/2\rceil, 
      \end{align*}
      so both terms do have the same length.

      Now, for
      \(1\leq i\leq \lceil m/2\rceil\),
      \( \atom{(x0)\lhalf}^m_i = \atom{x0}^m_i = \atom{x}^m_i \)
      and
      \begin{align*}
         &\!\atom{x\lhalf\lchop\rhalf x\C\big(x\lhalf,
            x\lhalf\cat((\rhalf x\cat 0)\rchop\rhalf x),
            x\lhalf\cat((\rhalf x\cat 0)\rchop\rhalf x)\big)}^m_i \\
         &\liff\begin{cases}
               \atom{x\lhalf}^m_i
                & \text{if } \lceil m/2\rceil=\lfloor m/2\rfloor \\
               \atom{x\lhalf\cat((\rhalf x\cat 0)\rchop\rhalf x)}^m_i
                & \text{if } \lceil m/2\rceil>\lfloor m/2\rfloor 
            \end{cases} \\
         &\liff\begin{cases}
               \atom{x}^m_i
                & \text{if } \lceil m/2\rceil=\lfloor m/2\rfloor \\
               \begin{cases}
                  \atom{x\lhalf}^m_i
                   & \text{if } i\leq\lfloor m/2\rfloor \\
                  \atom{((\rhalf x\cat 0)\rchop\rhalf x)}^m_1
                   & \text{if } i>\lfloor m/2\rfloor 
               \end{cases}
                & \text{if } \lceil m/2\rceil>\lfloor m/2\rfloor 
            \end{cases} \\
         &\liff\begin{cases}
               \atom{x}^m_i
                & \text{if } \lceil m/2\rceil=\lfloor m/2\rfloor \\
               \begin{cases}
                  \atom{x}^m_i
                   & \text{if } i\leq\lfloor m/2\rfloor \\
                  \atom{x}^m_i
                   & \text{if } i=\lceil m/2\rceil 
               \end{cases}
                & \text{if } \lceil m/2\rceil>\lfloor m/2\rfloor 
            \end{cases} \\
         &\liff\atom{x}^m_i 
      \end{align*}
      since
      \( \atom{((\rhalf x\cat 0)\rchop\rhalf x)}^m_1
          = \atom{x}^m_{1+\lfloor m/2\rfloor} \).
      The case for $x1$ is identical.

   \item
   \label{Fproof:halves2}
      Note that
      \( \tlen_{\rhalf(x0)}(m) = \lceil(m+1)/2\rceil
          = \lfloor m/2\rfloor+1 \)
      and
      \begin{align*}
         &\!\tlen_{x\lhalf\lchop\rhalf x\C(\rhalf x\cat 0,
            1\lchop(\rhalf x\cat 0),1\lchop(\rhalf x\cat 0))}(m) \\
         &= \symbf{if } \lceil m/2\rceil\dotminus\lfloor m/2\rfloor=0
            \symbf{ then } \lceil m/2\rceil+1
            \symbf{ else } (\lceil m/2\rceil+1)-1 \\
         &= \lfloor m/2\rfloor+1, 
      \end{align*}
      so both terms do have the same length.

      Now, for \(1\leq i\leq \lfloor m/2\rfloor+1\),
      \[
         \atom{\rhalf(x0)}^m_i = \atom{x0}^m_{i+\lfloor(m+1)/2\rfloor}
          = \begin{cases}
               \atom{x}^m_{i+\lceil m/2\rceil}
                 & \text{if } i\leq\lfloor m/2\rfloor \\
               \atom{0}_1 & \text{if } i=\lfloor m/2\rfloor+1 
            \end{cases}
      \]
      and
      \begin{align*}
         &\!\atom{x\lhalf\lchop\rhalf x\C\big(\rhalf x\cat 0,
            1\lchop(\rhalf x\cat 0),1\lchop(\rhalf x\cat 0)\big)}^m_i \\
         &\liff\begin{cases}
               \atom{\rhalf x\cat 0}^m_i
                & \text{if } \lceil m/2\rceil=\lfloor m/2\rfloor \\
               \atom{1\lchop(\rhalf x\cat 0)}^m_i
                & \text{if } \lceil m/2\rceil>\lfloor m/2\rfloor 
            \end{cases} \\
         &\liff\begin{cases}
               \begin{cases}
                  \atom{\rhalf x}^m_i
                   & \text{if } i\leq\lfloor m/2\rfloor \\
                  \atom{0}_1
                   & \text{if } i>\lfloor m/2\rfloor 
               \end{cases}
                & \text{if } \lceil m/2\rceil=\lfloor m/2\rfloor \\
               \atom{\rhalf x\cat 0}^m_{i+1}
                & \text{if } \lceil m/2\rceil>\lfloor m/2\rfloor 
            \end{cases} \\
         &\liff\begin{cases}
               \begin{cases}
                  \atom{\rhalf x}^m_i
                   & \text{if } i\leq\lfloor m/2\rfloor \\
                  \atom{0}_1
                   & \text{if } i>\lfloor m/2\rfloor 
               \end{cases}
                & \text{if } \lceil m/2\rceil=\lfloor m/2\rfloor \\
               \begin{cases}
                  \atom{\rhalf x}^m_{i+1}
                   & \text{if } i\leq\lfloor m/2\rfloor \\
                  \atom{0}_1
                   & \text{if } i>\lfloor m/2\rfloor 
               \end{cases}
                & \text{if } \lceil m/2\rceil>\lfloor m/2\rfloor 
            \end{cases} \\
         &\liff\begin{cases}
               \atom{x}^m_{i+\lceil m/2\rceil}
                 & \text{if } i\leq\lfloor m/2\rfloor \\
               \atom{0}_1 & \text{if } i=\lfloor m/2\rfloor+1 
            \end{cases} 
      \end{align*}
      The case for $x1$ is identical.

   \item
      Similarly to~\ref{Fproof:halves1}.

   \item
      Similarly to~\ref{Fproof:halves2}.

   \end{enumerate}

\item 
   \( \atom{[\fhead\tpl x\fdot t](\tpl x)}^{\tpl m}_i =
      \atom{t[\tpl x/\tpl x]}^{\tpl m'}_i =
      \atom{t}^{\tpl m'}_i \)
   for all \(1\leq i\leq \tlen_t(\tpl m')\).

\item 
   \begin{enumerate}

   \item
   \label{Fproof:lCRN}
      Since \( \tlen_{\lCRN[h](\E,\tpl y)}(\tpl n) = 0 \),
      \( \form{\lCRN[h](\E,\tpl y)=\E}^{\tpl n} = \ltrue \).

      Also, because
      \( \tlen_{(z\cat 0)\rchop z}(p) = p+1\dotminus p = 1 \),
      we have that
      \begin{align*}
         \atom{\lCRN[h](0x,\tpl y)}^{m,\tpl n}_1
         &= \atom{h(z,\tpl y)\cat 0}^{m+1,\tpl n}_1
            \big[\atom{0x}^m_j/\atom{z}^{m+1}_j\big]_{1\leq j\leq m+1} \\
         &= \atom{h(0x,\tpl y)\cat 0}^{m,\tpl n}_1 \\
         &= \atom{(h(0x,\tpl y)\cat 0)\rchop h(0x,\tpl y)}
            ^{m,\tpl n}_1 \\
         &= \atom{\big((h(0x,\tpl y)\cat 0)\rchop h(0x,\tpl y)\big)
               \cat\lCRN[h](x,\tpl y)}^{m,\tpl n}_1 
      \end{align*}
      (similarly for $1x$),

      and for \(1<i\leq m+1\),
      \begin{align*}
         \atom{\lCRN[h](0x,\tpl y)}^{m,\tpl n}_i
         &= \atom{h(z,\tpl y)\cat 0}^{m+2-i,\tpl n}_1
            \big[\atom{0x}^m_{j+i-1}/\atom{z}^{m+2-i}_j\big]
            _{1\leq j\leq m+2-i} \\
         &= \atom{h(z,\tpl y)\cat 0}^{m+1-(i-1),\tpl n}_1 \\*
         &\pheq \quad
            \big[\atom{x}^m_{j+(i-1)-1}/\atom{z}^{m+1-(i-1)}_j\big]
            _{1\leq j\leq m+1-(i-1)} \\
         &= \atom{\lCRN[h](x,y)}^{m,\tpl n}_{i-1} \\
         &= \atom{\big((h(0x,\tpl y)\cat 0)\rchop h(0x,\tpl y)\big)
               \cat\lCRN[h](x,\tpl y)}^{m,\tpl n}_i 
      \end{align*}
      (similarly for $1x$).

   \item
      Similarly to~\ref{Fproof:lCRN}.

   \end{enumerate}

\item 
   For all \(1\leq i\leq
      \tlen_{\TRN[g,h,h_\ell,h_r](x,z,\tpl y)}(m,p,\tpl n)\),
   and by the remark above,
   \begin{align*}
      \atom{x\rchop 1\C(g(x,z,\tpl y),t,t)}^{m,p,\tpl n}_i
      &= \begin{cases}
            \atom{g(x,z,\tpl y)}^{p,\tpl n}_i
             & \text{if } m\dotminus 1=0 \\
            \bigl(\lnot\atom{x\rchop 1}^m_{m\dotminus 1}\land
               \atom{{}_0(t\rchop t)\cat t}^{m,p,\tpl n}_i\bigr) \\
            \lor \bigl(\atom{x\rchop 1}^m_{m\dotminus 1}\land
               \atom{{}_0(t\rchop t)\cat t}^{m,p,\tpl n}_i\bigr)
             & \text{if } m\dotminus 1>0 
         \end{cases} \\
      &\liff\begin{cases}
            \atom{g(x,z,\tpl y)}^{p,\tpl n}_i
             & \text{if } m\dotminus 1=0 \\
            \atom{t}^{m,p,\tpl n}_i
             & \text{if } m\dotminus 1>0 
         \end{cases} 
   \end{align*}
   where \( t = h\bigl(x,z,\tpl y,
         \TRN[g,h,h_\ell,h_r](x\lhalf,h_\ell(z),\tpl y),
         \TRN[g,h,h_\ell,h_r](\rhalf x,h_r(z),\tpl y)\bigr) \).

\end{enumerate}

\subsection{Rules of inference}
\label{sec:sim_rules}

 For all the rules in Definition~\ref{def:T1rules}, if one of the
 premises contains an equation of the form \(\form{t=u}^{\tpl m}\)
 that degenerates to $\lfalse$ because
 \(\tlen_t(\tpl m_0)\neq\tlen_u(\tpl m_1)\), then the rule becomes
 trivial.  We therefore assume that none of the propositional
 translations of atomic formulas of $T_1$ are degenerate cases.
 Also, when we use the notation ``$\bigand$'' with no subscript, we
 implicitly assume that the conjunction is over all relevant values of
 the index of the term formulas involved.
\begin{enumerate}\setcounter{enumi}{-1}

\item
   Any standard, complete set of rules for the propositional calculus
   can be p-simulated within any $\F$-system.

\item
   We have short $\F$-proofs of \(\form{A}^{m,\tpl n}\).  Substituting
   \(\atom{t}^{\tpl p}_i\) for \(\atom{x}^m_i\) throughout these
   proofs yield short $\F$-proofs of
   \(\form{A[t/x]}^{\tpl p,\tpl n}\).

\item
   \begin{enumerate}

   \item
      First, a few observations.  Let
      \( F = (x=\E \lor x=0\cat\ldel x \lor x=1\cat\ldel x) \).
      Then,
      \begin{align*}
         \form{x=\E}^m &= \begin{cases}
               \ltrue & \text{if } m=0 \\
               \lfalse & \text{if } m>0 
            \end{cases} \\
         \form{x=0\cat\ldel x}^m &=
            \bigand \atom{x}^m_i\liff\atom{0\cat\ldel x}^m_i \\
         &= \atom{x}^m_1\liff\atom{0}_1 \land
            \bigand_{1<i} \atom{x}^m_i\liff\atom{\ldel x}^m_{i-1} \\
         &= \atom{x}^m_1\liff\lfalse \land
            \bigand_{1<i} \atom{x}^m_i\liff\atom{x}^m_i \\
         \form{x=1\cat\ldel x}^m &= \atom{x}^m_1\liff\ltrue \land
            \bigand_{1<i} \atom{x}^m_i\liff\atom{x}^m_i 
      \end{align*}
      so that
      \begin{equation*}
      \begin{aligned}
         \form{F}^m &= \begin{cases}
               \ltrue \lor \lfalse \lor \lfalse & \text{if } m=0 \\
               \lfalse \lor \bigl(\atom{x}^m_1\liff\lfalse \land
                  \bigand_{1<i} \atom{x}^m_i\liff\atom{x}^m_i\bigr) \\
               \lor \bigl(\atom{x}^m_1\liff\ltrue \land
                  \bigand_{1<i} \atom{x}^m_i\liff\atom{x}^m_i\bigr)
             & \text{if } m>0 
            \end{cases} \\
         &\liff\begin{cases}
               \ltrue & \text{if } m=0 \\
               \bigl( \bigl( \atom{x}^m_1\liff\lfalse
                  \lor \atom{x}^m_1\liff\ltrue \bigr) \land
                  \bigand_{1<i} \atom{x}^m_i\liff\atom{x}^m_i
               \bigr) & \text{if } m>0 
            \end{cases} \\
         &\liff\begin{cases}
               \ltrue & \text{if } m=0 \\
               \lnot\atom{x}^m_1\lor\atom{x}^m_1
             & \text{if } m>0 
            \end{cases} 
      \end{aligned}
      \end{equation*}
      Therefore, \(\form{F}^m\) has linear-size $\F$-proofs.

      Now, we have short $\F$-proofs of
      \begin{align*}
         A_\E &= \form{A[\E]}^{\tpl n}, \\
         A_0 &= \form{A[x]}^{m,\tpl n} \limp
            \form{A[0x]}^{m,\tpl n}, \\
         \text{and } A_1 &= \form{A[x]}^{m,\tpl n} \limp
            \form{A[1x]}^{m,\tpl n}. 
      \end{align*}
      If \(m=0\), then by Axiom~\ref{axiom:fcneq}, there are short
      proofs of
      \[
         \form{x=\E}^m \limp \bigl(\form{A}^{m,\tpl n}\liff
         \form{A[\E]}^{\tpl n}\bigr),
      \]
      which shows that there are short proofs of
      \begin{equation}
         \form{x=\E}^m \limp \bigl(\form{A[\E]}^{\tpl n}\limp
         \form{A}^{m,\tpl n}\bigr).
      \label{eq0}
      \end{equation}
      If \(m>0\), then substituting $1\lchop x$ for $x$ in $A_0$ gives
      short proofs of
      \[
         \form{A[1\lchop x]}^{m,\tpl n} \limp
         \form{A[0\cat\ldel x]}^{m,\tpl n};
      \]
      moreover, by Axiom~\ref{axiom:fcneq}, there are short proofs of
      \[
         \form{x=0\cat\ldel x}^m \limp
         \bigl(\form{A}^{m,\tpl n}\liff
            \form{A[0\cat\ldel x]}^{m,\tpl n}\bigr),
      \]
      and therefore, by transitivity, of
      \[
         \form{x=0\cat\ldel x}^m \limp
         \bigl(\form{A[1\lchop x]}^{m,\tpl n}\limp
            \form{A}^{m,\tpl n}\bigr).
      \]
      A similar argument shows that there are short proofs of
      \[
         \form{x=1\cat\ldel x}^m \limp
         \bigl(\form{A[1\lchop x]}^{m,\tpl n}\limp
            \form{A}^{m,\tpl n}\bigr),
      \]
      which, together with (\ref{eq0}), implies that there are short
      proofs of
      \[
         \form{F}^m \limp \bigl(\form{A[1\lchop x]}^{m,\tpl n}\limp
         \form{A}^{m,\tpl n}\bigr).
      \]
      Applying modus ponens to this and \(\form{F}^m\) gives short
      proofs of
      \begin{equation}
         \form{A[1\lchop x]}^{m,\tpl n} \limp \form{A}^{m,\tpl n}.
      \label{eq1}
      \end{equation}

      Repeated substitutions of $1\lchop x$ for $x$ in the proof of
      (\ref{eq1}) give short proofs of
      \begin{align*}
         \form{A[11\lchop x]}^{m,\tpl n} &\limp
            \form{A[1\lchop x]}^{m,\tpl n} \\
         &\ \:\vdots \\
         \form{A[\cst{m}\lchop x]}^{m,\tpl n} &\limp
         \form{A[\cst{m-1}\lchop x]}^{m,\tpl n} 
      \end{align*}
      (where we remind the reader that ``$\cst k$'' is a shorthand for
      $\overbrace{1\dotsm 1}^k$).

      Since \(\cst{m}\lchop x=\E\), using Axiom~\ref{axiom:fcneq}
      and modus ponens gives short proofs of
      \[
         \form{A[\E]}^{\tpl n} \limp
         \form{A[\cst{m}\lchop x]}^{m,\tpl n},
      \]
      and using transitivity $m+1$ times now gives short proofs of
      \[
         \form{A[\E]}^{\tpl n} \limp \form{A}^{m,\tpl n}.
      \]
      A final application of modus ponens with $A_\E$ gives the short
      proofs of \(\form{A}^{m,\tpl n}\) we wanted: as can easily be
      seen, the size of this proof is $\bigOh(m\cdot p(m,\tpl n))$,
      where $p(m,\tpl n)$ was the maximum size of the proofs of
      $A_\E,A_0,A_1$.

   \item
      The same reasoning as for part (a) applies.

   \end{enumerate}

\item
   We have short proofs of \(\form{A[\E,z]}^{p,\tpl n}\),
   \(\form{A[0,z]}^{p,\tpl n}\), \(\form{A[1,z]}^{p,\tpl n}\), and
   \begin{equation*}
      A' = \bigl(\form{A[x\lhalf,h_\ell(z)]}^{m,p,\tpl n}\land
               \form{A[\rhalf x,h_r(z)]}^{m,p,\tpl n}\bigr)
            \limp \form{A[x,z]}^{m,p,\tpl n}. 
   \end{equation*}
   If \(m=0\), using modus ponens twice on formula (\ref{eq0}) gives a
   short proof of \(\form{A}^{m,p,\tpl n}\).  If \(m>0\), then by
   repeatedly substituting first $x\lhalf,h_\ell(z)$ and then
   $\rhalf x,h_r(z)$ for $x,z$ in the proof of $A'$, we get a binary
   tree of short proofs of formulas of the form of $A'$, where the
   formula at the root is \(\form{A[x,z]}^{m,p,\tpl n}\), the formula
   at each node is implied by the conjunction of the formulas at its
   children nodes, and at the leaves, the terms being substituted for
   $u$ can all be proved to be equal to $0$ or $1$ (single bits of
   $x$).

   For example, if \(m=3\), the tree would have the form depicted in
   Figure~\ref{fig:prooftree} (where we've indicated only the
   consequent of the formula being proved at each node, so that a node
   $B$ with children $C$ and $D$ represents a proof of the formula
   \((C\land D) \limp B\) and a node $B$ with one child $C$ represents
   a proof of the formula \(C \limp B\)).

   \begin{figure}[!ht]
   \centering
      \begin{picture}(340,131)(-170,-125)
         \put(0,0){\makebox(0,0){\(A[x,z]\)}}
         \put(-120,-40){\makebox(0,0){\(A[x\lhalf,h_\ell(z)]\)}}
         \put(60,-40){\makebox(0,0){\(A[\rhalf x,h_r(z)]\)}}
         \put(0,-80){\makebox(0,0){\small
            \(A\big[(\rhalf x)\lhalf,h_\ell(h_r(z))\big]\)}}
         \put(120,-80){\makebox(0,0){\small
            \(A\big[\rhalf\rhalf x,h_r(h_r(z))\big]\)}}
         \put(-120,-80){\makebox(0,0){\(A[0,z']\land A[1,z']\)}}
         \put(0,-120){\makebox(0,0){\(A[0,z']\land A[1,z']\)}}
         \put(120,-120){\makebox(0,0){\(A[0,z']\land A[1,z']\)}}
         \put(-10,-10){\line(-5,-1){100}}
         \put(10,-10){\line(2,-1){40}}
         \put(45,-50){\line(-2,-1){40}}
         \put(75,-50){\line(2,-1){40}}
         \put(-120,-50){\line(0,-1){20}}
         \put(0,-90){\line(0,-1){20}}
         \put(120,-90){\line(0,-1){20}}
      \end{picture}
   \caption{Proof tree for $m=3$.}
   \label{fig:prooftree}
   \end{figure}
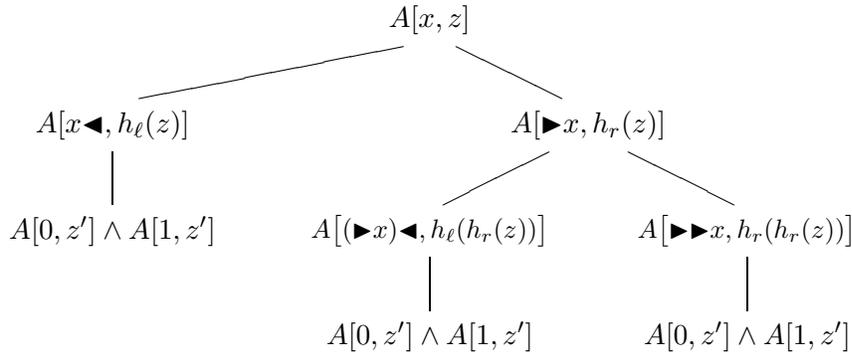

   Therefore, the proofs of \(\form{A[0,z]}^{p,\tpl n}\) and
   \(\form{A[1,z]}^{p,\tpl n}\) can be used with Rule~\ref{rule:sub}
   (substituting the right terms for $z$) and modus ponens to prove
   the formulas at the first level, and going up level by level using
   modus ponens, we obtain a proof of the consequent of the formula at
   the root of the tree, \ie, \(\form{A}^{m,p,\tpl n}\).  Moreover, if
   $q(m,p,\tpl n)$ is the maximum proof size of the premises and $c$
   is a constant satisfying \(|h_\ell(z)|\leq 2^c|z|\) and
   \(|h_r(z)|\leq 2^c|z|\), then the size of this proof is
   $\bigOh(m\cdot q(m,p\cdot m^c,\tpl n))$ since the tree has depth no
   more than $\lceil\log_2m\rceil$ and thus size no more than $2m$.

\end{enumerate}

\begin{remark}
   Note that the estimates on the size of $\F$-proofs for $T_1$'s
   theorems given above might be used to get more precise upper bounds
   than are currently known for the size of $\F$-proofs of certain
   families of tautologies.  For example, the family of tautologies
   arising from the pigeonhole principle was first shown to have
   polysize $\F$-proofs by Buss~\cite{Buss87b} (whose estimate of the
   size of the proofs was $\bigOh(n^{20})$); a careful analysis of the
   $T_1$-proof given in Chapter~\ref{chap:T1} together with the
   results of this chapter could provide a better bound (the details
   would be somewhat tedious but straightforward), and the same could
   be done for the other families of tautologies mentionned at the end
   of Chapter~\ref{chap:T1}.
\end{remark}

%% file: chapter5.tex
\chapter{$T_1$ Proves the Soundness of $\F$}
\label{chap:soundness}

 In this chapter, we will show how to formalize a particular
 $\F$-system in $T_1$, how to formalize Buss's algorithm for the
 \emph{Boolean Sentence Value Problem} ($\BSVP$)~\cite{Buss93} in
 $T_1$, and how to use the $\BSVP$ algorithm to prove the soundness
 of the given $\F$-system in $T_1$.  Then, we show that $\F$ provably
 $p$-simulates any proof system $S$ whose soundness can be proved in
 $T_1$.

\section{Formalizing $\F$-systems}

 Because any two $\F$-systems $p$-simulate each other, we will focus
 on the particular $\F$-system below:
\begin{description}
\item[language:]
   variables $p_1,p_2,\dots$, constants $\ltrue$ and $\lfalse$,
   connective $\limp$, brackets $($ $)$
\item[formulas:]
   \(\ltrue\), \(\lfalse\), \(p_i\) (for any \(i\geq 1\)), and
   recursively, \((A\limp B)\) for any formulas $A$ and $B$
\item[axiom schemes]
   (for any formulas $A$, $B$, $C$):
   \begin{enumerate}
   \item
      \( ( A \limp (B\limp A) ) \)
   \item
      \( \big( (A\limp(B\limp C)) \limp ((A\limp B)\limp(A\limp C))
         \big) \)
   \item
      \( \big( ((B\limp\lfalse)\limp(A\limp\lfalse))
         \limp (A\limp B) \big) \)
   \item
      \( \ltrue \)
   \end{enumerate}
\item[rule]
   (modus ponens):
   \( A,(A\limp B) \lded B \)
\end{description}

\subsection{Formulas}

 Given a formula $A$ of $\F$, we will encode $A$ into a string
 $\code_\ell A$ in the following way:  Let
 \(w_A=\max\{i:p_i\text{ appears in }A\}\) and
 \(\ell_A=1+\lfloor\lg(w_A+1)\rfloor\) (one more than the binary
 length of $w_A$).  Then, for \(2^\ell\geq\ell_A+2\),
 $\code_{2^\ell}A$ is obtained from $A$ by using the following
 G\"odel-numbering scheme (where
 \((i)_2^{2^\ell-2}\in\{0,1\}^{2^\ell-2}\) represents $i$ in binary
 using exactly $2^\ell-2$ bits---note that because of our choice of
 $2^\ell$, this will always start with a $0$).
\[
\begin{array}{c|c}
   \text{symbol} &    \text{code}     \\
\hline
        p_i      & 00(i)_2^{2^\ell-2} \\
      \lfalse    &  0010^{2^\ell-3}   \\
      \ltrue     &  0011^{2^\ell-3}   \\
         (       &   100^{2^\ell-2}   \\
       \limp     &   110^{2^\ell-2}   \\
         )       &   010^{2^\ell-2}   
\end{array}
\]
 In fact, because we can count in $T_1$, it is possible to define a
 slightly more complicated encoding where the codes for ``$($'',
 ``$\limp$'', and ``$)$'' include information about the logical depth
 of the symbol (making sure that $\ell_A$ is adjusted to be the
 maximum of its old value and the logical depth of $A$).  In what
 follows, we will use ``$(_j$'', ``$\limp_j$'', and ``$)_j$'' to
 represent the corresponding symbol at a logical depth of $j$.

 {}For example, the formula
\( A = \big(p_2\limp\big((p_{10}\limp\lfalse)\limp p_2\big)\big) \)
 can be rewritten as
\( A = (_0p_2\limp_0(_1(_2p_{10}\limp_2\lfalse)_2\limp_1p_2)_1)_0 \)
 by including the depth information for each symbol except logical
 variables and constants, which would be encoded as follows (where the
 string was split in two to fit on the page, and a little bit of space
 was added between blocks of bits representing each symbol, for
 readability):
\begin{align*}
   \code_8A &= \,10000000\;00000010\;11000000\;10000001\;
      10000010\;00001010\dotsm \\
   &\dotsm11000010\;00100000\;01000010\;11000001\;00000010\;
      01000001\;01000000. 
\end{align*}

 {}With this encoding, it is easy to write a function
 $\fcn{formula}(x,z)$ in $T_1$ that returns $1$ if
 \(x=\code_{2^\ell}A\) for some formula $A$ such that
 \(|\lenfcn{pow}(z)|=2^\ell\geq\ell_A+2\), or $0$ otherwise.  The
 function is defined by $\rpowCRN$ and simply checks that $x$ is one
 of ``$\code_{2^\ell}\lfalse$'', ``$\code_{2^\ell}\ltrue$'',
 ``$\code_{2^\ell}p_i$'', or that $x$ has the form ``$(_a\dotsb)_a$''
 and that each symbol in $x$ is preceded by a valid string of symbols,
 according to the following simple rules (using counting and masking
 operations):
\begin{itemize}
\item
   ``$p_i$'', ``$\lfalse$'', and ``$\ltrue$'' must immediately follow
   either ``$($'' or ``$\limp$'';
\item
   ``$(_j$'' must immediately follow either ``$(_{j-1}$'' or
   ``$\limp_{j-1}$'';
\item
   ``$\limp_j$'' must either immediately follow ``$(_jp_i$'',
   ``$(_j\lfalse$'', or ``$(_j\ltrue$'', or it must follow, in order,
   ``$(_j(_{j+1}\dotsb)_{j+1}$'' (where everything between the
   parentheses has depth at least $j+1$);
\item
   ``$)_j$'' must either immediately follow ``$\limp_jp_i$'',
   ``$\limp_j\lfalse$'', or ``$\limp_j\ltrue$'', or it must follow, in
   order, ``$\limp_j(_{j+1}\dotsb)_{j+1}$'' (where everything between
   the parentheses has depth at least $j+1$).
\end{itemize}

\subsection{Proofs}

 Now, we can encode $\F$-proofs easily.  A proof $A_1,\dots,A_k$ is
 encoded by a pair of strings:
 $\tuple{\code_{2^\ell}A_1,\dots,\code_{2^\ell}A_k}_k$ and
 $\tuple{j_1,\dots,j_k}_k$, where
 \(2^\ell\geq\max_{1\leq i\leq k}\{\ell_{A_i}\}+2\) and
\[
   j_i =
      \begin{cases}
         \tuple{0,m} & \text{if $A_i$ is an instance of axiom $m$}, \\
         \tuple{k_1,k_2} & \text{if } A_{k_2}=A_{k_1}\limp A_i. 
      \end{cases}
\]
 Again, it is straightforward to write a function
 $\fcn{proof}(x,y,z,w)$ in $T_1$ that returns $1$ if $x,y$ encode an
 $\F$-proof for \(2^\ell=|\lenfcn{pow}(z)|\) and \(k=|w|\), or $0$
 otherwise, by using simple masking operations and $\rpowCRN$.
 (Technically speaking, the encodings of the formulas in a proof must
 be padded so they all have the same length, but it is a simple matter
 to take care of.)

 {}Finally, we can define in $T_1$ the following function:
\[
   F(x,y,z,w) =
      \begin{cases}
         \proj^{|w|}_{|w|}(x) & \text{if } \fcn{proof}(x,y,z,w)=1, \\
         \code_{|\lenfcn{pow}(z)|}\ltrue & \text{otherwise}, 
      \end{cases}
\]
 that returns the tautology proved by $x,y$ (or some fixed tautology
 if $x,y$ is not an $\F$-proof).

\section{Buss's algorithm for the $\BSVP$}

 For reference purposes, we will now summarize Buss's most recent
 published $\ALT$ algorithm for the $\BSVP$~\cite{Buss93}.  Actually,
 we present a slight variation of his algorithm applied to formulas
 containing only the connective ``$\limp$'' as opposed to ``$\land$''
 and ``$\lor$''.  Given such a Boolean sentence, we can represent it
 as a binary tree with $2^{d+1}-1$ leaves for some $d$ (we can pad any
 sentence so that it meets this condition by preceding it with enough
 copies of ``$\ltrue\limp\dotsb$''), where each leaf stores either
 ``$\ltrue$'' or ``$\lfalse$'' and each interior node represents the
 connective ``$\limp$''.  For two nodes $U$ and $V$ in this tree, we
 write ``\(U\anc V\)'' to mean that $U$ is an ancestor of $V$, and
 ``\(U\anceq V\)'' to mean \(U=V\) or \(U\anc V\).  The least common
 ancestor of $U$ and $V$ is denoted $\lca(U,V)$.  By convention, we
 draw trees with the root at the top and the leaves at the bottom, so
 that ``above'' and ``below'' correspond to ``ancestor'' and
 ``descendant'', respectively.  Also, we define a \emph{scarred
 sentence} as a binary tree whose leaves store $\ltrue$ or $\lfalse$
 and that contains exactly one internal node with only one child (the
 missing child is called the \emph{scar}).  The ``value'' of a scarred
 sentence is defined to be a pair of truth-values
 $(t_\ltrue,t_\lfalse)$, where $t_\ltrue$ is the value of the Boolean
 sentence obtained when the scar is replaced by $\ltrue$, and
 similarly for $t_\lfalse$.

 {}The algorithm will be described as a \emph{pebbling game} on the
 formula's tree between two players: the \emph{Pebbler} and the
 \emph{Challenger}, and it proceeds in \emph{rounds}.  During each
 round, the Pebbler places \emph{pebbles} labelled with a truth-value
 ``$0$'' or ``$1$'' on nodes of the tree, representing assertions by
 the Pebbler that the subformulas rooted at those nodes have the
 indicated truth-values.  Following the Pebbler's move, the Challenger
 \emph{challenges} one of the pebbled positions $U$, representing an
 assertion by the Challenger that the pebble value at $U$ is incorrect
 (and implicitly, that every pebbled position below $U$ is correct).

 {}Intuitively, the essential feature of the pebbling-challenging game
 is to break up the work by creating scarred subsentences and
 evaluating them at the same time as their scar, instead of performing
 the evaluation sequentially.  (For example, we could evaluate
 $(A\limp B)\limp C$ by evaluating $(A\limp\scar)\limp C$ in parallel
 with $B$, where ``$\scar$'' indicates the position of the scar.)
 Together with Buss's innovative technique for finding scar positions
 in a semi-oblivious fashion through distinguished leaves, this
 feature of the algorithm allows even unbalanced sentences to be
 evaluated in a logarithmic number of steps.

 {}The game is designed so that the Pebbler has a winning strategy if
 the value of the sentence is ``$1$''; otherwise, the Challenger has a
 winning strategy.  Many of the rules of the game might seem somewhat
 arbitrary and strict, but they are designed so that a play of the
 game can be evaluated in $\ALT$ while preserving the property that
 the correct player has a winning strategy.  For example, the game
 will never last more than $d$ rounds (when there are $2^{d+1}-1$
 leaves), and since specifying arbitrary pebble positions would
 require $\bigOh(d)$ bits (which would take us outside $\ALT$), there
 must be a ``semi-oblivious'' way of specifying pebble positions using
 only a constant number of bits per round.

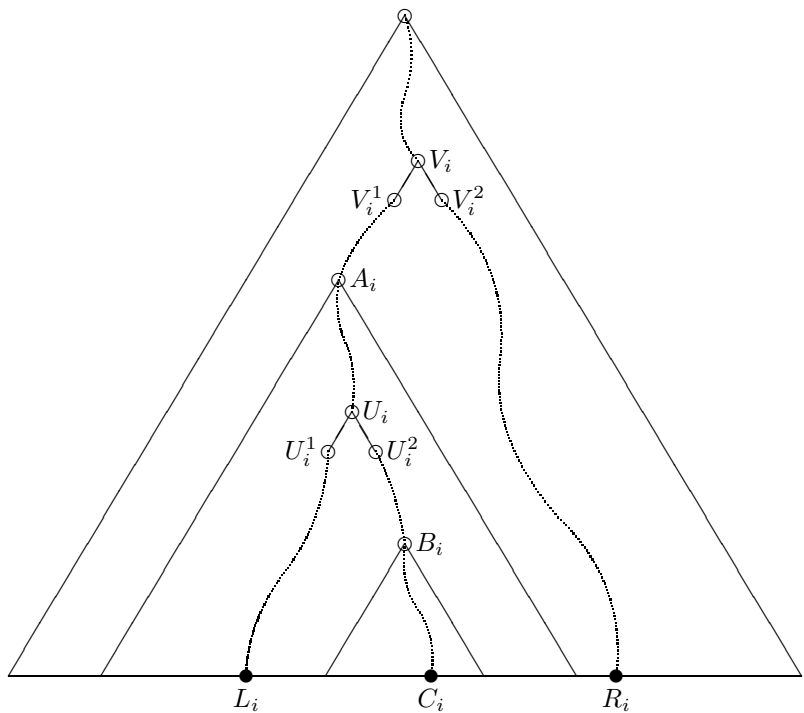
\begin{figure}[!ht]
\centering\small
   \begin{picture}(300,265)(-150,-262.5)
      \put(-150,-250){\line(1,0){300}}
      \put(0,0){\circle{5}}
      \put(0,0){\line(-3,-5){150}}
      \put(0,0){\line(3,-5){150}}
      \put(5,-55){\circle{5}}
      \put(9,-55){\makebox(0,0)[l]{$V_i$}}
      \put(5,-55){\line(-3,-5){9}}
      \put(5,-55){\line(3,-5){9}}
      \put(-4,-70){\circle{5}}
      \put(-8,-70){\makebox(0,0)[r]{$V_i^1$}}
      \put(14,-70){\circle{5}}
      \put(18,-70){\makebox(0,0)[l]{$V_i^2$}}
      \put(-25,-100){\circle{5}}
      \put(-21,-100){\makebox(0,0)[l]{$A_i$}}
      \put(-25,-100){\line(-3,-5){90}}
      \put(-25,-100){\line(3,-5){90}}
      \put(-20,-150){\circle{5}}
      \put(-16,-150){\makebox(0,0)[l]{$U_i$}}
      \put(-20,-150){\line(-3,-5){9}}
      \put(-20,-150){\line(3,-5){9}}
      \put(-29,-165){\circle{5}}
      \put(-33,-165){\makebox(0,0)[r]{$U_i^1$}}
      \put(-11,-165){\circle{5}}
      \put(-7,-165){\makebox(0,0)[l]{$U_i^2$}}
      \put(0,-200){\circle{5}}
      \put(4,-200){\makebox(0,0)[l]{$B_i$}}
      \put(0,-200){\line(-3,-5){30}}
      \put(0,-200){\line(3,-5){30}}
      \put(10,-250){\circle*{5}}
      \put(10,-255){\makebox(0,0)[t]{$C_i$}}
      \put(-60,-250){\circle*{5}}
      \put(-60,-255){\makebox(0,0)[t]{$L_i$}}
      \put(80,-250){\circle*{5}}
      \put(80,-255){\makebox(0,0)[t]{$R_i$}}
      \qbezier[20](0,0)(5,-15)(0,-30)
      \qbezier[20](0,-30)(-5,-45)(5,-55)
      \qbezier[30](-4,-70)(-20,-85)(-25,-100)
      \qbezier[50](14,-70)(40,-95)(36,-130)
      \qbezier[50](36,-130)(32,-160)(58,-190)
      \qbezier[50](58,-190)(84,-215)(80,-250)
      \qbezier[20](-25,-100)(-27,-115)(-22,-125)
      \qbezier[20](-22,-125)(-18,-135)(-20,-150)
      \qbezier[40](-29,-165)(-30,-190)(-45,-210)
      \qbezier[40](-45,-210)(-60,-230)(-60,-250)
      \qbezier[30](-11,-165)(-3,-180)(0,-200)
      \qbezier[20](0,-200)(-2,-215)(5,-225)
      \qbezier[20](5,-225)(12,-235)(10,-250)
   \end{picture}
\caption{Distinguished nodes (triangles delineate subtrees; dotted
   lines indicate paths).  Note that this illustrates only \emph{one}
   of the many configurations possible.}
\label{fig:BSVPtree}
\end{figure}

 {}Before giving the details of the pebbling game, we need to
 introduce a bit more notation.  First, leaves will be numbered from
 left to right, starting with $1$, and assigned a \emph{rank} equal to
 the largest integer $k$ such that $2^k$ divides the leaf number.
 Next, in each round \(i\geq 1\), there will be distinguished leaves
 $L_i$, $C_i$, and $R_i$ (for ``left'', ``center'', and ``right'',
 respectively) and distinguished nodes $A_i$ and $B_i$ (for ``above''
 and ``below'', respectively), satisfying the following conditions
 (see Figure~\ref{fig:BSVPtree} for a picture).

\begin{enumerate}
\item
   \( A_i \anceq B_i \anceq C_i \), with \( A_i = B_i \) only if
   \( B_i = C_i \);
\item
   $A_i$ is the lowest (and latest) challenged node, while $B_i$ is
   the highest pebbled position satisfying the first condition---or
   \(B_i=C_i\) if there is no pebbled node below $A_i$ (informally,
   the players have ``agreed'' at $B_i$ but ``disagree'' at $A_i$);
\item
   $L_i$ and $R_i$ are distinct leaves of rank $d-i$ and $C_i$ is of
   rank greater than $d-i$;
\item
   every leaf in the subtree rooted at $A_i$ but outside the subtree
   rooted at $B_i$ has number in the range
   \[
      (L_i-2^{d-i},L_i+2^{d-i}) \cup (R_i-2^{d-i},R_i+2^{d-i})
   \]
   (where the intervals are open).
\end{enumerate}

 {}The pebbling game proceeds as follows.  In round $0$, the Pebbler
 must pebble the root node with value ``$1$'', and the Challenger must
 challenge the pebble at the root.  In preparation for round $1$, set
 $A_1$ to be the root, $B_1=C_1$ to be the leaf numbered $2^d$, and
 $L_1$ and $R_1$ to be the leaves numbered $2^{d-1}$ and
 $2^d+2^{d-1}$, respectively (see Figure~\ref{fig:BSVP} for an example
 where each leaf's number and rank is indicated).  In round
 \(i\geq 1\), let \(U_i=\lca(L_i,C_i)\) and \(V_i=\lca(C_i,R_i)\)
 (note that $U_i$ and $V_i$ are distinct because $L_i$ and $R_i$ are
 distinct).

\begin{figure}[!ht]
\centering\small\setlength{\unitlength}{1.3125pt}
   \begin{picture}(337,128)(-228,-118)
      \put(0,0){\circle{3}}
      \put(0,0){\line(-6,-1){60}}
      \put(0,0){\line(5,-1){50}}
      \put(-60,-10){\circle{3}}
      \put(-60,-10){\line(-2,-1){20}}
      \put(-60,-10){\line(2,-1){20}}
      \put(50,-10){\circle{3}}
      \put(50,-10){\line(-3,-1){30}}
      \put(50,-10){\line(3,-1){30}}
      \put(-80,-20){\circle{3}}
      \qbezier(-80,-20)(-130,-25)(-180,-30)
      \put(-80,-20){\line(1,-1){10}}
      \put(-40,-20){\circle{3}}
      \put(-40,-20){\line(-1,-1){10}}
      \put(-40,-20){\line(3,-1){30}}
      \put(20,-20){\circle{3}}
      \put(20,-20){\line(-1,-1){10}}
      \put(20,-20){\line(2,-1){20}}
      \put(80,-20){\circle{3}}
      \put(80,-20){\line(-2,-1){20}}
      \put(80,-20){\line(2,-1){20}}
      \put(-180,-30){\circle{3}}
      \put(-180,-30){\line(-1,-1){10}}
      \put(-180,-30){\line(2,-1){20}}
      \put(-70,-30){\circle{3}}
      \put(-70,-30){\line(-1,-2){5}}
      \put(-70,-30){\line(1,-2){5}}
      \put(-50,-30){\circle{3}}
      \put(-50,-30){\line(-1,-2){5}}
      \put(-50,-30){\line(1,-2){5}}
      \put(-10,-30){\circle{3}}
      \put(-10,-30){\line(-2,-1){20}}
      \put(-10,-30){\line(1,-2){5}}
      \put(10,-30){\circle{3}}
      \put(10,-30){\line(-1,-2){5}}
      \put(10,-30){\line(1,-2){5}}
      \put(40,-30){\circle{3}}
      \put(40,-30){\line(-1,-1){10}}
      \put(40,-30){\line(1,-2){5}}
      \put(60,-30){\circle{3}}
      \put(60,-30){\line(-1,-2){5}}
      \put(60,-30){\line(1,-1){10}}
      \put(100,-30){\circle{3}}
      \put(100,-30){\line(-1,-1){10}}
      \put(100,-30){\line(1,-2){5}}
      \put(-190,-40){\circle{3}}
      \put(-190,-40){\line(-1,-2){5}}
      \put(-190,-40){\line(1,-2){5}}
      \put(-160,-40){\circle{3}}
      \put(-160,-40){\line(-1,-1){10}}
      \put(-160,-40){\line(3,-1){30}}
      \put(-30,-40){\circle{3}}
      \put(-30,-40){\line(-1,-2){5}}
      \put(-30,-40){\line(1,-1){10}}
      \put(30,-40){\circle{3}}
      \put(30,-40){\line(-1,-2){5}}
      \put(30,-40){\line(1,-2){5}}
      \put(70,-40){\circle{3}}
      \put(70,-40){\line(-1,-2){5}}
      \put(70,-40){\line(1,-2){5}}
      \put(90,-40){\circle{3}}
      \put(90,-40){\line(-1,-2){5}}
      \put(90,-40){\line(1,-2){5}}
      \put(-170,-50){\circle{3}}
      \put(-170,-50){\line(-1,-2){5}}
      \put(-170,-50){\line(1,-2){5}}
      \put(-130,-50){\circle{3}}
      \put(-130,-50){\line(-1,-1){10}}
      \put(-130,-50){\line(3,-1){30}}
      \put(-20,-50){\circle{3}}
      \put(-20,-50){\line(-1,-2){5}}
      \put(-20,-50){\line(1,-2){5}}
      \put(-140,-60){\circle{3}}
      \put(-140,-60){\line(-1,-1){10}}
      \put(-140,-60){\line(1,-2){5}}
      \put(-100,-60){\circle{3}}
      \put(-100,-60){\line(-2,-1){20}}
      \put(-100,-60){\line(1,-1){10}}
      \put(-150,-70){\circle{3}}
      \put(-150,-70){\line(-1,-2){5}}
      \put(-150,-70){\line(1,-2){5}}
      \put(-120,-70){\circle{3}}
      \put(-120,-70){\line(-1,-2){5}}
      \put(-120,-70){\line(1,-1){10}}
      \put(-90,-70){\circle{3}}
      \put(-90,-70){\line(-1,-2){5}}
      \put(-90,-70){\line(1,-2){5}}
      \put(-110,-80){\circle{3}}
      \put(-110,-80){\line(-1,-2){5}}
      \put(-110,-80){\line(1,-2){5}}
      \put(-195,-50){\circle*{3}}
      \put(-185,-50){\circle*{3}}
      \put(-175,-60){\circle*{3}}
      \put(-165,-60){\circle*{3}}
      \put(-155,-80){\circle*{3}}
      \put(-145,-80){\circle*{3}}
      \put(-135,-70){\circle*{3}}
      \put(-125,-80){\circle*{3}}
      \put(-115,-90){\circle*{3}}
      \put(-105,-90){\circle*{3}}
      \put(-95,-80){\circle*{3}}
      \put(-85,-80){\circle*{3}}
      \put(-75,-40){\circle*{3}}
      \put(-65,-40){\circle*{3}}
      \put(-55,-40){\circle*{3}}
      \put(-45,-40){\circle*{3}}
      \put(-35,-50){\circle*{3}}
      \put(-25,-60){\circle*{3}}
      \put(-15,-60){\circle*{3}}
      \put(-5,-40){\circle*{3}}
      \put(5,-40){\circle*{3}}
      \put(15,-40){\circle*{3}}
      \put(25,-50){\circle*{3}}
      \put(35,-50){\circle*{3}}
      \put(45,-40){\circle*{3}}
      \put(55,-40){\circle*{3}}
      \put(65,-50){\circle*{3}}
      \put(75,-50){\circle*{3}}
      \put(85,-50){\circle*{3}}
      \put(95,-50){\circle*{3}}
      \put(105,-40){\circle*{3}}
      \qbezier[25](-195,-50)(-195,-75)(-195,-100)
      \qbezier[25](-185,-50)(-185,-75)(-185,-100)
      \qbezier[20](-175,-60)(-175,-80)(-175,-100)
      \qbezier[20](-165,-60)(-165,-80)(-165,-100)
      \qbezier[10](-155,-80)(-155,-90)(-155,-100)
      \qbezier[10](-145,-80)(-145,-90)(-145,-100)
      \qbezier[15](-135,-70)(-135,-85)(-135,-100)
      \qbezier[5](-125,-90)(-125,-95)(-125,-100)
      \qbezier[5](-115,-90)(-115,-95)(-115,-100)
      \qbezier[5](-105,-90)(-105,-95)(-105,-100)
      \qbezier[10](-95,-80)(-95,-90)(-95,-100)
      \qbezier[10](-85,-80)(-85,-90)(-85,-100)
      \qbezier[30](-75,-40)(-75,-70)(-75,-100)
      \qbezier[30](-65,-40)(-65,-70)(-65,-100)
      \qbezier[30](-55,-40)(-55,-70)(-55,-100)
      \qbezier[21](-45,-58)(-45,-79)(-45,-100)
      \qbezier[25](-35,-50)(-35,-75)(-35,-100)
      \qbezier[20](-25,-60)(-25,-80)(-25,-100)
      \qbezier[20](-15,-60)(-15,-80)(-15,-100)
      \qbezier[30](-5,-40)(-5,-70)(-5,-100)
      \qbezier[30](5,-40)(5,-70)(5,-100)
      \qbezier[30](15,-40)(15,-70)(15,-100)
      \qbezier[25](25,-50)(25,-75)(25,-100)
      \qbezier[20](35,-60)(35,-80)(35,-100)
      \qbezier[30](45,-40)(45,-70)(45,-100)
      \qbezier[30](55,-40)(55,-70)(55,-100)
      \qbezier[25](65,-50)(65,-75)(65,-100)
      \qbezier[25](75,-50)(75,-75)(75,-100)
      \qbezier[25](85,-50)(85,-75)(85,-100)
      \qbezier[25](95,-50)(95,-75)(95,-100)
      \qbezier[30](105,-40)(105,-70)(105,-100)
      \put(-200,-105){\makebox(0,0)[r]{\vphantom{q}number:}}
      \put(-195,-105){\makebox(0,0){1}}
      \put(-185,-105){\makebox(0,0){2}}
      \put(-175,-105){\makebox(0,0){3}}
      \put(-165,-105){\makebox(0,0){4}}
      \put(-155,-105){\makebox(0,0){5}}
      \put(-145,-105){\makebox(0,0){6}}
      \put(-135,-105){\makebox(0,0){7}}
      \put(-125,-105){\makebox(0,0){8}}
      \put(-115,-105){\makebox(0,0){9}}
      \put(-105,-105){\makebox(0,0){10}}
      \put(-95,-105){\makebox(0,0){11}}
      \put(-85,-105){\makebox(0,0){12}}
      \put(-75,-105){\makebox(0,0){13}}
      \put(-65,-105){\makebox(0,0){14}}
      \put(-55,-105){\makebox(0,0){15}}
      \put(-45,-105){\makebox(0,0){16}}
      \put(-35,-105){\makebox(0,0){17}}
      \put(-25,-105){\makebox(0,0){18}}
      \put(-15,-105){\makebox(0,0){19}}
      \put(-5,-105){\makebox(0,0){20}}
      \put(5,-105){\makebox(0,0){21}}
      \put(15,-105){\makebox(0,0){22}}
      \put(25,-105){\makebox(0,0){23}}
      \put(35,-105){\makebox(0,0){24}}
      \put(45,-105){\makebox(0,0){25}}
      \put(55,-105){\makebox(0,0){26}}
      \put(65,-105){\makebox(0,0){27}}
      \put(75,-105){\makebox(0,0){28}}
      \put(85,-105){\makebox(0,0){29}}
      \put(95,-105){\makebox(0,0){30}}
      \put(105,-105){\makebox(0,0){31}}
      \put(-200,-115){\makebox(0,0)[r]{\vphantom{q}rank:}}
      \put(-195,-115){\makebox(0,0){0}}
      \put(-185,-115){\makebox(0,0){1}}
      \put(-175,-115){\makebox(0,0){0}}
      \put(-165,-115){\makebox(0,0){2}}
      \put(-155,-115){\makebox(0,0){0}}
      \put(-145,-115){\makebox(0,0){1}}
      \put(-135,-115){\makebox(0,0){0}}
      \put(-125,-115){\makebox(0,0){3}}
      \put(-115,-115){\makebox(0,0){0}}
      \put(-105,-115){\makebox(0,0){1}}
      \put(-95,-115){\makebox(0,0){0}}
      \put(-85,-115){\makebox(0,0){2}}
      \put(-75,-115){\makebox(0,0){0}}
      \put(-65,-115){\makebox(0,0){1}}
      \put(-55,-115){\makebox(0,0){0}}
      \put(-45,-115){\makebox(0,0){4}}
      \put(-35,-115){\makebox(0,0){0}}
      \put(-25,-115){\makebox(0,0){1}}
      \put(-15,-115){\makebox(0,0){0}}
      \put(-5,-115){\makebox(0,0){2}}
      \put(5,-115){\makebox(0,0){0}}
      \put(15,-115){\makebox(0,0){1}}
      \put(25,-115){\makebox(0,0){0}}
      \put(35,-115){\makebox(0,0){3}}
      \put(45,-115){\makebox(0,0){0}}
      \put(55,-115){\makebox(0,0){1}}
      \put(65,-115){\makebox(0,0){0}}
      \put(75,-115){\makebox(0,0){2}}
      \put(85,-115){\makebox(0,0){0}}
      \put(95,-115){\makebox(0,0){1}}
      \put(105,-115){\makebox(0,0){0}}
      \put(-45,-43){\makebox(0,0)[t]{$C_1$}}
      \put(-125,-83){\makebox(0,0)[t]{$L_1$}}
      \put(35,-53){\makebox(0,0)[t]{$R_1$}}
      \put(-45,-50){\makebox(0,0)[t]{$B_1$}}
      \put(0,-3){\makebox(0,0)[t]{$A_1$}}
      \put(0,3){\makebox(0,0)[b]{$V_1$}}
      \put(-60,0){\makebox(0,0)[b]{$V_1^1$}}
      \put(50,-7){\makebox(0,0)[b]{$V_1^2$}}
      \put(-60,-7){\makebox(0,0)[b]{$U_1$}}
      \put(-80,-17){\makebox(0,0)[b]{$U_1^1$}}
      \put(-40,-17){\makebox(0,0)[b]{$U_1^2$}}
   \end{picture}
\caption{Labelled example of the $\BSVP$ algorithm at round 1.}
\label{fig:BSVP}
\end{figure}

 {}In each round, the Pebbler uses six bits of information (one per
 node) to pebble $U_i$, $V_i$, and their two immediate children
 ($U_i^1$, $V_i^1$ to the left and $U_i^2$, $V_i^2$ to the right,
 respectively).  In addition, the Pebbler must use three bits of
 information to specify the relative positions of $A_i$, $U_i$, and
 $V_i$ (\ie, which nodes are ancestors of which ones---since all three
 nodes are ancestors of $C_i$, they all lie on the path from $C_i$ to
 the root).

 {}The Challenger then challenges one node from among $A_i$, $U_i$,
 $U_i^1$, $U_i^2$, $V_i$, $V_i^1$, or $V_i^2$ (using three bits to
 specify which one), subject to the conditions that the node
 challenged must be in the subtree rooted at $A_i$ and outside the
 subtree rooted at $B_i$.

 {}For round $i+1$, $A_{i+1}$ is set to the node just challenged,
 $B_{i+1}$ is set to the highest pebbled node below $A_{i+1}$ (or to
 $C_{i+1}$ if there is no pebbled node below $A_{i+1}$), and
 $L_{i+1}$, $C_{i+1}$, $R_{i+1}$ are set according to
 Table~\ref{table:BSVP}.

\begin{table}[!ht]
\begin{center}
   \vspace*{.5ex}
   \begin{tabular}{|c||c|c|}
   \hline
      Challenged
    & Pebbler says
    & Pebbler says \\[-.5ex]
      Node
    & \( U_i \anc V_i \)
    & \( V_i \anc U_i \) \\
   \hline
      
    & \tableline{C_{i+1}}{L_i}
    & \tableline{C_{i+1}}{L_i} \\
      $U_i^1$
    & \tableline{R_{i+1}}{L_i+2^{d-i-1}}
    & \tableline{R_{i+1}}{L_i+2^{d-i-1}} \\
      
    & \tableline{L_{i+1}}{L_i-2^{d-i-1}}
    & \tableline{L_{i+1}}{L_i-2^{d-i-1}} \\
   \hline
      
    & \tableline{C_{i+1}}{C_i}
    & \tableline{C_{i+1}}{C_i} \\
      $U_i^2$
    & \tableline{R_{i+1}}{R_i+2^{d-i-1}}
    & \tableline{R_{i+1}}{R_i-2^{d-i-1}} \\
      
    & \tableline{L_{i+1}}{L_i+2^{d-i-1}}
    & \tableline{L_{i+1}}{L_i+2^{d-i-1}} \\
   \hline
      
    & \tableline{C_{i+1}}{C_i}
    & \tableline{C_{i+1}}{C_i} \\
      $V_i^1$
    & \tableline{R_{i+1}}{R_i-2^{d-i-1}}
    & \tableline{R_{i+1}}{R_i-2^{d-i-1}} \\
      
    & \tableline{L_{i+1}}{L_i+2^{d-i-1}}
    & \tableline{L_{i+1}}{L_i-2^{d-i-1}} \\
   \hline
      
    & \tableline{C_{i+1}}{R_i}
    & \tableline{C_{i+1}}{R_i} \\
      $V_i^2$
    & \tableline{R_{i+1}}{R_i+2^{d-i-1}}
    & \tableline{R_{i+1}}{R_i+2^{d-i-1}} \\
      
    & \tableline{L_{i+1}}{R_i-2^{d-i-1}}
    & \tableline{L_{i+1}}{R_i-2^{d-i-1}} \\
   \hline
      $U_i$ or $V_i$
    & Game Ends
    & Game Ends \\
   \hline
   \end{tabular} \\[4.5ex]
   \begin{tabular}{|c|c|c|c|}
   \hline
      \multicolumn{4}{|c|}{Challenged Node: $A_i$} \\
   \hline
   \hline
      Pebbler says
    & Pebbler says
    & Pebbler says
    & Pebbler says \\[-.5ex]
      \( A_i \anc U_i,V_i \)
    & \( U_i,V_i \anceq A_i \)
    & \( U_i \anceq A_i \anc V_i \)
    & \( V_i \anceq A_i \anc U_i \) \\
   \hline
      \tableline{C_{i+1}}{C_i}
    & \tableline{C_{i+1}}{C_i}
    & \tableline{C_{i+1}}{C_i}
    & \tableline{C_{i+1}}{C_i} \\
      \tableline{R_{i+1}}{R_i+2^{d-i-1}}
    & \tableline{R_{i+1}}{R_i-2^{d-i-1}}
    & \tableline{R_{i+1}}{R_i+2^{d-i-1}}
    & \tableline{R_{i+1}}{R_i-2^{d-i-1}} \\
      \tableline{L_{i+1}}{L_i-2^{d-i-1}}
    & \tableline{L_{i+1}}{L_i+2^{d-i-1}}
    & \tableline{L_{i+1}}{L_i+2^{d-i-1}}
    & \tableline{L_{i+1}}{L_i-2^{d-i-1}} \\
   \hline
   \end{tabular} \\
\end{center}
\caption{Next leaf nodes in Buss's $\BSVP$ algorithm.}
\label{table:BSVP}
\end{table}

 {}Now, it is easy to show by induction on the number of rounds played
 that properties 1--4 are preserved for the duration of the algorithm:
 it is simply a matter of checking case-by-case each possibility in
 Table~\ref{table:BSVP} for the values of $C_i$, $L_i$, and $R_i$.
 For example, suppose that $V_1^2$ is challenged in
 Figure~\ref{fig:BSVP}, then for round 2, \(A_2=V_1^2\),
 \(B_2=C_2=R_1=24\), \(L_2=R_1-4=20\), and \(R_2=R_1+4=28\) (we refer
 to leaves by their number), so \(A_2\anc B_2 = C_2\), $A_2$ is the
 lowest challenged node and \(B_2=C_2\), $L_2$ and $R_2$ have rank
 $2=4-2$ and $C_2$ has rank $3>2$, and every leaf in the subtree
 rooted at $A_2$ but outside the subtree rooted at $B_2$ has number in
 the range
\( (20-2^{4-2},20+2^{4-2})\cup(28-2^{4-2},28+2^{4-2}) =
   (16,24)\cup(24,32) \).

 {}The game ends as soon as one of the players makes an ``obvious''
 mistake, \ie, one from the following list.  (Note that by property 4,
 the game must end by round number $d$ because
\( (L_d-2^{d-d},L_d+2^{d-d})\cup(R_d-2^{d-d},R_d+2^{d-d}) = 
   \{L_d,R_d\} \);
 it is easy to see that in that case, one of the two players will be
 forced to make a mistake from the list below.)
\begin{itemize}
\item
   Pebbler: when the input nodes of a gate are either leaves or are
   pebbled and the output node is pebbled incompatibly.
\item
   Challenger: when the output of a gate whose input nodes are either
   leaves or pebbled is challenged, even though it is correctly
   pebbled.
\item
   Pebbler: when a leaf is incorrectly pebbled.
\item
   Challenger: when a correctly pebbled leaf is challenged.
\item
   Pebbler: when a node is pebbled with both ``$0$'' and ``$1$''.
\item
   Pebbler: when an incorrect assertion is made about whether
   \(U_i\anc V_i\), \(A_i\anc U_i\), \(A_i\anc V_i\).
\item
   Challenger: when the challenged node is above a previously
   challenged node.
\item
   Challenger: when the challenged node is at or below a previously
   agreed upon pebble value (a pebble is ``agreed upon'' if it was
   placed in an earlier round and in that round, the Challenger
   challenged an ancestor of that pebble).
\end{itemize}

 {}It is straightforward to see that the game produces the correct
 result: if the value of the sentence is ``true'', the Pebbler can win
 the game by simply pebbling every node with its correct value and
 making assertions compatible with the structure of the sentence,
 while if the value of the sentence is ``false'', the Challenger can
 win the game by always challenging the lowest incorrectly pebbled
 node that is not below a previously agreed upon node.

 {}Moreover, the game can be translated into an $\ALT$ algorithm, as
 follows: First, simulate possible plays of the game using existential
 moves for the Pebbler and universal moves for the Challenger.  Then,
 for each such game, existentially guess the first mistake made and
 universally verify that no earlier mistake was made.  Note that from
 a play of the game, it is easy to determine the last round when $L_j$
 was computed from $R_j$ and to compute the appropriate sum of powers
 of $2$ to add to $R_j$ in order to get $L_i$ (the same goes for
 $R_i$).  As for $C_i$, simply find the last round when $C_j$ was
 equal to $L_j$ or $R_j$ and we know that \(C_i=C_j\).  Finally,
 because it is possible to count in $\ALT$, ancestors can readily be
 computed to find $U_i$, $V_i$ and thus determine $A_i$ and $B_i$.

\section{Formalizing the $\BSVP$}

 The algorithm that we will use to solve the $\BSVP$ in $T_1$ is
 simply a formalization inside our theory of the algorithm described
 in the previous section.  To formalize this algorithm inside $T_1$,
 we will define a function $\fcn{BSVP}$ that decomposes and evaluates
 the sentence, using $\TRN$ to perform the work in parallel,
 disjunction ($\fcn{OR}$) over all possible Pebbler guesses to emulate
 existential moves, and composition and implication to emulate
 Challenger's universal moves.

 {}First, it is easy to define by $\rpowCRN$ a function
 $\fcn{sentence}(v,x,z)$ that takes as argument a truth-value
 assignment $v$ (represented simply by a string whose $1^{\text{st}}$
 bit is the value of $p_1$, whose $2^{\text{nd}}$ bit is the value of
 $p_2$, etc.) and the encoding of a formula
 \(x=\code_{|\lenfcn{pow}(z)|}A\), and returns the sentence obtained
 by substituting the given truth-values for the variables in $x$.

 {}Next, we define the function $\fcn{BSVP}(d,h,m,s)$ that does the
 work according to Buss's algorithm.  There will be one variation:
 because we want the function to apply to arbitrary sentences, but a
 sentence must have a power of $2$ minus $1$ leaves in the algorithm,
 we will pad sentences so they have $2^{d+1}-1$ leaves and remember
 the position of the root of the original sentence inside the padded
 version as ``$M$''.  (The algorithm needs to be changed so that it
 takes the distinguished node $M$ into account at the same time as
 $A$, $B$, $L$, $C$, $R$, but the changes are easy to make since $M$
 remains fixed for the duration of the algorithm.)  In what follows,
 the parameter $s$ is fixed and encodes a superformula of the Boolean
 sentence we are evaluating (padded so it always has a power of two
 minus one leaves), the parameter $m$ is fixed and indicates the root
 $M$ of the subformula of $s$ whose value we are interested in, the
 parameter $h$ varies and represents the history of the game so far,
 as a sequence of blocks $b_1a_1\dotsm b_ia_i$ (each block a
 constant-length string encoding Pebbler's guesses on the relative
 positions of the nodes $A,U,V,B$, and $M$ (in $b_j$), as well as
 Challenger's chosen node (in $a_j$)), and $d$ varies and represents
 $2$ to the power of the current round number, in unary.  The function
 $\fcn{BSVP}(d,h,m,s)$ returns a \emph{truth-triplet}
 $(c,t_\ltrue,t_\lfalse)$, where $c$ is a check-bit indicating whether
 $h$ is a valid description of the structure of $s$ or not, and
 $(t_\ltrue,t_\lfalse)$ is the value of the possibly scarred
 subformula of $M$ picked out by $h$.  The function will be defined by
 $\TRN$ on $d$, following Buss's algorithm, but first, we must specify
 how truth-triplets can be combined in various ways.

 {}We generalize disjunction and implication to truth-triplets, and
 define composition of truth-triplets, as follows:
\begin{align*}
   (c^1,t_\ltrue^1,t_\lfalse^1)\tlor(c^2,t_\ltrue^2,t_\lfalse^2)
   &= \big(c^1\lor c^2,(c^1\land t_\ltrue^1)\lor(c^2\land t_\ltrue^2),
        (c^1\land t_\lfalse^1)\lor(c^2\land t_\lfalse^2)\big), \\
   (c^1,t_\ltrue^1,t_\lfalse^1)\tlimp(c^2,t_\ltrue^2,t_\lfalse^2)
   &= \big(c^1\land c^2,t_\ltrue^1\limp t_\ltrue^2,
         t_\lfalse^1\limp t_\lfalse^2\big), \\
   (c^1,t_\ltrue^1,t_\lfalse^1)\tcomp(c^2,t_\ltrue^2,t_\lfalse^2)
   &= \big(c^1\land c^2,t_{t_\ltrue^2}^1,t_{t_\lfalse^2}^1\big) 
\end{align*}
 (where $t_{t_\ltrue^2}^1$ is equal to $t_\ltrue^1$ if
 \(t_\ltrue^2=\ltrue\) and $t_\lfalse^1$ if \(t_\ltrue^2=\lfalse\),
 and similarly for $t_{t_\lfalse^2}^1$).

\noindent
 Then, for \(h=b_1a_1\dotsm b_ia_i\), we can define
 $\fcn{BSVP}(d,h,m,s)$ as follows:
\[
   \fcn{BSVP}(d,h,m,s) = \bigor_{\substack
        {\text{all Pebbler}\\
         \text{guesses $b$}}}\left[\text{\parbox{3.75in}{
         composition of $\fcn{BSVP}(\rhalf d,hba,m,s)$'s using
         $\tcomp$ and $\tlimp$, based on structure induced by $b$ and
         where $a$ picks out different subformulas at the current
         round
      }}\right].
\]
 Because $b$ has a fixed length, the disjunction actually represents a
 fixed number of cases, each one of which has a unique structure
 determined by the value of $b$.  We will not list all possible cases
 here (they can easily be written down from the description of Buss's
 algorithm and Table~\ref{table:BSVP}), but we give two illustrative
 examples based on the sentence depicted in Figure~\ref{fig:BSVP}.
\begin{enumerate}
\item
   Consider the sentence depicted in Figure~\ref{fig:BSVP1}, where we
   have filled-in the unique interior node that represents ``$m$'' and
   each leaf that falls inside the correct intervals around $L_1$ and
   $R_1$.  At round 1, we have that
   \( \fcn{BSVP}(d,h,m,s) = \fcn{BSVP}(\rhalf d,hba_{U^1},m,s)
      \tlimp\fcn{BSVP}(\rhalf d,hba_{U^2},m,s) \),
   where ``$a_{U^1}$'' and ``$a_{U^2}$'' are fixed-length strings
   representing which subsentence is selected, and ``$b$'' is the
   unique fixed-length string representing the correct structure of
   the formula.  The other parts of the formula (under $V_1^2$) are of
   no interest because they fall outside ``$m$''.

   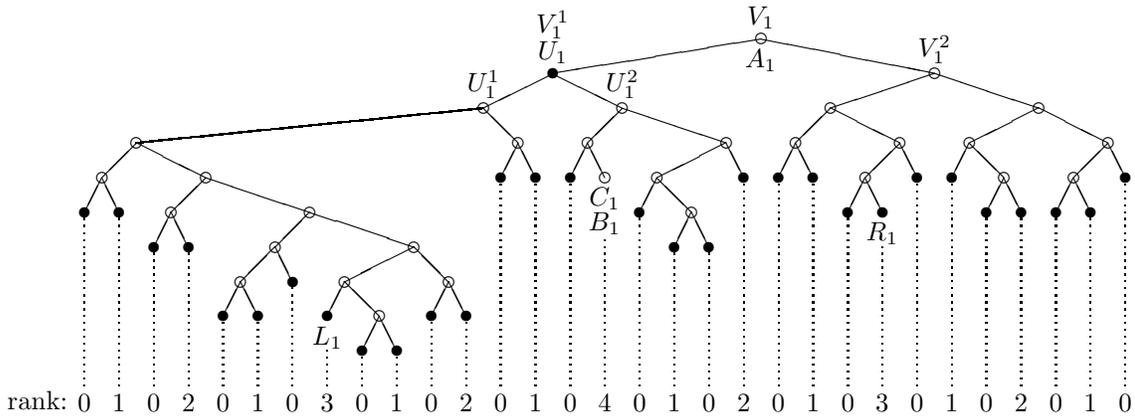
\begin{figure}[!ht]
   \centering\small\setlength{\unitlength}{1.3125pt}
      \begin{picture}(327,118)(-218,-108)
         \put(0,0){\circle{3}}
         \put(0,0){\line(-6,-1){60}}
         \put(0,0){\line(5,-1){50}}
         \put(-60,-10){\circle*{3}}
         \put(-60,-10){\line(-2,-1){20}}
         \put(-60,-10){\line(2,-1){20}}
         \put(50,-10){\circle{3}}
         \put(50,-10){\line(-3,-1){30}}
         \put(50,-10){\line(3,-1){30}}
         \put(-80,-20){\circle{3}}
         \qbezier(-80,-20)(-130,-25)(-180,-30)
         \put(-80,-20){\line(1,-1){10}}
         \put(-40,-20){\circle{3}}
         \put(-40,-20){\line(-1,-1){10}}
         \put(-40,-20){\line(3,-1){30}}
         \put(20,-20){\circle{3}}
         \put(20,-20){\line(-1,-1){10}}
         \put(20,-20){\line(2,-1){20}}
         \put(80,-20){\circle{3}}
         \put(80,-20){\line(-2,-1){20}}
         \put(80,-20){\line(2,-1){20}}
         \put(-180,-30){\circle{3}}
         \put(-180,-30){\line(-1,-1){10}}
         \put(-180,-30){\line(2,-1){20}}
         \put(-70,-30){\circle{3}}
         \put(-70,-30){\line(-1,-2){5}}
         \put(-70,-30){\line(1,-2){5}}
         \put(-50,-30){\circle{3}}
         \put(-50,-30){\line(-1,-2){5}}
         \put(-50,-30){\line(1,-2){5}}
         \put(-10,-30){\circle{3}}
         \put(-10,-30){\line(-2,-1){20}}
         \put(-10,-30){\line(1,-2){5}}
         \put(10,-30){\circle{3}}
         \put(10,-30){\line(-1,-2){5}}
         \put(10,-30){\line(1,-2){5}}
         \put(40,-30){\circle{3}}
         \put(40,-30){\line(-1,-1){10}}
         \put(40,-30){\line(1,-2){5}}
         \put(60,-30){\circle{3}}
         \put(60,-30){\line(-1,-2){5}}
         \put(60,-30){\line(1,-1){10}}
         \put(100,-30){\circle{3}}
         \put(100,-30){\line(-1,-1){10}}
         \put(100,-30){\line(1,-2){5}}
         \put(-190,-40){\circle{3}}
         \put(-190,-40){\line(-1,-2){5}}
         \put(-190,-40){\line(1,-2){5}}
         \put(-160,-40){\circle{3}}
         \put(-160,-40){\line(-1,-1){10}}
         \put(-160,-40){\line(3,-1){30}}
         \put(-30,-40){\circle{3}}
         \put(-30,-40){\line(-1,-2){5}}
         \put(-30,-40){\line(1,-1){10}}
         \put(30,-40){\circle{3}}
         \put(30,-40){\line(-1,-2){5}}
         \put(30,-40){\line(1,-2){5}}
         \put(70,-40){\circle{3}}
         \put(70,-40){\line(-1,-2){5}}
         \put(70,-40){\line(1,-2){5}}
         \put(90,-40){\circle{3}}
         \put(90,-40){\line(-1,-2){5}}
         \put(90,-40){\line(1,-2){5}}
         \put(-170,-50){\circle{3}}
         \put(-170,-50){\line(-1,-2){5}}
         \put(-170,-50){\line(1,-2){5}}
         \put(-130,-50){\circle{3}}
         \put(-130,-50){\line(-1,-1){10}}
         \put(-130,-50){\line(3,-1){30}}
         \put(-20,-50){\circle{3}}
         \put(-20,-50){\line(-1,-2){5}}
         \put(-20,-50){\line(1,-2){5}}
         \put(-140,-60){\circle{3}}
         \put(-140,-60){\line(-1,-1){10}}
         \put(-140,-60){\line(1,-2){5}}
         \put(-100,-60){\circle{3}}
         \put(-100,-60){\line(-2,-1){20}}
         \put(-100,-60){\line(1,-1){10}}
         \put(-150,-70){\circle{3}}
         \put(-150,-70){\line(-1,-2){5}}
         \put(-150,-70){\line(1,-2){5}}
         \put(-120,-70){\circle{3}}
         \put(-120,-70){\line(-1,-2){5}}
         \put(-120,-70){\line(1,-1){10}}
         \put(-90,-70){\circle{3}}
         \put(-90,-70){\line(-1,-2){5}}
         \put(-90,-70){\line(1,-2){5}}
         \put(-110,-80){\circle{3}}
         \put(-110,-80){\line(-1,-2){5}}
         \put(-110,-80){\line(1,-2){5}}
         \put(-195,-50){\circle*{3}}
         \put(-185,-50){\circle*{3}}
         \put(-175,-60){\circle*{3}}
         \put(-165,-60){\circle*{3}}
         \put(-155,-80){\circle*{3}}
         \put(-145,-80){\circle*{3}}
         \put(-135,-70){\circle*{3}}
         \put(-125,-80){\circle*{3}}
         \put(-115,-90){\circle*{3}}
         \put(-105,-90){\circle*{3}}
         \put(-95,-80){\circle*{3}}
         \put(-85,-80){\circle*{3}}
         \put(-75,-40){\circle*{3}}
         \put(-65,-40){\circle*{3}}
         \put(-55,-40){\circle*{3}}
         \put(-45,-40){\circle{3}}
         \put(-35,-50){\circle*{3}}
         \put(-25,-60){\circle*{3}}
         \put(-15,-60){\circle*{3}}
         \put(-5,-40){\circle*{3}}
         \put(5,-40){\circle*{3}}
         \put(15,-40){\circle*{3}}
         \put(25,-50){\circle*{3}}
         \put(35,-50){\circle*{3}}
         \put(45,-40){\circle*{3}}
         \put(55,-40){\circle*{3}}
         \put(65,-50){\circle*{3}}
         \put(75,-50){\circle*{3}}
         \put(85,-50){\circle*{3}}
         \put(95,-50){\circle*{3}}
         \put(105,-40){\circle*{3}}
         \qbezier[25](-195,-50)(-195,-75)(-195,-100)
         \qbezier[25](-185,-50)(-185,-75)(-185,-100)
         \qbezier[20](-175,-60)(-175,-80)(-175,-100)
         \qbezier[20](-165,-60)(-165,-80)(-165,-100)
         \qbezier[10](-155,-80)(-155,-90)(-155,-100)
         \qbezier[10](-145,-80)(-145,-90)(-145,-100)
         \qbezier[15](-135,-70)(-135,-85)(-135,-100)
         \qbezier[5](-125,-90)(-125,-95)(-125,-100)
         \qbezier[5](-115,-90)(-115,-95)(-115,-100)
         \qbezier[5](-105,-90)(-105,-95)(-105,-100)
         \qbezier[10](-95,-80)(-95,-90)(-95,-100)
         \qbezier[10](-85,-80)(-85,-90)(-85,-100)
         \qbezier[30](-75,-40)(-75,-70)(-75,-100)
         \qbezier[30](-65,-40)(-65,-70)(-65,-100)
         \qbezier[30](-55,-40)(-55,-70)(-55,-100)
         \qbezier[21](-45,-58)(-45,-79)(-45,-100)
         \qbezier[25](-35,-50)(-35,-75)(-35,-100)
         \qbezier[20](-25,-60)(-25,-80)(-25,-100)
         \qbezier[20](-15,-60)(-15,-80)(-15,-100)
         \qbezier[30](-5,-40)(-5,-70)(-5,-100)
         \qbezier[30](5,-40)(5,-70)(5,-100)
         \qbezier[30](15,-40)(15,-70)(15,-100)
         \qbezier[25](25,-50)(25,-75)(25,-100)
         \qbezier[20](35,-60)(35,-80)(35,-100)
         \qbezier[30](45,-40)(45,-70)(45,-100)
         \qbezier[30](55,-40)(55,-70)(55,-100)
         \qbezier[25](65,-50)(65,-75)(65,-100)
         \qbezier[25](75,-50)(75,-75)(75,-100)
         \qbezier[25](85,-50)(85,-75)(85,-100)
         \qbezier[25](95,-50)(95,-75)(95,-100)
         \qbezier[30](105,-40)(105,-70)(105,-100)
         \put(-200,-105){\makebox(0,0)[r]{\vphantom{q}rank:}}
         \put(-195,-105){\makebox(0,0){0}}
         \put(-185,-105){\makebox(0,0){1}}
         \put(-175,-105){\makebox(0,0){0}}
         \put(-165,-105){\makebox(0,0){2}}
         \put(-155,-105){\makebox(0,0){0}}
         \put(-145,-105){\makebox(0,0){1}}
         \put(-135,-105){\makebox(0,0){0}}
         \put(-125,-105){\makebox(0,0){3}}
         \put(-115,-105){\makebox(0,0){0}}
         \put(-105,-105){\makebox(0,0){1}}
         \put(-95,-105){\makebox(0,0){0}}
         \put(-85,-105){\makebox(0,0){2}}
         \put(-75,-105){\makebox(0,0){0}}
         \put(-65,-105){\makebox(0,0){1}}
         \put(-55,-105){\makebox(0,0){0}}
         \put(-45,-105){\makebox(0,0){4}}
         \put(-35,-105){\makebox(0,0){0}}
         \put(-25,-105){\makebox(0,0){1}}
         \put(-15,-105){\makebox(0,0){0}}
         \put(-5,-105){\makebox(0,0){2}}
         \put(5,-105){\makebox(0,0){0}}
         \put(15,-105){\makebox(0,0){1}}
         \put(25,-105){\makebox(0,0){0}}
         \put(35,-105){\makebox(0,0){3}}
         \put(45,-105){\makebox(0,0){0}}
         \put(55,-105){\makebox(0,0){1}}
         \put(65,-105){\makebox(0,0){0}}
         \put(75,-105){\makebox(0,0){2}}
         \put(85,-105){\makebox(0,0){0}}
         \put(95,-105){\makebox(0,0){1}}
         \put(105,-105){\makebox(0,0){0}}
         \put(-45,-43){\makebox(0,0)[t]{$C_1$}}
         \put(-125,-83){\makebox(0,0)[t]{$L_1$}}
         \put(35,-53){\makebox(0,0)[t]{$R_1$}}
         \put(-45,-50){\makebox(0,0)[t]{$B_1$}}
         \put(0,-3){\makebox(0,0)[t]{$A_1$}}
         \put(0,3){\makebox(0,0)[b]{$V_1$}}
         \put(-60,0){\makebox(0,0)[b]{$V_1^1$}}
         \put(50,-7){\makebox(0,0)[b]{$V_1^2$}}
         \put(-60,-7){\makebox(0,0)[b]{$U_1$}}
         \put(-80,-17){\makebox(0,0)[b]{$U_1^1$}}
         \put(-40,-17){\makebox(0,0)[b]{$U_1^2$}}
      \end{picture}
   \caption{Labelled example of the $\BSVP$ algorithm at round 1.}
   \label{fig:BSVP1}
   \end{figure}

\item
   If we look at the first recursive call of $\fcn{BSVP}$ in the
   preceding case, we have the situation depicted in
   Figure~\ref{fig:BSVP2}, in which case it is easy to see that
   \( \fcn{BSVP}(d,h,m,s) = \fcn{BSVP}(\rhalf d,hba_{A},m,s)\tcomp
      \Big(\fcn{BSVP}(\rhalf d,hba_{U^1},m,s)\tlimp \\ \mbox{}\hfill
      \Big(\fcn{BSVP}(\rhalf d,hba_{U^2},m,s)\tcomp
      \big(\fcn{BSVP}(\rhalf d,hba_{V^1},m,s)\tlimp
         \fcn{BSVP}(\rhalf d,hba_{V^2},m,s)\big)\Big)\Big) \).

   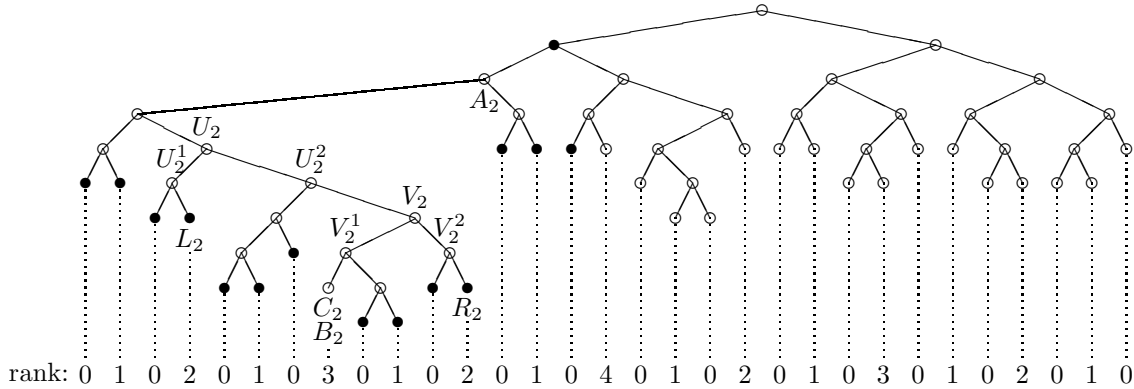
\begin{figure}[!ht]
   \centering\small\setlength{\unitlength}{1.3125pt}
      \begin{picture}(327,118)(-218,-108)
         \put(0,0){\circle{3}}
         \put(0,0){\line(-6,-1){60}}
         \put(0,0){\line(5,-1){50}}
         \put(-60,-10){\circle*{3}}
         \put(-60,-10){\line(-2,-1){20}}
         \put(-60,-10){\line(2,-1){20}}
         \put(50,-10){\circle{3}}
         \put(50,-10){\line(-3,-1){30}}
         \put(50,-10){\line(3,-1){30}}
         \put(-80,-20){\circle{3}}
         \qbezier(-80,-20)(-130,-25)(-180,-30)
         \put(-80,-20){\line(1,-1){10}}
         \put(-40,-20){\circle{3}}
         \put(-40,-20){\line(-1,-1){10}}
         \put(-40,-20){\line(3,-1){30}}
         \put(20,-20){\circle{3}}
         \put(20,-20){\line(-1,-1){10}}
         \put(20,-20){\line(2,-1){20}}
         \put(80,-20){\circle{3}}
         \put(80,-20){\line(-2,-1){20}}
         \put(80,-20){\line(2,-1){20}}
         \put(-180,-30){\circle{3}}
         \put(-180,-30){\line(-1,-1){10}}
         \put(-180,-30){\line(2,-1){20}}
         \put(-70,-30){\circle{3}}
         \put(-70,-30){\line(-1,-2){5}}
         \put(-70,-30){\line(1,-2){5}}
         \put(-50,-30){\circle{3}}
         \put(-50,-30){\line(-1,-2){5}}
         \put(-50,-30){\line(1,-2){5}}
         \put(-10,-30){\circle{3}}
         \put(-10,-30){\line(-2,-1){20}}
         \put(-10,-30){\line(1,-2){5}}
         \put(10,-30){\circle{3}}
         \put(10,-30){\line(-1,-2){5}}
         \put(10,-30){\line(1,-2){5}}
         \put(40,-30){\circle{3}}
         \put(40,-30){\line(-1,-1){10}}
         \put(40,-30){\line(1,-2){5}}
         \put(60,-30){\circle{3}}
         \put(60,-30){\line(-1,-2){5}}
         \put(60,-30){\line(1,-1){10}}
         \put(100,-30){\circle{3}}
         \put(100,-30){\line(-1,-1){10}}
         \put(100,-30){\line(1,-2){5}}
         \put(-190,-40){\circle{3}}
         \put(-190,-40){\line(-1,-2){5}}
         \put(-190,-40){\line(1,-2){5}}
         \put(-160,-40){\circle{3}}
         \put(-160,-40){\line(-1,-1){10}}
         \put(-160,-40){\line(3,-1){30}}
         \put(-30,-40){\circle{3}}
         \put(-30,-40){\line(-1,-2){5}}
         \put(-30,-40){\line(1,-1){10}}
         \put(30,-40){\circle{3}}
         \put(30,-40){\line(-1,-2){5}}
         \put(30,-40){\line(1,-2){5}}
         \put(70,-40){\circle{3}}
         \put(70,-40){\line(-1,-2){5}}
         \put(70,-40){\line(1,-2){5}}
         \put(90,-40){\circle{3}}
         \put(90,-40){\line(-1,-2){5}}
         \put(90,-40){\line(1,-2){5}}
         \put(-170,-50){\circle{3}}
         \put(-170,-50){\line(-1,-2){5}}
         \put(-170,-50){\line(1,-2){5}}
         \put(-130,-50){\circle{3}}
         \put(-130,-50){\line(-1,-1){10}}
         \put(-130,-50){\line(3,-1){30}}
         \put(-20,-50){\circle{3}}
         \put(-20,-50){\line(-1,-2){5}}
         \put(-20,-50){\line(1,-2){5}}
         \put(-140,-60){\circle{3}}
         \put(-140,-60){\line(-1,-1){10}}
         \put(-140,-60){\line(1,-2){5}}
         \put(-100,-60){\circle{3}}
         \put(-100,-60){\line(-2,-1){20}}
         \put(-100,-60){\line(1,-1){10}}
         \put(-150,-70){\circle{3}}
         \put(-150,-70){\line(-1,-2){5}}
         \put(-150,-70){\line(1,-2){5}}
         \put(-120,-70){\circle{3}}
         \put(-120,-70){\line(-1,-2){5}}
         \put(-120,-70){\line(1,-1){10}}
         \put(-90,-70){\circle{3}}
         \put(-90,-70){\line(-1,-2){5}}
         \put(-90,-70){\line(1,-2){5}}
         \put(-110,-80){\circle{3}}
         \put(-110,-80){\line(-1,-2){5}}
         \put(-110,-80){\line(1,-2){5}}
         \put(-195,-50){\circle*{3}}
         \put(-185,-50){\circle*{3}}
         \put(-175,-60){\circle*{3}}
         \put(-165,-60){\circle*{3}}
         \put(-155,-80){\circle*{3}}
         \put(-145,-80){\circle*{3}}
         \put(-135,-70){\circle*{3}}
         \put(-125,-80){\circle{3}}
         \put(-115,-90){\circle*{3}}
         \put(-105,-90){\circle*{3}}
         \put(-95,-80){\circle*{3}}
         \put(-85,-80){\circle*{3}}
         \put(-75,-40){\circle*{3}}
         \put(-65,-40){\circle*{3}}
         \put(-55,-40){\circle*{3}}
         \put(-45,-40){\circle{3}}
         \put(-35,-50){\circle{3}}
         \put(-25,-60){\circle{3}}
         \put(-15,-60){\circle{3}}
         \put(-5,-40){\circle{3}}
         \put(5,-40){\circle{3}}
         \put(15,-40){\circle{3}}
         \put(25,-50){\circle{3}}
         \put(35,-50){\circle{3}}
         \put(45,-40){\circle{3}}
         \put(55,-40){\circle{3}}
         \put(65,-50){\circle{3}}
         \put(75,-50){\circle{3}}
         \put(85,-50){\circle{3}}
         \put(95,-50){\circle{3}}
         \put(105,-40){\circle{3}}
         \qbezier[25](-195,-50)(-195,-75)(-195,-100)
         \qbezier[25](-185,-50)(-185,-75)(-185,-100)
         \qbezier[20](-175,-60)(-175,-80)(-175,-100)
         \qbezier[15](-165,-70)(-165,-85)(-165,-100)
         \qbezier[10](-155,-80)(-155,-90)(-155,-100)
         \qbezier[10](-145,-80)(-145,-90)(-145,-100)
         \qbezier[15](-135,-70)(-135,-85)(-135,-100)
         \qbezier[1](-125,-98)(-125,-99)(-125,-100)
         \qbezier[5](-115,-90)(-115,-95)(-115,-100)
         \qbezier[5](-105,-90)(-105,-95)(-105,-100)
         \qbezier[10](-95,-80)(-95,-90)(-95,-100)
         \qbezier[5](-85,-90)(-85,-95)(-85,-100)
         \qbezier[30](-75,-40)(-75,-70)(-75,-100)
         \qbezier[30](-65,-40)(-65,-70)(-65,-100)
         \qbezier[30](-55,-40)(-55,-70)(-55,-100)
         \qbezier[30](-45,-40)(-45,-70)(-45,-100)
         \qbezier[25](-35,-50)(-35,-75)(-35,-100)
         \qbezier[20](-25,-60)(-25,-80)(-25,-100)
         \qbezier[20](-15,-60)(-15,-80)(-15,-100)
         \qbezier[30](-5,-40)(-5,-70)(-5,-100)
         \qbezier[30](5,-40)(5,-70)(5,-100)
         \qbezier[30](15,-40)(15,-70)(15,-100)
         \qbezier[25](25,-50)(25,-75)(25,-100)
         \qbezier[25](35,-50)(35,-75)(35,-100)
         \qbezier[30](45,-40)(45,-70)(45,-100)
         \qbezier[30](55,-40)(55,-70)(55,-100)
         \qbezier[25](65,-50)(65,-75)(65,-100)
         \qbezier[25](75,-50)(75,-75)(75,-100)
         \qbezier[25](85,-50)(85,-75)(85,-100)
         \qbezier[25](95,-50)(95,-75)(95,-100)
         \qbezier[30](105,-40)(105,-70)(105,-100)
         \put(-200,-105){\makebox(0,0)[r]{\vphantom{q}rank:}}
         \put(-195,-105){\makebox(0,0){0}}
         \put(-185,-105){\makebox(0,0){1}}
         \put(-175,-105){\makebox(0,0){0}}
         \put(-165,-105){\makebox(0,0){2}}
         \put(-155,-105){\makebox(0,0){0}}
         \put(-145,-105){\makebox(0,0){1}}
         \put(-135,-105){\makebox(0,0){0}}
         \put(-125,-105){\makebox(0,0){3}}
         \put(-115,-105){\makebox(0,0){0}}
         \put(-105,-105){\makebox(0,0){1}}
         \put(-95,-105){\makebox(0,0){0}}
         \put(-85,-105){\makebox(0,0){2}}
         \put(-75,-105){\makebox(0,0){0}}
         \put(-65,-105){\makebox(0,0){1}}
         \put(-55,-105){\makebox(0,0){0}}
         \put(-45,-105){\makebox(0,0){4}}
         \put(-35,-105){\makebox(0,0){0}}
         \put(-25,-105){\makebox(0,0){1}}
         \put(-15,-105){\makebox(0,0){0}}
         \put(-5,-105){\makebox(0,0){2}}
         \put(5,-105){\makebox(0,0){0}}
         \put(15,-105){\makebox(0,0){1}}
         \put(25,-105){\makebox(0,0){0}}
         \put(35,-105){\makebox(0,0){3}}
         \put(45,-105){\makebox(0,0){0}}
         \put(55,-105){\makebox(0,0){1}}
         \put(65,-105){\makebox(0,0){0}}
         \put(75,-105){\makebox(0,0){2}}
         \put(85,-105){\makebox(0,0){0}}
         \put(95,-105){\makebox(0,0){1}}
         \put(105,-105){\makebox(0,0){0}}
         \put(-125,-83){\makebox(0,0)[t]{$C_2$}}
         \put(-165,-63){\makebox(0,0)[t]{$L_2$}}
         \put(-85,-83){\makebox(0,0)[t]{$R_2$}}
         \put(-125,-90){\makebox(0,0)[t]{$B_2$}}
         \put(-80,-23){\makebox(0,0)[t]{$A_2$}}
         \put(-160,-37){\makebox(0,0)[b]{$U_2$}}
         \put(-170,-47){\makebox(0,0)[b]{$U_2^1$}}
         \put(-130,-47){\makebox(0,0)[b]{$U_2^2$}}
         \put(-100,-57){\makebox(0,0)[b]{$V_2$}}
         \put(-120,-67){\makebox(0,0)[b]{$V_2^1$}}
         \put(-90,-67){\makebox(0,0)[b]{$V_2^2$}}
      \end{picture}
   \caption{Labelled example of the $\BSVP$ algorithm at round 2.}
   \label{fig:BSVP2}
   \end{figure}

\end{enumerate}
 At the last round (when \(d=1\)), the history $h$ is analyzed and
 compared with the actual structure of the sentence, and the check-bit
 $c$ returned is $\ltrue$ iff they agree.  The check-bit can be
 obtained by taking the $\fcn{OR}$ of a bit-string computed by
 $\rpowCRN$ on $h$, where the bit output for each block $b$ in $h$ is
 $1$ iff $b$ is correct.  Moreover, each of these bits can be computed
 by finding the positions of the $L$, $C$, and $R$ leaves from the
 first part of the history $h$ (which is easy to do by $\TRN$) and
 then finding least common ancestors of these leaves (which again can
 be done easily by $\TRN$).

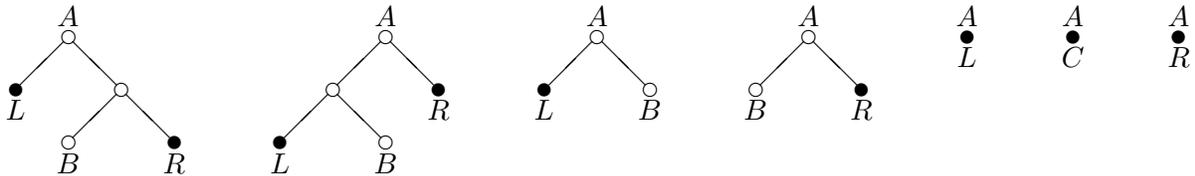
\begin{figure}[!ht]
\centering
   \begin{picture}(446,64)(-23,-52)
      \put(0,0){\circle{5}}
      \put(0,4){\makebox(0,0)[b]{$A$}}
      \put(-1.77,-1.77){\line(-1,-1){16.46}}
      \put(1.77,-1.77){\line(1,-1){16.46}}
      \put(-20,-20){\circle*{5}}
      \put(-20,-24){\makebox(0,0)[t]{$L$}}
      \put(20,-20){\circle{5}}
      \put(18.23,-21.77){\line(-1,-1){16.46}}
      \put(21.77,-21.77){\line(1,-1){16.46}}
      \put(0,-40){\circle{5}}
      \put(0,-44){\makebox(0,0)[t]{$B$}}
      \put(40,-40){\circle*{5}}
      \put(40,-44){\makebox(0,0)[t]{$R$}}
      \put(120,0){\circle{5}}
      \put(120,4){\makebox(0,0)[b]{$A$}}
      \put(118.23,-1.77){\line(-1,-1){16.46}}
      \put(121.77,-1.77){\line(1,-1){16.46}}
      \put(100,-20){\circle{5}}
      \put(98.23,-21.77){\line(-1,-1){16.46}}
      \put(101.77,-21.77){\line(1,-1){16.46}}
      \put(80,-40){\circle*{5}}
      \put(80,-44){\makebox(0,0)[t]{$L$}}
      \put(120,-40){\circle{5}}
      \put(120,-44){\makebox(0,0)[t]{$B$}}
      \put(140,-20){\circle*{5}}
      \put(140,-24){\makebox(0,0)[t]{$R$}}
      \put(200,0){\circle{5}}
      \put(200,4){\makebox(0,0)[b]{$A$}}
      \put(198.23,-1.77){\line(-1,-1){16.46}}
      \put(201.77,-1.77){\line(1,-1){16.46}}
      \put(180,-20){\circle*{5}}
      \put(180,-24){\makebox(0,0)[t]{$L$}}
      \put(220,-20){\circle{5}}
      \put(220,-24){\makebox(0,0)[t]{$B$}}
      \put(280,0){\circle{5}}
      \put(280,4){\makebox(0,0)[b]{$A$}}
      \put(278.23,-1.77){\line(-1,-1){16.46}}
      \put(281.77,-1.77){\line(1,-1){16.46}}
      \put(260,-20){\circle{5}}
      \put(260,-24){\makebox(0,0)[t]{$B$}}
      \put(300,-20){\circle*{5}}
      \put(300,-24){\makebox(0,0)[t]{$R$}}
      \put(340,0){\circle*{5}}
      \put(340,4){\makebox(0,0)[b]{$A$}}
      \put(340,-4){\makebox(0,0)[t]{$L$}}
      \put(380,0){\circle*{5}}
      \put(380,4){\makebox(0,0)[b]{$A$}}
      \put(380,-4){\makebox(0,0)[t]{$C$}}
      \put(420,0){\circle*{5}}
      \put(420,4){\makebox(0,0)[b]{$A$}}
      \put(420,-4){\makebox(0,0)[t]{$R$}}
   \end{picture}
\caption{Base cases for $\fcn{BSVP}$.}
\label{fig:BSVPbase}
\end{figure}

 {}At the same time, the actual sentence left at the last round will
 have one of the simple forms shown in Figure~\ref{fig:BSVPbase},
 which can all be evaluated trivially since we know the values of $L$,
 $C$, and $R$, and we can easily check when $B=C$.

 {}There remains one technical detail that needs to be taken care of
 in the definition of $\fcn{BSVP}$: as given, the definition is not a
 proper application of $\TRN$ since the value of $\fcn{BSVP}(d,h,m,s)$
 is recursively defined in terms of $\fcn{BSVP}(\rhalf d,h',m,s)$ for
 more than one value of $h'$.  The situation can easily be remedied,
 in the following way.  First, the bounded ``$\bigor$'' over all
 values of $b$ can be implemented with a subtree whose depth is equal
 to the number of bits in $b$ (which is a constant), where a single
 bit is added to $b$ at every level (to get all possible values of $b$
 at the bottom) and $\tlor$ is used to combine the values at each node
 in the subtree.  Next, the composition of values based on the
 structure determined by $b$ can similarly be carried out step-by-step
 using a subtree of constant depth, where at each step, the value of
 $b$ is used to determine which function ($\tcomp$ or $\tlimp$) to
 apply to combine the results, and which bits to add to $a$ to select
 subformulas.  Both of these steps only require increasing the length
 of $d$ by a constant factor, and using $\lenfcn{powmod}$ to determine
 which level is being evaluated in the subtrees.

 {}Finally, we can define a function $\fcn{VALUE}(v,x,z)$ that
 computes the truth-value of the formula encoded by $(x,z)$ under $v$:
\[
   \fcn{VALUE}(v,x,z) = \fcn{AND}\big(\fcn{BSVP}
      \big(2^k\times\fcn{leaves}(x,z),\E,\fcn{mask}(x,z),
         \fcn{sentence}(v,\fcn{pad}(x,z),z)\big)\big),
\]
 where $\fcn{pad}(x,z)$ pads the formula encoded by $x$ and $z$,
 adding enough copies of ``$\ltrue\limp\dotsb$'' to the left so that
 it has $2^{\delta+1}-1$ leaves for some integer $\delta\geq 0$,
 $\fcn{mask}(x,z)$, returns the position of the root of the formula
 encoded by $x$ inside $\fcn{pad}(x,z)$, using a bitmask (where the
 root is indicated by its main connective ``$\limp$''),
 $\fcn{leaves}(x,z)$ returns a string of length $2^\delta$ (for the
 same $\delta$ as above), and $k$ is the fixed number of bits in one
 block ``$ba$'' of the history.  All these functions are easily
 defined in $T_1$ using $\rpowCRN$ and $\TRN$, as follows.

 {}First, note that a formula with $n$ leaves always contains exactly
 $4n-3 = n + 3(n-1)$ symbols (1 for each leaf, and 3 for each
 connective: two parentheses and one connective), each one encoded by
 a block of length $|\lenfcn{pow}(z)|$.  This means that the function
\[
   \fcn{numleaves}(x,z) = \lenfcn{powdiv}
      \big((x\cat(3\times\lenfcn{pow}(z)))\lhalf\lhalf,z\big)
\]
 returns the number of leaves of the formula encoded by $x$, in unary.
 Also, we can check whether a formula has a power of $2$ minus $1$
 leaves or not with the function $\lenfcn{ispow}$.  Hence, if we define
\( \fcn{leaves}(x,z) = \fcn{sevael}(x\cat(7\times\lenfcn{pow}(z)),z) \),
 where
\[
   \fcn{sevael}(x,z) =
      \begin{cases}
         1 & \text{if } x\lenrel\leq 8\times\lenfcn{pow}(z), \\
         \fcn{sevael}(x\lhalf,z)\cat\fcn{sevael}(\rhalf x,z)
          & \text{otherwise}, 
      \end{cases}
\]
 then the string $x\cat(7\times\lenfcn{pow}(z))$ contains
 $4n-3+7=4(n+1)$ blocks of bits of length $\lenfcn{pow}(z)$ so that
 $\fcn{leaves}(x,z)$ contains exactly $2^\delta$ bits for any formula
 $x$ that contains between $2^\delta$ and $2^{\delta+1}-1$ leaves,
 inclusive.

 {}Next, we define by $\TRN$ a function $\fcn{padding}(y,w,z)$ that
 returns a balanced sentence containing exactly $|y|$ leaves, each one
 having the value $\ltrue$, where the logical depth of each symbol is
 at least $|w|$ and the length of each symbol's encoding is
 $|\lenfcn{pow}(z)|$:
\[
   \fcn{padding}(y,w,z) =
   \begin{cases}
      y\zlC\big(\E,\code_{|\lenfcn{pow}(z)|}\ltrue\big)
       & \text{if } y=\E,0,1, \\
      \code_{|\lenfcn{pow}(z)|}(_{|w|}\cat\fcn{padding}(y\lhalf,w1,z)
      \cat\code_{|\lenfcn{pow}(z)|}\limp_{|w|} \\
      \hphantom{\code_{|\lenfcn{pow}(z)|}(_{|w|}}\cat
      \fcn{padding}(\rhalf y,w1,z)\cat\code_{|\lenfcn{pow}(z)|})_{|w|}
       & \text{otherwise}. 
   \end{cases}
\]
 Now, we can easily define
\begin{align*}
   \fcn{pad}(x,z) &= x \qquad
      \text{(if $x$ has a power of $2$ minus $1$ leaves)}, \\
   \fcn{pad}(x,z) &= \code_{|\lenfcn{pow}(z)|}(_0\cat
      \fcn{padding}(\fcn{numleaves}(x,z)\lchop\fcn{leaves}(x,z),1,z) \\*
   &\pheq \hphantom{\code_{|\lenfcn{pow}(z)|}(_0}\cat
      \code_{|\lenfcn{pow}(z)|}\limp_0\cat\fcn{deepen}(x,z)\cat
      \code_{|\lenfcn{pow}(z)|})_0 \qquad\text{(otherwise)}, 
\end{align*}
 where $\fcn{deepen}(x,z)$ is easily defined by $\rpowCRN$ to add $1$
 to the logical depth of every symbol in the formula $x$.  Finally, we
 can define $\fcn{mask}$ by $\rpowCRN$:
\begin{align*}
   \fcn{mask}(x,z) &=
   \text{the unique connective $\limp$ at logical depth $0$ inside }
      \fcn{pad}(x,z) \\*
   &\pheq 
      \text{(if $x$ has a power of $2$ minus $1$ leaves)}, \\
   \fcn{mask}(x,z) &=
   \text{the second connective $\limp$ at logical depth $1$ inside }
      \fcn{pad}(x,z) 
   \qquad \text{(otherwise)}. 
\end{align*}

\subsection{Proof of correctness in $T_1$}

\begin{theorem}
\label{thm:VALUE}
   $T_1$ proves
   \( \fcn{VALUE}(v,\code_{|\lenfcn{pow}(z)|}\ltrue,z) = 1 \),
   \( \fcn{VALUE}(v,\code_{|\lenfcn{pow}(z)|}\lfalse,z) = 0 \), and
   for arbitrary formulas $M$ and $N$,
   \[         
      \fcn{VALUE}(v,\code_{|\lenfcn{pow}(z)|}(M\limp N),z) =
         \fcn{VALUE}(v,\code_{|\lenfcn{pow}(z)|} M,z)\bitbin\limp
         \fcn{VALUE}(v,\code_{|\lenfcn{pow}(z)|} N,z)
   \]
   (\ie, $\fcn{VALUE}$ is intensional).
\end{theorem}
\begin{proof}
 The first two statements follow directly from the definition of the
 functions involved.  The third statement follows from
 Claim~\ref{claim:BSVP} below.
\end{proof}

\begin{claim}\mbox{}\nopagebreak
\label{claim:BSVP}
\begin{enumerate}
\item
\label{claim:BSVP1}
   If $h$ picks out a supersentence of \((M\limp N)\), possibly
   scarred at $B$, in the sentence encoded by $s$, then
   \( \fcn{BSVP}(d,h,\fcn{lca}(m,n),s) =
      \fcn{BSVP}(d,h,m,s)\tlimp\fcn{BSVP}(d,h,n,s) \)
   (where $\fcn{lca}(m,n)$ is a mask indicating the position of the
   least common ancestor of the nodes masked by $m$ and $n$ in the
   sentence $s$).
\item
\label{claim:BSVP2}
   \( \fcn{BSVP}(d,\E,m,s) = \fcn{BSVP}(d',\E,m',s') \)
   for all values of the parameters that represent the same sentence,
   \ie, given \(x=\code_\ell A\), for any $s,s'$ that are
   supersentences of $x$, where $m,m'$ and $d,d'$ are defined
   appropriately.
\end{enumerate}
\end{claim}
\begin{proof}\mbox{}\nopagebreak
\begin{enumerate}

\item
   By induction on $d$.  When \(d=1\), only the first four cases of
   Figure~\ref{fig:BSVPbase} apply.  Suppose we are in case 4; then,
   \(A=(M\limp N)\) and
   \( \fcn{BSVP}(d,h,\fcn{lca}(m,n),s) = (c,r,\ltrue) \)
   (where $r$ is the value of node $R$),
   \( \fcn{BSVP}(d,h,m,s) = (c,\ltrue,\lfalse) \)
   (since \(M=B\) is the scar), and
   \( \fcn{BSVP}(d,h,n,s) = (c,r,r) \),
   so the statement is true.  The other three cases are similar.

   Now, if $h$ picks out a supersentence of \((M\limp N)\), possibly
   scarred at $B$, then consider the following cases.
   \begin{enumerate}
   \item
      If \( \fcn{BSVP}(d,h,\fcn{lca}(m,n),s) =
         \fcn{BSVP}(\rhalf d,hba,\fcn{lca}(m,n),s) \), then
      $hba$ also picks out a supersentence of $(M\limp N)$, possibly
      scarred at $B$, and the statement is true by the induction
      hypothesis.
   \item
      If \(U>V=(M\limp N)\), then
      \begin{align*}
         &\ \fcn{BSVP}(d,h,\fcn{lca}(m,n),s) \\
         &= \fcn{BSVP}(\rhalf d,hba_{V^1},\fcn{lca}(m,n),s)\tlimp
            \fcn{BSVP}(\rhalf d,hba_{V^2},\fcn{lca}(m,n),s).
      \end{align*}
      Also,
      \[
         \fcn{BSVP}(d,h,m,s) =
         \fcn{BSVP}(\rhalf d,hba_{V^1},m,s) =
         \fcn{BSVP}(\rhalf d,hba_{V^1},\fcn{lca}(m,n),s)
      \]
      and
      \[
         \fcn{BSVP}(d,h,n,s) =
         \fcn{BSVP}(\rhalf d,hba_{V^2},n,s) =
         \fcn{BSVP}(\rhalf d,hba_{V^2},\fcn{lca}(m,n),s)
      \]
      by the induction hypothesis.  Hence, the statement is true.
      (The case when \(V>U=(M\limp N)\) is similar.)
   \item
      If \((M\limp N)=U>V\), then
      \begin{align*}
         &\ \fcn{BSVP}(d,h,\fcn{lca}(m,n),s) =
            \fcn{BSVP}(\rhalf d,hba_{U^1},\fcn{lca}(m,n),s)\tlimp \\
         &\ \big(\fcn{BSVP}(\rhalf d,hba_{U^2},\fcn{lca}(m,n),s)\tcomp
               \fcn{BSVP}(\rhalf d,hba_{V},\fcn{lca}(m,n),s)\big).
      \end{align*}
      Also,
      \[
         \fcn{BSVP}(d,h,m,s) =
         \fcn{BSVP}(\rhalf d,hba_{U^1},m,s)
      \]
      and
      \[
         \fcn{BSVP}(d,h,n,s) =
         \fcn{BSVP}(\rhalf d,hba_{U^2},n,s)\tcomp
         \fcn{BSVP}(\rhalf d,hba_{V},n,s),
      \]
      where
      \begin{align*}
         \fcn{BSVP}(\rhalf d,hba_{V},\fcn{lca}(m,n),s)
         &= \fcn{BSVP}(\rhalf d,hba_{V},n,s) \\
         &= \fcn{BSVP}(\rhalf d,hba_{V^1},n,s)\tlimp
            \fcn{BSVP}(\rhalf d,hba_{V^2},n,s).
      \end{align*}
      Since \(t\tlimp(t_1\tcomp t_2)=(t\tlimp t_1)\tcomp t_2\) for any
      truth-triplets $t$, $t_1$, and $t_2$ such that $t$ is unscarred
      (\ie, \(t_\ltrue=t_\lfalse\)), the statement follows immediately
      by the induction hypothesis.  (The case when \((M\limp N)=V>U\)
      is similar.)
   \item
      If \(U>(M\limp N)>V\), then assuming \(M>V\) (the other cases
      being similar), we have that
      \begin{align*}
         &\ \fcn{BSVP}(d,h,\fcn{lca}(m,n),s) \\
         &= \fcn{BSVP}(\rhalf d,hba_{U^2},\fcn{lca}(m,n),s)\tcomp
            \fcn{BSVP}(\rhalf d,hba_{V},\fcn{lca}(m,n),s).
      \end{align*}
      Also,
      \[
         \fcn{BSVP}(d,h,m,s) =
         \fcn{BSVP}(\rhalf d,hba_{U^2},m,s)\tcomp
         \fcn{BSVP}(\rhalf d,hba_{V},m,s)
      \]
      and
      \[
         \fcn{BSVP}(d,h,n,s) =
         \fcn{BSVP}(\rhalf d,hba_{U^2},n,s),
      \]
      where
      \begin{align*}
         \fcn{BSVP}(\rhalf d,hba_{V},\fcn{lca}(m,n),s)
         &= \fcn{BSVP}(\rhalf d,hba_{V},m,s) \\
         &= \fcn{BSVP}(\rhalf d,hba_{V^1},m,s)\tlimp
            \fcn{BSVP}(\rhalf d,hba_{V^2},m,s).
      \end{align*}
      Since \((t_1\tcomp t_2)\tlimp t=(t_1\tlimp t)\tcomp t_2\) for
      any truth-triplets $t$, $t_1$, and $t_2$ such that $t$ is
      unscarred, the statement follows immediately by the induction
      hypothesis.
      (The case when \(U>(M\limp N)>V\) is similar.)
   \item
      The case when \((M\limp N)>U,V\) is similar to the last one, and 
      all cases can be suitably simplified when \(B>U\) or \(B>V\).
   \end{enumerate}

\item
   We will actually prove that for all supersentences $s$ of $x$,
   given a mask $m$ for $x$ in $s$ and a suitable value for $d$,
   \( \fcn{BSVP}(d,\E,m,s) = \fcn{BSVP}(d_x,\E,m_x,s_x) \),
   where $s_x$, $m_x$, and $d_x$ are the default values for $x$.  This
   will be proved by induction on the number of leaves of the sentence
   encoded by $x$.  If $x$ encodes a sentence with only one leaf, then
   $x$ either encodes ``$\ltrue$'' or ``$\lfalse$'', in which case
   there is a unique history $h$ that picks out $x$ from any given
   sentence $s$ containing $x$.  For that history, it is easy to see
   that
   \( \fcn{BSVP}(d,\E,m,s) = \fcn{BSVP}(1,h,m,s) \)
   which is equal to the value of $x$, and the same is true for the
   default values of $s$, $m$, and $d$.

   Now, suppose that $x$ encodes a sentence \(M'=(M\limp N)\) with
   \(k>1\) leaves, and let $s$ be any supersentence of $x$.  Then,
   \begin{align*}
      \fcn{BSVP}(d,\E,m',s)
      &= \fcn{BSVP}(d,\E,m,s)\tlimp\fcn{BSVP}(d,\E,n,s) \\
      &= \fcn{BSVP}(d_x,\E,m_x,s_x)\tlimp\fcn{BSVP}(d_x,\E,n_x,s_x) \\
      &= \fcn{BSVP}(d_x,\E,m'_x,s_x) 
   \end{align*}
   where the first and last equality are true by
   Claim~\ref{claim:BSVP}(\ref{claim:BSVP1}), and the middle equality
   is true by the induction hypothesis. \quad\QED

\end{enumerate}
\end{proof}

\section{The soundness proof}

\subsection{Preliminaries}

 If we let
 \( \fcn{TRUE}(v,x,z) =
    \fcn{formula}(x,z)\bitbin\land\fcn{VALUE}(v,x,z) \),
 then we can express the fact that our $\F$-system is sound with the
 following statement in $T_1$:
 \(\fcn{TRUE}(v,F(x,y,z,w),z)=1\).

 {}We will show that $T_1$ can prove this statement, by induction on
 the parameter $w$ (which indicates the number of lines in the proof
 encoded by $x,y$).  For this, though, we will have to define a form
 of ``strong induction'' in $T_1$.  First, we need to define a notion
 of prefix for strings: ``$y\pref x$'', defined below, returns $1$
 when $y$ is a prefix of $x$, $0$ otherwise.
\[
   y\pref x = y\strrel=\fcn{lc}(x,y).
\]
 Next, we will formalize the notion of ``part-of'' quantifiers in
 $T_1$.  More precisely, we will show how to represent the part-of
 quantifications \(\bigand_{y\pref x}A[y]\) and
 \(\bigor_{y\pref x}A[y]\) for any fixed formula $A$.  Since we have
 shown that $\strrel=$ is equivalent to $=$ in $T_1$ and that the
 connectives of $T_1$ are equivalent to their functional
 counterparts, we can replace $=$ with $\strrel=$, $\lnot$ with
 $\bitbin\lnot$, etc. inside $A[y]$ to obtain a term $\term{A}(y)$ for
 which we know $T_1$ can prove \(A[y]\liff\term{A}(y)=1\), and this
 for each value of $y$.  Then, if we define
\[
   \fcn{cat}_A(yi) = \fcn{cat}_A(y)\cat\term{A}(yi)
\]
 by $\CRN$, it is immediately clear that
\begin{align*}
   \bigand_{y\pref x}A[y] &\liff \fcn{AND}(\fcn{cat}_A(x))=1, \\
    \bigor_{y\pref x}A[y] &\liff \fcn{OR}(\fcn{cat}_A(x))=1. 
\end{align*}

 {}Now, suppose that for a particular formula $A$, we can prove in
 $T_1$ that \(A[\E]\) and \(\bigand_{y\pref x}A[y]\limp A[xi]\).  Can
 we conclude that \(A\) is true?  An easy $\NIND$ on $x$ proves
 \(\bigand_{y\pref x}A[y]\) in $T_1$:
\begin{enumerate}
\item
   \(\bigand_{y\pref\E}A[y]=A[\E]\), which we can prove by assumption;
\item
   assuming that \(\bigand_{y\pref x}A[y]\), an application of modus
   ponens gives us \(A[xi]\), which implies that
   \(\bigand_{y\pref xi}A[y]\) holds by properties of $\fcn{AND}$ and
   the definition of $\fcn{cat}_A$.
\end{enumerate}
 Hence, $T_1$ proves \(\bigand_{y\pref x}A[y]\), which implies, in
 particular, that \(A[x]\) holds.

\subsection{The proof}

\begin{theorem}
\label{thm:TRUE-F}
   $T_1$ proves \(\fcn{TRUE}(v,F(x,y,z,w),z)=1\).
\end{theorem}
 We use the ``strong induction'' described above to prove the theorem.
 The proof itself will be quite short.

 {}First, \(F(x,y,z,\E)=\code_{|\lenfcn{pow}(z)|}\ltrue\), and
 \(\fcn{TRUE}(v,\code_{|\lenfcn{pow}(z)|}\ltrue,z)\) is obviously
 equal to $1$ by Theorem~\ref{thm:VALUE}.  From now on, we will
 implicitly assume that $F(x,y,z,w)$ is not equal to
 $\code_{|\lenfcn{pow}(z)|}\ltrue$.

 {}Next, assume that
 \( \bigand_{u\pref w}\fcn{TRUE}(v,F(x,y,z,u),z) = 1 \),
 and consider the following cases, based on the value of $y_{|w1|}$,
 for the value of \(\fcn{TRUE}(v,F(x,y,z,w1),z)\).
\begin{itemize}
\item
   If \(y_{|w1|}=\tuple{0,1}\), then $F(x,y,z,w1)$ is the encoding of
   a formula of the form \((A\limp(B\limp A))\), in which case
   Theorem~\ref{thm:VALUE} implies that
   \begin{multline*}
      \fcn{TRUE}(v,F(x,y,z,w1),z) = \\
      \fcn{TRUE}(v,\code_{|\lenfcn{pow}(z)|}A,z)\bitbin\limp
      \big(\fcn{TRUE}(v,\code_{|\lenfcn{pow}(z)|}B,z)\bitbin\limp
         \fcn{TRUE}(v,\code_{|\lenfcn{pow}(z)|}A,z)\big), 
   \end{multline*}
   which is obviously equal to $1$ in $T_1$, being a simple
   tautology.
\item
   The other three axioms can easily be dealt with similarly.
\item
   If \(y_{|w1|}=\tuple{k_1,k_2}\), then we know that
   \(F(x,y,z,\cst{k_2})=\proj^{|w1|}_{k_2}(x)\) encodes a formula
   \(A_{k_2}=A_{k_1}\limp A_{|w1|}\), where $A_{k_1}$ is the formula
   encoded by \(F(x,y,z,\cst{k_1})=\proj^{|w1|}_{k_1}(x)\) and
   $A_{|w1|}$ is the formula encoded by $F(x,y,z,w1)$.  But then, by
   the induction hypothesis, we know that
   \( \fcn{TRUE}(v,F(x,y,z,\cst{k_1}),z) = 1 \) and
   \[
      \fcn{TRUE}(v,F(x,y,z,\cst{k_2}),z) =
         \fcn{TRUE}(v,F(x,y,z,\cst{k_1}),z)\bitbin\limp
         \fcn{TRUE}(v,F(x,y,z,w1),z) = 1,
   \]
   so it immediately follows that \(\fcn{TRUE}(v,F(x,y,z,w1),z)=1\).
\end{itemize}

\section{Simulation results}

 In this section, we show that $\F$ can $p$-simulate any proof system
 $S$ whose soundness can be proved in $T_1$.  A similar result was
 first proved for $\PV$ by Cook~\cite{Cook75} (Kraj\'\i\v{c}ek gives a
 more detailed proof in his book~\cite[Theorem~9.3.17]{Kraj95}), but
 to our knowledge, this is the first theory of $\ALT$ reasoning for
 which such a result is shown.

 {}Intuitively, the proof hinges on the fact that for any formula $A$,
 the propositional translations of the $T_1$ equation
 ``\( \fcn{TRUE}(v,\code_{2^\ell}A,\una l) = 1 \)''
 can be proven equivalent to a substitution instance of $A$ itself, so
 that if $T_1$ proves the equation, then $A$ is a tautology with
 short $\F$-proofs.

 {}More precisely, recall from Chapter~\ref{chap:polyproofs} that for
 any equation $t=u$ of $T_1$, we defined a family of propositional
 translations $\form{t=u}^{\tpl m,\tpl n}$ that have polysize
 $\F$-proofs whenever $t=u$ is a theorem of $T_1$.  Hence, for any
 formula $A$ and string $\una l$ such that
 \(|\una l|=2^\ell\geq\ell_A+2\), if $T_1$ proves
\( \fcn{TRUE}(v,\code_{2^\ell}A,\una l) = 1 \),
 then there are polysize $\F$-proofs of the corresponding
 propositional tautologies
\( \form{\fcn{TRUE}(v,\code_{2^\ell}A,\una l)=1}^k \)
 (where $k$ is the length of the free variable $v$), and these
 tautologies are defined as
\( \atom{\fcn{TRUE}(v,\code_{2^\ell}A,\una l)}^k_1\liff\ltrue \),
 which is equivalent to
\( \atom{\fcn{TRUE}(v,\code_{2^\ell}A,\una l)}^k_1 \)
 (the term formula describing the first and only bit of the term
 $\fcn{TRUE}(v,\code_{2^\ell}A,\una l)$ as a function of the bits of
 its free variable $v$).  Also, since $T_1$ can prove
\( \fcn{TRUE}(v,\code_{2^\ell}(A\limp B),\una l) =
   \fcn{TRUE}(v,\code_{2^\ell}A,\una l)\bitbin\limp
   \fcn{TRUE}(v,\code_{2^\ell}B,\una l) \)
 for any formulas $A$ and $B$, there are polysize $\F$-proofs that the
 corresponding propositional translations are equivalent, \ie,
\[
   \atom{\fcn{TRUE}(v,\code_{2^\ell}(A\limp B),\una l)}^k_1 \liff
   \left(\atom{\fcn{TRUE}(v,\code_{2^\ell}A,\una l)}^k_1\limp
      \atom{\fcn{TRUE}(v,\code_{2^\ell}B,\una l)}^k_1\right),
\]
 so that if $T_1$ proves
\( \fcn{TRUE}(v,\code_{2^\ell}(A\limp B),\una l) = 1 \),
 then there are polysize $\F$-proofs of
\[
   \atom{\fcn{TRUE}(v,\code_{2^\ell}A,\una l)}^k_1\limp
   \atom{\fcn{TRUE}(v,\code_{2^\ell}B,\una l)}^k_1.
\]
 Applying this reasoning recursively shows that if $T_1$ proves
\( \fcn{TRUE}(v,\code_{2^\ell}A,\una l) = 1 \),
 then there are polysize $\F$-proofs (call them $\Pi_k$) of
\( A\big[ \atom{\fcn{TRUE}(v,\code_{2^\ell}p_i,\una l)}^k_1
    / p_i \big] \)
 (\ie, the formula $A$ where each propositional variable $p_i$ has
 been replaced by the formula
\( \atom{\fcn{TRUE}(v,\code_{2^\ell}p_i,\una l)}^k_1 \)).
 Since the subformulas
 $\atom{\fcn{TRUE}(v,\code_{2^\ell}p_i,\una l)}^k_1$
 are never ``broken up'' inside $\Pi_k$, we can just substitute $p_i$
 for
 $\atom{\fcn{TRUE}(v,\code_{2^\ell}p_i,\una l)}^k_1$
 throughout to get polysize $\F$-proofs of $A$.

 {}Now, since $\F$ can ``evaluate'' sentences (\ie, given a
 propositional formula with no variables, $\F$ has polysize proofs
 that it is equivalent to its truth-value), for any function
 $f(x_1,\dots,x_n)$ definable in $T_1$, and any tuple of strings
 \(a_1\in\{0,1\}^*,\dots,a_n\in\{0,1\}^*\), there are polysize
 $\F$-proofs that
\( \atom{f(\tpl x)}^{\tpl m}_i
   \big[\atom{a_j}_{i_j}/\atom{x_j}^{m_j}_{i_j}\big]
    \liff \atom{f(\tpl a)}_i \)
 (where $\atom{a_j}_i$ is equal to $\ltrue$ or $\lfalse$ depending on
 the value of bit number $i$ of $a_j$).  In particular, if $S$ is a
 proof system formalizable in $T_1$ as a function symbol $S(x,z)$,
 then for any particular $S$-proof $(a,b)$ of a formula $A$, there are
 polysize $\F$-proofs that $S(a,b)$ is equivalent to the encoding of
 $A$.

 {}Putting these two facts together, we have that if $S(x,z)$ is a
 proof system whose soundness can be proved in $T_1$ (\ie, for which
 $T_1$ can prove \(\fcn{TRUE}(v,S(x,z),z)=1\)), then for any
 particular $S$-proof $(a,b)$, there are polysize $\F$-proofs of the
 formula encoded by $S(a,b)$, \ie, $\F$ $p$-simulates $S$.  Moreover,
 it appears that the translation from $S$-proofs to $\F$-proofs can be
 carried out in $\NC^1$ and thus formalized in $T_1$, where the
 simulation proof can also be formalized (although we do not carry
 this out, we do not expect any technical difficulties to arise in the
 details of such a formalization).

\begin{remark}
   Note that Theorems~\ref{thm:simulation} and~\ref{thm:TRUE-F}
   immediately give an alternative proof that $\F$-systems have
   polysize proofs of their own partial consistency (when suitably
   expressed), a fact first proved directly by Buss~\cite{Buss91}.
   The partial self-consistency statements obtained through $T_1$
   would be different from the ones Buss considered, but it should be
   possible to prove that they are equivalent.
\end{remark}

%% file: chapter6.tex
\chapter{Related Work}
\label{chap:related}

 In this chapter, we show that Arai's $\AID$ is equivalent to $QT_1$
 (a suitably defined quantified version of $T_1$), and briefly
 discuss the relationship between $T_1$ and Clote's $\ALV$ or
 $\ALV'$.  We will keep the discussion at a high level, with few
 technical details.

\section{$T_1$ and $\AID$}

 If we define $QT_1$ to be a first-order theory whose non-logical
 symbols are those of $T_1$ and whose axioms are the universal
 closures of the axioms of $T_1$, together with axiom schemes
 corresponding to $\NIND$ and $\TIND$, then we can show that
\begin{itemize}
\item
   $QT_1$ is a conservative extension of $T_1$ (the proof is similar
   to Cook's proof in~\cite{Cook98} that $Q\ALV$ is conservative over
   $\ALV'$);
\item
   for every $\Sigma^b_0$-formula $B$ in $QT_1$, there exists a
   function symbol $\term{B}$ in $T_1$ such that $QT_1$ proves
   \( B[\tpl x] \liff \term{B}(\tpl x)=1 \)
   (sharply bounded quantifiers, \eg ``$\forall x\leq|t|$'', are
   easy to represent functionally since we already have ``part-of''
   quantifiers from the end of Chapter~\ref{chap:soundness} and
   \( \forall x\leq|t| B[x] \liff \bigand_{x\pref t}B[|x|] \),
   for example);
\item
   $QT_1$ proves the scheme of $\LIND[\Sigma^b_0]$ (with a
   straightforward application of $\NIND$).
\end{itemize}

\noindent
 Next, the primitive functions of $\AID$ are all easily defined in
 $T_1$ (all treating their inputs ``numerically'', \ie, ignoring
 leading zeroes), and their defining axioms can be proven without
 difficulty.  Also, for every inductively defined predicate
 $A^{\ell,B,\tpl D, I}$ in $\AID$, we can define a $\{0,1\}$-valued
 function $\term{A}$ in $T_1$ such that the equation
 \(\term{A}(\tpl x,p)=1\) provably satisfies the defining axioms
 (A.0)--(A.2) of $A$.  This can be done by $\TRN$ in a relatively
 straightforward manner, except for two technicalities that we discuss
 now.

 {}First, the recursive definition of $A$ in $\AID$ involves computing
 the values of predicates $D_1(\tpl x,p),\dots,D_m(\tpl x,p)$ at every
 level of the recursion, even though these computations can only be
 represented in $T_1$ by function symbols of rank $1$ (\ie, the
 computations are in $\NC^1$).  Since functions of rank $1$ are not
 allowed in the recursive part of a definition by $\TRN$, we need to
 precompute the values of the $\tpl D$ predicates for every level and
 extract the correct values during the recursive definition by $\TRN$.
 This is accomplished by first computing the concatenation of all the
 values of $p$ for which the $\tpl D$ predicates need to be computed,
 then using $\CRN$ to compute the concatenation of the $\tpl D$
 predicates for each value, and finally breaking up this string in the
 appropriate way during the recursion (so the order in which the
 values are listed is important and must be chosen carefully).

 {}Second, the depth of the recursion is controlled by the ``linear
 form'' $\ell||\tpl x||$, which is represented in $T_1$ by a
 numerical function, \ie, one whose actual string length could be
 arbitrarily longer than its numerical length because of leading
 $0$'s.  By simply extending the precomputation carried out above for
 the $\tpl D$ predicates so that the values of the $B$ predicate are
 also computed for each value of $p$ at the ``bottom'' of the
 recursion, we can easily define the term $\term{A}$ by $\TRN$ so that
 the recursion terminates early once the base cases are reached, so
 that the exact string length of the parameter controlling the $\TRN$
 does not matter, as long as it is long enough.

 {}In the other direction, every primitive function of $T_1$ can
 easily be defined in $\AID$, using Arai's representation of strings
 (a string $b_{k-1}\dotsm b_0$ is represented by the integer
 $1b_{k-1}\dotsm b_0$), since the primitive ``part'' function of
 $\AID$ can be used directly to extract arbitrary substrings.  From
 the properties of ``part'', the defining axioms for the
 $T_1$-functions can be proven without difficulty in $\AID$.  Also,
 function symbols defined by $\lCRN$ or $\rCRN$ in $T_1$ can be
 defined in $\AID$ in a straightforward manner using Comprehension,
 and the defining axioms for these functions follows directly from the
 Comprehension axiom in $\AID$.  Finally, functions defined by $\TRN$
 in $T_1$ can be defined in $\AID$ using an inductively defined
 predicate $A^{\ell,B,\tpl D,I}$ that uses its parameter $p$ to keep
 track of a \emph{path} through the recursion tree and that computes
 the appropriate substring of $x$ and the appropriate function of $z$
 at each level using the predicates $\tpl D$.  Moreover, the defining
 axioms of such a function symbol follow directly from the axioms for
 $A$ in $\AID$.

 {}Now, for every formula $B$ in $\AID$, we let $\term{B}$ denote the
 formula in $QT_1$ obtained from $B$ by replacing each primitive
 function symbol by its definition in $T_1$ and each inductively
 defined predicate symbol $A$ by its definition $\term{A}$ in $T_1$.
 Similarly, for every formula $B$ in $QT_1$, we let $\formaid{B}$
 denote the formula in $\AID$ obtained from $B$ by replacing each
 function symbol of $T_1$ by its definition in $\AID$.

 {}These translations allow us to show that if a formula $B$ is
 provable in $\AID+\CA[\Sigma^b_0]$, then $QT_1$ can prove $\term{B}$
 (since $QT_1$ proves $\LIND[\Sigma^b_0]$ and $\CA[\Sigma^b_0]$ can
 be defined using $\CRN$ and proven using $\NIND$), and that if a
 formula $B$ is provable in $QT_1$, then $\AID+\CA[\Sigma^b_0]$ can
 prove $\formaid{B}$ (since $\AID+\CA[\Sigma^b_0]$ can prove $\NIND$
 using Comprehension, and $\TIND$ in a way similar to Arai's proof of
 ``tree induction'').

 {}Hence, $\AID$ is equivalent to $QT_1$, which implies that $\AID$
 is conservative over $T_1$.  Moreover, since Arai proves in his
 paper that $\AID$ is equivalent to $Q\ALV$ for $\Sigma^b_1$-formulas,
 this also implies that $\ALV$ is equivalent to $T_1$ (modulo the
 translations from numbers to strings and from strings to numbers
 given above).  Unfortunately, the corresponding result for $\ALV'$
 and $T_1$ is not known.

 {}Now, even though $\AID$ is conservative over $T_1$ (through the
 appropriate translations between strings and numbers), $T_1$ appears
 to be more natural and easier to reason with for a variety of
 reasons.
\begin{itemize}
\item
   $T_1$ reasons directly with functions in $\FALT$, whereas $\AID$
   reasons only with predicates (functions have to be defined
   implicitly).
\item
   $T_1$'s scheme of $\TRN$ is simpler than $\AID$'s inductive
   definitions, in the sense that a function defined by $\TRN$ carries
   out only simple computations at each level of the recursion (\ie,
   the $h$, $h_\ell$, and $h_r$ functions can be computed by
   constant-depth circuits), unlike the computations requiring
   logdepth circuits carried out by the ``$D$'' predicates at each
   level of an inductive definition.
\item
   It seems quite tedious to work out precise estimates on the size of
   $\F$-proofs for the propositional translations of the
   $\Sigma^b_1$-theorems of $\AID$, whereas the corresponding task for
   $T_1$ is straightforward (so in a sense, $T_1$ is ``closer'' to
   $\F$-systems than $\AID$).
\end{itemize}

\section{$T_1$ and $\PV$}

 Based on a similar result of Buss~\cite{Buss94}, which uses
 Kraj\'\i\v{c}ek, Pudl\'ak \& Takeuti's Herbrand-type witnessing
 theorem~\cite{KrPuTa91}, Cook~\cite{Cook98} has argued that if $Q\PV$
 is conservative over $Q\ALV$, then \(\DP=\ALT\), where $Q\PV$ is the
 appropriately defined quantified theory corresponding to $\PV$ and
 $Q\ALV$ is a quantified theory based on Clote's $\ALV'$.  A similar
 result should hold with $QT_1$ in place of $Q\ALV$, given a suitable
 interpretation of strings as numbers and numbers as strings.

 {}As for the quantifier-free theories $\PV$ and $T_1$, it is easy to
 see that if $\PV$ is conservative over $T_1$ (through an appropriate
 translation between numbers and strings, such as the one used above),
 then $T_1$ proves the soundness of $\eF$ (since $\PV$ proves the
 soundness of $\eF$ and $\eF$ can be defined in $T_1$), so that $\F$
 $p$-simulates $\eF$.  Unfortunately, the converse is not known, and
 this has no known implication for the complexity classes $\DP$ and
 $\NC^1$.

%% file: chapter7.tex
\chapter{Conclusion}
\label{chap:concl}

\section{Summary}

 As Chapter~\ref{chap:L1} has shown, $L_1$ is an elegant and natural
 recursive characterization of $\NC^1$: simple functions are easy to
 define and even for more complex functions, the definitions are not
 unnecessarily complicated.  The only exception to that statement
 might be the ``numerical'' functions, but even there, the definitions
 are straightforward and correspond quite closely to the computation
 of these functions by circuit families.  Also, our scheme of $\TRN$
 seems to capture the computational power of $\ALT$ in the most
 natural way, as evidenced by the short proof that
 $\FALT\subseteq L_1$.  It would be interesting to prove the other
 direction also (that $L_1\subseteq\FALT$) by using computations by
 $\ATM$s as opposed to uniform circuit families, so that both
 directions of the proof are similar, but time constraints did not
 allow us to carry out such a proof.

 {}The theory $T_1$ based on $L_1$ has the desirable property that
 its appropriately translated theorems have short $\F$-proofs, and the
 proof of that fact is quite simple (especially when compared to the
 corresponding proofs for other theories of $\ALT$ reasoning in the
 literature).  In fact, the structure of the $\F$-proofs is
 straightforward enough that we get precise estimates on their size
 (as a function of the lengths of the variables).  Also, considering
 the inherent complexity of evaluating Boolean sentences in $\ALT$,
 the $T_1$-proof of the soundness of a particular $\F$-system is
 straightforward, consisting mainly in the formalization of Buss's
 $\BSVP$ algorithm and the proof of its properties.  Finally, the fact
 that $\F$ $p$-simulates any proof system $S$ whose soundness can be
 proved in $T_1$ is also straightforward to prove, and $T_1$ is the
 first theory of $\ALT$ reasoning for which this result has been
 shown.  All these facts strongly support our claim that $T_1$ is one
 of the most natural theories available for $\ALT$ reasoning, even
 though it is based on strings instead of numbers (unlike most of the
 other theories for polytime or $\ALT$ reasoning).

 {}To conclude, it might seem that any algebraic characterization of a
 complexity class could be used to define a quantifier-free theory
 like $T_1$, by simply having function symbols defined recursively
 and induction rules based on the recursion operators.  Although such
 a theory would undoubtedly reason on functions in the desired
 complexity class, we would still need to show that the type of
 reasoning that can be carried out in this theory also falls within
 the desired complexity class, and there is no clear way of doing this
 for arbitrary complexity classes.  Also, as evidenced by Clote's
 theories $\ALV$ and $\ALV'$, it is not an easy task to get a theory
 that is natural and simple enough to be useful in practice.

\section{Future work}

 First, an obvious generalization of $L_0$ and $L_1$ suggests
 itself: for \(i\geq 1\), let $L_{i+1}$ be the closure of $L_i$
 under $\COMP$, $\CRN$, and $\TRN\restrict^{L_{i+1}}_{L_i}$, defined
 recursively.  A study of these classes (or of a similar extension for
 the theory $T_1$) might be interesting.  (One fact about $L_i$
 which is relatively easy to prove is that it is a subset of the class
 of functions computable by uniform circuit families of
 $\bigOh(\log^i)$ depth, but it is unknown if this is a proper
 containment.  It might be interesting to try to prove better results,
 maybe that the $L_i$'s exactly captures these circuit families, or
 that the union of the $L_i$ hierarchy defines the class of functions
 computable by uniform circuit families of polylog depth.)

 {}Next, it would be interesting to compare $QT_1$ to Takeuti and
 Clote's $\TNC^0$, maybe to show that the two first-order theories are
 equivalent.  Also, from Arai's results on $\AID$ and $\ALV$, we have
 concluded in Chapter~\ref{chap:related} that $T_1$ and $\ALV$ are
 equivalent, but it is unknown whether or not $\ALV'$ is equivalent to
 $\ALV$ (or to $T_1$).

 {}It would also be interesting to see if ``tree recursion'' and
 ``tree induction'' can be adapted to define a similar quantifier-free
 theory for uniform $\TC^0$ reasoning, hopefully as natural as $T_1$
 is natural for $\ALT$ reasoning (such a theory would correspond to
 bounded-depth $\F$-systems with threshold gates in the same way that
 $T_1$ corresponds to $\F$-systems).  Note that there is already a
 first-order theory $\bar R^0_2$ for $\TC^0$ reasoning, defined by
 Johannsen~\cite{Joha96}.

 {}Similarly, there should be a way to extend $L_0$ and $L_1$ to
 capture all of $\NC$ (based on Bloch's characterization of $\NC$).
 This could possibly be used to define a quantifier-free theory for
 $\NC$ reasoning, which might lead to a natural propositional proof
 system that reasons in $\NC$.

 {}Finally, fully relating conservativity results between logical
 theories for $\DP$ and $\ALT$ reasoning to equivalence results
 between $\eF$ and $\F$ systems to collapse results between $\DP$ and
 $\NC^1$ remains a central open problem in this area.

%% file: appendix.tex
\chapter{Details of Proofs in the Formal Development of $T_1$}
\label{app:proofs}

 This appendix contains most of the proofs missing from the formal
 development of $T_1$ given in Chapter~\ref{chap:T1}.  It is included
 here mainly for the sake of completeness, so the style will be quite
 terse.  In particular, most proofs that consist only in a
 straightforward application of $\NIND$ will be omitted.

\heading{On generalizations of $\NIND$}

\begin{claim*}{\ref{claim:cuts}}
\mbox{}\nopagebreak
   \begin{enumerate}
   \item
      \leftright
     {\( z\lchop yx = z\lchop y\cat(z\rchop y)\lchop x \)}
     {\( xy\rchop z = x\rchop(y\lchop z)\cat y\rchop z \)}
   \item
      \leftright
     {\( y\rdel\lchop x = \fcn{lb}(x,y)\cat y\lchop x \)}
     {\( x\rchop\ldel y = x\rchop y\cat\fcn{rb}(x,y) \)}
   \item
      \leftright
     {\( y\lchop((x\rchop y)\lchop x) = \E \)}
     {\( (x\rchop(y\lchop x))\rchop y = \E \)}
   \item
      \leftright
     {\( \fcn{lc}(\fcn{rc}(x,y),y) = \fcn{rc}(x,y) \)}
     {\( \fcn{rc}(\fcn{lc}(x,y),y) = \fcn{lc}(x,y) \)}
   \item
      \leftright
     {\( \lbit(y\lchop x) =
         y\lchop x\zlC\big(\E,(x\rchop\ldel(y\lchop x))\rbit\big) \)}
     {\( (x\rchop y)\rbit =
         x\rchop y\zlC\big(\E,\lbit((x\rchop y)\rdel\lchop x)\big) \)}
   \item
      \leftright
     {\( \fcn{lc}(x,yi) = \fcn{lc}(x,y)\cat\fcn{lb}(x,yi) \)}
     {\( \fcn{rc}(x,iy) = \fcn{rb}(x,iy)\cat\fcn{rc}(x,y) \)}
   \item
      \leftright
     {\( \fcn{lc}(x,y)\cat y\lchop x = x \)}
     {\( x = x\rchop y\cat\fcn{rc}(x,y) \)}
   \end{enumerate}
\end{claim*}
\begin{proof}
\mbox{}\nopagebreak
   \begin{enumerate}
   \item
      (L) By $\NIND$ on $z$, Axioms~\ref{axiom:lchopE}
      and~\ref{axiom:rchopE}, and Claim~\ref{claim:chops}:
      \( \E\lchop yx = yx = \E\lchop y\cat\E\lchop x
          = \E\lchop y\cat(\E\rchop y)\lchop x \),
      \( iz\lchop yx = \ldel(z\lchop yx)
          = \ldel(z\lchop y\cat(z\rchop y)\lchop x)
          = z\lchop y\zlC\big(\ldel((z\rchop y)\lchop x),
            \ldel(z\lchop y)\cat(z\rchop y)\lchop x\big)
          = z\lchop y\zlC\big(\E\cat(i\cat z\rchop y)\lchop x,
            iz\lchop y\cat\E\lchop x\big)
          = z\lchop y\zlC\big(iz\lchop y\cat(iz\rchop y)\lchop x,
            iz\lchop y\cat(iz\rchop y)\lchop x\big)
          = iz\lchop y\cat(iz\rchop y)\lchop x \).
   \item
      (L) By $\NIND$ on $y$:
      \( \E\rdel\lchop x = x = \E\cat x = \fcn{lb}(x,\E)\cat\E\lchop x\),
      \( (yi)\rdel\lchop x = y\lchop x
          = \lbit(y\lchop x)\cat\ldel(y\lchop x)
          = \lbit((yi)\rdel\lchop x)\cat((yi)\lchop x)
          = \fcn{lb}(x,yi)\cat(yi)\lchop x \).
   \item
      (L) By $\NIND$ on $y$, and preceding claims:
      \( \E\lchop((x\rchop\E)\lchop x) = x\lchop x = \E \),
      \begin{align*}
         yi\lchop((x\rchop yi)\lchop x)
         &= \ldel\big(y\lchop((x\rchop y)\rdel\lchop x)\big) \\
         &= y\lchop\ldel(\fcn{lb}(x,x\rchop y)\cat
               (x\rchop y)\lchop x) \\
         &= y\lchop((x\rchop y)\lchop x) = \E 
      \end{align*}
      (Note that in this proof, we did not explicitly deal with the
      cases when \(x\rchop y=\E\) or when
      \((x\rchop y)\rdel\lchop x=\E\).  However, it can easily be seen
      that both cases make the statement trivially true.)
   \item
      (L) By preceding claims:
      \( \fcn{lc}(\fcn{rc}(x,y),y) = ((x\rchop y)\lchop x)\rchop
            \big(y\lchop((x\rchop y)\lchop x)\big)
          = \fcn{rc}(x,y)\rchop\E = \fcn{rc}(x,y) \).
   \item
      (L) By $\NIND$ on $x$ and preceding claims:
      \( \lbit(y\lchop\E) = \E = y\lchop\E\zlC(\E,\E) \),
      \( \lbit(y\lchop xi) = y\lchop xi\zlC(\E,\lbit(y\lchop x\cat i))
          = y\lchop xi\zlC\big(\E,y\lchop x\zlC(i,\lbit(y\lchop x))\big)
          = y\lchop xi\zlC\big(\E,y\lchop x\zlC
            \big(i,(x\rchop\ldel(y\lchop x))\rbit\big)\big)
          = y\lchop xi\zlC\big(\E,(xi\rchop(y\lchop x))\rbit\big)
          = y\lchop xi\zlC\big(\E,(xi\rchop\ldel(y\lchop xi))\rbit\big) \).
   \item
      (L) By preceding claims:
      \( \fcn{lc}(x,yi) = x\rchop(yi\lchop x)
          = x\rchop\ldel(y\lchop x)
          = x\rchop(y\lchop x)\cat y\lchop x\zlC
            \big(\E,(x\rchop\ldel(y\lchop x))\rbit\big)
          = x\rchop(y\lchop x)\cat\lbit(y\lchop x)
          = \fcn{lc}(x,y)\cat\fcn{lb}(x,yi) \).
      \quad\QED
   \end{enumerate}
\end{proof}

\heading{On propositional reasoning}

\begin{theorem*}{\ref{thm:prop}}
\mbox{}\nopagebreak
   \begin{enumerate}
   \item
      \( \bit\lsg x=1 \lor \bit\lsg x=0 \)
   \item
      \( \bit\lsg\bit\lsg x = \bit\lsg x \)
   \item
      \( \bit\lnot x=1 \liff \lnot(\bit\lsg x=1) \)
   \item
      \( x\bitbin\land y=1 \liff (\bit\lsg x=1\land\bit\lsg y=1) \)
   \item
      \( x\bitbin\lor y=1 \liff (\bit\lsg x=1\lor\bit\lsg y=1) \)
   \item
      \( x\bitbin\limp y=1 \liff (\bit\lsg x=1\limp\bit\lsg y=1) \)
   \item
      \( x\bitbin\liff y=1 \liff (\bit\lsg x=1\liff\bit\lsg y=1) \)
   \item
      \( x\bitbin\lxor y=1 \liff (\bit\lsg x=1\lxor\bit\lsg y=1) \)
   \end{enumerate}
\end{theorem*}
\begin{proof}
   The first three can be proved by straightforward $\NIND$ on $x$,
   all the others with a simple and direct application of Derived
   Rule~\ref{derule:2NIND}.
\end{proof}

\heading{On ``$\fcn{AND}$'' and ``$\fcn{OR}$''}

\begin{lemma*}{\ref{lem:AND}}
\mbox{}\nopagebreak
   \begin{enumerate}
   \item
      \leftright
     {\( x\neq\E\land x\neq 0\land x\neq 1 \limp
         (\ldel x\cat j)\lhalf = \ldel(x\lhalf)\cat\lbit\rhalf x \)\\}
     {\( x\neq\E\land x\neq 0\land x\neq 1 \limp
         \rhalf(\ldel x\cat j) = \ldel\rhalf x\cat j \)}
   \item
      \leftright
     {\( \lbit x\bitbin\land\fcn{AND}(\ldel x\cat j)
          = \fcn{AND}(x)\bitbin\land j \)}
     {\( j\bitbin\land\fcn{AND}(x)
          = \fcn{AND}(j\cat x\rdel)\bitbin\land x\rbit \)}
   \item
      \( \lbit x\bitbin\land\fcn{AND}(\ldel x) = \fcn{AND}(x)
          = \fcn{AND}(x\rdel)\bitbin\land x\rbit \) \quad
      for \(x\neq\E,0,1\)
   \end{enumerate}
\end{lemma*}
\begin{proof}
\mbox{}\nopagebreak
   \begin{enumerate}
   \item
      (L) By Derived Rule~\ref{derule:cases}:
      since \(\E=\E\), the statement is trivially true for \(x=\E\).
      Assuming \(ix\neq 0\land ix\neq 1\), \ie, \(x\neq\E\),
      \begin{align*}
         (\ldel(ix)\cat j)\lhalf
         &= (xj)\lhalf \\
         &= x\elC(x\lhalf,x\lhalf\cat\lbit(\rhalf x\cat j)) \\
         &= x\elC(x\lhalf,x\lhalf\cat\lbit\rhalf x) \\
         &= x\elC(x\lhalf\rdel\cat x\lhalf\rbit,
            \ldel(i\cat x\lhalf)\cat\lbit\rhalf x) \\
         &= x\elC\big(\ldel(i\cat x\lhalf)\rdel\cat
               \lbit((i\cat x\lhalf)\rbit\cat\rhalf x),
            \ldel(i\cat x\lhalf)\cat\lbit\rhalf x\big) \\
         &= \ldel((ix)\lhalf)\cat\lbit\rhalf(ix). 
      \end{align*}
      (A similar proof shows the same theorem with $i\cat x\rdel$ in
      place of $\ldel x\cat i$.)
   \item
      (L) By the preceding lemma and by $\TIND$ on $x$:
      \( \lbit\E\bitbin\land\fcn{AND}(\ldel\E\cat j) = 0
          = \fcn{AND}(\E)\bitbin\land j \),
      \( \lbit i\bitbin\land\fcn{AND}(\ldel i\cat j)
          = i\bitbin\land\fcn{AND}(j) = \fcn{AND}(i)\bitbin\land j \),
      and assuming that
      \( \lbit(x\lhalf)\bitbin\land\fcn{AND}(\ldel(x\lhalf)\cat j)
          = \fcn{AND}(x\lhalf)\bitbin\land j \)
      and
      \( \lbit(\rhalf x)\bitbin\land\fcn{AND}(\ldel(\rhalf x)\cat j)
          = \fcn{AND}(\rhalf x)\bitbin\land j \),
      then, for $x$ such that \(\ldel x\neq\E\) (the case when it is
      equal being trivial),
      \begin{align*}
         \lbit x\bitbin\land\fcn{AND}(\ldel x\cat j)
         &= \lbit x\bitbin\land\fcn{AND}((\ldel x\cat j)\lhalf)\bitbin\land
            \fcn{AND}(\rhalf(\ldel x\cat j)) \\
         &= \lbit(x\lhalf)\bitbin\land
            \fcn{AND}(\ldel(x\lhalf)\cat\lbit\rhalf x)\bitbin\land
            \fcn{AND}(\ldel\rhalf x\cat j) \\
         &= \fcn{AND}(x\lhalf)\bitbin\land\lbit\rhalf x\bitbin\land
            \fcn{AND}(\ldel\rhalf x\cat j) \\
         &= \fcn{AND}(x\lhalf)\bitbin\land\fcn{AND}(\rhalf x)\bitbin\land j \\
         &= \fcn{AND}(x)\bitbin\land j. 
      \end{align*}
   \item
      By preceding lemmas and by Derived Rule~\ref{derule:TINDE} on
      $x$:
      \( \lbit(ij)\bitbin\land\fcn{AND}(\ldel(ij))
          = i\bitbin\land\fcn{AND}(j) = \fcn{AND}(ij)
          = \fcn{AND}(i)\bitbin\land j
          = \fcn{AND}((ij)\rdel)\bitbin\land(ij)\rbit \),
      \( \lbit(ikj)\bitbin\land\fcn{AND}(\ldel(ikj))
          = i\bitbin\land\fcn{AND}(kj) = \fcn{AND}(ikj)
          = \fcn{AND}(ik)\bitbin\land j
          = \fcn{AND}((ikj)\rdel)\bitbin\land(ikj)\rbit \),
      and under the induction hypotheses that
      \( \lbit(x\lhalf)\bitbin\land\fcn{AND}(\ldel(x\lhalf)) = \fcn{AND}(x)
          = \fcn{AND}((x\lhalf)\rdel)\bitbin\land(x\lhalf)\rbit \)
      and
      \( \lbit(\rhalf x)\bitbin\land\fcn{AND}(\ldel(\rhalf x))
          = \fcn{AND}(x)
          = \fcn{AND}((\rhalf x)\rdel)\bitbin\land(\rhalf x)\rbit \)
      for \(x\lhalf\lenrel>1\), then
      \begin{align*}
         \lbit x\bitbin\land\fcn{AND}(\ldel x)
         &= \lbit x\bitbin\land\fcn{AND}((\ldel x)\lhalf)\bitbin\land
            \fcn{AND}(\rhalf(\ldel x)) \\
         &= x\elC\big(\lbit(x\lhalf)\bitbin\land\fcn{AND}(\ldel(x\lhalf))
               \bitbin\land\fcn{AND}(\rhalf x), \big.\\*
         &\pheq \hphantom{x\elC\big(\big.}\big. \lbit(x\lhalf)\bitbin\land
               \fcn{AND}(\ldel(x\lhalf)\cat\lbit(\rhalf x))\bitbin\land
               \fcn{AND}(\ldel(\rhalf x))\big) \\
         &= x\elC\big(\fcn{AND}(x\lhalf)\bitbin\land\fcn{AND}(\rhalf x),
               \fcn{AND}(x\lhalf)\bitbin\land\lbit(\rhalf x)\bitbin\land
               \fcn{AND}(\ldel(\rhalf x))\big) \\
         &= x\elC\big(\fcn{AND}(x\lhalf)\bitbin\land\fcn{AND}(\rhalf x),
               \fcn{AND}(x\lhalf)\bitbin\land\fcn{AND}(\rhalf x)\big) \\
         &= x\elC(\fcn{AND}(x),\fcn{AND}(x)) \\
         &= \fcn{AND}(x) \displaybreak[0]\\
         &= x\elC(\fcn{AND}(x),\fcn{AND}(x)) \\
         &= x\elC\big(\fcn{AND}(x\lhalf)\bitbin\land\fcn{AND}(\rhalf x),
               \fcn{AND}(x\lhalf)\bitbin\land\fcn{AND}(\rhalf x)\big) \\
         &= x\elC\big(\fcn{AND}((x\lhalf)\rdel)\bitbin\land(x\lhalf)\rbit
               \bitbin\land\fcn{AND}(\rhalf x),\fcn{AND}(x\lhalf)
               \bitbin\land\fcn{AND}(\rhalf x)\big) \\
         &= x\elC\big(\fcn{AND}((x\lhalf)\rdel)\bitbin\land
               \fcn{AND}((x\lhalf)\rbit\cat(\rhalf x)\rdel)\bitbin\land
               (\rhalf x)\rbit, \big.\\*
         &\pheq \hphantom{x\elC\big(\big.}\big.
            \fcn{AND}(x\lhalf)\bitbin\land
               \fcn{AND}((\rhalf x)\rdel)\bitbin\land(\rhalf x)\rbit\big) \\
         &= \fcn{AND}((x\rdel)\lhalf)\bitbin\land\fcn{AND}(\rhalf(x\rdel))
            \bitbin\land x\rbit \\
         &= \fcn{AND}(x\rdel)\bitbin\land x\rbit. 
         \quad\QED
      \end{align*}
   \end{enumerate}
\end{proof}

\heading{On generalizations of $\CRN$---part I}

\begin{theorem*}{\ref{thm:_jchop}}
   \( {}_jx={}_jy \liff x\lchop y=\E=x\rchop y \)
\end{theorem*}
\begin{proof}
   By Derived Rule~\ref{derule:2NIND}:
   \( {}_j\E={}_jy \liff \E=y \liff \E\lchop y=\E=\E\rchop y \),
   \( {}_jx={}_j\E \liff x=\E \liff x\lchop\E=\E=x\rchop\E \),
   \( {}_j(xi)={}_j(ky) \liff {}_jx\cat j=j\cat{}_jy
      \liff {}_jx={}_jy
      \liff x\lchop y=\E=x\rchop y
      \liff xi\lchop ky=\E=xi\rchop ky \).
\end{proof}

\begin{claim*}{\ref{claim:_jCRN}}
   For \(f=\rCRN[h]\),
   \begin{enumerate}
   \item
      \( {}_j(f(x,\tpl y)) = {}_jx \)
   \item
      \( f(x,\tpl y)\rchop z = f(x\rchop z,\tpl y) \)
   \item
      \( \fcn{lb}(f(x,\tpl y),z) = x\lchop z\zlC
         \big((0\cat h(\fcn{lc}(x,z),\tpl y))\rbit,\E\big) \)
      \quad for \(x,z\neq\E\)
   \item
      \( \fcn{lc}(f(x,\tpl y),z) = f(\fcn{lc}(x,z),\tpl y) \)
   \end{enumerate}
\end{claim*}
\begin{proof}
\mbox{}\nopagebreak
   \begin{enumerate}
   \item
      By $\NIND$ on $x$:
      \( {}_j(f(\E,\tpl y)) = {}_j\E \),
      \( {}_j(f(xi,\tpl y)) = {}_j\big(f(x,\tpl y)\cat
         (0\cat h(xi,\tpl y))\rbit\big) = {}_jx\cat j = {}_j(xi) \).
   \item
      By Derived Rule~\ref{derule:2NIND}:
      \( f(\E,\tpl y)\rchop z = \E\rchop z = f(\E\rchop z,\tpl y) \),
      \( f(x,\tpl y)\rchop\E = f(x,\tpl y) = f(x\rchop\E,\tpl y) \),
      \( f(xi,\tpl y)\rchop zj
         = \big(f(x,\tpl y)\cat (0\cat h(xi,\tpl y))\rbit\big)\rdel
           \rchop z = f(x,\tpl y)\rchop z = f(x\rchop z,\tpl y)
         = f(xi\rchop zj,\tpl y) \).
   \item
      First, a straightforward proof by $\NIND$ on $x$ shows that
      \( \lbit f(x,\tpl y) = (0\cat h(\lbit x,\tpl y))\rbit \).
      Now, by $\NIND$ on $z\neq\E$ and the claims above:
      \( \fcn{lb}(f(x,\tpl y),j)
          = \lbit\big(\E\lchop f(x,\tpl y)\big)
          = (0\cat h(\lbit x,\tpl y))\rbit
          = (0\cat h(\fcn{lc}(x,j),\tpl y))\rbit \),
      \begin{align*}
         \fcn{lb}(f(x,\tpl y),zi)
         &= \lbit\big(z\lchop f(x,\tpl y)\big) \\
         &= z\lchop f(x,\tpl y)\zlC\big(\E,
               \big(f(x,\tpl y)\rchop\ldel(z\lchop f(x,\tpl y))\big)
            \rbit\Big) \\
         &= z\lchop x\zlC\big(\E,
               \big(f(x,\tpl y)\rchop\ldel(z\lchop x)\big)\rbit\big) \\
         &= z\lchop x\zlC\big(\E,f(x\rchop\ldel(z\lchop x),\tpl y)\rbit
            \big) \\
         &= z\lchop x\zlC\big(\E,
               f(x\rchop(z\lchop x)\cat\lbit(z\lchop x),\tpl y)\rbit
            \big) \\
         &= z\lchop x\zlC\big(\E,(0\cat
               h(x\rchop(z\lchop x)\cat\lbit(z\lchop x),\tpl y)
            )\rbit\big) \\
         &= x\lchop zi\zlC\big(
               (0\cat h(\fcn{lc}(x,z)\cat\fcn{lb}(x,zi),\tpl y))\rbit,
            \E\big) \\
         &= x\lchop zi\zlC\big((0\cat h(\fcn{lc}(x,zi),\tpl y))\rbit,
            \E\big). 
      \end{align*}
   \item
      By $\NIND$ on $z$:
      \( \fcn{lc}(f(x,\tpl y),\E) = \E
          = f(\fcn{lc}(x,\E),\tpl y) \), and
      \begin{align*}
         \fcn{lc}(f(x,\tpl y),zi)
         &= \fcn{lc}(f(x,\tpl y),z)\cat
            \fcn{lb}(f(x,\tpl y),zi) \\
         &= f(\fcn{lc}(x,z),\tpl y)\cat x\lchop zi\zlC
            \big((0\cat h(\fcn{lc}(x,zi),\tpl y))\rbit,\E\big) \\
         &= z\lchop x\zlC\big(f(\fcn{lc}(x,z),\tpl y)\cat\E,
            f(\fcn{lc}(x,z),\tpl y)\cat
               (0\cat h(\fcn{lc}(x,zi),\tpl y))\rbit\big) \\
         &= z\lchop x\zlC\big(f(\fcn{lc}(x,zi),\tpl y),
            f(\fcn{lc}(x,zi),\tpl y)\big) \\
         &= f(\fcn{lc}(x,zi),\tpl y). 
         \quad\QED
      \end{align*}
   \end{enumerate}
\end{proof}

\begin{lemma*}{\ref{lem:chopchop}}
\mbox{}\nopagebreak
   \begin{enumerate}
   \item
      \leftright
     {\( y\lchop x = x\lchop y\zlC(y\lchop x,\E) \)}
     {\( x\rchop y = x\lchop y\zlC(x\rchop y,\E) \)}
   \item
      \leftright
     {\( (y\rchop(x\lchop y))\lchop x = y\lchop x \)}
     {\( x\rchop((y\rchop x)\lchop y) = x\rchop y \)}
   \end{enumerate}
\end{lemma*}
\begin{proof}
\mbox{}\nopagebreak
   \begin{enumerate}
   \item
      (L) Immediate from the fact that
      \( x\lchop y\neq\E \limp y\lchop x=\E \).
   \item
      (L) By $\NIND$ on $y$:
      \( (\E\rchop(x\lchop\E))\lchop x = (\E\rchop\E)\lchop x
          = \E\lchop x \),
      \begin{align*}
         (yj\rchop(x\lchop yj))\lchop x
         &= x\lchop yj\zlC\big((yj\rchop\E)\lchop x,
               (yj\rchop((x\lchop y)\cat j))\lchop x\big) \\
         &= x\lchop yj\zlC\big(yj\lchop x,
               (y\rchop(x\lchop y))\lchop x\big) \\
         &= x\lchop yj\zlC(yj\lchop x,y\lchop x) 
          = x\lchop yj\zlC(yj\lchop x,\E) \\
         &= x\lchop yj\zlC(yj\lchop x,yj\lchop x) 
          = yj\lchop x. 
         \quad\QED
      \end{align*}
   \end{enumerate}
\end{proof}

\begin{claim*}{\ref{claim:lpmaxl}}
\mbox{}\nopagebreak
   \begin{enumerate}
   \item
      \leftright
     {\( \fcn{lp}_j(x,\lenfcn{max}(x,y)) = \fcn{lp}_j(x,y) \)}
     {\( \fcn{rp}_j(x,\lenfcn{max}(x,y)) = \fcn{rp}_j(x,y) \)}
   \item
      \leftright
     {\( \fcn{lb}\big(\fcn{lp}_0(xi,\lenfcn{max}(xi,yj)),
      \lenfcn{max}(xi,yj)\big) = i \)\\}
     {\( \fcn{rb}\big(\fcn{rp}_0(ix,\lenfcn{max}(ix,jy)),
      \lenfcn{max}(ix,jy)\big) = i \)}
   \end{enumerate}
\end{claim*}
\begin{proof}
\mbox{}\nopagebreak
   \begin{enumerate}
   \item
      (L):
      \( \fcn{lp}_j(x,\lenfcn{max}(x,y))
          = x\lchop y\zlC\big({}_j(x\rchop x)\cat x,
               {}_j(y\rchop x)\cat x\big)
          = x\lchop y\zlC\big(\E\cat x,{}_j(y\rchop x)\cat x\big)
          = x\lchop y\zlC\big({}_j(y\rchop x)\cat x,
               {}_j(y\rchop x)\cat x\big)
          = {}_j(y\rchop x)\cat x
          = \fcn{lp}_j(x,y) \).
   \item
      (L):
      \begin{align*}
         &\!\fcn{lb}\big(\fcn{lp}_0(xi,\lenfcn{max}(xi,yj)),
            \lenfcn{max}(xi,yj)\big) \\
         &= \fcn{lb}\big(\fcn{lp}_0(x,\lenfcn{max}(x,y))\cat i,
               \lenfcn{max}(x,y)\cat x\lchop y\zlC(i,j)\big) \\
         &= \lbit\big(\lenfcn{max}(x,y)\lchop
               ({}_0(\lenfcn{max}_x(x,y)\rchop x)\cat x\cat i)\big) \\
         &= x\lchop y\zlC
            \big(\lbit\big(x\lchop({}_0(x\rchop x)\cat xi)\big),
               \lbit\big(y\lchop({}_0(y\rchop x)\cat xi)\big)\big) \\
         &= x\lchop y\zlC\big(\lbit(x\lchop(\E\cat xi)),
               \lbit\big(y\lchop(x\lchop{}_0y)\cat
               (y\rchop(x\lchop{}_0y))\lchop xi\big)\big) \\
         &= x\lchop y\zlC\big(\lbit i,
               \lbit(\E\cat(y\rchop(x\lchop y))\rchop x\cat i)\big) \\
         &= x\lchop y\zlC\big(i,\lbit(y\lchop x\cat i)\big) \\
         &= x\lchop y\zlC\big(i,\lbit(\E\cat i)\big) 
            = x\lchop y\zlC(i,i) = i
         \quad\QED
      \end{align*}
   \end{enumerate}
\end{proof}

\begin{theorem*}{\ref{thm:CRNm}}
   \begin{align*}
      &\!\lCRN_m[h](x_1,\dots,x_m,\tpl y) \\*
      &= x_1\cat\dots\cat x_m\zlC\Big(\E,\lbit
         \big(h\big(\fcn{lp}_0(x_1,\lenfcn{max}_m(\tpl x_m)),\dots,
         \fcn{lp}_0(x_m,\lenfcn{max}_m(\tpl x_m)),\tpl y\big)\cat
         0\big) \\*
      &\pheq \hphantom{x_1\cat\dotsb\cat x_m\zlC\Big(\E, } \cat
         \lCRN_m[h]\big(\ldel\fcn{lp}_0(x_1,\lenfcn{max}_m(\tpl x_m)),
         \dots,\ldel\fcn{lp}_0(x_m,\lenfcn{max}_m(\tpl x_m)),
         \tpl y\big)\Big) \\
      &\!\rCRN_m[h](x_1,\dots,x_m,\tpl y) \\*
      &= x_1\cat\dots\cat x_m\zlC\Big(\E,\rCRN_m[h]
         \big(\fcn{rp}_0(x_1,\lenfcn{max}_m(\tpl x_m))\rdel,\dots,
         \fcn{rp}_0(x_m,\lenfcn{max}_m(\tpl x_m))\rdel,\tpl y\big) \\*
      &\pheq \hphantom{x_1\cat\dotsb\cat x_m\zlC\Big(\E, } \cat
         \big(0\cat h\big(\fcn{rp}_0(x_1,\lenfcn{max}_m(\tpl x_m)),
         \dots,\fcn{rp}_0(x_m,\lenfcn{max}_m(\tpl x_m)),
         \tpl y\big)\big)\rbit\Big) 
   \end{align*}
\end{theorem*}
\begin{proof}
   From the fact that
   \( \lenfcn{max}_m(\tpl x_m)=\E \liff x_1\cat\dots\cat x_m=\E \)
   (easily proved by Derived Rule~\ref{derule:2NIND}), the theorem
   follows from Claim~\ref{claim:lpmaxl} by a straightforward
   application of Derived Rule~\ref{derule:2NIND}, generalized to $m$
   variables.
\end{proof}

\pagebreak[2]

\begin{claim*}{\ref{claim:bitand}}
\mbox{}\nopagebreak
   \begin{enumerate}
   \item
      \( \bitfcn{and}_m(x_1i_1,\dots,x_mi_m) =
         \bitfcn{and}_m(\tpl x_m)\cat\bitfcn{and}_m(\tpl i_m) \)
   \item
      \( {}_jx_1=\dots={}_jx_m\land{}_jy_1=\dots={}_jy_m \limp
         \bitfcn{and}_m(x_1y_1,\dots,x_my_m)=
         \bitfcn{and}_m(\tpl x_m)\cat\bitfcn{and}_m(\tpl y_m) \).
   \item
      \( \bitfcn{and}_m(\tpl x_m)\rdel =
         \bitfcn{and}_m(x_1\rdel,\dots,x_m\rdel) \)
   \item
      \( \bitfcn{and}_m(\tpl x_m)\rchop y =
         \bitfcn{and}_m(x_1\rchop y,\dots,x_m\rchop y) \)
   \end{enumerate}
\end{claim*}
\begin{proof}
\mbox{}\nopagebreak
   \begin{enumerate}
   \item
      By a straightforward application of Derived
      Rule~\ref{derule:2NIND}, generalized to $m$ variables.
   \item
      Again, by a straightforward application of Derived
      Rule~\ref{derule:2NIND}, generalized to $m$ variables, together
      with the previous claim.
   \item
      From the first claim, by a straightforward generalized $\NIND$.
   \item
      Directly from the preceding claim, with a straightforward
      $\NIND$ on $y$.
      \quad\QED
   \end{enumerate}
\end{proof}

\begin{theorem*}{\ref{thm:ANDnotOR}}
   \( \bit\lnot\fcn{AND}(x) = \fcn{OR}(\bitfcn{not}(x)) \)
   \quad and \quad
   \( \bit\lnot\fcn{OR}(x) = \fcn{AND}(\bitfcn{not}(x)) \)
   \quad for \(x\neq\E\)
\end{theorem*}
\begin{proof}
   (We prove only the first statement, the second one being almost
   identical.)  By $\TIND$ on $x$:
   \( \bit\lnot\fcn{AND}(i) = \bit\lnot i
       = \fcn{OR}(\bit\lnot i) = \fcn{OR}(\bitfcn{not}(i)) \),
   \begin{align*}
      \bit\lnot\fcn{AND}(x)
      &= \bit\lnot\big(\fcn{AND}(x\lhalf)\bit\land
         \fcn{AND}(\rhalf x)\big) \\
      &= \bit\lnot\fcn{AND}(x\lhalf)\bit\lor
         \bit\lnot\fcn{AND}(\rhalf x) \\
      &= \fcn{OR}(\bitfcn{not}(x\lhalf))\bit\lor
         \fcn{OR}(\bitfcn{not}(\rhalf x)) \\
      &= \fcn{OR}(\bitfcn{not}(x)\lhalf)\bit\lor
         \fcn{OR}(\rhalf\bitfcn{not}(x))
       = \fcn{OR}(\bitfcn{not}(x)). 
      \quad\QED
   \end{align*}

\end{proof}

\heading{On generalizations of $\CRN$---part II}

\begin{claim*}{\ref{claim:pow}}
\mbox{}\nopagebreak
   \begin{enumerate}
   \item
      \( \rhalf\lenfcn{pow}(y) = \lenfcn{pow}(\rhalf y)
          = \lenfcn{pow}(y)\lhalf \)
      \quad (for \(y\neq\E,0,1\))
   \item
      \( {}_1\lenfcn{pow}(y) = \lenfcn{pow}({}_1y)
          = \lenfcn{pow}(y) \)
   \item
      \( \lenfcn{pow}(\lenfcn{pow}(y)) = \lenfcn{pow}(y) \)
   \item
      \( \lenfcn{pow}(y)\lchop y = \E \)
   \item
      \( \lenfcn{ispow}(\lenfcn{pow}(y)) = \E \)
   \end{enumerate}
\end{claim*}
\begin{proof}
   All can be proved by very simple applications of $\TIND$.  We give
   the proof of the third statement as an illustration:
   \( \lenfcn{pow}(\lenfcn{pow}(\E)) = \lenfcn{pow}(\E) \),
   \( \lenfcn{pow}(\lenfcn{pow}(i)) = \lenfcn{pow}(1) =
      \lenfcn{pow}(i) \),
   \( \lenfcn{pow}(\lenfcn{pow}(y))
       = \lenfcn{pow}(\rhalf\lenfcn{pow}(y))\cat
         \lenfcn{pow}(\rhalf\lenfcn{pow}(y))
       = \lenfcn{pow}(\lenfcn{pow}(\rhalf y))\cat
         \lenfcn{pow}(\lenfcn{pow}(\rhalf y))
       = \lenfcn{pow}(\rhalf y)\cat\lenfcn{pow}(\rhalf y)
       = \lenfcn{pow}(y) \).
\end{proof}

\begin{claim*}{\ref{claim:smsh}}
\mbox{}\nopagebreak
   \begin{enumerate}
   \item
      \( \E\smsh y = \E = x\smsh\E \)
   \item
      \leftright
     {\( {}_1(xi\smsh y) = {}_1(x\smsh y)\cat{}_1y \)}
     {\( {}_1(x\smsh yi) = {}_1(x\smsh y)\cat{}_1x \)}
   \item
      \leftright
     {\( {}_1((x\cat y)\smsh z) = {}_1(x\smsh z)\cat{}_1(y\smsh z) \)}
     {\( {}_1(x\smsh(y\cat z)) = {}_1(x\smsh y)\cat{}_1(x\smsh z) \)}
   \item
      \( {}_1(x\smsh y) = {}_1(y\smsh x) \)
   \end{enumerate}
\end{claim*}
\begin{proof}
   Again, all these statements can be proved by very simple
   applications of $\NIND$ or $\TIND$; we give the proof of the third
   statement (for the left case) as an illustration:
   \( {}_1((x\cat\E)\smsh z) = {}_1(x\smsh z) = {}_1(x\smsh z)\cat
         {}_1(\E\smsh z) \),
   \( {}_1((x\cat yi)\smsh z) = {}_1((x\cat y)\smsh z)\cat{}_1z
       = {}_1(x\smsh z)\cat{}_1(y\smsh z)\cat{}_1z
       = {}_1(x\smsh z)\cat{}_1(yi\smsh z) \).
\end{proof}

\begin{claim*}{\ref{claim:powdiv}}
\mbox{}\nopagebreak
   \begin{enumerate}
   \item
      \( \lenfcn{powdiv}(x,y) = \lenfcn{powdiv}({}_1x,{}_1y) \)
   \item
      \( \lenfcn{powdiv}(x,y) = \lenfcn{powdiv}(x,\lenfcn{pow}(y)) \)
   \item
      \( x\lchop\lenfcn{pow}(y)\neq\E \limp \lenfcn{powdiv}(x,y)=\E \)
   \item
      \( \lenfcn{powdiv}(\lenfcn{pow}(y)\smsh z,y) = y\zlC(\E,{}_1z) \)
   \end{enumerate}
\end{claim*}
\begin{proof}
\mbox{}\nopagebreak
   \begin{enumerate}
   \item
      By a simple $\TIND$ on $y$.
   \item
      By a simple $\TIND$ on $y$.
   \item
      First, we prove that
      \( x\lchop\lenfcn{pow}(y)\neq\E \limp
         x\lhalf\lchop\lenfcn{pow}(\rhalf y)\neq\E \)
      by proving the contrapositive by $\TIND$ on $y$:
      \( x\lchop\lenfcn{pow}(\E)=\E \)
      (so the statement is vacuously true),
      \( x\lhalf\lchop\lenfcn{pow}(\rhalf 1)=\E \limp
         x\lhalf\neq\E \limp x\neq\E \limp x\lchop\lenfcn{pow}(1)=\E \),
      and assuming that
      \( x\lhalf\lchop\lenfcn{pow}(\rhalf y) = \E \), then
      \begin{align*}
         x\lchop y
         &= (x\lhalf\cat\rhalf x)\lchop
            (\lenfcn{pow}(y)\lhalf\cat\rhalf\lenfcn{pow}(y)) \\
         &= \rhalf x\lchop(x\lhalf\lchop
            (\lenfcn{pow}(\rhalf y)\cat\lenfcn{pow}(\rhalf y))) \\
         &= \rhalf x\lchop((x\lhalf\lchop\lenfcn{pow}(\rhalf y))\cat
               (x\lhalf\rchop\lenfcn{pow}(\rhalf y))\lchop
               \lenfcn{pow}(\rhalf y)) \displaybreak[0]\\
         &= \rhalf x\lchop((x\lhalf\rchop\lenfcn{pow}(\rhalf y))\lchop
            \lenfcn{pow}(\rhalf y)) \\
         &= (x\lhalf\rchop\lenfcn{pow}(\rhalf y))\lchop
            (\rhalf x\lchop\lenfcn{pow}(\rhalf y)) \\
         &= (x\lhalf\rchop\lenfcn{pow}(\rhalf y))\lchop\E
          = \E 
      \end{align*}
      (where we have used the fact that
      \( (z\cat y)\lchop x = y\lchop(z\lchop x) = z\lchop(y\lchop x) \),
      which is easy to prove by $\NIND$ on $z$).
      The result then follows by a simple application of $\TIND$.
   \item
      By $\TIND$ on $y$:
      \( \lenfcn{powdiv}(\lenfcn{pow}(\E)\smsh z,\E)
          = \E = \E\zlC(\E,{}_1z) \),
      \( \lenfcn{powdiv}(\lenfcn{pow}(i)\smsh z,i)
          = \lenfcn{powdiv}(z,i) = {}_1z = i\zlC(\E,{}_1z) \),
      \( \lenfcn{powdiv}(\lenfcn{pow}(y)\smsh z,y)
          = \lenfcn{powdiv}((\lenfcn{pow}(y)\smsh z)\lhalf,\rhalf y)
          = \lenfcn{powdiv}(\lenfcn{pow}(y)\lhalf\smsh z,\rhalf y)
          = \lenfcn{powdiv}(\lenfcn{pow}(\rhalf y)\smsh z,\rhalf y)
          = {}_1z = y\zlC(\E,{}_1z) \).
      \quad\QED
   \end{enumerate}
\end{proof}

\begin{claim*}{\ref{claim:powdivmod}}
\mbox{}\nopagebreak
   \begin{enumerate}
   \item
      \( x\lchop(\lenfcn{pow}(y)\smsh\lenfcn{powdiv}(x,y)) = \E \)
   \item
      \( y\neq\E \limp
         \lenfcn{powmod}(x,y)\lchop\lenfcn{pow}(y)\neq\E \)
   \item
      \( y\neq\E \limp \lenfcn{powdiv}(x1,y)=\lenfcn{powdiv}(x,y)\cat
         \big((\lenfcn{powmod}(x,y)\cat 1)\lchop\lenfcn{pow}(y)
            \zlC(1,\E)\big) \) \\
      \( y\neq\E \limp \lenfcn{powmod}(x1,y)=
         (\lenfcn{powmod}(x,y)\cat 1)\lchop\lenfcn{pow}(y)\zlC
         \big(\E,\lenfcn{powmod}(x,y)\cat 1\big) \)
   \item
      \( \lenfcn{powdiv}((\lenfcn{pow}(y)\smsh z)\cat x,y) =
         y\zlC(\E,{}_1z)\cat\lenfcn{powdiv}(x,y) \land{} \) \\
      \( \lenfcn{powmod}((\lenfcn{pow}(y)\smsh z)\cat x,y) =
         \lenfcn{powmod}(x,y) \)
   \item
      \( y\neq\E \land x\lchop\lenfcn{pow}(y)=\E \limp
         \lenfcn{powdiv}(x,y)=
            \lenfcn{powdiv}(x\rchop\lenfcn{pow}(y),y)\cat 1 \) \\
      \( y\neq\E \land x\lchop\lenfcn{pow}(y)=\E \limp
         \lenfcn{powmod}(x,y)=
            \lenfcn{powmod}(x\rchop\lenfcn{pow}(y),y) \)
   \end{enumerate}
\end{claim*}
\begin{proof}
\mbox{}\nopagebreak
   \begin{enumerate}
   \item
      By $\TIND$ on $y$ (with $h_\ell=h_r=\lhalf$):
      \( x\lchop(\lenfcn{pow}(\E)\smsh\lenfcn{powdiv}(x,\E))
          = x\lchop\E = \E \),
      \( x\lchop(\lenfcn{pow}(i)\smsh\lenfcn{powdiv}(x,i))
          = x\lchop(1\smsh{}_1x) = \E \),
      \begin{align*}
         &\!x\lchop(\lenfcn{pow}(y)\smsh\lenfcn{powdiv}(x,y)) \\
         &= x\lchop(\lenfcn{pow}(y)\smsh\lenfcn{powdiv}
               (x\lhalf,\rhalf y)) \\
         &= (x\lhalf\cat\rhalf x)\lchop\big((\lenfcn{pow}(\rhalf y)
               \cat\lenfcn{pow}(\rhalf y))\smsh\lenfcn{powdiv}
               (x\lhalf,\rhalf y)\big) \\
         &= \rhalf x\lchop\Big(\big(x\lhalf\lchop(\lenfcn{pow}(\rhalf y)
                  \smsh\lenfcn{powdiv}(x\lhalf,\rhalf y))\big)\cat \\*
         &\pheq \hphantom{\rhalf x\lchop\Big( }
               \big(x\lhalf\rchop(\lenfcn{pow}(\rhalf y)
                  \smsh\lenfcn{powdiv}(x\lhalf,\rhalf y))\big)\lchop
               (\lenfcn{pow}(\rhalf y)\smsh\lenfcn{powdiv}
                  (x\lhalf,\rhalf y))\Big) \\
         &= (x\lhalf\rchop(\lenfcn{pow}(\rhalf y)\smsh
               \lenfcn{powdiv}(x\lhalf,\rhalf y)))\lchop
               \big(\rhalf x\lchop(\lenfcn{pow}(\rhalf y)\smsh
                  \lenfcn{powdiv}(x\lhalf,\rhalf y))\big) \\
         &= (x\lhalf\rchop(\lenfcn{pow}(\rhalf y)\smsh
               \lenfcn{powdiv}(x\lhalf,\rhalf y)))\lchop\E
          = \E. 
      \end{align*}
   \item
      By $\TIND$ on $y\neq\E$:
      \( \lenfcn{powmod}(x,1)\lchop\lenfcn{pow}(1) = \E\lchop 1
         \neq \E \),
      \begin{align*}
         &\!\lenfcn{powmod}(x,y)\lchop\lenfcn{pow}(y) \\
         &= \big((\lenfcn{pow}(y)\smsh\lenfcn{powdiv}(x,y))\lchop
               {}_1x\big)\lchop\lenfcn{pow}(y) \\
         &= \Big(\big((\lenfcn{pow}(\rhalf y)\cat
                  \lenfcn{pow}(\rhalf y))\smsh
               \lenfcn{powdiv}(x\lhalf,\rhalf y)\big)
               \lchop({}_1x\lhalf\cat{}_1\rhalf x)\Big) \\*
         &\pheq \hphantom{\Big( } \lchop
            (\lenfcn{pow}(\rhalf y)\cat\lenfcn{pow}(\rhalf y)) \\
         &= \Big(\big((\lenfcn{pow}(\rhalf y)\smsh
                  \lenfcn{powdiv}(x\lhalf,\rhalf y))\cat
               (\lenfcn{pow}(\rhalf y)\smsh
                  \lenfcn{powdiv}(x\lhalf,\rhalf y))\big) \\*
         &\pheq \hphantom{\Big( } \lchop
            ({}_1x\lhalf\cat{}_1x\lhalf\cat(x\elC(\E,1)))\Big)\lchop
            (\lenfcn{pow}(\rhalf y)\cat\lenfcn{pow}(\rhalf y)) \\
         &= \Big(\big((\lenfcn{pow}(\rhalf y)\smsh
                  \lenfcn{powdiv}(x\lhalf,\rhalf y))\lchop
                  {}_1x\lhalf\big)\cat \\*
         &\pheq \hphantom{\Big( } \big((\lenfcn{pow}(\rhalf y)\smsh
                  \lenfcn{powdiv}(x\lhalf,\rhalf y))\lchop
                  {}_1x\lhalf\cat(x\elC(\E,1))\big)\Big)\lchop
            (\lenfcn{pow}(\rhalf y)\cat\lenfcn{pow}(\rhalf y)) \\
         &= \Big(\lenfcn{powmod}(x\lhalf,\rhalf y)\cat
               \big(\lenfcn{powmod}(x\lhalf,\rhalf y)\cat
                  (x\elC(\E,1))\big)\Big)\lchop
            (\lenfcn{pow}(\rhalf y)\cat\lenfcn{pow}(\rhalf y)) \\
         &= \lenfcn{powmod}(x\lhalf,\rhalf y)\lchop
               \lenfcn{pow}(\rhalf y)\cat
            (\lenfcn{powmod}(x\lhalf,\rhalf y)\cat(x\elC(\E,1)))\lchop
               \lenfcn{pow}(\rhalf y) \neq \E 
      \end{align*}
      (where we have used the fact that
      \( y\rchop x=\E \land w\rchop z=\E \limp
         wy\lchop{}_1z{}_1x=w\lchop{}_1z\cat y\lchop{}_1x \),
      which is a direct consequence of Claim~\ref{claim:cuts} and the
      fact that \(zy\lchop x=y\lchop(z\lchop x)\)).
   \item
      We prove the first statement by $\TIND$ on $y\neq\E$:
      \( \lenfcn{powdiv}(x1,1) = {}_1x1 = {}_1x\cat
            (1\lchop 1\zlC(1,\E)) = \lenfcn{powdiv}(x,1)\cat
            ((\lenfcn{powmod}(x,1)\cat 1)\lchop 1\zlC(1,\E)) \),
      and before proving the inductive case, we can show by $\TIND$ on
      $y\neq\E$ that
      \begin{align*}
         &\!(\lenfcn{powmod}(x,y)\cat 1)\lchop\lenfcn{pow}(y)\zlC(1,\E) \\
         &= x\elC\Big(\big((\lenfcn{powmod}(x\lhalf,\rhalf y)\cat 1)
               \lchop\lenfcn{pow}(\rhalf y)\cat
            \lenfcn{powmod}(x\lhalf,\rhalf y)\lchop
               \lenfcn{pow}(\rhalf y)\big)\zlC(1,\E), \\*
         &\pheq\quad
            \big((\lenfcn{powmod}(x\lhalf,\rhalf y)\cat 1)\lchop
               \lenfcn{pow}(\rhalf y)\cat
            (\lenfcn{powmod}(x\lhalf,\rhalf y)\cat 1)\lchop
               \lenfcn{pow}(\rhalf y)\big)\zlC(1,\E)\Big) \\
         &= x\elC\Big(\E,(\lenfcn{powmod}(x\lhalf,\rhalf y)\cat 1)
               \lchop\lenfcn{pow}(\rhalf y)\zlC(1,\E)\Big) 
      \end{align*}
      (the proof is similar to that of the preceding claim), so that
      \begin{align*}
         &\!\lenfcn{powdiv}(x1,y) \\
         &= \lenfcn{powdiv}((x1)\lhalf,\rhalf y) \\
         &= x\elC\big(\lenfcn{powdiv}(x\lhalf,\rhalf y),
            \lenfcn{powdiv}(x\lhalf\cat 1,\rhalf y)\big) \\
         &= x\elC\big(\lenfcn{powdiv}(x,y),
            \lenfcn{powdiv}(x\lhalf,\rhalf y)\cat
               (\lenfcn{powmod}(x\lhalf,\rhalf y)\cat 1)\lchop
               \lenfcn{pow}(\rhalf y)\zlC(1,\E)\big) \\
         &= \lenfcn{powdiv}(x,y)\cat\big(x\elC\big(\E,
               (\lenfcn{powmod}(x\lhalf,\rhalf y)\cat 1)\lchop
               \lenfcn{pow}(\rhalf y)\zlC(1,\E)\big)\big) \\
         &= \lenfcn{powdiv}(x,y)\cat\big((\lenfcn{powmod}(x,y)\cat 1)
               \lchop\lenfcn{pow}(y)\zlC(1,\E)\big). 
      \end{align*}

      As for the second statement, it follows directly from the first
      by the definition of $\lenfcn{powmod}$ and the fact that
      \( \lenfcn{powmod}(x,y)\lchop\lenfcn{pow}(y)\neq\E \limp
         \lenfcn{pow}(y)\lchop(\lenfcn{powmod}(x,y)\cat 1)=\E \).
   \item
      By $\NIND$ on $x$:
      \( \lenfcn{powdiv}((\lenfcn{pow}(y)\smsh z)\cat\E,y) =
         y\zlC(\E,{}_1z) = y\zlC(\E,{}_1z)\cat\lenfcn{powdiv}(\E,y)
         \land \lenfcn{powmod}((\lenfcn{pow}(y)\smsh z)\cat\E,y) =
         \E = \lenfcn{powmod}(\E,y) \),
      \begin{align*}
         &\!\lenfcn{powdiv}((\lenfcn{pow}(y)\smsh z)\cat xi,y) \\
         &= \lenfcn{powdiv}((\lenfcn{pow}(y)\smsh z)\cat x,y)\cat \\*
         &\pheq\big((\lenfcn{powmod}((\lenfcn{pow}(y)\smsh z)\cat x,y)
               \cat 1)\lchop\lenfcn{pow}(y)\zlC(1,\E)\big) \\
         &= (y\zlC(\E,{}_1z))\cat\lenfcn{powdiv}(x,y)\cat
            \big((\lenfcn{powmod}(x,y)\cat 1)\lchop\lenfcn{pow}(y)
               \zlC(1,\E)\big) \\
         &= (y\zlC(\E,{}_1z))\cat\lenfcn{powdiv}(x1,y), \displaybreak[0]\\
         &\!\lenfcn{powmod}((\lenfcn{pow}(y)\smsh z)\cat xi,y) \\
         &= \big(\lenfcn{pow}(y)\smsh
               \lenfcn{powdiv}((\lenfcn{pow}(y)\smsh z)\cat xi,y)\big)
            \lchop{}_1\big((\lenfcn{pow}(y)\smsh z)\cat xi\big) \\
         &= \big(\lenfcn{pow}(y)\smsh
               ((y\zlC(\E,{}_1z))\cat\lenfcn{powdiv}(xi,y))\big)
            \lchop{}_1(\lenfcn{pow}(y)\smsh z)\cat{}_1(xi) \\
         &= (\lenfcn{pow}(y)\smsh(y\zlC(\E,{}_1z)))\cat
               (\lenfcn{pow}(y)\smsh\lenfcn{powdiv}(xi,y))
            \lchop(\lenfcn{pow}(y)\smsh{}_1z)\cat{}_1(xi) \\
         &= \lenfcn{pow}(y)\smsh\lenfcn{powdiv}(xi,y)\lchop{}_1(xi) \\
         &= \lenfcn{powmod}(xi,y). 
      \end{align*}
   \item
      By $\NIND$ on $x$:
      \( y\neq\E \land \E\lchop\lenfcn{pow}(y)=\E \)
      is false so the statement is vacuously true, and
      \( y\neq\E \land xi\lchop\lenfcn{pow}(y)=\E \limp
         (y\neq\E \land x\lchop\lenfcn{pow}(y)=\E) \lor
         (y\neq\E \land x\lchop\lenfcn{pow}(y)=1) \),
      so we prove the statement by cases.

      First, if
      \( y\neq\E \land x\lchop\lenfcn{pow}(y)=\E \)
      (which implies that
      \( {}_1(x1\rchop\lenfcn{pow}(y)) =
         {}_1(x\rchop\lenfcn{pow}(y))\cat 1 \)),
      \begin{align*}
         &\!\lenfcn{powdiv}(x1,y) \\
         &= \lenfcn{powdiv}(x,y)\cat\big((\lenfcn{powmod}(x,y)\cat 1)
               \lchop\lenfcn{pow}(y)\zlC(1,\E)\big) \\
         &= \lenfcn{powdiv}(x\rchop\lenfcn{pow}(y),y)\cat 1\cat
            \big((\lenfcn{powmod}(x\rchop\lenfcn{pow}(y))\cat 1)\lchop
               \lenfcn{pow}(y)\zlC(1,\E)\big) \\
         &= \lenfcn{powdiv}((x\rchop\lenfcn{pow}(y))\cat 1,y)\cat 1
          = \lenfcn{powdiv}(x1\rchop\lenfcn{pow}(y),y)\cat 1, 
            \displaybreak[0]\\
         &\!\lenfcn{powmod}(x1,y) \\
         &= (\lenfcn{powmod}(x,y)\cat 1)\lchop\lenfcn{pow}(y)\zlC
            (\E,\lenfcn{powmod}(x,y)\cat 1) \\
         &= (\lenfcn{powmod}(x\rchop\lenfcn{pow}(y),y)\cat 1)\lchop
               \lenfcn{pow}(y)\zlC(\E,\lenfcn{powmod}
                  (x\rchop\lenfcn{pow}(y),y)\cat 1) \\
         &= \lenfcn{powmod}((x\rchop\lenfcn{pow}(y))\cat 1,y)
          = \lenfcn{powmod}(x1\rchop\lenfcn{pow}(y),y). 
      \end{align*}

      Second, if
      \( y\neq\E \land x\lchop\lenfcn{pow}(y)=1 \)
      (which implies that
      \( {}_1(x1\rchop\lenfcn{pow}(y)) = \E \)),
      \begin{align*}
         \lenfcn{powdiv}(x1,y)
         &= \lenfcn{powdiv}(x,y)\cat\big((\lenfcn{powmod}(x,y)\cat 1)
               \lchop\lenfcn{pow}(y)\zlC(1,\E)\big) \\
         &= \E\cat\big(({}_1x1)\lchop\lenfcn{pow}(y)\zlC(1,\E)\big) \\
         &= 1 = \lenfcn{powdiv}(\E,y)\cat 1 \\
         &= \lenfcn{powdiv}(x1\rchop\lenfcn{pow}(y),y)\cat 1, 
            \displaybreak[0]\\
         \lenfcn{powmod}(x1,y)
         &= (\lenfcn{powmod}(x,y)\cat 1)\lchop\lenfcn{pow}(y)\zlC
            (\E,\lenfcn{powmod}(x,y)\cat 1) \\
         &= {}_1x1\lchop\lenfcn{pow}(y)\zlC(\E,{}_1x1) \\
         &= \E = \lenfcn{powmod}(\E,y) \\
         &= \lenfcn{powmod}(x1\rchop\lenfcn{pow}(y),y). 
         \quad\QED
      \end{align*}
   \end{enumerate}
\end{proof}

\heading{On ``$\numrel=$'' and ``$\numrel<$''}

\begin{theorem*}{\ref{thm:=n}}
\mbox{}\nopagebreak
   \begin{enumerate}
   \item
      \( x\numrel=\E \liff x={}_0x \)
   \item
      \( x\numrel=y \liff \fcn{lp}_0(x,y)=\fcn{lp}_0(y,x) \)
   \end{enumerate}
\end{theorem*}
\begin{proof}
\mbox{}\nopagebreak
   \begin{enumerate}
   \item
      By a straightforward $\NIND$ on $x$.
   \item
      By the preceding property and Derived Rule~\ref{derule:2NIND}:
      \( x\numrel=\E \liff x={}_0 x \liff
         \fcn{lp}_0(x,\E)=\fcn{lp}_0(\E,x) \)
      (and similarly for $\E\numrel=y$),
      \( xi\numrel=yj \liff x\numrel=y\bitbin\land(i\bitbin\liff j)
         \liff \fcn{lp}_0(x,y)=\fcn{lp}_0(y,x)\land i=j
         \liff \fcn{lp}_0(x,y)\cat i=\fcn{lp}_0(y,x)\cat j
         \liff \fcn{lp}_0(xi,yj)=\fcn{lp}_0(yj,xi) \).
      \quad\QED
   \end{enumerate}
\end{proof}

\begin{claim*}{\ref{claim:=ncat}}
\mbox{}\nopagebreak
   \begin{enumerate}
   \item
      \( x0\numrel=y0 \liff x\numrel=y \liff x1\numrel=y1 \)
   \item
      \( \bit\lnot(x0\numrel=y1) \)
   \item
      \( \bit\lnot(x1\numrel=y0) \)
   \end{enumerate}
\end{claim*}
\begin{proof}
   Directly from Theorem~\ref{thm:=n}:
   \( xi\numrel=yj \liff \fcn{lp}_0(xi,yj)=\fcn{lp}_0(yj,xi)
      \liff \fcn{lp}_0(x,y)\cat i=\fcn{lp}_0(y,x)\cat j
      \liff \fcn{lp}_0(x,y)=\fcn{lp}_0(y,x)\land i=j
      \liff x\numrel=y\land i=j \).
\end{proof}

\begin{claim*}{\ref{claim:<ncat}}
\mbox{}\nopagebreak
   \begin{enumerate}
   \item
      \( x0\numrel<y0
          = x\numrel<y\bitbin\lor(x\numrel=y\bitbin\land 0\bitrel<0)
          = x\numrel<y \)
   \item
      \( x0\numrel<y1
          = x\numrel<y\bitbin\lor(x\numrel=y\bitbin\land 0\bitrel<1)
          = x\numrel\leq y \)
   \item
      \( x1\numrel<y0
          = x\numrel<y\bitbin\lor(x\numrel=y\bitbin\land 1\bitrel<0)
          = x\numrel<y \)
   \item
      \( x1\numrel<y1
          = x\numrel<y\bitbin\lor(x\numrel=y\bitbin\land 1\bitrel<1)
          = x\numrel<y \)
   \end{enumerate}
\end{claim*}
\begin{proof}
   From Theorem~\ref{thm:CRNm}, we have that
   \( xi\numrel<yj =
      x\numrel<y\bitbin\lor(x\numrel=y\bitbin\land i\bitbin<j) \),
   so the theorem follows directly from the preceding claim.
\end{proof}

\begin{claim*}{\ref{claim:<n}}
\mbox{}\nopagebreak
   \begin{enumerate}
   \item
      \( \bit\lnot(x\numrel<\E) \)
   \item
      \( \bit\lnot(x\numrel<x) \)
   \item
      \( \bit\lnot(\E\numrel<{}_0x) \)
   \end{enumerate}
\end{claim*}
\begin{proof}
   A straightforward proof by $\NIND$, using the preceding claims.
\end{proof}

\begin{lemma*}{\ref{lem:cat=n}}
\mbox{}\nopagebreak
   \begin{enumerate}
   \item
      \( x\numrel=y = 0x\numrel=y = x\numrel=0y = 0x\numrel=0y \)
   \item
      \( x\numrel<y = 0x\numrel<0y \)
   \item
      \( x\numrel<y = 0x\numrel<y = x\numrel<0y \)
   \end{enumerate}
\end{lemma*}
\begin{proof}
\mbox{}\nopagebreak
   \begin{enumerate}
   \item
      Direct from the fact that
      \( (\fcn{lp}_0(0x,y) = \fcn{lp}_0(x,y)) \ \lor
       \ (\fcn{lp}_0(0x,y) = 0\cat\fcn{lp}_0(x,y)) \)
      (which can easily be proved by cases depending on the length of
      $x\lchop y$), and by Derived Rule~\ref{derule:2NIND}.
   \item
      By a simple application of Derived Rule~\ref{derule:2NIND} and
      the first claim (we show only the inductive step, the base cases
      being just as simple):
      \( 0xi\numrel<0yj
          = 0x\numrel<0y\bitbin\lor(0x\numrel=0y\bitbin\land i\bitbin<j)
          = x\numrel<y\bitbin\lor(x\numrel=y\bitbin\land i\bitbin<j)
          = xi\numrel<yj \).
   \item
      By the second claim:
      \begin{align*}
         0x\numrel<y
         &= 0x\rchop y\zlC\big({}_0(y\rchop0x)\cat 0x\numrel<y,
            0x\numrel<{}_0(0x\rchop y)\cat y\big) \\
         &= x\lchop y\zlC\big(0x\numrel<0\cat{}_0(x\rchop y)\cat y,
            {}_0(y\rchop x)\rdel\cat 0\cat x\numrel<y\big) \\
         &= x\lchop y\zlC\big(x\numrel<{}_0(x\rchop y)\cat y,
            {}_0(y\rchop x)\cat x\numrel<y\big) \\
         &= x\numrel<y 
      \end{align*}
      (and similarly for $x\numrel<0y$).
      \quad\QED
   \end{enumerate}
\end{proof}

\begin{theorem*}{\ref{thm:=n<ntrans}}
   \( x\numrel=y \land y\numrel<z \limp x\numrel<z \)
   \quad and \quad
   \( x\numrel=y \land y\numrel>z \limp x\numrel>z \)
\end{theorem*}
\begin{proof}
   Follows directly from the (already proven) facts that
   \( x\numrel=y \liff \fcn{lp}_0(x,\lenfcn{max}_3(x,y,z))=
         \fcn{lp}_0(y,\lenfcn{max}_3(x,y,z)) \)
   and
   \( y\numrel<z \liff \fcn{lp}_0(y,\lenfcn{max}_3(x,y,z))\numrel<
         \fcn{lp}_0(z,\lenfcn{max}_3(x,y,z)) \).
\end{proof}
\begin{corollary*}{\ref{cor:=nleqntrans}}
   \( x\numrel=y \land y\numrel\leq z \limp x\numrel\leq z \)
   \quad and \quad
   \( x\numrel=y \land y\numrel\geq z \limp x\numrel\geq z \)
\end{corollary*}
\begin{proof}
   Directly from the theorem.
\end{proof}

\begin{theorem*}{\ref{thm:<ntrans}}
   \( x\numrel<y \land y\numrel<z \limp x\numrel<z \)
   \quad and \quad
   \( x\numrel>y \land y\numrel>z \limp x\numrel>z \)
\end{theorem*}
\begin{proof}
   By Derived Rule~\ref{derule:2NIND}: the base cases for \(z=\E\) and
   \(y=\E\) are trivially true since \(\E\numrel{\not<}x\), while the
   third base case
   \( \E\numrel<y \land y\numrel<z \limp \E\numrel<z \)
   is itself proved by Derived Rule~\ref{derule:2NIND}: the two base
   cases are again trivially true, and
   \( \E\numrel<yj\bitbin\land yj\numrel<zk
       = \big(\E\numrel<y\bitbin\lor
            (\E\numrel=y\bitbin\land 0\bitbin<j)\big)\bitbin\land
         \big(y\numrel<z\bitbin\lor
            (y\numrel=z\bitbin\land j\bitbin<k)\big)
       = (\E\numrel<y\bitbin\land y\numrel<z)\bitbin\lor
         (\E\numrel<y\bitbin\land y\numrel=z\bitbin\land j\bitbin<k)\bitbin\lor
         (\E\numrel=y\bitbin\land 0\bitbin<j\bitbin\land y\numrel<z)\bitbin\lor
         (\E\numrel=y\bitbin\land 0\bitbin<j\bitbin\land
            y\numrel=z\bitbin\land j\bitbin<k)
       = \E\numrel<z\bitbin\lor(\E\numrel=z\bitbin\land 0\bitbin<k)
       = \E\numrel<zk \)
   (the general inductive step is almost identical).
\end{proof}
\begin{corollary*}{\ref{cor:leqn<ntrans}}
   \( x\numrel\leq y \land y\numrel<z \limp x\numrel<z \)
   \quad and \quad
   \( x\numrel\geq y \land y\numrel>z \limp x\numrel>z \)
\end{corollary*}
\begin{proof}
   Directly from the theorem.
\end{proof}
\begin{corollary*}{\ref{cor:leqntrans}}
   \( x\numrel\leq y \land y\numrel\leq z \limp x\numrel\leq z \)
   \quad and \quad
   \( x\numrel\geq y \land y\numrel\geq z \limp x\numrel\geq z \)
\end{corollary*}
\begin{proof}
   Directly from the theorem.
\end{proof}

\heading{On ``$|\mathord\cdot|$'' and ``$\numfcn{succ}$''}

\begin{claim*}{\ref{claim:succcat}}
\label{thm:succ}\mbox{}\nopagebreak
   \begin{align*}
      \numfcn{succ}(\E) &= 1 \\
      \numfcn{succ}(x0) &= 0x1 \\
      \numfcn{succ}(x1) &= \numfcn{succ}(x)\cat 0 
   \end{align*}
\end{claim*}
\begin{proof}
    Simple proofs by $\NIND$ (proving the relevant properties first
    for the auxiliary function $\numfcn{cuss}$, and then for
    $\numfcn{succ}$).
\end{proof}

\begin{claim*}{\ref{claim:catsucc}}
\mbox{}\nopagebreak
   \begin{align*}
      \numfcn{succ}(0x) &= 0\cat\numfcn{succ}(x) \\
      \numfcn{succ}(1x) &= \lbit\numfcn{succ}(x)\cat
         \bit\lnot\lbit\numfcn{succ}(x)\cat\ldel\numfcn{succ}(x) \\
      \lbit\numfcn{succ}(x) &= \fcn{AND}(1x) \\
      \numfcn{succ}(x) &= \fcn{AND}(1x)\bC
         \big(1\cat{}_0x,0\cat\ldel\numfcn{succ}(x)\big) 
   \end{align*}
\end{claim*}
\begin{proof}
   All the proofs are simple, but we will illustrate them by proving
   the second property, by $\NIND$ on $x$:
   \( \numfcn{succ}(1\E) = 10 = 1\cat\bit\lnot 1\cat\ldel 1 =
      \lbit\numfcn{succ}(\E)\cat\bit\lnot\lbit\numfcn{succ}(\E)\cat
         \ldel\numfcn{succ}(\E) \),
   \( \numfcn{succ}(1x0) = 01x1
       = \lbit(0x1)\cat\bit\lnot\lbit(0x1)\cat\ldel(0x1)
       = \lbit\numfcn{succ}(x0)\cat\bit\lnot\lbit\numfcn{succ}(x0)
         \cat\ldel\numfcn{succ}(x0) \),
   \( \numfcn{succ}(1x1) = \numfcn{succ}(1x)\cat 0
       = \lbit\numfcn{succ}(x)\cat\bit\lnot\lbit\numfcn{succ}(x)
         \cat\ldel\numfcn{succ}(x)\cat 0
       = \lbit(\numfcn{succ}(x)\cat 0)\cat
         \bit\lnot\lbit(\numfcn{succ}(x)\cat 0)\cat
         \ldel(\numfcn{succ}(x)\cat 0)
       = \lbit\numfcn{succ}(x1)\cat\bit\lnot\lbit\numfcn{succ}(x1)
         \cat\ldel\numfcn{succ}(x1) \).
\end{proof}

\begin{theorem*}{\ref{thm:<nsucc>n}}
\mbox{}\nopagebreak
   \begin{enumerate}
   \item
      \( x \numrel< \numfcn{succ}(x) \)
   \item
      \( x\numrel>y = x\numrel\geq\numfcn{succ}(y) \)
   \end{enumerate}
\end{theorem*}
\begin{proof}
\mbox{}\nopagebreak
   \begin{enumerate}
   \item
      A simple direct proof by $\NIND$:
      \( \E\numrel<\numfcn{succ}(\E) = \E\numrel<1 \),
      \( x0\numrel<\numfcn{succ}(x0) = x0\numrel<0x1 = x\numrel\leq 0x \), 
      \( x1\numrel<\numfcn{succ}(x1) = x1\numrel<\numfcn{succ}(x)\cat 0
          = x\numrel<\numfcn{succ}(x) \).
   \item
      By Derived Rule~\ref{derule:2NIND}:
      \( \E\numrel>y = 0 = \E\numrel\geq\numfcn{succ}(y) \) (since
      \(\numfcn{succ}(y)\numrel\neq\E\)),
      \( x\numrel>\E = x\numrel\geq\numfcn{succ}(\E) \) (proved by an
      easy $\NIND$ on $x$),
      \begin{align*}
         xi\numrel>y0
         &= x\numrel>y\bitbin\lor(x\numrel=y\bitbin\land i\bitrel>0) \\
         &= x\numrel>y\bitbin\lor(x\numrel=y\bitbin\land i) \\
         &= x\numrel>0y\bitbin\lor(x\numrel=0y\bitbin\land i\bitrel>1)
            \bitbin\lor(x\numrel=0y\bitbin\land i) \\
         &= xi\numrel>0y1\bitbin\lor xi\numrel=0y1 \\
         &= xi\numrel\geq 0y1 
          = xi\numrel\geq\numfcn{succ}(y0), \displaybreak[0]\\
         xi\numrel>y1
         &= x\numrel>y\bitbin\lor(x\numrel=y\bitbin\land i\bitrel>1) \\
         &= x\numrel>y \\
         &= x\numrel\geq\numfcn{succ}(y) \\
         &= x\numrel>\numfcn{succ}(y)\bitbin\lor
            (x\numrel=y\bitbin\land i\bitrel\geq 0) \\
         &= x\numrel>\numfcn{succ}(y)\bitbin\lor
            (x\numrel=y\bitbin\land i\bitrel>0)\bitbin\lor
            (x\numrel=y\bitbin\land i\bitrel=0) \\
         &= xi\numrel>\numfcn{succ}(y)\cat 0\bitbin\lor
            xi\numrel=\numfcn{succ}(y)\cat 0 \\
         &= xi\numrel\geq\numfcn{succ}(y)\cat 0 
          = xi\numrel\geq\numfcn{succ}(y1).
         \quad\QED
      \end{align*}
   \end{enumerate}
\end{proof}

\begin{claim*}{\ref{claim:|_j|}}
   \( |x| = |{}_jx| \)
\end{claim*}
\begin{proof}
   By $\TIND$ on $x$: \( |\E| = \E = |{}_j\E| \),
   \( |i| = 1 = |1| = |{}_1i| \),
   \( |x| = x\elC\big(|x\lhalf|\cat 0,|x\lhalf|\cat 1\big)
       = {}_jx\elC\big(|{}_jx\lhalf|\cat 0,|{}_jx\lhalf|\cat 1\big)
       = |{}_jx| \).
\end{proof}

\begin{theorem*}{\ref{thm:succ||}}
   \( |xi| \numrel= \numfcn{succ}(|x|) \)
\end{theorem*}
\begin{proof}
   By $\TIND$ on $x$: \( |\E i| = 1 = \numfcn{succ}(|\E|) \),
   \( |ji| = 10 = \numfcn{succ}(1) = \numfcn{succ}(|j|) \),
   \begin{align*}
      |xi|
      &= xi\elC\big(|(xi)\lhalf|\cat 0,|(xi)\lhalf|\cat 1\big) \\
      &= x\elC\big(|(xi)\lhalf|\cat 1,|(xi)\lhalf|\cat 0\big) \\
      &= x\elC\big(|x\lhalf|\cat 1,
            |x\lhalf\cat\lbit(\rhalf x\cat i)|\cat 0\big) \\
      &= x\elC\big(|x\lhalf|\cat 1,|x\lhalf\cat j|\cat 0\big) \\
      &\numrel= x\elC\big(\numfcn{succ}(|x\lhalf|\cat 0),
            \numfcn{succ}(|x\lhalf|)\cat 0\big) \\
      &= x\elC\big(\numfcn{succ}(|x\lhalf|\cat 0),
            \numfcn{succ}(|x\lhalf|\cat 1)\big) \\
      &= \numfcn{succ}(|x|) 
   \end{align*}
   (where the fifth equality, where ``$\numrel=$'' is introduced,
   holds by the induction hypothesis).
\end{proof}

\heading{On ``masking'' functions}

\begin{theorem*}{\ref{thm:firstj}}
\mbox{}\nopagebreak
   \begin{alignat*}{2}
      \fcn{first}_0(0x) &= 1\cat{}_0x &\qquad
      \fcn{first}_0(1x) &= 0\cat\fcn{first}_0(x) \\
      \fcn{first}_1(0x) &= 0\cat\fcn{first}_1(x) &\qquad
      \fcn{first}_1(1x) &= 1\cat{}_0x 
   \end{alignat*}
\end{theorem*}
\begin{proof}
   By a straightforward $\NIND$.
\end{proof}

\begin{corollary*}{\ref{cor:first01}}
   \( \fcn{first}_0(x) = \fcn{first}_1(\bitfcn{not}(x)) \)
\end{corollary*}
\begin{proof}
   By a straightforward $\NIND$, from the preceding theorem.
\end{proof}

\heading{On binary addition}

\begin{theorem*}{\ref{thm:+comm}}
   \( x\numbin+y = y\numbin+x \)
\end{theorem*}
\begin{proof}
   Direct from the commutativity of the functions involved (\ie,
   $\bitfcn{xor}$ and $\bitfcn{and}$).
\end{proof}

\begin{lemma*}{\ref{lem:carry1}}
\mbox{}\nopagebreak
   \begin{align*}
      \numfcn{carry}(x0,1)
      &= \numfcn{carry}(x,\E)\cat 0 = {}_0x0 \\
      \numfcn{carry}(x1,1)
      &= x\zlC\big(1,\numfcn{carry}(x,1)\cat 1\big) 
   \end{align*}
\end{lemma*}
\begin{proof}
   (We prove only the second statement, the first is a simple
   application of $\NIND$ on $x$.)  By $\NIND$ on $x$:
   \( \numfcn{carry}(\E\cat 1,1) = 1
       = \E\zlC(1,\numfcn{carry}(\E,1)\cat 1) \),
   \begin{align*}
      \numfcn{carry}(0x1,1)
      &= 0\cat\numfcn{carry}(x1,1) \\
      &= 0\cat x\zlC\big(1,\numfcn{carry}(x,1)\cat 1\big) \\
      &= x\zlC\big(0\cat 1,
         0\cat\numfcn{carry}(x,1)\cat 1\big) \\
      &= x\zlC\big(\numfcn{carry}(0,1)\cat 1,
         \numfcn{carry}(0x,1)\cat 1\big) \\
      &= \numfcn{carry}(0x,1)\cat 1, \displaybreak[0]\\
      \numfcn{carry}(1x1,1)
      &= \lbit\numfcn{carry}(x1,1)\cat\numfcn{carry}(x1,1) \\
      &= x\zlC\big(\lbit(1)\cat 1,
         \lbit(\numfcn{carry}(x,1)\cat 1)\cat
            \numfcn{carry}(x,1)\cat 1\big) \\
      &= x\zlC\big(11,\lbit\numfcn{carry}(x,1)\cat
         \numfcn{carry}(x,1)\cat 1\big) \\
      &= x\zlC\big(\numfcn{carry}(1,1)\cat 1,
         \numfcn{carry}(1x,1)\cat 1\big) \\
      &= \numfcn{carry}(1x,1)\cat 1. 
      \quad\QED
   \end{align*}

\end{proof}

\begin{theorem*}{\ref{thm:succ=+1}}
   \( x\numbin+1 =
      x\zlC\big(0\cat\numfcn{succ}(x),\numfcn{succ}(x)\big)
      \numrel= \numfcn{succ}(x) \)
\end{theorem*}
\begin{proof}
   By $\NIND$ on $x$:
   \( \E\numbin+1 = 01 =
      \E\zlC\big(0\cat\numfcn{succ}(\E),\numfcn{succ}(\E)\big) \),
   \begin{align*}
      x0\numbin+1
      &= \bitfcn{xor}_3(\numfcn{carry}(x0,1)\cat 0,x0,1) \\
      &= \bitfcn{xor}_3(0{}_0x0,0x0,0{}_0x1) \\
      &= 0x1 \\
      &= \numfcn{succ}(x0), \displaybreak[0]\\
      x1\numbin+1
      &= \bitfcn{xor}_3(\numfcn{carry}(x1,1)\cat 0,x1,1) \\
      &= \bitfcn{xor}_3(\numfcn{carry}(x,1)\cat 10,0x1,0{}_0x1) \\
      &= \bitfcn{xor}_3(\numfcn{carry}(x,1)\cat 1,0x,{}_0x0)
         \cat 0 \\
      &= \bitfcn{xor}_3(\numfcn{carry}(x,1)\cat 0,0x,{}_0x1)
         \cat 0 \\
      &= \bitfcn{xor}_3(\numfcn{carry}(x,1)\cat 0,x,1)\cat 0 \\
      &= (x\numbin+1)\cat 0 \\
      &\numrel= \numfcn{succ}(x)\cat 0 \\
      &= \numfcn{succ}(x1). 
      \quad\QED
   \end{align*}

\end{proof}

\begin{claim*}{\ref{claim:catcarry}}
   For ${}_jx={}_jy$,
   \begin{align*}
      \numfcn{carry}(x,\E) &= 0\cat{}_0x \\
      \numfcn{carry}(x,x) &= x0 \\
      \numfcn{carry}(0x,0y) &= 0\cat\numfcn{carry}(x,y) \\
      \numfcn{carry}(1x,0y)
      &= \lbit\numfcn{carry}(x,y)\cat\numfcn{carry}(x,y) \\
      \numfcn{carry}(1x,1y) &= 1\cat\numfcn{carry}(x,y) 
   \end{align*}
\end{claim*}
\begin{proof}
   All these properties can be proved with a simple application of
   Derived Rule~\ref{derule:len2NIND}, or directly from the definition
   of $\numfcn{carry}$.  The last three depend on the following facts.
   \begin{align*}
      \fcn{maskbit}\big(\bitfcn{and}_2(0x,0y),
         \fcn{first}_0(\bitfcn{xor}_2(0x,0y))\big)
      &= \fcn{maskbit}\big(0\cat\bitfcn{and}_2(x,y),
         \fcn{first}_0(0\cat\bitfcn{xor}_2(x,y))\big) \\
      &= \fcn{maskbit}\big(0\cat\bitfcn{and}_2(x,y),
         1\cat{}_0\bitfcn{xor}_2(x,y)\big) \\
      &= 0 \displaybreak[0]\\
      \fcn{maskbit}\big(\bitfcn{and}_2(1x,0y),
         \fcn{first}_0(\bitfcn{xor}_2(1x,0y))\big)
      &= \fcn{maskbit}\big(0\cat\bitfcn{and}_2(x,y),
         \fcn{first}_0(1\cat\bitfcn{xor}_2(x,y))\big) \\
      &= \fcn{maskbit}\big(0\cat\bitfcn{and}_2(x,y),
         0\cat\fcn{first}_0(\bitfcn{xor}_2(x,y))\big) \\
      &= \fcn{maskbit}\big(\bitfcn{and}_2(x,y),
         \fcn{first}_0(\bitfcn{xor}_2(x,y))\big) \displaybreak[0]\\
      \fcn{maskbit}\big(\bitfcn{and}_2(1x,1y),
         \fcn{first}_0(\bitfcn{xor}_2(1x,1y))\big)
      &= \fcn{maskbit}\big(1\cat\bitfcn{and}_2(x,y),
         \fcn{first}_0(0\cat\bitfcn{xor}_2(x,y))\big) \\
      &= \fcn{maskbit}\big(1\cat\bitfcn{and}_2(x,y),
         1\cat{}_0\bitfcn{xor}_2(x,y)\big) \\
      &= 1
      \quad\QED
   \end{align*}

\end{proof}

\begin{claim*}{\ref{claim:cat+}}
   For ${}_jx={}_jy$,
   \begin{align*}
      x\numbin+\E &= 0x \\
      x\numbin+x &= x0 \\
      0x\numbin+0y &= 0\cat(x\numbin+y) \\
      1x\numbin+0y &= \lbit(x\numbin+y)\cat\bit\lnot\lbit(x\numbin+y)\cat
         \ldel(x\numbin+y) \\
      1x\numbin+1y &= 1\cat(x\numbin+y) 
   \end{align*}
\end{claim*}
\begin{proof}
   Directly from the corresponding properties for $\numfcn{carry}$,
   where we have used the fact that
   \( \lbit(x\numbin+y) = \lbit\numfcn{carry}(x,y) \).
   Note that the second property implies that
   \( \ldel\numfcn{succ}(x\numbin+x) = x1 \).
\end{proof}

\begin{claim*}{\ref{claim:+cat}}
\mbox{}\nopagebreak
   \begin{align*}
      x0\numbin+y0 &= (x\numbin+y)\cat 0 \\
      x1\numbin+y0 &= (x\numbin+y)\cat 1 \\
      x1\numbin+y1 &= \ldel\numfcn{succ}(x\numbin+y)\cat 0 
   \end{align*}
\end{claim*}
\begin{proof}
   The first two properties follow directly from the Claim above by
   Derived Rule~\ref{derule:len2NIND}.  We prove the third property
   because it is more involved.  First, note that
   \(\bit\lnot\fcn{AND}(x\numbin+y)\) can be proved directly from
   Claim~\ref{claim:cat+} by Derived Rule~\ref{derule:len2NIND}.  This
   implies that \( \lbit\numfcn{succ}(x\numbin+y) = 0 \), which in turn
   implies that
   \( \numfcn{succ}(1\cat(x\numbin+y))
       = 01\cat\ldel\numfcn{succ}(x\numbin+y)\).
   Now, we prove the third property by Derived
   Rule~\ref{derule:len2NIND}:
   \( 1\numbin+1 = 10 = \ldel(01)\cat 0 = \ldel\numfcn{succ}(0)\cat 0
       = \ldel\numfcn{succ}(\E\numbin+\E)\cat 0 \),
   \begin{align*}
      &\!0x1\numbin+0y1 \\
      &= 0\cat(x1\numbin+y1) \\
      &= 0\cat\ldel\numfcn{succ}(x\numbin+y)\cat 0 \\
      &= \numfcn{succ}(x\numbin+y)\cat 0 \\
      &= \ldel(0\cat\numfcn{succ}(x\numbin+y))\cat 0 \\
      &= \ldel\numfcn{succ}(0\cat(x\numbin+y))\cat 0 \\
      &= \ldel\numfcn{succ}(0x\numbin+0y)\cat 0, \displaybreak[0]\\
      &\!1x1\numbin+1y1 \\
      &= 1\cat(x1\numbin+y1) \\
      &= 1\cat\ldel\numfcn{succ}(x\numbin+y)\cat 0 \\
      &= \ldel(01\cat\ldel\numfcn{succ}(x\numbin+y))\cat 0 \\
      &= \ldel\numfcn{succ}(1\cat(x\numbin+y))\cat 0 \\
      &= \ldel\numfcn{succ}(1x\numbin+1y)\cat 0, \displaybreak[0]\\
      &\!1x1\numbin+0y1 \\
      &= \lbit(x1\numbin+y1)\cat\bit\lnot\lbit(x1\numbin+y1)\cat
         \ldel(x1\numbin+y1) \\
      &= \lbit(\ldel\numfcn{succ}(x\numbin+y)\cat 0)\cat
         \bit\lnot\lbit(\ldel\numfcn{succ}(x\numbin+y)\cat 0)\cat
         \ldel(\ldel\numfcn{succ}(x\numbin+y)\cat 0) \\
      &= \lbit\ldel\numfcn{succ}(x\numbin+y)\cat
         \bit\lnot\lbit\ldel\numfcn{succ}(x\numbin+y)\cat
         \ldel\ldel\numfcn{succ}(x\numbin+y)\cat 0 \\
      &= \lbit(x\numbin+y)\zlC\Big(
         \lbit\ldel\numfcn{succ}(1\ldel(x\numbin+y))\cat
            \bit\lnot\lbit\ldel\numfcn{succ}(1\ldel(x\numbin+y))\cat
            \ldel\ldel\numfcn{succ}(1\ldel(x\numbin+y))\cat 0, \Big.\\*
      &\pheq \hphantom{\lbit(x\numbin+y)\zlC\Big(\Big.}\Big.
         \lbit\ldel\numfcn{succ}(0\ldel(x\numbin+y))\cat
            \bit\lnot\lbit\ldel\numfcn{succ}(0\ldel(x\numbin+y))\cat
            \ldel\ldel\numfcn{succ}(0\ldel(x\numbin+y))\cat 0\Big) \\
      &= \lbit(x\numbin+y)\zlC\Big(
         \bit\lnot\lbit\numfcn{succ}(\ldel(x\numbin+y))\cat
            \lbit\numfcn{succ}(\ldel(x\numbin+y))\cat
            \ldel\numfcn{succ}(\ldel(x\numbin+y))\cat 0, \Big.\\*
      &\pheq \hphantom{\lbit(x\numbin+y)\zlC\Big(\Big.}\Big.
         \lbit\numfcn{succ}(\ldel(x\numbin+y))\cat
            \bit\lnot\lbit\numfcn{succ}(\ldel(x\numbin+y))\cat
            \ldel\numfcn{succ}(\ldel(x\numbin+y))\cat 0\Big) \\
      &= \lbit(x\numbin+y)\zlC\Big(
         \bit\lnot\fcn{AND}(1\ldel(x\numbin+y))\cat
            \numfcn{succ}(\ldel(x\numbin+y))\cat 0,
         \numfcn{succ}(1\ldel(x\numbin+y))\cat 0\Big) \\
      &= \lbit(x\numbin+y)\zlC\Big(
         1\cat\numfcn{succ}(\ldel(x\numbin+y))\cat 0,
         \numfcn{succ}(1\ldel(x\numbin+y))\cat 0\Big) \\
      &= \lbit(x\numbin+y)\zlC\Big(
         \ldel(01\cat\numfcn{succ}(\ldel(x\numbin+y)))\cat 0,
         \ldel(0\cat\numfcn{succ}(1\ldel(x\numbin+y)))\cat 0\Big) \\
      &= \lbit(x\numbin+y)\zlC\Big(
         \ldel\numfcn{succ}(10\ldel(x\numbin+y))\cat 0,
         \ldel\numfcn{succ}(01\ldel(x\numbin+y))\cat 0\Big) \\
      &= \ldel\numfcn{succ}\big(\lbit(x\numbin+y)\cat
            \bit\lnot\lbit(x\numbin+y)\cat\ldel(x\numbin+y)\big)\cat 0 \\
      &= \ldel\numfcn{succ}(1x\numbin+0y)\cat 0. 
      \quad\QED
   \end{align*}

\end{proof}

\begin{theorem*}{\ref{thm:+}}
\mbox{}\nopagebreak
   \begin{enumerate}
   \item
      \( x\numbin+\numfcn{succ}(y) \numrel= \numfcn{succ}(x\numbin+y) \)
   \item
      \( x\numbin+(y\numbin+z) \numrel= (x\numbin+y)\numbin+z \)
   \item
      \( y\numrel<z \liff x\numbin+y\numrel<x\numbin+z \)
   \item
      \( x\numrel=y \land z\numrel=w \limp x\numbin+z\numrel=y\numbin+w \)
   \item
      \( x\numrel=y \land z\numrel<w \limp x\numbin+z\numrel<y\numbin+w \)
   \item
      \( x\numrel<y \land z\numrel<w \limp x\numbin+z\numrel<y\numbin+w \)
   \end{enumerate}
\end{theorem*}
\begin{proof}
\mbox{}\nopagebreak
   \begin{enumerate}
   \item
      By Derived Rule~\ref{derule:2NIND}:
      \( \E\numbin+\numfcn{succ}(y) \numrel= \numfcn{succ}(y)
         \numrel= \numfcn{succ}(\E\numbin+y) \),
      \( x\numbin+\numfcn{succ}(\E) = x\numbin+1 \numrel= \numfcn{succ}(x)
         \numrel= \numfcn{succ}(x\numbin+\E) \),
      \begin{align*}
         x0\numbin+\numfcn{succ}(y0)
         &\numrel= x0\numbin+y1 \\
         &\numrel= (x\numbin+y)\cat 1 \\
         &\numrel= \numfcn{succ}((x\numbin+y)\cat 0) \\
         &\numrel= \numfcn{succ}(x0\numbin+y0), \displaybreak[0]\\
         x0\numbin+\numfcn{succ}(y1)
         &\numrel= x0\numbin+\numfcn{succ}(y)\cat 0 \\
         &\numrel= (x\numbin+\numfcn{succ}(y))\cat 0 \\
         &\numrel= \numfcn{succ}(x\numbin+y)\cat 0 \\
         &\numrel= \numfcn{succ}((x\numbin+y)\cat 1) \\
         &\numrel= \numfcn{succ}(x0\numbin+y1), \displaybreak[0]\\
         x1\numbin+\numfcn{succ}(y0)
         &\numrel= x1\numbin+y1 \\
         &\numrel= \numfcn{succ}(x\numbin+y)\cat 0 \\
         &\numrel= \numfcn{succ}((x\numbin+y)\cat 1) \\
         &\numrel= \numfcn{succ}(x1\numbin+y0), \displaybreak[0]\\
         x1\numbin+\numfcn{succ}(y1)
         &\numrel= x1\numbin+\numfcn{succ}(y)\cat 0 \\
         &\numrel= (x\numbin+\numfcn{succ}(y))\cat 1 \\
         &\numrel= \numfcn{succ}(x\numbin+y)\cat 1 \\
         &\numrel= \numfcn{succ}(\numfcn{succ}(x\numbin+y)\cat 0) \\
         &\numrel= \numfcn{succ}(x1\numbin+y1). 
      \end{align*}
   \item
      By Derived Rule~\ref{derule:2NIND}, generalized to three
      variables:
      \( \E\numbin+(y\numbin+z) \numrel= y\numbin+z
         \numrel= (\E\numbin+y)\numbin+z \)
      (and similarly for $x,\E,z$ and $x,y,\E$),
      \( x0\numbin+(y0\numbin+z0)
         \numrel= x0\numbin+(y\numbin+z)0
         \numrel= (x\numbin+(y\numbin+z))0
         \numrel= ((x\numbin+y)\numbin+z)0
         \numrel= (x\numbin+y)0\numbin+z0
         \numrel= (x0\numbin+y0)\numbin+z0 \),
      \( x1\numbin+(y0\numbin+z0)
         \numrel= x1\numbin+(y\numbin+z)0
         \numrel= (x\numbin+(y\numbin+z))1
         \numrel= ((x\numbin+y)\numbin+z)1
         \numrel= (x\numbin+y)1\numbin+z0
         \numrel= (x1\numbin+y0)\numbin+z0 \)
      (and similarly for $x0,y1,z0$ and $x0,y0,z1$),
      \( x0\numbin+(y1\numbin+z1)
         \numrel= x0\numbin+\numfcn{succ}(y\numbin+z)0
         \numrel= (x\numbin+\numfcn{succ}(y\numbin+z))0
         \numrel= \numfcn{succ}(x\numbin+(y\numbin+z))0
         \numrel= \numfcn{succ}((x\numbin+y)\numbin+z)0
         \numrel= (x\numbin+y)1\numbin+z1
         \numrel= (x0\numbin+y1)\numbin+z1 \)
      (and similarly for $x1,y0,z1$ and $x1,y1,z0$),
      \( x1\numbin+(y1\numbin+z1)
         \numrel= x1\numbin+\numfcn{succ}(y\numbin+z)0
         \numrel= (x\numbin+\numfcn{succ}(y\numbin+z))1
         \numrel= \numfcn{succ}(x\numbin+(y\numbin+z))1
         \numrel= \numfcn{succ}((x\numbin+y)\numbin+z)1
         \numrel= (\numfcn{succ}(x\numbin+y)\numbin+z)1
         \numrel= \numfcn{succ}(x\numbin+y)0\numbin+z1
         \numrel= (x1\numbin+y1)\numbin+z1 \).
   \item[3--6.]
      Similar to the last case, straightforward, case-by-case proofs
      by Derived Rule~\ref{derule:2NIND}.
      \quad\QED
   \end{enumerate}
\end{proof}

\heading{On iterated sums}

\begin{claim*}{\ref{claim:CS}}
\mbox{}\nopagebreak
   \begin{gather*}
      \fcn{CScar}_3(x0,y0,z0) = \fcn{CScar}_3(x,y,z)\cat 0 \\
      \fcn{CScar}_3(x1,y0,z0) = \fcn{CScar}_3(x0,y1,z0)
       = \fcn{CScar}_3(x1,y0,z1) = \fcn{CScar}_3(x,y,z)\cat 0 \\
      \fcn{CScar}_3(x0,y1,z1) = \fcn{CScar}_3(x1,y0,z1)
       = \fcn{CScar}_3(x1,y1,z0) = \fcn{CScar}_3(x,y,z)\cat 1 \\
      \fcn{CScar}_3(x1,y1,z1) = \fcn{CScar}_3(x,y,z)\cat 1 \\
      {}_0\fcn{CScar}_3(x,y,z)\cat 0 = {}_0\fcn{CSadd}_3(x,y,z)
       = 0\cat{}_0\lenfcn{max}_3(x,y,z)
       = 0\cat\lenfcn{max}_3({}_0x,{}_0y,{}_0z) \\
      0\cat\lenfcn{max}_3({}_0x,{}_0y,{}_0z)
       = \fcn{CScar}_3({}_0x,{}_0y,{}_0z)\cat 0
       = \fcn{CSadd}_3({}_0x,{}_0y,{}_0z) \\
      {}_0\fcn{CScar}(x,y,z,w) = {}_0\fcn{CSadd}(x,y,z,w)
       = 00\cat{}_0\lenfcn{max}_4(x,y,z,w) 
       = 00\cat\lenfcn{max}_4({}_0x,{}_0y,{}_0z,{}_0w) \\
      00\cat\lenfcn{max}_4({}_0x,{}_0y,{}_0z,{}_0w)
       = \fcn{CScar}({}_0x,{}_0y,{}_0z,{}_0w)
       = \fcn{CSadd}({}_0x,{}_0y,{}_0z,{}_0w) 
   \end{gather*}
\end{claim*}
\begin{proof}
   All can be proved with a very simple application of Derived
   Rule~\ref{derule:len2NIND}, generalized to three variables,
   directly from the definitions of the functions involved.
\end{proof}

\begin{lemma*}{\ref{lem:CSsucc}}
\mbox{}\nopagebreak
   \begin{multline*}
      (\fcn{CScar}_3(\numfcn{succ}(x),y,z)\cat 0)\numbin+
         \fcn{CSadd}_3(\numfcn{succ}(x),y,z) \\*
      \numrel= \numfcn{succ}\big((\fcn{CScar}_3(x,y,z)\cat 0)\numbin+
         \fcn{CSadd}_3(x,y,z)\big) 
   \end{multline*}
\end{lemma*}
\begin{proof}
   The proof is a straightforward, if tedious, application of Derived
   Rule~\ref{derule:2NIND} (generalized to three variables).  First,
   the base cases for $\E,y,z$ and $x,\E,z$ and $x,y,\E$ are proved
   (each one with another application of Derived
   Rule~\ref{derule:2NIND}), and then the eight cases from $x0,y0,z0$
   to $x1,y1,z1$ are proved from the assumption that the lemma holds
   for $x,y,z$.  We do not show the full proof here as it is not
   particularly instructive; instead, we give parts of the proof for
   two illustrative cases.  First, in the proof of the base case
   \( (\fcn{CScar}_3(\numfcn{succ}(\E),y,z)\cat 0)\numbin+
         \fcn{CSadd}_3(\numfcn{succ}(\E),y,z)
      \numrel= \numfcn{succ}\big((\fcn{CScar}_3(\E,y,z)\cat 0)\numbin+
         \fcn{CSadd}_3(\E,y,z)\big) \)
   by Derived Rule~\ref{derule:2NIND}, we show the case for $y1,z1$.
   \begin{align*}
      &\!(\fcn{CScar}_3(\numfcn{succ}(\E),y1,z1)\cat 0)\numbin+
         \fcn{CSadd}_3(\numfcn{succ}(\E),y1,z1) \\
      &\numrel= (\fcn{CScar}_3(1,y1,z1)\cat 0)\numbin+
         \fcn{CSadd}_3(1,y1,z1) \\
      &\numrel= (\fcn{CScar}_3(\E,y,z)\cat 10)\numbin+
         (\fcn{CSadd}_3(\E,y,z)\cat 1) \\
      &\numrel= \big((\fcn{CScar}_3(\E,y,z)\cat 1)\numbin+
         \fcn{CSadd}_3(\E,y,z)\big)\cat 1 \\
      &\numrel=
         \numfcn{succ}\big(\big((\fcn{CScar}_3(\E,y,z)\cat 1)\numbin+
         \fcn{CSadd}_3(\E,y,z)\big)\cat 0\big) \\
      &\numrel=
         \numfcn{succ}\big((\fcn{CScar}_3(\E,y1,z1)\cat 0)\numbin+
         \fcn{CSadd}_3(\E,y1,z1)\big) 
   \end{align*}
   Second, in the inductive step, we show the case for $x0,y1,z0$.
   \begin{align*}
      &\!(\fcn{CScar}_3(\numfcn{succ}(x0),y1,z0)\cat 0)\numbin+
         \fcn{CSadd}_3(\numfcn{succ}(x0),y1,z0) \\
      &\numrel= (\fcn{CScar}_3(x1,y1,z0)\cat 0)\numbin+
         \fcn{CSadd}_3(x1,y1,z0) \\
      &\numrel= (\fcn{CScar}_3(x,y,z)\cat 10)\numbin+
         (\fcn{CSadd}_3(x,y,z)\cat 0) \\
      &\numrel= \big(\numfcn{succ}(\fcn{CScar}_3(x,y,z)\cat 0)
         \numbin+\fcn{CSadd}_3(x,y,z)\big)\cat 0 \\
      &\numrel= \numfcn{succ}\big(\big((\fcn{CScar}_3(x,y,z)\cat 0)
         \numbin+\fcn{CSadd}_3(x,y,z)\big)\cat 1\big) \\
      &\numrel= \numfcn{succ}\big((\fcn{CScar}_3(x,y,z)\cat 00)
         \numbin+(\fcn{CSadd}_3(x,y,z)\cat 1)\big) \\
      &\numrel= \numfcn{succ}\big((\fcn{CScar}_3(x0,y1,z0)\cat 0)
         \numbin+\fcn{CSadd}_3(x0,y1,z0)\big) 
      \quad\QED
   \end{align*}

\end{proof}

\begin{theorem*}{\ref{thm:CS+}}
   \( \fcn{CScar}(x,y,z,w)\numbin+\fcn{CSadd}(x,y,z,w)
      \numrel= x\numbin+y\numbin+z\numbin+w \)
\end{theorem*}
\begin{proof}
   As for the preceding lemma, the proof is a straightforward, if
   tedious, application of Derived Rule~\ref{derule:2NIND}
   (generalized to four variables).  First, the base cases for
   $\E,y,z,w$ and $x,\E,z,w$ and $x,y,\E,w$ and $x,y,z,\E$ are proved
   (each one with another application of Derived
   Rule~\ref{derule:2NIND}), and then the sixteen cases from
   $x0,y0,z0,w0$ to $x1,y1,z1,w1$ are proved from the assumption that
   the lemma holds for $x,y,z,w$.  We do not show the full proof here
   as it is not particularly instructive; instead, we give parts of
   the proof for one illustrative case.  In the inductive step, we
   show the case for $x1,y1,z1,w1$.
   \begin{align*}
      &\!\fcn{CScar}(x1,y1,z1,w1)\numbin+\fcn{CSadd}(x1,y1,z1,w1) \\
      &\numrel= \fcn{CScar}_3\big(
            \fcn{CScar}_3(x,y,z)\cat 10,
            \fcn{CSadd}_3(x,y,z)\cat 1,w1\big)\cat 0\numbin+ \\*
      &\phanrel{\numrel=} \fcn{CSadd}_3\big(
            \fcn{CScar}_3(x,y,z)\cat 10,
            \fcn{CSadd}_3(x,y,z)\cat 1,w1\big) \\
      &\numrel= \fcn{CScar}_3\big(
            \numfcn{succ}(\fcn{CScar}_3(x,y,z)\cat 0),
            \fcn{CSadd}_3(x,y,z),w\big)\cat 10\numbin+ \\*
      &\phanrel{\numrel=} \fcn{CSadd}_3\big(
            \numfcn{succ}(\fcn{CScar}_3(x,y,z)\cat 0),
            \fcn{CSadd}_3(x,y,z),w\big)\cat 0 \\
      &\numrel= \numfcn{succ}\Big(\fcn{CScar}_3\big(
            \numfcn{succ}(\fcn{CScar}_3(x,y,z)\cat 0),
            \fcn{CSadd}_3(x,y,z),w\big)\cat 0\numbin+\Big. \\*
      &\phanrel{\numrel=} \hphantom{\numfcn{succ}\Big(\Big.}
         \Big.\fcn{CSadd}_3\big(
            \numfcn{succ}(\fcn{CScar}_3(x,y,z)\cat 0),
            \fcn{CSadd}_3(x,y,z),w\big)\Big)\cat 0 \\
      &\numrel= \numfcn{succ}\big(\numfcn{succ}\big(
            \fcn{CScar}(x,y,z,w)\numbin+\fcn{CSadd}(x,y,z,w)
         \big)\big)\cat 0 \\
      &\numrel= \numfcn{succ}\big(\numfcn{succ}\big(
            x\numbin+y\numbin+z\numbin+w\big)\big)\cat 0 \\
      &\numrel= \numfcn{succ}(x\numbin+y)\cat 0\numbin+
            \numfcn{succ}(z\numbin+w)\cat 0 \\
      &\numrel= x1\numbin+y1\numbin+z1\numbin+w1 
      \quad\QED
   \end{align*}

\end{proof}

\begin{claim*}{\ref{claim:CARADD_0}}
\mbox{}\nopagebreak
   \begin{enumerate}
   \item
     \( \fcn{CARADD}({}_0x) \numrel= 0 \)
   \item
     \( \fcn{sum}({}_0x) = \fcn{CAR}({}_0x)\numbin+\fcn{ADD}({}_0x)
        \numrel= 0\numbin+0 \numrel= 0 \)
   \end{enumerate}
\end{claim*}
\begin{proof}
\mbox{}\nopagebreak
   \begin{enumerate}
   \item
      By $\TIND$ on $x$:
      \( \fcn{CARADD}({}_0\E) = \E \numrel= 0 \),
      \( \fcn{CARADD}({}_0i) = 00 \numrel= 0 \),
      \begin{align*}
         \fcn{CARADD}({}_0x)
         &= \fcn{CScar}(\fcn{CAR}({}_0x\lhalf),\fcn{ADD}({}_0x\lhalf),
               \fcn{CAR}(\rhalf{}_0x),\fcn{ADD}(\rhalf{}_0x))\cat \\*
         &\pheq \fcn{CSadd}(\fcn{CAR}({}_0x\lhalf),\fcn{ADD}({}_0x\lhalf),
               \fcn{CAR}(\rhalf{}_0x),\fcn{ADD}(\rhalf{}_0x)) \\
         &\numrel= \fcn{CScar}(0,0,0,0)\cat\fcn{CSadd}(0,0,0,0) \\
         &\numrel= 0\cat 0 \numrel= 0. 
      \end{align*}
   \item
      Direct corollary of the first claim.
      \quad\QED
   \end{enumerate}
\end{proof}

\begin{theorem*}{\ref{thm:sumhalf}}
   \( \fcn{sum}(x) \numrel=
      \fcn{sum}(x\lhalf)\numbin+\fcn{sum}(\rhalf x) \)
\end{theorem*}
\begin{proof}
   \begin{align*}
      \fcn{sum}(x)
      &= \fcn{CScar}(\fcn{CAR}(x\lhalf),\fcn{ADD}(x\lhalf),
            \fcn{CAR}(\rhalf x),\fcn{ADD}(\rhalf x))\numbin+ \\*
      &\pheq \fcn{CSadd}(\fcn{CAR}(x\lhalf),\fcn{ADD}(x\lhalf),
            \fcn{CAR}(\rhalf x),\fcn{ADD}(\rhalf x)) \\
      &\numrel= \fcn{CAR}(x\lhalf)\numbin+\fcn{ADD}(x\lhalf)\numbin+
            \fcn{CAR}(\rhalf x)\numbin+\fcn{ADD}(\rhalf x) \\
      &= \fcn{sum}(x\lhalf)\numbin+\fcn{sum}(\rhalf x) 
      \quad\QED
   \end{align*}

\end{proof}
 From this theorem, it is possible to prove that
\( \fcn{sum}(xy) \numrel= \fcn{sum}(x)\numbin+\fcn{sum}(y) \)
 with a sequence of lemmas and theorems similar to the ones used to
 show that
\( \fcn{AND}(xy) = \fcn{AND}(x)\bitbin\land\fcn{AND}(y) \).
 In particular, we have that
\( \fcn{sum}(x0) \numrel= \fcn{sum}(x)\numbin+0
   \numrel= \fcn{sum}(x) \)
 and
\( \fcn{sum}(x1) \numrel= \fcn{sum}(x)\numbin+1
   \numrel= \numfcn{succ}(\fcn{sum}(x)) \).

\begin{theorem*}{\ref{thm:sum_1}}
   \( \fcn{sum}(x) \numrel\leq \fcn{sum}({}_1x) \numrel= |x| \)
\end{theorem*}
\begin{proof}
   By $\TIND$ on $x$:
   \( \fcn{sum}(\E) = \fcn{sum}({}_1\E) = 0 \numrel= |\E| \),
   \( \fcn{sum}(i) \numrel= i \numrel\leq 1
      \numrel= \fcn{sum}(1) \numrel= |1| \),
   \( \fcn{sum}(x)
      \numrel= \fcn{sum}(x\lhalf)\numbin+\fcn{sum}(\rhalf x)
      \numrel\leq \fcn{sum}({}_1x\lhalf)\numbin+\fcn{sum}(\rhalf{}_1x)
      \numrel= \fcn{sum}({}_1x)
      \numrel= \fcn{sum}({}_1x\lhalf)\numbin+\fcn{sum}(\rhalf{}_1x)
      \numrel= |{}_1x\lhalf|\numbin+|\rhalf{}_1x| \numrel= |x| \).
   To complete the inductive case for \(\fcn{sum}({}_1x)=|x|\), we
   need to prove that \( |x| \numrel= |x\lhalf|\numbin+|\rhalf x| \) by
   $\TIND$ on $x$: the base cases are trivial, and
   \begin{align*}
      |x\lhalf|\numbin+|\rhalf x|
      &\numrel= x\elC\big(|{}_1x\lhalf|\numbin+|{}_1x\lhalf|,
            |{}_1x\lhalf|\numbin+|{}_1x\lhalf\cat 1|\big) \\
      &\numrel= x\elC\big(|{}_1x\lhalf|\cat 0,
            |x\lhalf|\numbin+\numfcn{succ}(|x\lhalf|)\big) \\
      &\numrel= x\elC\big(|x\lhalf|\cat 0,
            \numfcn{succ}(|x\lhalf|\numbin+|x\lhalf|)\big) \\
      &\numrel= x\elC\big(|x\lhalf|\cat 0,|x\lhalf|\cat 1\big) \\
      &\numrel= |x|. 
      \quad\QED
   \end{align*}

\end{proof}

\heading{Lemmas for the proof of $\PHP$}

\begin{lemma*}{A.1}
   \( \fcn{AND}\big(\bitfcn{or}_2(\bitfcn{not}(x),y)\big)\bitbin\limp
      \fcn{sum}(x)\numrel\leq\fcn{sum}(y) \)
\end{lemma*}
\begin{proof}
   By Derived Rule~\ref{derule:2NIND}:
   \begin{align*}
      &\!\fcn{AND}\big(\bitfcn{or}_2(\bitfcn{not}(\E),y)\big)\bitbin\limp
         \fcn{sum}(\E)\numrel\leq\fcn{sum}(y) 
       = \fcn{AND}(y)\bitbin\limp\E\numrel\leq\fcn{sum}(y) = 1,
         \displaybreak[0]\\
      &\!\fcn{AND}\big(\bitfcn{or}_2(\bitfcn{not}(x),\E)\big)\bitbin\limp
         \fcn{sum}(x)\numrel\leq\fcn{sum}(\E) \\
      &= \fcn{AND}(\bitfcn{not}(x))\bitbin\limp\fcn{sum}(x)\numrel=\E =
         \bit\lnot\fcn{OR}(x)\bitbin\limp x\strrel={}_0x = 1,
         \displaybreak[0]\\
      &\!\fcn{AND}\big(\bitfcn{or}_2(\bitfcn{not}(xi),yj)\big)
         \bitbin\limp\fcn{sum}(xi)\numrel\leq\fcn{sum}(yj) \\
      &= \fcn{AND}\big(\bitfcn{or}_2(\bitfcn{not}(x),y)\big)
         \bitbin\land(\bit\lnot i\bitbin\lor j)\bitbin\limp
         \fcn{sum}(x)\numbin+i\numrel\leq\fcn{sum}(y)\numbin+j \\
      &= \fcn{AND}\big(\bitfcn{or}_2(\bitfcn{not}(x),y)\big)
         \bitbin\land(\bit\lnot i\bitbin\lor j)\bitbin\limp
         \fcn{sum}(x)\numrel\leq\fcn{sum}(y)\bitbin\land
         \bit\lnot(i\bitbin\land\bit\lnot j) = 1. 
      \quad\QED
   \end{align*}

\end{proof}

\begin{lemma*}{A.2}
   \( \fcn{AND}\big(\bitfcn{or}_2(\bitfcn{not}(x),y)\big)\bitbin\land
      \fcn{OR}\big(\bitfcn{and}_2(\bitfcn{not}(x),y)\big) \limp
      \fcn{sum}(x)\numrel<\fcn{sum}(y) \)
\end{lemma*}
\begin{proof}
   By Derived Rule~\ref{derule:2NIND}:
   \begin{align*}
      &\!\fcn{AND}\big(\bitfcn{or}_2(\bitfcn{not}(\E),y)\big)\bitbin\land
      \fcn{OR}\big(\bitfcn{and}_2(\bitfcn{not}(\E),y)\big) \bitbin\limp
      \fcn{sum}(\E)\numrel<\fcn{sum}(y) \\
      &= \fcn{AND}(y)\bitbin\land\fcn{OR}({}_0y) \bitbin\limp
         \E\numrel<\fcn{sum}(y) \\
      &= 0\bitbin\limp\E\numrel<\fcn{sum}(y) = 1, \displaybreak[0]\\
      &\!\fcn{AND}\big(\bitfcn{or}_2(\bitfcn{not}(x),\E)\big)\bitbin\land
      \fcn{OR}\big(\bitfcn{and}_2(\bitfcn{not}(x),\E)\big) \bitbin\limp
      \fcn{sum}(x)\numrel<\fcn{sum}(\E) \\
      &= \fcn{AND}(\bitfcn{not}(x))\bitbin\land\fcn{OR}({}_0x)
         \bitbin\limp \fcn{sum}(x)\numrel<\E \\
      &= 0\bitbin\limp 0 = 1, \displaybreak[0]\\
      &\!\fcn{AND}\big(\bitfcn{or}_2(\bitfcn{not}(xi),yj)\big)\bitbin\land
      \fcn{OR}\big(\bitfcn{and}_2(\bitfcn{not}(xi),yj)\big) \bitbin\limp
      \fcn{sum}(xi)\numrel<\fcn{sum}(yj) \\
      &= \big(\fcn{AND}\big(\bitfcn{or}_2(\bitfcn{not}(x),y)\big)
         \bitbin\land(\bit\lnot i\bitbin\land j)\big)\bitbin\land
         \big(\fcn{OR}\big(\bitfcn{and}_2(\bitfcn{not}(x),y)\big)
         \bitbin\lor(\bit\lnot i\bitbin\land j)\big) \bitbin\limp \\
      &\pheq \fcn{sum}(x)\numbin+i\numrel<\fcn{sum}(y)\numbin+j 
   \end{align*}
   (and a simple check of all four cases for the possible values of
   $i$ and $j$ shows that the property holds in each one).
\end{proof}

\begin{lemma*}{A.3}
   \( \fcn{maskbit}(x,\fcn{first}_1(x)) = \fcn{OR}(x) \)
\end{lemma*}
\begin{proof}
   By $\NIND$ on $x$:
   \( \fcn{maskbit}(\E,\fcn{first}_1(\E)) = \fcn{markbit}(\E,\E)
       = \fcn{OR}(\bitfcn{and}_2(\E,\E)) = \fcn{OR}(\E) \), and
   \( \fcn{maskbit}(xi,\fcn{first}_1(xi))
       = \fcn{maskbit}(x,\fcn{first}_1(x))\bitbin\lor
         (i\bitbin\land\fcn{OR}(x)\bC(0,i))
       = \fcn{OR}(x)\bitbin\lor\fcn{OR}(x)\bC(0,i)
       = \fcn{OR}(x)\bC(1,i)
       = \fcn{OR}(x)\bitbin\lor i
       = \fcn{OR}(xi) \).
\end{proof}

\begin{lemma*}{A.4}
\mbox{}\nopagebreak\\
   \leftright
  {\( \bit\lsg\fcn{lb}(x,y) =
      \fcn{maskbit}\big(x,(1\cat{}_0(y\lchop x1))\rdel\big) \) \\}
  {\( \bit\lsg\fcn{rb}(x,y) =
      \fcn{maskbit}\big(x,(1\cat{}_0y)\rdel\big) \)}
\end{lemma*}
\begin{proof}
   (L) By Derived Rule~\ref{derule:2NIND}:
   \begin{align*}
      &\!\fcn{maskbit}\big(\E,(1\cat{}_0(y\lchop 1))\rdel\big) \\
      &= y\zlC\big(\fcn{maskbit}(\E,1),\fcn{maskbit}(\E,\E)\big) \\
      &= y\zlC(0,0) = 0 = \bit\lsg\E = \bit\lsg\fcn{lb}(\E,y),
         \displaybreak[0]\\
      &\!\fcn{maskbit}\big(x,(1\cat{}_0(\E\lchop x1))\rdel\big) \\
      &= \fcn{maskbit}(x,1\cat{}_0x) \\
      &= \fcn{OR}(\bitfcn{and}_2(0x,1\cat{}_0x)) \\
      &= (0\bitbin\land 1)\bitbin\lor\fcn{OR}(\bitfcn{and}_2(x,{}_0x)) \\
      &= 0\bitbin\lor\fcn{OR}({}_0x) \\
      &= 0 = \bit\lsg\E = \bit\lsg\fcn{lb}(x,\E), \displaybreak[0]\\
      &\!\fcn{maskbit}\big(ix,(1\cat{}_0(yj\lchop ix1))\rdel\big) \\
      &= \fcn{maskbit}\big(ix,(1\cat{}_0(y\lchop x1))\rdel\big) \\
      &= y\zlC\Big(\fcn{maskbit}(ix,1\cat{}_0x), \Big.\\*
      &\pheq \hphantom{y\zlC\Big(\Big.}\Big.\fcn{maskbit}\big(ix,
            {}_0\big(ix\rchop(1\cat{}_0(y\lchop x1))\rdel\big)\cat
            (1\cat{}_0(y\lchop x1))\rdel\big)\Big) \\
      &= y\zlC\Big((i\bitbin\land 1)\bitbin\lor\fcn{maskbit}(x,{}_0x),
         \Big.\\*
      &\pheq \hphantom{y\zlC\Big(\Big.}\Big.\fcn{maskbit}\big(ix,
            {}_0(ix\rchop(y\lchop x1))\cat
            (1\cat{}_0(y\lchop x1))\rdel\big)\Big) \\
      &= y\zlC\Big(i\bitbin\lor 0,\fcn{maskbit}\big(ix,
            0\cat{}_0(x\rchop(y\lchop x1))\cat
            (1\cat{}_0(y\lchop x1))\rdel\big)\Big) \\
      &= y\zlC\Big(i,(i\bitbin\land 0)\bitbin\lor
         \fcn{maskbit}\big(x,{}_0(x\rchop(y\lchop x1))\cat
            (1\cat{}_0(y\lchop x1))\rdel\big)\Big) \\
      &= y\zlC\Big(i,\fcn{maskbit}\big(x,
            (1\cat{}_0(y\lchop x1))\rdel\big)\Big) \\
      &= y\zlC(i,\bit\lsg\fcn{lb}(x,y)) \\
      &= \bit\lsg\fcn{lb}(ix,yj). 
      \quad\QED
   \end{align*}

\end{proof}

\begin{lemma*}{A.5}
   \( \bit\lnot\fcn{OR}(\fcn{delfirst}_1(x))\bitbin\land\fcn{lb}(x,y)
      \bitbin\limp \bit\lnot\fcn{OR}(\fcn{lc}(x,y\rdel)) \)
\end{lemma*}
\begin{proof}
   By $\NIND$ on $x$:
   \( \bit\lnot\fcn{OR}(\fcn{delfirst}_1(\E))\bitbin\land\fcn{lb}(\E,y)
      \bitbin\limp \bit\lnot\fcn{OR}(\fcn{lc}(\E,y\rdel)) =
      \bit\lnot\fcn{OR}(\E)\bitbin\land\E \bitbin\limp
      \bit\lnot\fcn{OR}(\E) = 0 \bitbin\limp 0 = 1 \),
   If \(y=\E\) or \(y\lenrel>xi\), then \(\fcn{lb}(xi,y)=\E\), which
   makes the antecedent of $\bitbin\limp$ false and the entire statement
   trivially true.  Hence, we prove the inductive step under the
   implicit assumption that \(y\neq\E\) and \(y\lenrel\leq xi\).
   \begin{align*}
      &\!\bit\lnot\fcn{OR}(\fcn{delfirst}_1(xi))
         \bitbin\land\fcn{lb}(xi,y)\bitbin\limp\bit\lnot\fcn{OR}
            (\fcn{lc}(xi,y\rdel)) \\
      &= y\lenrel=xi\zlC\Big(\bit\lnot(i\bitbin\land
            \bit\lnot(\fcn{OR}(x)\bC(0,i)))\bitbin\land
         \bit\lnot\fcn{OR}(\fcn{delfirst}_1(x))\bitbin\land i
         \bitbin\limp \bit\lnot\fcn{OR}(\fcn{lc}(xi,(xi)\rdel)), \Big.\\*
      &\pheq \quad\quad\quad \Big.\bit\lnot(i
            \bitbin\land\bit\lnot(\fcn{OR}(x)\bC(0,i)))\bitbin\land
          \bit\lnot\fcn{OR}(\fcn{delfirst}_1(x))\bitbin\land\fcn{lb}(x,y)
          \bitbin\limp\bit\lnot\fcn{OR}(\fcn{lc}(xi,y\rdel))\Big) \\
      &= y\lenrel=xi\zlC\Big(\bit\lnot(i\bitbin\land
            \fcn{OR}(x)\bC(1,\bit\lnot i))\bitbin\land i\bitbin\land
         \bit\lnot\fcn{OR}(\fcn{delfirst}_1(x))\bitbin\limp
         \bit\lnot\fcn{OR}(\fcn{lc}(xi,x)),1\Big) \\
      &= y\lenrel=xi\zlC\Big(\bit\lnot(\fcn{OR}(x)\bC(i,0))\bitbin\land
         \fcn{OR}(x)\bC(i,i)\bitbin\land\bit\lnot
            \fcn{OR}(\fcn{delfirst}_1(x))\bitbin\limp
         \bit\lnot\fcn{OR}(x),1\Big) \\
      &= y\lenrel=xi\zlC\big(\fcn{OR}(x)\bC(0,i)\bitbin\limp
         \fcn{OR}(x)\bC(0,1),1\big) \\
      &= y\lenrel=xi\zlC(1,1) = 1
      \quad\QED
   \end{align*}

\end{proof}

%% file: biblio.tex
\addcontentsline{toc}{chapter}{Bibliography}
\bibliographystyle{plain}
\bibliography{thesis}